\documentclass[11pt,ngerman]{scrbook}
\usepackage{ngerman}
\usepackage{amsmath}
\usepackage{amssymb}
\usepackage{amsfonts}
\usepackage{textcomp}
\usepackage{dcolumn}
\usepackage[T1]{fontenc}
\usepackage[latin1]{inputenc}
\usepackage[dvips]{graphics}
\usepackage{epsfig}
\usepackage{bm}

\begin{document}

\pagenumbering{Roman}


\begin{titlepage}

\vspace*{1 cm}

\begin{center}

\textbf{\huge Über die Beziehung der begrifflichen Grundlagen der
Quantentheorie und der Allgemeinen Relativitätstheorie}\\

\vspace{4.5 cm}

{\Large Dissertation\\ zur Erlangung des Doktorgrades\\ der Naturwissenschaften}\\

\vspace{1cm}

{\Large vorgelegt beim Fachbereich Physik\\
der Johann Wolfgang Goethe-Universität\\
in Frankfurt am Main}\\

\begin{figure}[h]
\centering
\epsfig{figure=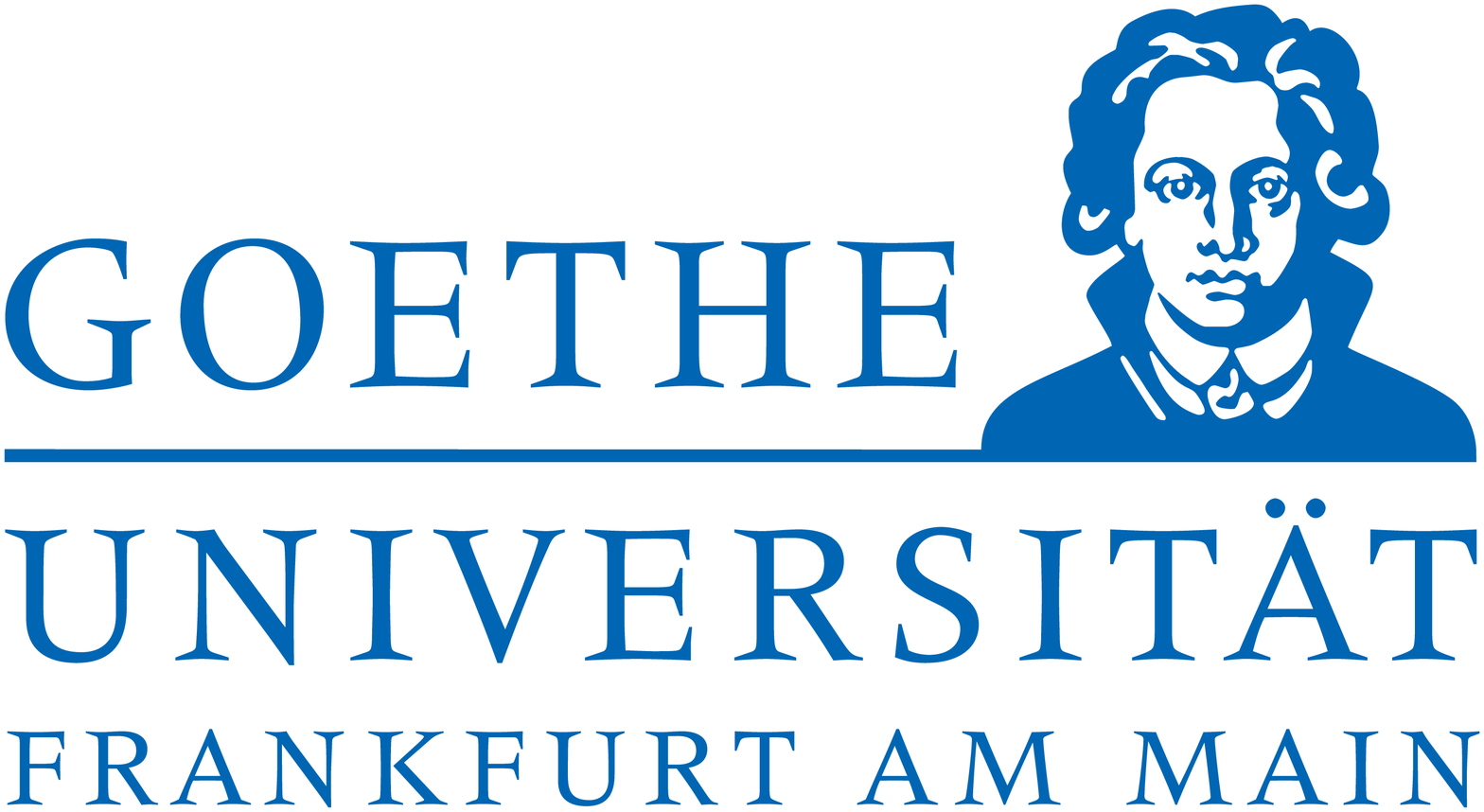,width=4cm}
\end{figure}

\vspace{1 cm}

{\Large von \\
Martin Kober\\
aus Wiesbaden}

\vspace{1.5 cm}

{\Large Frankfurt am Main 2010\\ (D 30)}
\end{center}

\newpage

\pagestyle{empty}

\vspace*{7 cm}

\noindent
{\Large Vom Fachbereich Physik der\\
Johann Wolfgang Goethe-Universität als Dissertation angenommen.}

\vspace{7 cm}

{\Large Dekan: ........................................................................}

\vspace{1.5 cm}

{\Large Gutachter: ..................................................................}

\vspace{1.5 cm}

{\Large Datum der Disputation: .............................................}

\end{titlepage}

\pagestyle{empty}

\cleardoublepage

\vspace*{7.5 cm}

\begin{quote}
\textbf{\Large "`Ich bin Physiker, ich bin Naturwissenschaftler, und unter den Naturwissenschaftlern der Neuzeit ist derjenige,
den ich wohl am meisten liebe, Johannes Kepler. Johannes Kepler war ein Christlicher Neuplatoniker, und er verstand die
Naturwissenschaft als das Nachdenken der Schöpfungsgedanken Gottes, welches dem Menschen möglich ist, weil er von Gott 
nach dem Bilde Gottes geschaffen ist. Also als Gottesdienst."'\\
\\
Carl Friedrich von Weizsäcker, in einem Vortrag auf dem Evangelischen Kirchentag, 1987}
\end{quote}

\cleardoublepage

\pagestyle{plain}

\tableofcontents

\cleardoublepage

\chapter*{Vorwort}
\addcontentsline{toc}{chapter}{Vorwort}
\pagenumbering{arabic}

Die vorliegende Dissertation kann als das vorläufige Ergebnis meines etwa vor zehn Jahren begonnenen Versuches angesehen
werden, die Quantentheorie, die Allgemeine Relativitätstheorie und deren Beziehung zueinander zu verstehen. Als ich zum
ersten Mal begann, über die Quantentheorie und die Allgemeine Relativitätstheorie zu lesen, hielt ich es für beinahe
unmöglich, jemals auch nur in die Nähe eines Verständnisses dieser abstrakten und zunächst fremdartig erscheinenden Realität
zu gelangen, welche durch sie beschrieben wird. Es erschien mir, und erscheint es auch heute noch, als ein Wunder, dass es
den großen Theoretischen Physikern des zwanzigsten Jahrhunderts gelungen ist, durch Interpretation von Phänomenen,
Anwendung abstrakter Mathematik und begrifflich-philosophische Analyse in einen Bereich der Wirklichkeit geistig vorzudringen,
der die für Menschen sinnlich erlebbare Ebene der Realität aus prinzipiellen Gründen in grundsätzlicher Weise übersteigt.

Albert Einstein, Niels Bohr, Werner Heisenberg und Carl Friedrich von Weizsäcker erschienen mir in Bezug auf die Analyse der
begrifflichen Grundlagen der Theoretischen Physik immer als die bedeutendsten Denker des zwanzigsten Jahrhunderts. Sie haben
nach meinem Empfinden den vielleicht wichtigsten Beitrag zur Entstehung eines vollkommen neuen physikalischen Weltbildes
geliefert. Albert Einstein hat die beiden Relativitätstheorien begründet, womit er die klassische Vorstellung von Raum und Zeit
in grundlegender Weise verändert hat, und auf die Entwicklung der Quantentheorie entscheidenden Einfluss genommen.
Er war wahrscheinlich nicht nur der größte Physiker des zwanzigsten Jahrhunderts sondern sogar aller Zeiten. Seine
geistesgeschichtliche Bedeutung besteht darin, dass es ihm als erstem Menschen gelang, im Rahmen der Physik als basalster
der konkreten, empirischen Wissenschaften die Grundstrukturen unseres Denkens zu überwinden und eine neue Ebene der
Realität zu eröffnen.
Niels Bohr und Werner Heisenberg haben die von Max Planck und Albert Einstein begonnene Entwicklung der Quantentheorie gemeinsam
mit Wolfgang Pauli, Paul Adrien Maurice Dirac, Erwin Schrödinger, Louis de Broglie, Max Born, Pascual Jordan und anderen zur
Vollendung geführt.
Niels Bohr erkannte als erster in vollem Umfang, dass die Beschreibung der Atome nicht nur neue Modelle, sondern eine neue
Auffassung der physikalischen Realität notwendig machte und hat die Quantentheorie wohl am gründlichsten interpretiert und am
tiefsten verstanden. Sein Schüler Werner Heisenberg begründete die endgültige Gestalt der Quantentheorie als vollendeter Theorie
und leitete aus ihr mit der Unbestimmtheitsrelation eine konkrete Darstellung der Grenze der Anwendbarkeit klassischer Begriffe
im atomaren und subatomaren Bereich her.
Damit haben Niels Bohr und Werner Heisenberg im Rahmen eines kollektiven Werkes, an dem sehr viele Physiker entscheidenden
Anteil hatten, eine Revolution des physikalischen Denkens hervorgerufen, welche der in der Einzelleistung Albert Einsteins
bestehenden gleichkommt. 
Carl Friedrich von Weizsäcker hat zwar nicht an der eigentlichen Entstehung der Theorien mitgewirkt.
Er hat die Quantentheorie aber von Niels Bohr und Werner Heisenberg ausgehend und unter Einbeziehung
der Klassischen Philosophie in einem größeren philosophischen Zusammenhang gedeutet und mit seiner hieraus hervorgehenden
bisher primär auf philosophischer Ebene behandelten Quantentheorie der Ur-Alternativen, wie mir scheint, bereits den Weg zu einer
möglichen neuen wissenschaftlichen Revolution gewiesen, welche Quantentheorie, Elementarteilchenphysik und Allgemeine
Relativitätstheorie in eine innere Einheit bringen könnte. Er gehört im vollen Sinne sowohl der Physik als konkreter Wissenschaft
als auch der die Interpretation der großen Zusammenhänge betreffenden Philosophie gleichermaßen an und kann daher als wirklicher
Universalgelehrter gelten. Es erfüllt mich mit tiefer Dankbarkeit, dass ich den wunderbaren Schriften dieser großen Theoretischen
Physiker begegnet bin und so zu ihren grundlegenden Erkenntnissen Zugang erhalten konnte.

Der bedeutende Wissenschaftsjournalist, Psychiater und Philosoph Hoimar v. Ditfurth hat mir die existentiell-philosophische
Dimension der Naturwissenschaft eröffnet. Dem Studium seiner Schriften verdanke ich unendlich viel und von ihm ausgehend wurde
ich überhaupt erst in die Lage versetzt, den Sinn wissenschaftlicher Theorien und philosophischer Anschauungen verstehen zu
können. Hätte ich als Jugendlicher seine Schriften nicht kennengelernt, wäre ich wahrscheinlich mein Leben lang geistig
"`blind"' geblieben.

Für mein Verständnis der Quantentheorie und der Allgemeinen Relativitätstheorie waren weiter die Schriften von Carlo Rovelli,
Holger Lyre und Peter Mittelstaedt entscheidend. Durch die Lektüre des Buches von Carlo Rovelli habe ich zum ersten Mal das
Gefühl bekommen, die Allgemeine Relativitätstheorie wirklich verstanden zu haben. Holger Lyre verdanke ich ebenfalls wichtige
Einsichten in Bezug auf die Allgemeine Relativitätstheorie und deren eichtheoretische Formulierung. In der Lektüre von Peter
Mittelstaedts Buch "`Philosophische Probleme der modernen Physik"' habe ich die einzige Behandlung des quantentheoretischen
Messproblems gefunden, welche mich einigermaßen zufrieden gestellt hat. Meine eigene in dieser Arbeit dargelegte Haltung zu
dieser Problematik entspricht in weiten Teilen derjenigen Mittelstaedts.

\newpage

\chapter*{Einleitung}
\addcontentsline{toc}{chapter}{Einleitung}

Auf der Ebene der Erkenntnis, die in der zeitgenössischen Theoretischen Physik erreicht ist, existieren zwei fundamentale
Theorien zur Beschreibung der basalen Zusammenhänge in der Natur. Es handelt sich um die Quantentheorie und die Allgemeine
Relativitätstheorie. Die Quantentheorie stellt ein allgemeines Schema des dynamischen Verhaltens beliebiger Objekte dar, welche
durch abstrakte Zustände in einem Hilbert-Raum charakterisiert werden. In Gestalt relativistischer Quantenfeldtheorien bildet
sie damit die Basis der Beschreibung beliebiger Teilchen beziehungsweise Felder, welche in der Elementarteilchenphysik im Sinne des
Welle-Teilchen-Dualismus als fundamentale Entitäten der Natur angesehen werden. Die Aufgabe der auf der Quantentheorie
basierenden Elementarteilchenphysik ist es gerade, die real existierenden Objekte und deren Eigenschaften, insbesondere ihre
spezifischen Wechselwirkungen, auf ein ebenso einheitliches Schema zurückzuführen, wie es die Quantentheorie selbst darstellt,
die an sich zunächst nichts über die Existenz spezifischer Objekte aussagt. Letzteres ist bisher nicht vollständig gelungen.
Das Standardmodell der Elementarteilchenphysik liefert zwar empirisch höchst erfolgreiche mathematische Modelle der Beschreibung
der Elementarteilchen und ihrer Wechselwirkungen, die auch bis zu einem bestimmten Grade eine gewisse Geschlossenheit aufweisen.
Aber auf die Frage, warum gerade die bekannten Elementarteilchen mit den bekannten Symmetrien existieren, und warum sie gerade die
empirisch gefundenen Massen und Ladungen aufweisen, kann das Standardmodell der Elementarteilchenphysik in seiner heutigen Gestalt
keine Antwort geben. Unabhängig davon wird die Struktur der Raum-Zeit gemäß der Speziellen Relativitätstheorie in der heute
erreichten Formulierung relativistischer Quantenfeldtheorien schlicht als eine Grundannahme hingenommen.   
Die Allgemeine Relativitätstheorie liefert eine Theorie der Beschreibung von Raum und Zeit, oder spezifischer, ihrer
metrischen Struktur, die gemäß der Allgemeinen Relativitätstheorie selbst zu einer dynamischen Entität wird. Hierbei geht es um
die wechselseitige Wirkung von vorhandener Materie auf der Raum-Zeit mit der Struktur der Raum-Zeit selbst,
welche ihrerseits durch ein metrisches Feld charakterisiert wird. Gleichzeitig stellt sie auf diese Weise eine klassische
Beschreibung einer der vier bekannten fundamentalen Wechselwirkungen in der Natur dar, nämlich der Gravitation, und führt diese
damit auf die metrische Struktur der Raum-Zeit selbst zurück. 
Sie ermöglicht also die Behandlung makroskopischer Systeme, bei denen die Gravitation die dominierende Rolle spielt, wie etwa
beim Verhalten von Sternen. Darüber hinaus liefert sie als Theorie der Gestalt von Raum und Zeit die Basis der Beschreibung
des Universums als Ganzem, welche Gegenstand der Kosmologie ist.   
Eine angemessene quantentheoretische Fassung der Gravitation, aus der konkrete Vorhersagen abgeleitet werden könnten, ist
im Gegensatz zu den drei anderen Wechselwirkungen bisher nicht erreicht worden. 
\footnote{Die Frage, ob es sich bei der konsequenten quantentheoretischen Fassung der anderen Wechselwirkungen angesichts des
vielleicht unästhetisch und konstruiert erscheinenden effektiven Verfahrens der Renormalisierung, das als adäquates Mittel zum
Erhalt endlicher Werte bei der Vorhersage für konkrete empirische Befunde etabliert ist und dessen Zweckmäßigkeit auf dem
gegenwärtigen Erkenntnisstand auch unbestreitbar ist, wirklich bereits um eine vollends zufriedenstellende Lösung handelt, sei
hier nicht weiter diskutiert.}
Hierbei scheint es sich angesichts des oben Gesagten um ein prinzipielles Problem zu handeln, dass nicht nur ein neues
mathematisches Modell, sondern einen neuen begrifflichen Rahmen für die bezüglich ihrer Grundkonzepte so unterschiedlichen
Naturen der beiden Theorien erfordert. 
Dies kann man nun zunächst in wechselseitiger Hinsicht zum Ausdruck bringen. Auf der einen Seite muss die quantentheoretische
Art der Naturbeschreibung auf die Allgemeine Relativitätstheorie übertragen werden. Die Beschreibung der Dynamik des metrischen
Feldes muss im Sinne der Einheitlichkeit in Analogie zu den anderen Wechselwirkungen selbst im Rahmen einer Quantenfeldtheorie
erfolgen. Umgekehrt muss in die derzeitige Fassung relativistischer Quantenfeldtheorien, innerhalb welcher die Raum-Zeit-Struktur
der Speziellen Relativitätstheorie vorausgesetzt wird, eine allgemein-relativistische Raum-Zeit-Beschreibung eingeführt werden,
bei der also von keiner nicht-dynamischen metrischen Hintergrundstruktur mehr ausgegangen wird. Das Ziel ist also eine
Beschreibung der Gravitation als hintergrundunabhängiger Quantenfeldtheorie. Derjenige Ansatz zu einer solchen Theorie,
welcher diesem Ziel vielleicht bisher am nächsten gekommen ist, ist wohl die Theorie der Schleifenquantengravitation.

Nun sollte man aber im Sinne einer einheitlichen Beschreibung noch einen Schritt weiter gehen.
Die bisherigen Ansätze zu einer Quantentheorie der Gravitation versuchen, die Prinzipien der jeweils einen Theorie auf die
jeweils andere zu übertragen, stellen aber nicht wirklich eine Beziehung ihrer Begriffe zueinander her. In einer wirklich
einheitlichen Theorie würde man aber erwarten, dass sich sowohl die Raum-Zeit-Beschreibung einer quantentheoretisch formulierten
Allgemeinen Relativitätstheorie, als auch die bereits im Rahmen der Quantentheorie beschriebene Elementarteilchenphysik aus einem
einheitlichen begrifflichen Rahmen ergeben, in dem sie gewissermaßen aufgehen. Damit würde diesen beiden bisher getrennt
beschriebenen Aspekten der Realität also ein bestimmter Aspekt einer einzigen übergeordneten Theorie entsprechen. 
Hierzu muss zunächst noch einmal daran erinnert werden, dass die spezielle Form, in welcher die Quantentheorie im Rahmen der
konkreten Anwendungen in der Elementarteilchenphysik auftritt, nämlich als relativistischer Quantenfeldtheorie, einen
Spezialfall der Allgemeinen Relativitätstheorie im Sinne einer von außen an die Quantentheorie angehefteten Theorie
bereits integriert hat, nämlich die Spezielle Relativitätstheorie, und insofern nicht für sich alleine unabhängig von der
Relativitätstheorie besteht. Dies scheint umgekehrt zunächst sehr wohl der Fall zu sein. Denn die Allgemeine
Relativitätstheorie bedient sich zwar in der konkreten Anwendung auch unterschiedlicher Modelle für die Materie, die sie aus
anderen Theorien gewissermaßen ausleiht. Diese sind aber im Prinzip austauschbar, zumindest sind sie nicht für die Allgemeine
Relativitätstheorie als konkret angewandter Theorie konstitutiv.
Die Quantentheorie an sich in der allgemeinen Dirac-von Neumannschen Fassung enthält allerdings noch keine Annahme
über die Struktur der Raum-Zeit. Sie setzt noch nicht einmal die Existenz einer Raum-Zeit voraus. Nur ein mit der Zeit zu
identifizierender Parameter ist notwendig, um die quantentheoretische Dynamik in ihrer allgemeinen Fassung zu formulieren.
Damit ist die Quantentheorie in ihrer allgemeinen Form, im Gegensatz zu ihrer Manifestation im Rahmen relativistischer
Quantenfeldtheorien, aber eigentlich logisch unabhängig von der Allgemeinen Relativitätstheorie. Die Allgemeine
Relativitätstheorie hingegen handelt, wie die Quantentheorie, auch von der Beschreibung dynamischer Entitäten, spezifischer von
derjenigen des die Raum-Zeit-Struktur beschreibenden metrischen Feldes. Wenn man die Quantentheorie im Sinne der Einheit der
Physik in ihrer allgemeinen Formulierung als abstrakte Theorie des Hilbert-Raumes allerdings als allgemeines Schema der
Beschreibung der Dynamik beliebiger Objekte auffasst (alle gegenwärtig verfolgten Ansätze zu einer einheitlichen Theorie tun
dies, indem sie alle Dynamik der Quantentheorie zu unterwerfen versuchen), so unterliegt die Allgemeine Relativitätstheorie
jedoch sehr wohl der Quantentheorie im Sinne eines solchen allgemeinen Schemas, ist also logisch nicht unabhängig von ihr. Damit
eröffnet sich im Prinzip die Möglichkeit für die Vermutung, dass sowohl eine quantentheoretische Fassung der Allgemeinen
Relativitätstheorie als Theorie der Raum-Zeit als auch die auf der Relativitätstheorie basierende bisherige konkrete
Manifestation der Quantentheorie in Gestalt relativistischer Quantenfeldtheorien bereits eine spezielle Konsequenz der
als abstrakter Theorie aufgefassten Quantentheorie darstellen könnten.

Carl Friedrich von Weizsäcker hat im Rahmen eines sehr grundsätzlichen philosophischen Programmes den Versuch unternommen, die
universelle Gültigkeit der allgemeinen Quantentheorie durch prinzipielle Postulate bezüglich der Bedingungen der Möglichkeit
von Erfahrung zu begründen. Die Struktur der Quantentheorie wird hierbei aus endlichen Alternativen hergeleitet, deren
einzelnen Ereignissen Wahrheitswerte gemäß einer nicht-klassischen Logik zugeordnet werden, welche wiederum mit der Struktur der
Zeit in Verbindung gebracht wird. Letzteres rührt daher, dass man Aussagen, welche sich auf die Zukunft beziehen, nicht die
Wahrheitswerte wahr und falsch sondern nur Wahrscheinlichkeiten zuordnen kann. Demnach wird ein quantentheoretisches Objekt
durch die Information charakterisiert, die man von ihm gewinnen kann. 
Konstitutiv für dieses durch die Kantische Transzendentalphilosophie inspirierten Unternehmen ist die Tatsache, dass in seinem
Rahmen die Existenz eines physikalischen Ortsraumes, welcher mit der Zeit im Sinne der Speziellen Relativitätstheorie zu einem
vierdimensionalen Raum-Zeit-Kontinuum verknüpft ist, als eine Konsequenz und nicht als eine Voraussetzung erscheint.
Sie ergibt sich durch die Darstellbarkeit jedes Zustandes eines endlichdimensionalen Hilbert-Raumes durch die
tensorielle Kombination einer entsprechenden Anzahl zweidimensionaler komplexer Hilbert-Räume, deren Symmetrieeigenschaften
wiederum bereits die Symmetrieeigenschaften der realen Raum-Zeit widerspiegeln. Diese durch Zustände in zweidimensionalen
komplexen Hilbert-Räumen quantentheoretisch beschriebenen basalen Alternativen, welche die einfachsten in einer beliebigen
Quantentheorie überhaupt denkbaren Objekte darstellen, bezeichnet von Weizsäcker wegen ihres grundlegenden Charakters als
Ur-Alternativen und dieser Versuch einer Herleitung der bisher bekannten Physik mit den im Rahmen ihrer Beschreibung
auftretenden Entitäten aus einer bestimmten Darstellung der abstrakten Quantentheorie wird dementsprechend als Ur-Theorie
bezeichnet. Quantentheoretische Zustände in Hilbert-Räumen repräsentieren innerhalb dieser Theorie die einem Objekt
entsprechende Information, welche man durch binäre Alternativen ausdrücken kann und welche damit dem Materiebegriff übergeordnet
werden. Als sehr wichtig erscheint also zunächst die Tatsache, dass wenn dieser Ansatz zumindest im Prinzip Wahrheit enthielte,
die Spezielle Relativitätstheorie sich tatsächlich bereits als Konsequenz der abstrakten Quantentheorie herausstellen würde,
wie dies in der obigen Argumentation als zumindest denkbar erschien. Von entscheidender Bedeutung ist weiter,
dass die Annahme der Ur-Alternativen als fundamentaler dynamischer Entität, einerseits von einem nicht-räumlichen
grundlegenden Objekt in der Natur ausgeht und dass ein solches elementares Objekt zugleich wirklich das einfachste denkbare
Objekt ist, dass überhaupt in irgendeiner speziellen Theorie, sofern sie quantentheoretisch formuliert ist, vorkommen kann.     

Der spezifische Inhalt der vorliegenden Dissertation stellt sich nun in der folgenden Weise dar.
Zunächst sollen die wichtigsten Inhalte der Quantentheorie und der Allgemeinen Relativitätstheorie dargestellt werden.
Dies geschieht in den mit den jeweiligen Bezeichnungen der Theorien versehenen Teilen I und II der Arbeit. Hierbei kommt
einer naturphilosophischen Analyse der Grundbegriffe, welche im jeweils letzten Kapitel dieser beiden ersten jeweils drei
Kapitel umfassenden Teile des Buches geschehen soll, eine besondere Bedeutung zu. Damit sollen die Theorien für sich selbst
genommen einem tieferen Verständnis zugänglich gemacht werden.
Im Zusammenhang mit der Allgemeinen Relativitätstheorie spielt die mathematische Eigenschaft der Diffeomorphismeninvarianz
die entscheidende Rolle, während bezüglich der Quantentheorie von der Kopenhagener Deutung ausgehend
zunächst das Messproblem behandelt wird, ehe basierend auf der Analyse konkreter Phänomene und des mathematischen Apparates
der Quantentheorie die Nichtlokalität als eine wichtige, vielleicht ihre entscheidende, Eigenheit herausgestellt wird.
Dies bedeutet, dass beide Theorien, sowohl die Allgemeine Relativitätstheorie als auch die Quantentheorie, eine
relationalistische Anschauung der Natur des Raumes nahelegen, die allerdings in den konkreten Manifestationen der beiden
Theorien in gewissem Sinne nur implizit sichtbar ist, da sowohl die Allgemeine Relativitätstheorie als auch die konkrete
Formulierung relativistischer Quantenfeldtheorien vom Feldbegriff und damit von einer gegebenen Raum-Zeit ausgehen, auf der
diese Felder definiert sind.
Diese Analyse der begrifflichen Grundlagen der Quantentheorie und der Allgemeinen Relativitätstheorie in Bezug auf die Natur
des Raumes gewinnt erst unter Einbeziehung der Kantischen Philosophie und seiner Analyse der Begriffe Raum und Zeit als
Grundformen der Anschauung ihre volle Überzeugungskraft.
Im Teil III der Arbeit soll dann die oben geschilderte von Weizsäckersche Quantentheorie der Ur-Alternativen vorgestellt werden,
die wie bereits erwähnt als einzige bisher existierende Theorie auf der fundamentalen Ebene keinen physikalischen Ortsraum
voraussetzt, sondern stattdessen von nicht-räumlichen rein quantentheoretischen Objekten ausgeht, welche den Raum jedoch
konstituieren. Damit trägt sie eben diesem der Quantentheorie und der Allgemeinen Relativitätstheorie innewohnenden
Relationalismus bezüglich der Natur des Raumes Rechnung. Dies ist eines der entscheidenden Argumente dafür, dass es sich bei
der Quantentheorie der Ur-Alternativen unter den bisherigen Ansätzen zu einer einheitlichen Beschreibung der bisher bekannten
Physik um die wohl aussichtsreichste Theorie handelt. Dieses und weitere sehr gewichtige Argumente werden im letzten Abschnitt
des ersten Kapitels innerhalb dieses dritten Teiles der Arbeit in einer Liste zusammengestellt.
 
Schließlich wird im Teil IV der Arbeit der Versuch unternommen, diese in den ersten drei Teilen gewonnenen Erkenntnisse
konkret auf die Frage nach der quantentheoretischen Formulierung der Allgemeinen Relativitätstheorie anzuwenden. Zunächst
werden einige bisherige Versuche einer Quantisierung der Gravitation geschildert.
Danach wird eine von mir selbst entwickelte Theorie vorgestellt, welche eben im Sinne der obigen Betrachtungen einen Schritt
in die Richtung eines Ansatzes zu einer quantentheoretischen Beschreibung der Allgemeinen Relativitätstheorie darstellen soll,
welche auf fundamentaler Ebene keinen physikalischen Ortsraum voraussetzt. Sie ist es insofern, als sie zu Beginn keine von der
Beschreibung anderer physikalischer Objekte unabhängige Raum-Zeit-Metrik verwendet. Vielmehr versucht sie eine solche
Struktur von einem fundamentalen Spinorfeld ausgehend zu begründen, dass seinerseits die Elementarteilchen einheitlich
beschreiben soll. Hierbei wird die Heisenbergsche nichtlineare Spinorfeldtheorie als Ansatz zu einer einheitlichen
Quantenfeldtheorie der Elementarteilchen zugrunde gelegt. Im Unterschied zur ursprünglichen Fassung Heisenbergs wird aber eben
gerade keine Minkowski-Hintergrund-Metrik vorausgesetzt, sondern das Spinorfeld wird als auf einer vierdimensionalen
differenzierbaren Mannigfaltigkeit definiert angenommen, deren metrische Struktur aus dem Zusammenhang des fundamentalen
Spinorfeldes begründet wird, wobei die von Roger Penrose im Rahmen der sogenannten Twistortheorie entwickelte Mathematik
verwendet wird. Damit stellt sich die metrische Struktur der Raum-Zeit als eine direkt auf die Natur der
Elementarteilchen bezogene und nicht als eine durch einen unabhängigen Begriff beschriebene Realität dar. Dies bedeutet auch,
dass die Dualität zwischen einer Beschreibung der Raum-Zeit-Struktur einerseits und der auf ihr existierenden Materie, die ja
schon Albert Einstein im Rahmen seines Ansatzes zu einer einheitlichen Feldtheorie aufheben wollte, durch diese Theorie
zumindest abgeschwächt wird. Ein vorgestelltes hierauf basierendes Programm einer Quantisierung benutzt eine dementsprechende
Beschreibung des Zusammenhangs des elementaren Spinorfeldes durch zwei unabhängige Spinorfelder, deren direkte Quantisierung im
Sinne der üblichen Quantisierung eines fermionischen Feldes indirekt auch eine quantentheoretische Formulierung des aus ihnen
konstruierten metrischen Feldes zur Folge hat. Die sich hieraus ergebende quantentheoretische Beschreibung der Gravitation ist
allerdings bisher nur angedeutet. Diese Theorie stellt in jedem Falle nur eine vorläufige Stufe dar, denn es wird ja zu Beginn
noch von einer differenzierbaren Mannigfaltigkeit zur Beschreibung der Raum-Zeit ausgegangen.
Da sich aus der begrifflichen Analyse der Quantentheorie und der Allgemeinen Relativitätstheorie aber ergab, dass
die Natur auf der basalen Ebene im strengen Sinne nicht-räumlich ist, ist letztlich eine Theorie anzustreben, welche in ihren
Grundannahmen vollends ohne die Annahme einer bereits gegebenen Struktur eines physikalischen Ortsraumes auskommt. 
Die Quantentheorie der Ur-Alternativen geht aber als konsequent aus der begrifflichen Analyse der Quantentheorie erwachsende
Theorie gerade von einer auf der basalen Ebene nicht-räumlichen Beschreibung der Natur aus, indem sie abstrakte
quantentheoretische Alternativen als die fundamentalen Konstituenten der Naturbeschreibung postuliert.
Das bedeutet, dass das Ziel letztlich also darin bestehen muss, die oben beschriebene Theorie der Spinoren oder eine andere
quantentheoretische Formulierung der Allgemeinen Relativitätstheorie aus der Quantentheorie der Ur-Alternativen zu begründen,
welche somit den übergeordneten begrifflichen Rahmen einer solchen Theorie darstellen würde. Hierzu werden im letzten Kapitel
des vierten Teiles der Arbeit einige Grundgedanken geäußert, die aber in ihrem gegenwärtigen Stadium mehr als eine Art
Darstellung eines Forschungsprogrammes zu verstehen sind. Wichtig ist aber dennoch, dass der Idee, eine quantentheoretische
Formulierung der Allgemeinen Relativitätstheorie aus dieser in ihren Begriffsbildungen rein der Quantentheorie verhafteten und
damit den physikalischen Raumbegriff nicht voraussetzenden Theorie zu begründen, eine sehr hohe Bedeutung beigemessen
werden muss.

Konstitutiv für die gesamte Arbeit ist also die These, und darin besteht gewissermaßen ihr eigentlicher Sinn, dass die
Existenz der Raum-Zeit mit ihrer topologischen Struktur einerseits und ihrer metrischen Struktur im Sinne der Allgemeinen
Relativitätstheorie andererseits, letztlich die Konsequenz einer dahinterliegenden Darstellung der Quantentheorie ist, welche
im von Weizsäckerschen Sinne als abstrakte Theorie der Information angesehen wird. Die Raum-Zeit besteht demnach also nicht 
für sich, sondern ist nur eine Weise, die abstrakte dahinterstehende Realität der Quantentheorie als Theorie der Information zu
veranschaulichen. Die Dualität zwischen der durch die Allgemeine Relativitätstheorie beschriebenen Raum-Zeit und der durch die
konkrete Quantentheorie beschriebenen Elementarteilchen würde damit durch den der allgemeinen Quantentheorie entsprechenden
übergeordneten Begriff der Information aufgehoben und sich als Konsequenz verschiedener spezieller Manifestationen der
quantentheoretischen Darstellung physikalischer Information herausstellen. Um von der begrifflichen Seite her diese Deutung
rechtfertigen zu können, ist es unumgänglich, auch Überlegungen der Klassischen Philosophie, insbesondere solche
erkenntnistheoretischer Natur, konkret miteinzubeziehen. Im Speziellen handelt es sich hierbei in erster Linie um die bereits
erwähnte Kantische Erkenntnislehre, spezifischer seine epistemologische Interpretation von Raum und Zeit im Rahmen der
transzendentalen Ästhetik, dem ersten Teil der "`Kritik der reinen Vernunft"', die jedoch in sehr sorgfältiger Weise an die
Erkenntnisse der modernen Naturwissenschaft angepasst werden muss und dies unter Einbeziehung der Evolutionären
Erkenntnistheorie. Desweiteren ist ein Bezug auch auf die Platonische Philosophie von großer Wichtigkeit. Nur unter
Einbeziehung dieses weiteren philosophischen Rahmens können die konkreten physikalischen Darstellungen und Untersuchungen ihre
eigentliche Bedeutung und ihren eigentlichen Sinn voll entfalten. Daher beginnt die Arbeit mit einem philosophischen
Einleitungskapitel, das zumindest die Grundzüge der für die gezogenen physikalischen Schlüsse notwendigen Philosophie vorstellt
und den vier Hauptteilen der Arbeit als "`Erkenntnistheoretisches Prolegomena"' vorangestellt ist.\\
\\
Schließlich sollte noch ein Hinweis in Bezug auf die in dieser Arbeit verwendete Notation gegeben werden. Beim Verweis auf
Gleichungen innerhalb der Arbeit und auf Literatur im Literaturverzeichnis wird der üblichen Konvention gefolgt. In dieser
Arbeit wird aber zudem eine bestimmte Notation verwendet, um auf einen Abschnitt oder einen Unterabschnitt eines Kapitels
innerhalb der Arbeit zu verweisen. Hierbei steht die entsprechende Nummerierung innerhalb einer eckigen Klammer, wobei sowohl
die eckige Klammer als auch die Nummerierung fett geschrieben sind, also beispielsweise: \textbf{[1.1.1]}.

\chapter{Erkenntnistheoretisches Prolegomena}

\section{Einleitung}

Das Ziel aller Naturwissenschaft besteht letztlich natürlich darin, möglichst allgemeingültige Wahrheiten über die natürliche
Welt zu formulieren und zu verstehen. Im Alltag naturwissenschaftlicher Forschung stellt man sich im Allgemeinen nicht der
Frage, auf welche Weise und in welchem Maße es dem Menschen möglich ist, Wahrheit über die Welt zu erkennen. Es handelt sich
hierbei um eine philosophische Frage. Es ist der Sinn der Erkenntnistheorie als Teilgebiet der Philosophie, sich den Fragen nach
den Voraussetzungen und Quellen menschlicher Erkenntnis zu stellen. In der Tat ist es nicht notwendig, sich in Zusammenhang mit
konkreter Wissenschaft mit solchen Fragen zu beschäftigen, wenn man im Rahmen eines bestehenden begrifflichen Rahmens einer
Theorie oder eines Paradigmas, wie es der Wissenschaftstheoretiker Samuel Kuhn nennt, bestimmte Probleme löst. Möchte man aber
die Grundlagen einer Theorie an sich oder deren Verhältnis zu einer anderen Theorie verstehen, so wird es unumgänglich, sich
auch erkenntnistheoretischen Fragestellungen zu stellen, also dem Problem, wie der Mensch zu Erkenntnis über die Natur gelangt
und in welchem Verhältnis der menschliche Geist zur natürlichen Welt steht. Denn Erkenntnis einer Realität in der Welt setzt
eine geordnete Beziehung des erkennenden Geistes zur erkannten Realität voraus. Wenn man also neue Begriffe und Vorstellungen
zur Beschreibung der Realität verwendet, so muss man die Frage nach dem Verhältnis dieser Begriffe zur Realität an sich, wie
sie sich uns auf einer bestimmten Ebene in der Natur darstellt, neu stellen und versuchen, zu beantworten. Auf die Frage nach
der Natur dieser Beziehung sind in der Geschichte der Philosophie unabhängig von einer spezifischen Beschreibung eines
bestimmten Ausschnittes der Realität, wie ihn eine spezielle physikalische Theorie liefert, unterschiedliche Antworten gegeben
und dementsprechend unterschiedliche geistige Systeme entworfen worden. In Bezug auf die in dieser Dissertation behandelten
Fragen kommt insbesondere der Erkenntnistheorie Immanuel Kants \cite{Kant:1781},\cite{Kant:1783} eine besondere Bedeutung zu,
die auch in Bezug zur Evolutionären Erkenntnistheorie \cite{Lorenz:1970},\cite{Vollmer:1975} betrachtet werden soll.
Weiter scheint es wichtig, einige Grundelemente der Platonischen Philosophie
\cite{Platon:Politeia},\cite{Platon:Timaios},\cite{Platon:Parmenides} zu erwähnen.

\section{Kantische Philosophie und Evolutionäre Erkenntnistheorie}

\subsection{Die Erkenntnistheorie Kants} 

Grundsätzlich kann man in der Geschichte der Erkenntnistheorie zwischen zwei geistigen Strömungen unterscheiden. Es handelt
sich um diejenige des Rationalismus und diejenige des Empirismus. Der Rationalismus ist der Anschauung, dass der Urgrund
menschlicher Erkenntnis letztlich in der menschlichen Vernunft selbst liege, aus der heraus er die reale Welt deuten könne.
Der Empirismus glaubt, dass alles menschliche Wissen aus der Erfahrung stamme. Die Kantische Erkenntnistheorie stellt in
gewissem Sinne eine Synthese des Empirismus mit dem Rationalismus dar. Sie wird in dem ersten der drei Hauptwerke Immanuel
Kants entwickelt, der berühmten "`Kritik der reinen Vernunft"' \cite{Kant:1781},\cite{Kant:1783}.
Ein bedeutender Empirist in der Geschichte der Philosophie war David Hume. Hume war der Meinung, dass alle Annahmen des
menschlichen Geistes über die Welt, auch die grundlegendsten, aus der Erfahrung stammten. Die Verknüpfung zwischen Ursache und
Wirkung beispielsweise, also das Prinzip der Kausalität, erhielt für ihn seine Glaubwürdigkeit nur aus der Erfahrung. Wir
glauben laut Hume nur daran, dass jeder Wirkung eine Ursache voran geht, weil wir es bisher so erfahren haben. Unser Glaube an
die Kausalität beruht also auf einem Prozess der Gewöhnung. Dies führte ihn auf das berühmte Induktionsproblem, dass darin
besteht, dass aus der Erfahrung vergangener Ereignisse eigentlich keine Schlüsse für die Zukunft gezogen werden können,
zumindest nicht mit logischer Notwendigkeit. Kant wird zu seiner eigenen Philosophie durch Hume angeregt,
\footnote{Immanuel Kant:\ "`Ich gestehe frei: die Erinnerung des David Hume war eben dasjenige, was mir vor vielen Jahren
den dogmatischen Schlummer unterbrach, und meinen Untersuchungen im Felde der spekulativen Philosophie eine ganz andere
Richtung gab."'}
betritt aber eine grundsätzlich neue Reflexionsebene, indem er das Phänomen der Erfahrung nicht mehr als schlicht gegeben
hinnimmt, sondern es selbst hinterfragt. Er stellt also die Frage nach den Bedingungen der Möglichkeit von Erfahrung. Dabei
leugnet Kant keineswegs, dass alle konkrete Erkenntnis über die Welt auf Erfahrung basiere. Kant vertritt allerdings die
Auffassung, dass dem menschlichen Geist vor aller Erfahrung, a priori, bereits bestimmte Erkenntnisstrukturen inne wohnen,
welche ihm überhaupt erst ermöglichen, in diejenige Beziehung zur Welt zu treten, welche wir als Erfahrung bezeichnen. Und jene
dem menschlichen Geist a priori gegebenen Erkenntnisstrukturen bezeichnet er als transzendental.
\footnote{Dies ist ein eigens zu diesem Zwecke eingeführter rein erkenntnistheoretischer Begriff, der nicht mit dem
metaphysischen Begriff "`transzendent"' zu verwechseln ist.}
Zu der Überzeugung, dass bestimmte Erkenntnisstrukturen a priori gegeben sind, wird Kant dadurch geleitet, dass es in Bezug
auf entsprechende Grundgegebenheiten, die Bestandteil aller Erfahrung sind, unmöglich ist, sich auch nur vorzustellen, dass
diese anders beschaffen seien. Wenn eine solche Grundgegebenheit weiter eine Bedingung der Möglichkeit von Erfahrung darstellt,
so kann sie durch konkrete Erfahrung niemals widerlegt werden, da sie letztere eben erst ermöglicht und in diesem Sinne ist
sie a priori gewiss. Damit sind die Erkenntnisse, welche wir durch die Erfahrung gewinnen, eine Mischung aus den durch die
konkrete Erfahrung a posteriori angeregten spezielle Inhalten und dem a priori gegebenen allgemeinen Rahmen, innerhalb
dessen wir sie erfahren. 
Kant unterscheidet weiter zwischen zwei Instanzen innerhalb des menschlichen Geistes, nämlich der Anschauung und dem reinen
Verstand. Über die Anschauung erhält der Mensch Zugang zur Realität und mit Hilfe des reinen Verstandes werden die hierin
erfahrenen Erscheinungen geordnet und dadurch verstanden. Es ist nun von entscheidender Bedeutung, nicht zuletzt auch in Bezug
auf spätere Teile der vorliegenden Dissertation, dass Raum und Zeit bei Kant sogenannte Grundformen der Anschauung darstellen,
die a priori gegeben und damit erfahrungskonstitutiv sind. Diese Behauptung rechtfertigt Kant ganz im Sinne der obigen
allgemeinen Begründung einer Grundstruktur der Erfahrung als a priori gegeben dadurch, dass es zwar im Prinzip möglich ist, von
allen im Raum und Zeit befindlichen speziellen Gegebenheiten, also darin befindliche Objekten und deren Eigenschaften, zu
abstrahieren. Raum und Zeit an sich müssen aber in unserer Vorstellung bestehen bleiben.
Von Ihnen können wir nicht abstrahieren, denn sie sind konstitutiver Bestandteil unserer Anschauung. Innerhalb der Anschauung
unterscheidet Kant zwischen Empfindung und Erscheinung. Eine Empfindung ist die Wirkung der unabhängig vom menschlichen Geist
gegebenen (materiellen) Realität auf die menschliche Anschauung, eine Erscheinung ist die Art und Weise, wie ein Ding sich im
Rahmen der Grundformen der Anschauung darstellt, also in Raum und Zeit. 
Das "`Ding an sich"', wie Kant es nennt, ist für sich genommen völlig unerkennbar. Nur seine Manifestation innerhalb der
menschlichen Anschauung, in der es als Erscheinung auftritt und somit zum Objekt wird, ist dem Menschen zugänglich.
Zum Erfassen und Ordnen der in der Anschauung gegebenen Erscheinungen verwendet der menschliche Geist die im reinen Verstand
liegenden Begriffe. Diese beziehen sich entweder direkt auf eine Erscheinung innerhalb der Anschauung oder auf einen anderen
Begriff und werden miteinander zu Urteilen verknüpft. 
Eine Reihe konstitutiver Begriffe, welche eine Art Grundmuster des menschlichen Erkennens darstellen, bezeichnet er als
Kategorien. Zu diesen gehört beispielsweise die Kausalität. Wichtig ist nun weiter, dass den dem menschlichen Geist a priori
gegebenen Anschauungsformen und Kategorien keine ontologische Signifikanz zukommt. Das bedeutet, dass Raum, Zeit und Kausalität
zwar insofern real sind, als sie konstitutiv für menschliche Erfahrung sind und jeder menschliche Zugang zur Welt nur auf Ihnen
basieren kann, da sie Bedingungen der Möglichkeit von Erfahrung darstellen, sie aber unabhängig von menschlicher Erkenntnis
keine Realität besitzen. Ihnen kommt also keine ontische, sondern lediglich eine epistemische Realität zu.

Nun hat aber die Entwicklung der modernen Physik gezeigt, und dies auf empirischen Befunden basierend, dass Raum, Zeit und
Kausalität anders verstanden werden müssen, als dies in der Klassischen Physik geschieht.
\footnote{Bei der Kausalität ist dies meiner Meinung nach nur in einem rein epistemologischen Sinne der Fall. Dies wird im
Abschnitt über die Interpretation der Quantentheorie thematisiert werden.}
Die Kantische Erkenntnistheorie in in ihrer streng idealistischen Form würde eine solche Entwicklung aber als grundsätzlich
unmöglich erachten. Dennoch stellt sie auf der anderen Seite den entscheidenden Schlüssel zu einem wirklichen Verständnis
unseres Bezuges zu der neuartigen Realität dar, deren Beschreibung durch Relativitätstheorie und Quantentheorie eröffnet wird. 
Es besteht also die Notwendigkeit, die Kantische Erkenntnistheorie neu zu interpretieren. Und dies ist in der Tat in einer
Weise möglich, in welcher die entscheidende Aussage der Existenz von Bedingungen der Möglichkeit von Erfahrung ihre völlige
Gültigkeit behält, jedoch Raum und Zeit eine zweifache Bedeutung erhalten. Es besteht nämlich die Notwendigkeit zwischen dem
realen Raum und dem Raum unserer Anschauung begrifflich zu unterscheiden, wenn beide auch in einem bestimmten Verhältnis
zueinander stehen. Diese Uminterpretation der Kantischen Philosophie geschieht unter Einbeziehung der Evolutionstheorie und
führt zur Evolutionären Erkenntnistheorie, welche in diesem Sinne als Erweiterung der Kantischen Erkenntnistheorie verstanden
werden kann.

\subsection{Die zweite Kantische Antinomie}

Zunächst muss jedoch noch auf einen weiteren in Bezug auf die Theoretische Physik bedeutsamen Aspekt der Philosophie Kants
eingegangen werden. Kant kommt im Rahmen seiner erkenntnistheoretischen Untersuchungen unter anderem zu dem Schluss, dass der
menschliche Geist so beschaffen ist, dass er bezüglich bestimmter sehr grundsätzlicher Fragen zwangsläufig zu einander
widersprechenden Aussagen geführt wird. Es ist hier von den bekannten Kantischen Antinomien die Rede. In der "`Kritik der reinen
Vernunft"' werden vier solche Antinomien aufgeführt \cite{Kant:1781}. In Bezug auf die Grundfragen der Elementarteilchenphysik
ist die zweite Antinomie von großer Bedeutung. Sie bezieht sich auf die Frage nach der Existenz von kleinsten räumlichen
Objekten als Grundbestandteilen der Materie. Es geht also letztlich um den Atomismus. In diesem Zusammenhang argumentiert Kant
nun in zweierlei Weise:\\
\\
1) Die sich ständig verändernden Erscheinungsformen der Materie lassen sich nur dann verstehen, wenn sie eine implizite
unveränderliche Struktur aufweisen. Da sie selbst aber veränderlich sind, kann eine solche innere Ordnung nur unter der
Bedingung existieren, dass es einfache nicht veränderliche und daher substantielle Grundbestandteile gibt, welche an sich
selbst nicht veränderlich sind und die Veränderung des Zusammengesetzten durch den Wandel ihrer Anordnung begründen. Dies
führt zur Annahme von Atomen als kleinsten unteilbaren Einheiten des substantiell Zusammengesetzten.\\
\\
2) Nach Ren\'{e} Descartes ist jedes materielle Objekt seiner Natur nach räumlich ausgedehnt (res extensa)
\cite{Descartes:1641}. Jedes denkbare als fundamental angenommene räumliche Objekt lässt aber zumindest im Prinzip die
begriffliche Teilbarkeit in Teilvolumina zu, da sich jedes noch so kleine Volumen begrifflich noch weiter unterteilen lässt.
Man kann zwar die Existenz eines kleinstes Objekts postulieren, aber dieses bleibt dennoch zumindest auf begrifflicher Ebene
weiter unterteilbar und daher ist es in sich nicht völlig einfach. Damit stellt sich die Frage, was die verschiedenen Teile
zusammenhält. Das Objekt besitzt also aufgrund seiner Ausdehnung eine innere Struktur (die Beziehung der Teilvolumina
zueinander), welche selbst erklärungsbedürftig ist. Dies bedeutet, dass es also prinzipiell keine kleinsten räumlichen
Objekte geben kann.\\
\\
Diese beiden gegensätzlichen Schlüsse bezüglich der Existenz räumlich fundamentaler Objekte können bereits zur Annahme
führen, dass einfachste Objekte nur auf einer nicht-räumlichen Ebene sinnvoll definiert werden können und vielleicht auch,
dass die räumliche Anschauung als Grundmuster der Erkenntnis der Realität an sich ihre Grenzen aufweist. 
Dies ist eine Erkenntnis, die für das Verständnis der Quantentheorie und die weitere Argumentation in dieser Arbeit von
großer Bedeutung ist.  

\subsection{Bezug der Kantischen zur Evolutionären Erkenntnistheorie}

Die Kantische Erkenntnistheorie behauptet, dass es bestimmte a priori gegebene unserem Geist inhärierende Erkenntnisstrukturen
gibt, welche Bedingungen der Möglichkeit von Erfahrung darstellen und damit erfahrungskonstitutiv sind, denen jedoch an sich
keine ontische Realität zukommt. Die ursprünglich auf Konrad Lorenz zurückgehende Evolutionäre Erkenntnistheorie
\cite{Lorenz:1970},\cite{Vollmer:1975} geht nun davon aus, dass sich der menschliche Geist, auf dem Gehirn als biologischem
System basierend, in Analogie zu allen anderen biologischen Systemen, in Anpassung an die Umwelt nach dem Prinzip des
Überlebensvorteiles, dass er dem Menschen beziehungsweise seinen Vorfahren im Laufe der biologischen Evolution einbrachte,
entwickelt hat.
So können dann auch die seitens Kant als a priori dargestellten Erkenntnisstrukturen als das Ergebnis eines Äonen währenden
Anpassungsprozesses des menschlichen Gehirns mit seinen Sinnen an die Umwelt angesehen werden. Sie sind damit in einem Sinne
als a priori gegeben anzusehen und in einem anderen erweiterten Sinne als a posteriori erworben zu interpretieren.
Ontogenetisch gesehen sind diese grundlegenden Erkenntnisstrukturen als a priori gegeben zu verstehen, da sie dem einzelnen
Individuum angeboren sind und dieses nur in ihrem Rahmen spezielle Erfahrungen machen kann.
Bezieht man jedoch eine erweiterte phylogenetische Perspektive mit ein, so stellen sie erworbene Eigenschaften des menschlichen
Geistes dar und können damit als überindividuelle Erfahrung der ganzen Spezies Mensch interpretiert werden.
\footnote{Bei phylogenetisch älteren Bestandteilen des menschlichen ``Weltbildapparates'' (Konrad Lorenz) bezieht sich diese
überindividuelle Erfahrung analog auf diejenigen Lebensformen, aus denen sich der Mensch biologisch entwickelt hat.}
Da der Erwerb dieser Eigenschaften sich im Rahmen der biologischen Evolution vollzog, ist nun weiter entscheidend, dass
diese Erkenntnisstrukturen nur insoweit eine getreue Abbildung der Realität darstellen, als diese die Überlebenschancen der
biologischen Spezies, in diesem Falle des Menschen, signifikant erhöhten. So erscheint auch die Annahme plausibel, dass unsere
Erkenntnisstrukturen nicht deckungsgleich mit den Strukturen der realen Welt sind, und demnach versagen, sobald man es mit
Phänomenen zu tun hat, die weit jenseits der Dimensionen des natürlichen Erfahrungsbereiches des Menschen und seiner Vorfahren
liegen. Die Erkenntnisstrukturen unseres Geistes stellen zwar eine im Laufe der Evolution entstandene Abbildung der Realität
dar, aber es handelt sich eben nur um eine die Realität in einer gewissen Näherung korrekt beschreibende Abbildung. 

Das bedeutet auch, dass damit etwa die Struktur des realen Raumes als ontischer Realität seiner Natur nach von seiner
Abbildung als dem Individuum a priori gegebener Anschauungsform zu unterscheiden ist. Es besteht allerdings, zumindest
näherungsweise, eine Isomorphie zwischen der tatsächlichen Struktur des Raumes in der Realität und der Struktur des Raumes als
Anschauungsform unseres Geistes. Dies bedeutet, dass die durch die moderne Physik entdeckte Tatsache, dass die Raum-Zeit
anders beschaffen ist, als es die Klassische Physik voraussetzt, nicht im Widerspruch zur Kantischen Auffassung steht,
dass der Raum als Anschauungsform eine a priori gegebene Realität des menschlichen Geistes sei. Denn es ist dementsprechend
nur gezeigt worden, dass die Realität wirklich eine Struktur aufweist, welche durch den Raum als Anschauungsform abgebildet
werden kann, denn sonst hätten wir ja nicht durch indirekte aus der Erfahrung gezogene Schlüsse entdeckt, dass diese Struktur
allgemeiner ist, als sie in unserem Geist abgebildet wird. Es wird hierdurch aber in keiner Weise nahegelegt, dass diese
Anschauungsform an sich real sei, sondern nur, dass dies von der Struktur gilt, die sie näherungsweise widerspiegelt.
Damit behält die Kantische Argumentation, dass der Raum als Anschauungsform eine Realität unseres Geistes sei, ihre volle
Gültigkeit. Allgemein kann man dementsprechend sagen, dass es sinnvoll ist, davon auszugehen, dass unser Geist bestimmte in der
Realität wirklich existierende Strukturen in einer für ihn spezifischen Weise abbildet, aber dass die Isomorphie zwischen den
realen Strukturen und den Strukturen, wie sie der menschlichen Geistes abbildet, nur näherungsweise besteht.

Wenn man einerseits die Geschichte der biologischen Evolution anerkennt und andererseits das Gehirn oder generell ein
physikalisches System als die Voraussetzung des Auftretens von Intelligenz in der Welt ansieht, so kann man wohl nur schwer daran
zweifeln, dass der Evolutionären Erkenntnistheorie eine wichtige Rolle beim Verständnis der Beziehung unseres Geistes zur
Realität zukommt, wenn damit auch nichts darüber ausgesagt ist, inwiefern der menschliche Geist eine vom menschlichen Gehirn an
sich zu unterscheidende Realität darstellt.

\section{Platonismus}

\subsection{Allgemeine Ideenlehre}

Ein wesentliches Element der Platonischen Philosophie ist die berühmte Ideenlehre, wie sie beispielsweise in der Politeia
\cite{Platon:Politeia}, dem Dialog Timaios \cite{Platon:Timaios} oder dem Dialog Parmenides \cite{Platon:Parmenides}
behandelt wird. Die Platonische Vorstellung einer Idee entspringt der Frage nach dem Zusammenhang einer speziellen
Gegebenheit in der Erfahrung und der sich in ihr ausdrückenden allgemeinen Realität. So existieren in der Erfahrung etwa viele
konkrete Kreise. Bei Ihnen allen handelt es sich um voneinander unterscheidbare Objekte, die ferner nie das Ideal eines exakten
Kreises im Sinne der Mathematik erfüllen. Dennoch kann man in Ihnen nur deshalb Kreise erkennen, weil es eine abstrakte
mathematische Struktur gibt, welche wir einen idealen Kreis nennen.
Die Realität dieser Struktur inhäriert implizit allen wirklichen Kreisen. Gleichzeitig ist sie selbst natürlich nicht
Gegenstand einer natürlichen Erfahrung. Es stellt sich also das Problem des Realitätsstatus eines abstrakten Kreises oder
anderer Strukturen an sich. Platon verwendet in diesem Zusammenhang den Terminus der Idee. Die Ideen entsprechen einer für sich
bestehenden geistigen Realität. Die konkreten Gegenstände der Erfahrung sind in gewisser Weise konkrete Manifestationen der
Ideen und der menschliche Geist wird durch Erfahrung der konkreten Gegenstände an die Ideen erinnert, welche in der Erfahrung
nur unvollkommen realisiert sind. Hierdurch wird nach Platon Erkenntnis der Natur möglich.
Es ist diesbezüglich aber wichtig zu erwähnen, dass eine Idee im Sinne Platons nicht einem Begriff entspricht. Einem Begriff
kommt rein epistemische Realität zu. Er ist ein Instrument des menschlichen Geistes, mit Hilfe dessen dieser die Erfahrung
ordnet. Er mag zu einer realen Gegebenheit isomorph sein. Je genauer diese Isomorphie ist, desto besser ist der Begriff zur
Beschreibung der Realität geeignet. In jedem Falle ist er aber von der ihm entsprechenden Realität selbst zu unterscheiden. Ihm
kommt also an sich keine ontologische Bedeutung zu. Bei einer Idee im Platonischen Sinne ist dies jedoch grundsätzlich anders.
Eine Platonische Idee besitzt sowohl epistemische als auch ontische Realität, ist also auch nicht an den menschlichen Geist
gebunden, sondern existiert in einer Welt des objektiven Geistes. Dieser Welt des objektiven Geistes kommt nach Platon ein viel
höherer Realitätsstatus zu als der Erfahrungswelt, durch die sie immer nur in einer schwachen Weise hindurchschimmert.
Dementsprechend stellt eine bestimmte Idee, welche sich in der Struktur eines Objektes manifestiert, die eigentliche Realität
dar, von welcher das konkrete Objekt nur ein schwaches Abbild darstellt. Hier entsteht natürlich die Frage, in welcher
Beziehung ein konkretes Objekt zur allgemeinen Idee dahinter steht, inwiefern die spezielle Manifestation der Idee im Objekt
durch die Idee an sich Realität erhält. Dies ist der Kern des sogenannten Universalienproblems. Nach Platon hat das konkrete
Objekt Anteil an der Idee, erhält sogar durch sie erst seine Realität. 

\subsection{Die Natur der Materie in der Platonischen Philosophie}

Soweit wurde nun in sehr kurzer Form auf die Platonische Ideenlehre als grundlegendem Teil der Platonischen Erkenntnistheorie
im Allgemeinen eingegangen. Im Hinblick auf die Physik, genauer die Physik der Elementarteilchen, ist nun im Speziellen die im
Dialog Timaios dargelegte Vorstellung Platons vom Aufbau der Materie von Bedeutung \cite{Platon:Timaios}. Platon übernimmt hier
die ursprünglich durch Empedokles eingeführte Vorstellung der vier Elemente, nämlich Feuer, Erde, Wasser und Luft als
Grundbestandteile der Natur. Diese entsprechen aber im Rahmen der Platonischen Anschauung nun geometrischen Formen, genauer
gleichmäßigen Körpern. Das Feuer entspricht dem Tetraeder, die Erde dem Würfel, die Luft dem Oktaeder und das Wasser dem
Ikosaeder. Diese regulären Körper sind ihrerseits wiederum aus Dreiecken aufgebaut, welche letztlich aus einem elementaren
gleichseitigen Dreieck als fundamentalster Grundform bestehen. Dieses stellt die auf basaler Ebene eigentliche existierende
ontologische Entität dar. Die durch die Mathematik beschriebenen geometrischen Strukturen erhalten damit also selbst ontischen
Charakter. Und dies ist nicht etwa in dem Sinne zu verstehen, dass sie im Sinne der Dualität von Stoff und Form als Akzidenzien
ihnen entsprechenden Substanzen anhängen würden, die an sich materiell sind. Die mathematische Struktur besteht vielmehr für
sich selbst. Sie ist selbst Substanz und auf basaler Ebene als ontologische Entität die eigentliche Realität, welche die auf
ihr basierende der direkten Sinneswahrnehmung zugängliche materielle Realität erst konstituiert. Werner Heisenberg und 
Carl Friedrich von Weizsäcker haben diese Anschauung Platons zur Interpretation der Quantentheorie in Beziehung gesetzt.
Betrachtungen hierzu sind beispielsweise in \cite{Heisenberg:1969},\cite{Heisenberg:1979} beziehungsweise
\cite{Weizsaecker:1971},\cite{Weizsaecker:1981} vorzufinden.

\section{Heuristische Vorüberlegungen zur Theoretischen Physik}

\subsection{Die Entwicklung der Theoretischen Physik zur Einheit}

In der Geschichte der Theoretischen Physik gibt es eine Tendenz zu einer sich vergrößernden Einheitlichkeit der
Naturbeschreibung, die in gewisser Hinsicht sogar ihr eigentliches Bestreben darstellt. Es stellt sich die Frage, wie sich
dieser Prozess in der Vergangenheit vollzogen hat, und welche Bedeutung dies für die Suche nach neuen Theorien hat.
Samuel Kuhn unterscheidet in seiner Wissenschaftstheorie zwischen normaler Wissenschaft und wissenschaftlichen Revolutionen
\cite{Kuhn:1962}. In der normalen Wissenschaft werden laut Kuhn im Rahmen eines Paradigmas, also einer bestimmten
grundlegenden Vorstellung über die Natur in einem bestimmten Bereich und damit verbundenen allgemein akzeptierten Methoden,
spezielle Probleme gelöst. Den weitaus größten Teil der Zeit nehmen die Perioden einer Wissenschaft ein, in denen in diesem
Sinne normale Wissenschaft betrieben wird.
Es gibt allerdings auch Phasen, in denen eine Wissenschaft oder eine bestimmte Theorie in eine Krise gerät und sich prinzipielle
Schwierigkeiten bei der Behandlung der entsprechenden konkreten wissenschaftlichen Problemstellungen ergeben. Wenn dies
geschieht, so wird das der jeweiligen Wissenschaft zugrundeliegende Paradigma in Frage gestellt und nach einem neuen Paradigma
gesucht. Während der Suche nach einem neuen Paradigma vollzieht sich die Wissenschaft nicht in der gleichen schematischen Weise
wie dies während der Phasen der normalen Wissenschaft der Fall ist, in denen in kumulativer Weise eine Fragestellung nach der
anderen behandelt wird. Es handelt sich hingegen um einen nicht nach geordneten Regeln vor sich gehenden Prozess, an dessen
Ende sich jedoch schließlich ein grundlegend neues Paradigma durchsetzt. Kuhn bezeichnet die Phasen des Überganges daher als
wissenschaftliche Revolutionen. 

Heisenberg hat in diesem wissenschaftstheoretischen Zusammenhang in Bezug auf die Theoretische Physik den Begriff der
abgeschlossenen Theorie geprägt. Eine abgeschlossene Theorie besteht aus einem System basaler Begriffe, welche ihnen
entsprechenden dynamischen Entitäten zugeordnet werden, und aus einer Reihe grundlegender Aussagen über die Natur, welche diese
Begriffe miteinander verbinden, den sogenannten Axiomen. Alle speziellen Phänomene des Erfahrungsbereiches, auf den sich die
abgeschlossene Theorie bezieht, stellen sich im Rahmen einer solchen Theorie als Konsequenzen dieses konstitutiven
theoretischen Systems dar.
\footnote{Hierbei sollte erwähnt werden, dass die durch die Begriffe beschriebenen Entitäten bestimmte Strukturen aufweisen.
Die Beschreibung solcher Strukturen aber liefert die Mathematik. Daher spielt die Mathematik bei der Formulierung der Axiome
und der Herleitung aus diesen Axiomen folgender Konsequenzen eine zentrale Rolle.}
Verstehen eines Bereiches der Natur bedeutet für Heisenberg demnach, sehr viele verschiedene Erscheinungen im Rahmen einer
Theorie auf ein möglichst einfaches und einheitliches System von basalen Begriffen zurückzuführen. Eine solche Theorie ist
nicht endgültig, oder muss es zumindest nicht sein. Sie stellt jedoch ein in sich bestehendes System von großer innerer
Harmonie und Einheit dar und kann daher durch kleine Änderungen nicht mehr verbessert werden. Wenn sie durch eine auf sie
folgende noch allgemeinere und einheitlichere abgeschlossene Theorie relativiert wird, so definiert diese neue allgemeinere
Theorie die Gültigkeitsgrenzen der alten Theorie, welche in Bezug auf die neue Theorie in dem Sinne Wahrheit enthält, dass sie
in der allgemeineren Theorie implizit als Grenzfall enthalten ist.
Einer Kuhnschen wissenschaftlichen Revolution entspricht demgemäß in der Heisenbergschen Terminologie der Übergang zu einer
neuen abgeschlossenen Theorie. In einer neuen abgeschlossenen Theorie werden im allgemeinen mehrere frühere abgeschlossenen
Theorien miteinander vereinheitlicht. Dies bedeutet, dass sich die Theoretische Physik hin zu einer immer größeren inneren
Einheit bewegt, in der ein immer größerer Erfahrungsbereich im Rahmen einer einzigen Theorie beschrieben werden kann. Dies
bedeutet nicht, dass die Fülle der zu untersuchenden Detailfragen und Konsequenzen nicht immer mehr zunehmen würde, aber es
bedeutet, dass die Grundsysteme, auf denen alles aufbaut und aus denen heraus letztlich versucht wird, die Phänomene zu
erfassen, immer umfassender, allgemeiner und damit auch abstrakter werden.

\subsection{Die Entstehung neuer abgeschlossener Theorien}

Man kann nun in der Geschichte der Theoretischen Physik grundsätzlich zwischen zwei Arten der Entstehung neuer grundlegender
Theorien unterscheiden. Entweder wurden neue Phänomene entdeckt, die sich grundsätzlich nicht mehr in den alten begrifflichen
Rahmen integrieren ließen oder man hatte es mit mehreren in sich geschlossenen Theorien zu tun, deren basale Begriffssysteme
aber in Bezug aufeinander widersprüchlich erschienen. Meistens war es eine Kombination aus beidem.
Im Falle der Speziellen Relativitätstheorie lag das Michelson-Morley-Experiment vor, aus dem die universelle Konstanz der
Lichtgeschwindigkeit folgt. Zudem gab es die Tatsache, dass die Klassische Mechanik und die Klassische Elektrodynamik sich
zu widersprechen schienen, da in der Elektrodynamik und der aus ihr folgenden Wellengleichung für elektromagnetische Felder
die Lorentz-Gruppe als grundlegende Symmetriegruppe auftritt, welche die Lichtgeschwindigkeit konstant lässt. In der Klassischen
Mechanik hingegen war es die Galilei Gruppe, welche die grundlegende Symmetriegruppe darstellte.
Die Spezielle Relativitätstheorie löst diesen Widerspruch auf und trägt ebenso dem Michelson-Morley-Experiment Rechnung, indem
sie die Klassische Mechanik und das ihr inhärierende Relativitätsprinzip dem Prinzip der Konstanz der Lichtgeschwindigkeit
unterwirft und sie damit Lorentz-kovariant formuliert. Hierbei muss der Begriff der Gleichzeitigkeit aufgegeben beziehungsweise
neu definiert werden. In der Allgemeinen Relativitätstheorie, auf die später noch sehr viel ausführlicher eingegangen werden
wird, musste die Newtonsche Gravitationstheorie mit der Speziellen Relativitätstheorie vereinheitlicht werden.
Im Falle der Quantentheorie war der Ursprungspunkt das Spannungsverhältnis zwischen der Thermodynamik und der Klassischen
Elektrodynamik, welches nur durch die Einführung des Planckschen Wirkungsquantums aufgelöst werden konnte. Die weitere
Entwicklung zur endgültigen Fassung der Quantentheorie konnte dann nur unter der Inspiration durch neue empirische Phänomene
geschehen, wie etwa den Spektrallinien oder den Beugungs- und Interferenzeffekten.
Da man bei der Suche nach einer neuen abgeschlossenen Theorie also nach ganz neuen Begriffen zur Beschreibung der Realität
sucht, müssen dementsprechend grundsätzliche Fragen thematisiert werden. Und daher ergibt sich auch die Frage der Beziehung 
unseres Geistes mit seinen spezifischen Erkenntnisstrukturen zur Realität an sich, weshalb unter anderem eben Fragen der
Erkenntnistheorie miteinbezogen werden müssen.
So war für Albert Einstein nicht nur die Frage entscheidend, welche Bedeutung Raum und Zeit bei der Beschreibung natürlicher
Vorgänge wirklich spielen und in welcher Beziehung sie zu den anderen Begriffen stehen, die in der Physik verwendet werden,
sondern auch, wie wir überhaupt Wissen über räumliche und zeitliche Relationen gewinnen können. 
In der gleichen Weise war für Niels Bohr die Frage nach dem Grund der Stabilität der Materie, also der Tatsache, dass sich im
atomaren Bereich immer wieder die gleichen Strukturen herstellen, weshalb immer wieder die gleichen Atomsorten mit exakt den
gleichen für sie spezifischen Eigenschaften in Erscheinung treten, mit der kritischen Überlegung über die Grenzen der
Möglichkeit der Beschreibung des inneren der Atome mit auf klassischen Vorstellungen basierenden Anschauungen verbunden.  
Wichtig hierbei ist, dass während des Prozesses der Suche nach einer neuen abgeschlossenen Theorie der sorgfältigen Besinnung
auf die grundlegenden Begriffe und Aussagen der älteren Theorien und die mit ihnen in Zusammenhang stehenden bekannten
grundlegenden Phänomene in der Natur eine immense Bedeutung zukommt. Hierbei muss vor allem betont werden, dass eine
umfassendere Beschreibung der Natur, welche zwei oder mehrere ältere Theorien zu einer neuen begrifflichen Einheit führen soll,
welche neuen Begriffe sie auch einführen und welche neuen über das bisherige hinausgehende Annahmen über die Realität sie auch
zugrundelegen mag, in jedem Falle den jeweiligen Reflexionsgrad und die innere begriffliche Geschlossenheit der älteren
bewährten Theorien in sich Aufnehmen, gegebenenfalls erweitern muss, keinesfalls aber wieder aufbrechen darf. Denn diejenigen
Theorien, welche sich durchgesetzt und empirisch bewährt haben, beziehen ihre Glaubwürdigkeit daraus, dass sie in einer sehr
eleganten Weise durch die Einführung neuer Begriffe, viele Divergenzen und Inkonsistenzen älterer Theorien, die davor schon auf
einer darunterliegenden Reflexionsebene das Gleiche geleistet haben, in eine neue Harmonie geführt haben und ebendies auch mit
einem dazu parallelen empirischen Erfolg einherging. Hieraus ergibt sich aber die kategorische Notwendigkeit, alles, aber auch
wirklich alles daran zu setzen, die begrifflichen Neuerungen der älteren Theorien in sich zu verstehen, und neue Denkansätze
entweder hieraus hervorgehen zu lassen oder zumindest hiervon ausgehend zu entwickeln.
Eine neue Theorie, welche den ungeheuren Erkenntnisgewinn der älteren Theorien wirklich würdigen und eine neue Ebene der
Wahrheit erreichen soll, darf sich daher nicht damit begnügen, nur alte formale Methoden im Rahmen neuer Ansätze zu verwenden,
sondern sie muss vielmehr den eigentlichen Gehalt der Errungenschaften der älteren Theorien in sich aufnehmen und weiterführen.
Vor diesem Hintergrund sind auch ein Großteil der Bemühungen dieser Doktorarbeit zu sehen. Es geht also auch darum, zu
versuchen, in der Tiefe zu verstehen, worin der mit der Entstehung der Quantentheorie und der Allgemeinen Relativitätstheorie
einhergehende Erkenntnisgewinn in seinen grundlegenden Behauptungen über die Natur besteht und welche dementsprechenden
Konsequenzen in Bezug auf den physikalischen Realitätsbegriff daher aus ihnen zu ziehen sind.

\part{Die Allgemeine Relativitätstheorie}

\chapter{Grundbegriffe der Allgemeinen Relativitätstheorie}

\section{Einleitung}

\subsection{Die historische Rolle der Allgemeinen Relativitätstheorie}

Die Allgemeine Relativitätstheorie \cite{Einstein:1914bx},\cite{Einstein:1915by},\cite{Einstein:1916vd} kann als die Vollendung
der Klassischen Physik angesehen werden. Sie ist das alleinige Werk Albert Einsteins und von ihr glaubt man als einziger
Theorie, dass sie ohne ihren Begründer überhaupt nicht oder zumindest erst sehr viel später entdeckt worden wäre. Die
Gravitationsteorie Isaac Newtons wird durch sie mit der Speziellen Relativitätstheorie \cite{Einstein:1905} vereinheitlicht,
die selbst bereits die Klassische Mechanik Isaac Newtons und die Klassische Elektrodynamik Michael Faradays und James Clerk
Maxwells in sich beinhaltet und die ebenfalls bereits Einsteins Werk darstellt, auch wenn ihm hier im Gegensatz zur Allgemeinen
Relativitätstheorie schon wichtige Einsichten vorlagen, die sich insbesondere mit den Namen Hendrik Antoon Lorentz und Henri
Poincar\'{e} verbinden. Darüber hinaus liefert sie eine völlig neue Beschreibung von Raum und Zeit, indem sie deren metrische
Struktur selbst einer dynamischen Beschreibung zugänglich macht. Ihre Entstehung nahm insgesamt zehn Jahre in Anspruch. Die
Überlegungen Einsteins zur Allgemeinen Relativitätstheorie begannen wohl unmittelbar nach Vollendung der Speziellen
Relativitätstheorie im Jahre 1905, da dieser sehr früh im Sinne einer begrifflichen Konsistenz mit der Gravitation die
Notwendigkeit einer Verallgemeinerung sah. Die vollständige mathematische Fassung und im wesentlichen auch ihre
naturphilosophische Deutung, deren Kern mit der Diffeomorphismeninvarianz in Zusammenhang steht, lagen im Jahre 1915 vor.
Seitdem stellt sie die fundamentale Theorie der Gravitation dar, welche letztere als Konsequenz einer Beschreibung der
Raum-Zeit-Struktur darstellt. Erst durch sie wurde daher die Kosmologie als Theorie der Struktur des Universums zu einer
physikalischen Wissenschaft (siehe beispielsweise \cite{Weinberg:1972}). Alle anderen makroskopischen Phänomene auf
astrophysikalischer Ebene werden ebenfalls mit Hilfe der Allgemeinen Relativitätstheorie beschrieben. Mathematisch basiert sie
auf der Riemannschen Geometrie, welche Bernhard Riemann im neunzehnten Jahrhundert als Verallgemeinerung der Euklidischen
Geometrie für gekrümmte Räume formulierte.

\subsection{Vorbetrachtung der Grundideen der Allgemeinen Relativitätstheorie}

Wie die Bezeichnung bereits andeutet, stellt die Allgemeine Relativitätstheorie eine Verallgemeinerung der Speziellen
Relativitätstheorie dar und ist aus erneuten inhärenten begrifflichen Schwierigkeiten der Speziellen Relativitätstheorie als
Verallgemeinerung der Klassischen Mechanik erwachsen, welche die Elektrodynamik mit umfasst.
Es ergab sich einerseits die Frage nach der Bedeutung beliebiger Bezugssysteme. Die Spezielle Relativitätstheorie enthält
ebenso wie die Klassische Mechanik nur die Gleichberechtigung aller Inertialsysteme, also gleichmäßig zueinander bewegter
Bezugssysteme. In einer allgemeinen Theorie würde man aber die Gleichberechtigung beliebiger Bezugssysteme erwarten.
Desweiteren bestand die Schwierigkeit des Verhältnisses der Speziellen Relativitätstheorie zum Newtonschen Gravitationsgesetz.
Dieses beschreibt eine starre Kraft und enthält damit instantane Fernwirkungen. Wird ein Körper bewegt, so ändert sich die
durch ihn bewirkte Gravitationskraft instantan im gesamten Raum. Dies steht aber im Widerspruch zu der fundamentalen Aussage
der Speziellen Relativitätstheorie, dass die Lichtgeschwindigkeit die absolute Höchstgrenze für die Geschwindigkeit der
Informationsübertragung im Raum darstellt. Es bestand also die Schwierigkeit, eine feldtheoretische Formulierung der
Gravitation zu finden, wie sie für die Elektrodynamik bereits existierte, um diese Inkonsistenz zu beheben. Hierbei konnte sich
Einstein an der Faraday-Maxwellschen Elektrodynamik orientieren, aus welcher die Lichtgeschwindigkeit für die Ausbreitung
Elektromagnetischer Wellen ja gerade folgt. Diese stellt eine Verallgemeinerung des Coulombschen Kraftgesetzes dar. Die
elektrische Ladung tritt einerseits als Quelle des elektromagnetischen Feldes und andererseits als diejenige Größe auf, welche
bestimmt, wie sehr ein Körper auf das Elektromagnetische Feld reagiert. Im Falle der Gravitation ist es die schwere Masse,
welche die Stärke des Gravitationsfeldes und die Reaktion eines Körpers auf ein solches Feld bestimmt. Diese ist begrifflich
von der trägen Masse zu unterscheiden, welche den Widerstand beschreibt, welcher ein Körper einer Beschleunigung entgegensetzt.
Nun ist es aber ein empirisches Faktum, dass in der Natur die träge Masse der schweren Masse entspricht. Diese Äquivalenz
zwischen träger und schwerer Masse wird als Äquivalenzprinzip bezeichnet und entspricht der Tatsache, dass die Wirkung eines
beschleunigten Bezugssystems nicht von derjenigen eines Gravitationsfeldes zu unterscheiden ist. Dies führte Einstein zu der
Vermutung, dass das Gravitationsfeld eine Eigenschaft der Raum-Zeit selbst darstellt, da die Wirkung eines beschleunigten
Bezugssystems ja ein direkt auf die Eigenschaften der Raum-Zeit bezogenes Phänomen ist.

\section{Differentialgeometrie}

Die mathematische Formulierung der Allgemeinen Relativitätstheorie basiert auf der Riemannschen Geometrie, einer
Verallgemeinerung der Euklidischen Geometrie für gekrümmte Räume, in welcher das Parallelenaxiom nicht als gültig vorausgesetzt
wird. Letzteres besagt, dass es in einer Ebene zu einer bestimmten Geraden $g$ und einem nicht auf dieser Geraden liegenden
Punkt $P$ genau eine weitere Gerade gibt, welche durch den Punkt $P$ geht und die Gerade $g$ nicht schneidet, also parallel zu
ihr ist. Alle weiteren Konzepte der Riemannschen Geometrie bauen auf dem Begriff der differenzierbaren Mannigfaltigkeit auf.
Dieser wird daher im folgenden Unterabschnitt eingeführt werden, um anschließend die weiteren mathematischen Konzepte der
kovarianten Ableitung, der Krümmung und der metrischen Struktur zu behandeln (siehe auch \cite{Nakahara:1990}).

\subsection{Differenzierbare Mannigfaltigkeiten}

Es sei zunächst ein topologischer Raum gegeben, also eine Punktmenge $M$, auf der ein System offener Mengen definiert
ist, welches den Axiomen einer Topologie genügt. Desweiteren sei der Begriff der Karte eingeführt. Als Karte wird eine stetige
bijektive Abbildung $\varphi$, also ein Homöomorphismus, einer Teilmenge $U \subseteq M$ in eine Menge des $\mathbb{R}^d$
bezeichnet. Damit beschreibt eine Karte $\varphi$ Punkte $p \in U$ durch $d$ Koordinaten 

\begin{equation}
p \in U \quad,\quad \varphi:p\rightarrow x^\mu(p)\quad,\quad \mu=1...d. 
\end{equation}
Wenn zwei Karten $\varphi$ und $\varphi^{\prime}$ gegeben sind, welche sich auf zwei Teilmengen $U\subseteq M$ und
$U^{\prime}\subseteq M$ beziehen, welche eine nicht-leere Schnittmenge $U \cap U^{\prime}$ bilden, so kann jeder Punkt $p \in
\left(U \cap U^{\prime}\right)$ durch zweierlei Koordinaten, $x^\mu$ und $x^{\mu \prime}$, bezeichnet werden. Diese Koordinaten
lassen sich, da es sich bei $\varphi$ und $\varphi^{\prime}$ um Homöomorphismen handelt, in eindeutiger Weise aufeinander abbilden

\begin{equation}
x^\mu=A(x^{\nu \prime})\quad,\quad x^{\nu \prime}=A^{-1}(x^{\mu}).
\end{equation}
Wenn die Abbildung $A$ zwischen den beiden Karten $k$-mal stetig differenzierbar ist, so bezeichnet man die beiden Karten als
$C^k$-verknüpft. Ein Atlas ist eine Menge von $n$ Karten $\varphi_m$, $m=1...n$, deren Definitionsbereiche $U_m$ den gesamten 
Raum $M$ überdecken:

\begin{equation}
M=\bigcup_{m} U_m.
\end{equation}
\textbf{Definition:} Als eine $C^k$-Mannigfaltigkeit bezeichnet man einen topologischen Raum $M$, welcher mit einem Atlas versehen
ist, dessen Karten miteinander $C^k$-verknüpft sind.

Eine Kurve ist eine Abbildung $\kappa$ eines offenen Intervalls von $\mathbb{R}$ nach $M$. Sei $f$ eine in einer Umgebung von
$p \in M$ definierte reellwertige Funktion, $\kappa$ eine Kurve, welche durch $p$ läuft, und $\tau$ der Kurvenparameter, sodass
$\kappa(\tau_p)=p$, dann bezeichnet man die Ableitung der Funktion $f$ nach dem Kurvenparameter $\tau$ in $p$ als
Tangentialvektor $v$ der Kurve in $p$

\begin{equation}
v(f)=\frac{d}{d \tau}f[x^\mu(\tau)]|_{\tau=\tau_0}=\frac{\partial f(x^\mu)}{\partial x^\nu}|_{x^\mu(\tau_0)}
\frac{d x^\nu}{d\tau}|_{\tau=\tau_0}\quad,\quad v=\frac{d}{d \tau}=\frac{d x^\mu}{d \tau}\frac{\partial}{\partial x^\mu}.
\end{equation}
Ein Tangentialvektor beschreibt eine Richtung innerhalb der Mannigfaltigkeit. Desweiteren entspricht einem Tangentialvektor 
eine Äquivalenzklasse von Kurven, die in diesem Punkt den gleichen Tangentialvektor definieren. 
Da eine beliebige Linearkombination von Tangentialvektoren in einem Punkt wieder auf einen Tangentialvektor führt,
bilden alle Tangentialvektoren bezüglich eines Punktes einen Vektorraum, was die Bezeichnung Tangentialvektor rechtfertigt. 
Demnach wird die Menge aller Tangentialvektoren in einem Punkt $p$ einer Mannigfaltigkeit als Tangentialraum $\mathcal{T}_p$ 
bezeichnet. Die partiellen Ableitungen nach den Koordinaten einer Karte $\frac{\partial}{\partial x^\mu}$ bilden eine Basis des 
Tangentialraumes, während die Ableitungen dieser Koordinaten nach dem Kurvenparameter $\frac{d x^\mu}{d \tau}$ die Komponenten 
$v^\mu$ des zu dieser Kurve gehörigen Tangentialvektors $v$ darstellen. Natürlich gehorchen die Vektoren beim Übergang zu einem 
anderen Koordinatensystem den üblichen Transformationsregeln für Vektoren. Ein zugehöriger Kotangentialraum $\mathcal{T}_p^{*}$ 
lässt sich in der gewöhnlichen Weise als Menge der Linearformen $v^{*}$ auf dem Tangentialraum definieren. 
Die Vereinigung aller Tangentialvektorräume einer Mannigfaltigkeit bezeichnet man als Tangentialbündel $\mathcal{T}_M$: 

\begin{equation}
\mathcal{T}_M=\bigcup_{p \in M}\mathcal{T}_p.
\end{equation}
Ein Vektorfeld $V_M$ auf einer Mannigfaltigkeit stellt damit eine Abbildung aller Punkte $p$ der Mannigfaltigkeit $M$ in die 
entsprechenden Tangentialräume $\mathcal{T}_p$ dar 

\begin{equation}
V_M: M \rightarrow \mathcal{T}_M
\end{equation}
und wird als Schnitt im Tangentialbündel bezeichnet.

Man kann nun gemäß der Einführung von Vektoren und Kovektoren auch Tensoren einführen, welche sich auf Tangentialräume
beziehen. Sei $\mathcal{T}_p$ ein Tangentialraum und $\mathcal{T}_p^{*}$ der dazugehörige Kotangentialraum eines Punktes $p$ 
auf einer Mannigfaltigkeit, dann ist ein $(a|b)$-Tensor $T_{(a|b)}$ in Bezug auf diesen Tangentialraum der Mannigfaltigkeit
eine multilineare Abbildung der folgenden Form

\begin{equation}
T_{(a|b)}:\underbrace{\mathcal{T}_p^{*} \times ... \times \mathcal{T}_p^{*}}_{a-mal} \times
\underbrace{\mathcal{T}_p \times ... \times \mathcal{T}_p}_{b-mal} \rightarrow \mathbb{R}.
\end{equation}
Eine Abbildung, welche jedem Punkt der Mannigfaltigkeit einen $(a|b)$-Tensor zuordnet, bezeichnet man als ein
$(a|b)$-Tensorfeld. Dieses kann natürlich durch die Komponenten in Bezug auf eine bestimmte Koordinatenbasis 
in dem entsprechenden Tangentialraum dargestellt werden. Durch das Tensorprodukt $R_{(a|b)} \otimes S_{(c|d)}$
wird einem $R_{(a|b)}$- und einem $S_{(c|d)}$-Tensor ein $T_{(a+c|b+d)}$-Tensor in der Weise 
zugeordnet, sodass mit $T_M[n]=\underbrace{\mathcal{T}_p \times ... \times \mathcal{T}_p}_{n-mal}$ gilt:
$(R\otimes S)(\mathcal{T}_M^{*}[a+c] \times \mathcal{T}_M[b+d])=
R(\mathcal{T}_M^{*}[a] \times \mathcal{T}_M[b])S(\mathcal{T}_M^{*}[c] \times \mathcal{T}_M[d])$.

\subsection{Kovariante Ableitung und Krümmung}

Auf einer Mannigfaltigkeit ist die Verschiebung von Vektoren im Gegensatz zu einem affinen Raum im Allgemeinen nicht 
trivial. Es muss vielmehr durch einen Zusammenhang erst definiert werden, wann zwei Vektoren in sich auf verschiedene 
Punkte einer Mannigfaltigkeit beziehenden Tangentialräumen parallel zueinander stehen. Wenn man die Änderung eines Vektors 
entlang einer Kurve auf einer Mannigfaltigkeit bestimmen will, so benötigt man eine kovariante Ableitung, welche 
den Zusammenhang enthält und damit bei Anwendung auf einen Vektor dessen Änderung bestimmt.\\ 
\textbf{Definition:} Eine kovariante Ableitung $\nabla$ ist eine Abbildung, welche jedem $(a|b)$-Tensor einen $(a|b+1)$-Tensor 
zuordnet, wobei $\nabla$ additiv ist und der Produktregel genügt

\begin{equation}
\nabla(x+y)=\nabla x+\nabla y\quad,\quad \nabla(xy)=(\nabla x)y+x(\nabla y).
\end{equation}
Als Richtungsableitung $\nabla_v$ bezeichnet man die kovariante Ableitung $\nabla$ in Richtung eines Tangentialvektors $v$.
Dies bedeutet, dass die Richtungsableitung eines Vektorfeldes $x$ in Bezug auf einen Tangentialvektor $v$ folgende Gestalt hat

\begin{equation}
\nabla_v x=v^\nu \nabla_\nu x^\mu. 
\end{equation}
Bezeichne nun $e_\mu$ mit $\mu=1...d$ eine Basis in einem Tangentialraum $\mathcal{T}_p$, so sind die
Zusammenhangkoeffizienten $\Gamma_{\mu\nu}^\rho$ über die Ableitungen der verschiedenen Basisvektoren in Richtung der
verschiedenen Basisvektoren definiert

\begin{equation}
\nabla_{e_\mu}e_\nu=\Gamma_{\mu\nu}^\rho e_\rho. 
\end{equation}
Mit diesen Zusammenhangkoeffizienten lässt sich die Wirkung der kovarianten Ableitung auf ein Vektorfeld $x$ ausdrücken

\begin{equation}
\nabla_\mu x^\nu=\partial_\mu x^\nu+\Gamma_{\mu\rho}^\nu x^\rho.
\end{equation}
Die Auszeichnung der kovarianten Ableitung ermöglicht nun die Definition der Parallelverschiebung. Ein Vektor wird demgemäß 
genau dann parallel von einem Punkt zu einem anderen verschoben, wenn die Anwendung der kovarianten Ableitung auf diesen  
Vektor auf der Kurve, entlang derer der Vektor verschoben wird, null ergibt. 
Man bezeichnet eine Kurve $\kappa$ als geodätisch, wenn die kovariante Ableitung des Tangentialvektors entlang 
der Kurve verschwindet

\begin{equation}
\frac{d \kappa^\nu}{d\tau}\nabla_\nu \frac{d \kappa^\mu}{d\tau}=0,
\label{Geodaetengleichung1}
\end{equation}
die Kurve also in Bezug auf den definierten Zusammenhang ihre Richtung nicht ändert.

Einen Raum, der durch eine Mannigfaltigkeit, auf der ein beliebiger nicht-trivialer Zusammenhang ausgezeichnet ist, beschrieben
wird, bezeichnet man als gekrümmt. In einem solchen Raum wird die Richtung eines parallelverschobenen Vektors im Allgemeinen
vom Weg abhängig sein, auf dem er verschoben wurde. Diese Abhängigkeit der Richtung eines Vektors vom Weg, auf dem er von einem 
zum anderen Punkt verschoben wurde, ist ein Maß für die Krümmung des Raumes. Im infinitesimalen Grenzfall führt dies auf den
Kommutator der Komponenten der kovarianten Ableitung als Maß für die Krümmung in einem Punkt einer Mannigfaltigkeit. Demgemäß
ist der Riemannsche Krümmungstensor $R_{\mu\nu\rho}^{\ \ \ \ \sigma}$ über den Kommutator der kovarianten Ableitungen angewandt
auf ein Vektorfeld $x$ definiert

\begin{equation}
R_{\mu\nu\rho}^{\ \ \ \ \sigma}x^\rho=(\nabla_\mu \nabla_\nu-\nabla_\nu \nabla_\mu)x^\sigma.
\end{equation}
Er erfüllt die folgenden Identitäten: Antisymmetrie bezüglich der vorderen beiden Indizes

\begin{equation}
R_{\mu\nu\rho}^{\ \ \ \ \sigma}=-R_{\nu\mu\rho}^{\ \ \ \ \sigma},
\label{Antisymmetrie1}
\end{equation}
die zyklische Identität

\begin{equation}
R_{[\mu\nu\rho]}^{\ \ \ \ \sigma}=R_{\mu\nu\rho}^{\ \ \ \ \sigma}+R_{\nu\rho\mu}^{\ \ \ \ \sigma}
+R_{\rho\mu\nu}^{\ \ \ \ \sigma}=0
\label{Zyklische-Identitaet}
\end{equation}
und die Bianchi-Identität

\begin{equation}
\nabla_{[\lambda} R_{\mu\nu]\rho}^{\ \ \ \ \sigma}=0.
\label{Bianchi-Identitaet}
\end{equation}
Wenn man den Riemann-Tensor explizit in Abhängigkeit von den Zusammenhangkoeffizienten ausdrückt, so hat er folgende Gestalt

\begin{equation}
R_{\mu\nu\rho}^{\ \ \ \ \sigma}=\partial_\mu \Gamma_{\nu\rho}^{\sigma}-\partial_\nu
\Gamma_{\mu\rho}^{\sigma}+\Gamma_{\mu\lambda}^{\sigma}\Gamma_{\nu\rho}^{\lambda}
-\Gamma_{\nu\lambda}^{\sigma}\Gamma_{\mu\rho}^{\lambda}.
\label{Riemann-Tensor}
\end{equation}
Durch Kontraktion des zweiten mit dem vierten Index kann aus dem Riemann-Tensor der Ricci-Tensor gebildet werden

\begin{equation}
R_{\mu\nu}=R_{\mu\rho\nu}^{\ \ \ \ \rho},
\label{Ricci-Tensor}
\end{equation}
und aus diesem entsteht durch erneute Kontraktion der Ricci-Skalar

\begin{equation}
R=g^{\mu\nu}R_{\mu\nu}.
\label{Ricci-Skalar}
\end{equation}
Eine weitere wichtige Größe ist die der Torsion $T_{\mu\nu}^{\ \ \rho}$. Sie entspricht der Nichtkommutativität der kovarianten
Ableitungen in Bezug auf eine skalare Funktion $f$

\begin{equation}
T_{\mu\nu}^{\ \ \rho}\nabla_\rho f=\left(\nabla_\mu \nabla_\nu-\nabla_\nu \nabla_\mu\right)f.
\label{Torsion}
\end{equation}
In der ursprünglichen Formulierung der Allgemeinen Relativitätstheorie wird von einer verschwindenden Torsion ausgegangen,
was dazu führt, dass die Zusammenhangkoeffizienten bezüglich der unteren beiden Indizes symmetrisch sind

\begin{equation}
\left(\nabla_\mu \nabla_\nu-\nabla_\nu \nabla_\mu\right)f=0 \quad\Rightarrow\quad \Gamma_{\mu\nu}^\rho=\Gamma_{\nu\mu}^\rho.
\end{equation}

\subsection{Riemannsche Geometrie}

Als eine Riemannsche Mannigfaltigkeit bezeichnet man eine Mannigfaltigkeit, auf der eine Metrik definiert ist, welche einen
Abstandsbegriff konstituiert.\\
\textbf{Definition:} Ein metrisches Feld $g$ auf einer Mannigfaltigkeit ist ein $(0|2)$-Tensorfeld, das in Komponentendarstellung 
$g_{\mu\nu}$ symmetrisch bezüglich der beiden Indizes ist.\\
Da es zwei Vektoren eine reelle Zahl zuordnet, definiert es an jedem Punkt der Mannigfaltigkeit ein
inneres Produkt zweier Vektoren

\begin{equation}
\langle x,y \rangle=g_{\mu\nu}x^\mu y^\nu,
\end{equation}
wodurch ein Isomorphismus zwischen einem Tangentialraum $\mathcal{T}_p$ und dem dazugehörigen Kotangentialraum 
$\mathcal{T}_p^{*}$ konstituiert wird

\begin{equation}
x_\mu=g_{\mu\nu}x^\nu \quad,\quad x^\mu=g^{\mu\nu}x_\nu,
\end{equation}
wobei $g_{\mu\rho}g^{\rho\nu}=\delta_\mu^\nu$. Denn die Bildung des inneren Produktes eines Vektors $v$ mit einem 
anderen Vektor $w$ entspricht der Anwendung einer Linearform $w^{*}$ auf den Vektor $v$.
Desweiteren wird durch dieses innere Produkt die Orthogonalität zweier Vektoren definiert. Zwei Vektoren in einem 
Tangentialraum $\mathcal{T}_p$ sind genau dann orthogonal zueinander, wenn das innere Produkt der beiden Vektoren
verschwindet, das seinerseits durch den Wert des metrischen Feldes $g_{\mu\nu}$ im Punkt $p$ definiert ist:
$\langle x,y\rangle=0$. Die Länge eines Vektors ist durch die Wurzel seines inneren Produktes mit sich selbst
gegeben

\begin{equation}
\sqrt{\langle x,x \rangle}=\sqrt{g_{\mu\nu}x^\mu x^\nu}.
\end{equation}
Auf einer Riemannschen Mannigfaltigkeit kann man nun eine spezielle kovariante Ableitung in der Weise definieren, 
dass sie die Metrik konstant hält 

\begin{equation}
\nabla_\rho g_{\mu\nu}=0. 
\end{equation}
Das bedeutet, dass der entsprechende Zusammenhang über die Metrik definiert ist. Man bezeichnet einen solchen Zusammenhang als 
Levi-Civita-Zusammenhang und die entsprechenden Zusammenhangkoeffizienten als Christoffelsymbole. Sie haben folgende Gestalt

\begin{equation}
\Gamma^{\mu}_{\nu\rho}=\frac{1}{2}g^{\mu\sigma}\left(\partial_\nu g_{\sigma\rho}+\partial_\rho g_{\sigma\nu}
-\partial_\sigma g_{\nu\rho}\right).
\label{Christoffelsymbole}
\end{equation}
In einer solchen Geometrie beschreiben die in ($\ref{Geodaetengleichung1}$) definierten Geodäten die kürzesten Wege von einem
Punkt zu einem anderen. Hierauf wird im nächsten Abschnitt im Zusammenhang mit den Prinzipien der Allgemeinen
Relativitätstheorie noch näher eingegangen werden.
Der zu den Christoffelsymbolen gehörige spezielle Riemann-Tensor erfüllt neben den allgemeinen Eigenschaften 
($\ref{Antisymmetrie1}$),($\ref{Zyklische-Identitaet}$),($\ref{Bianchi-Identitaet}$) noch die weiteren Eigenschaften

\begin{equation}
R_{\mu\nu\rho\sigma}=-R_{\mu\nu\sigma\rho}
\label{Antisymmetrie2}
\end{equation}
und

\begin{equation}
R_{\mu\nu\rho\sigma}=R_{\rho\sigma\mu\nu},
\end{equation}
wobei letztere aus ($\ref{Antisymmetrie1}$),($\ref{Zyklische-Identitaet}$) und ($\ref{Antisymmetrie2}$) hergeleitet 
werden kann.

Man kann eine Verallgemeinerung der Riemannschen Geometrie in der Weise vornehmen, dass man einen verallgemeinerten 
Zusammenhang wie folgt definiert

\begin{equation}
\Gamma_{\mu\nu}^\rho=g^{\rho\lambda}\Delta_{\nu\mu\lambda}^{\alpha\beta\gamma}\left(\frac{1}{2}\partial_\alpha g_{\beta\gamma}
-g_{\gamma\delta}T_{\alpha\beta}^{\ \ \delta}+\frac{1}{2}Q_{\alpha\beta\gamma}\right),
\end{equation}
wobei $T_{\mu\nu}^{\ \ \rho}$ als der in ($\ref{Torsion}$) definierte Torsionstensor den antisymmetrischen Teil des 
Zusammenhangs beschreibt

\begin{equation}
T_{\mu\nu}^{\ \ \rho}=\frac{1}{2}\left(\Gamma_{\mu\nu}^\rho-\Gamma_{\nu\mu}^\rho\right)
\end{equation}
und $Q_{\mu\nu\rho}$ die Wirkung der kovarianten Ableitung auf die Metrik beschreibt 

\begin{equation}
Q_{\mu\nu\rho}=\nabla_\mu g_{\nu\rho}.
\end{equation}
Zusätzlich wurde $\Delta_{\nu\mu\rho}^{\alpha\beta\gamma}$ wie folgt definiert  

\begin{equation}
\Delta_{\nu\mu\rho}^{\alpha\beta\gamma}=\delta^\alpha_\nu \delta^\beta_\mu \delta^\gamma_\rho
+\delta^\alpha_\mu \delta^\beta_\rho \delta^\gamma_\nu+\delta^\alpha_\rho \delta^\beta_\nu \delta^\gamma_\mu.
\end{equation}
Eine vierdimensionale Raum-Zeit mit einer solchen Struktur bezeichnet man als $L_4$. Für $Q_{\mu\nu\rho}=0$ geht sie in einen
$U_4$ über, eine Riemann-Cartansche Raum-Zeit, die ihrerseits bei verschwindender Torsion $T_{\mu\nu}^{\ \ \rho}$ wieder in
eine Riemannsche Raum-Zeit $V_4$ übergeht, wie sie zur Formulierung der Allgemeinen Relativitätstheorie gewöhnlich verwendet
wird. Verschwindet auch noch die Krümmung, so bleibt die flache Minkowski-Raum-Zeit $R_4$ der Speziellen Relativitätstheorie

\begin{equation}
L_4 \xrightarrow{Q=0} U_4 \xrightarrow{T=0} V_4 \xrightarrow{R=0} R_4.
\end{equation}
Für die folgende Formulierung der Allgemeinen Relativitätstheorie wird eine Riemannsche Raum-Zeit vorausgesetzt. Es gibt
allerdings auch eine alternative Formulierung, in der die Torsion die entscheidende Rolle spielt. Dies wird im folgenden
Kapitel wichtig, indem es um die Formulierung der Allgemeinen Relativitätstheorie als Eichtheorie der Translationen gehen wird.

\section{Physikalische Prinzipien der Allgemeinen Relativitätstheorie}

\subsection{Grundpostulate der Allgemeinen Relativitätstheorie}

Die Allgemeine Relativitätstheorie geht davon aus, dass die reale Raum-Zeit eine (3+1)-dimensionale Riemannsche 
Mannigfaltigkeit mit einer pseudo-Euklidischen Metrik $g_{\mu\nu}$ darstellt, welche die Zeitkomponente bezüglich des 
Abstandsbegriffes mit einem negativen Vorzeichen belegt, was die Bezeichnung (3+1)-dimensional andeutet. Die metrische 
Struktur entspricht dem Gravitationsfeld, unterliegt selbst einer inneren Dynamik und wechselwirkt mit der in der 
Raum-Zeit vorhandenen Materie. Letztere Aussage führt in mathematischer Präzisierung auf die beiden grundlegenden 
Postulate, auf denen die Allgemeine Relativitätstheorie im Wesentlichen basiert:\\
\\
\textbf{1.} Objekte bewegen sich auf dem kürzesten Weg durch die Raum-Zeit. Dieser Sachverhalt wird mathematisch durch die
Geodätengleichung beschrieben: $\frac{d^2 X^\mu}{dt^2}+\Gamma^\mu_{\rho\sigma}\frac{dX^\rho}{dt}\frac{X^\sigma}{dt}=0$.
Es handelt sich hierbei um eine Verallgemeinerung des Galileischen Trägheitsgesetzes.\\
\\
\textbf{2.} Die durch die Krümmungsgrößen beschriebene Geometrie der Raum-Zeit steht in einer direkten Beziehung zur durch 
den Energie-Impuls-Tensor beschriebenen Materieverteilung in ihr. Dieser Sachverhalt wird in mathematisch exakter Weise 
durch die Einsteinsche Feldgleichung zum Ausdruck gebracht: $R_{\mu\nu}-\frac{1}{2}R g_{\mu\nu}=-8\pi G T_{\mu\nu}$. 
Für statische Gravitationsfelder bedeutet dies, dass die Materieverteilung die Geometrie der Raum-Zeit und damit das 
Gravitationsfeld bestimmt.\\
\\
Im Grenzfall eines konstanten metrischen Feldes $g_{\mu\nu}=\eta_{\mu\nu}$ geht die Allgemeine Relativitätstheorie in die 
Spezielle Relativitätstheorie über. 

\subsection{Äquivalenzprinzip und Geodätengleichung}

Der grundlegende Ausgangspunkt der Allgemeinen Relativitätstheorie ist das Äquivalenzprinzip, welches besagt, dass die
träge und die schwere Masse eines Körpers, welche begrifflich zunächst durchaus voneinander zu unterscheiden sind, physikalisch
äquivalent zueinander sind. Die schwere Masse bestimmt, wie sehr ein Körper auf ein Gravitationsfeld reagiert, also wie sehr
dieser von einem anderen Körper angezogen wird. Die träge Masse hingegen bestimmt, wie groß der Widerstand ist, den ein Körper
einer Änderung seines Bewegungszustandes (Beschleunigung) entgegensetzt. Sind träge und schwere Masse einander äquivalent, so
hat dies zur Folge, dass alle Körper sich bei gleicher Anfangsgeschwindigkeit und Abwesenheit sonstiger Einflüsse in einem
Gravitationsfeld exakt gleich verhalten. Dies drückt sich beispielsweise im Galileischen Fallgesetz aus, das besagt, dass
in einem Vakuum alle Körper gleich schnell zu Boden fallen. Dass dies keineswegs selbstverständlich ist, zeigt ein Blick auf
die Elektrodynamik. Hier ist es die elektrische Ladung, welche bestimmt, wie sehr ein Körper auf ein Elektromagnetisches Feld
reagiert und diese ist natürlich vollkommen unabhängig von der Trägheit des Körpers, weshalb Körper mit unterschiedlicher
Ladung in einem elektromagnetischen Feld unterschiedlich stark beschleunigt werden.
Da sich also alle Körper in einem Gravitationsfeld bei Ausbleiben sonstiger Einflüsse genau gleich verhalten, kann ein
beliebiges Gravitationsfeld im Prinzip durch Transformation in ein entsprechend beschleunigtes Bezugssystem zum Verschwinden
gebracht werden. Das bedeutet, dass die Existenz eines Gravitationsfeldes der Transformation in ein beschleunigtes Bezugssystem
entspricht. Weil demnach für jedes Gravitationsfeld ein Bezugssystem existiert, in Bezug auf das alle Körper träge Bewegungen 
ausführen, ergibt sich aber die Möglichkeit, die Existenz von Gravitationsfeldern als Eigenschaft der Raum-Zeit selbst zu
deuten. Als Konsequenz definiert das Gravitationsfeld dann sozusagen lokal, welche Bezugssysteme Inertialsysteme sind.

Geschwindigkeiten von Körpern werden mathematisch als Verschiebungsvektoren dargestellt. Wie oben beschrieben muss in einer
gekrümmten Raum-Zeit ein Zusammenhang definiert werden, welcher die Definition der Veränderung von Verschiebungsvektoren 
entlang einer Kurve liefert und damit den Vergleich von Verschiebungsvektoren an unterschiedlichen Raum-Zeit-Punkten
ermöglicht. Beschreibt also ein Verschiebungsvektor auf einer Mannigfaltigkeit die Geschwindigkeit eines Körpers, so entspricht
die Veränderung des Verschiebungsvektors dessen Beschleunigung, die deshalb über den Zusammenhang definiert ist.
Natürlich bestimmt im Falle eines sich in der Raum-Zeit bewegenden Körpers der seine Geschwindigkeit beschreibende
Verschiebungsvektor selbst lokal den Verlauf der Trajektorie des Körpers und entlang derer wird eben dieser 
Verschiebungsvektor selbst verschoben. Damit kann durch die Wahl eines entsprechenden Zusammenhangs auf der Raum-Zeit ein
beschleunigtes Bezugssystem definiert und hierdurch, da die Existenz eines Gravitationsfeldes dem Übergang in ein entsprechend
beschleunigtes Bezugssystem entspricht, ein beliebiges Gravitationsfeld dargestellt werden.
Der Zusammenhang ist gerade derjenige, welcher die Verschiebungsvektoren von Körpern in einem Gravitationsfeld konstant hält.
Man kann es nun umgekehrt auch so formulieren, dass wenn ein bestimmter Zusammenhang ein Gravitationsfeld beschreibt, sich 
die Körper in diesem Gravitationsfeld eben gerade so verhalten, dass die kovariante Ableitung mit dem entsprechenden das
Gravitationsfeld beschreibenden Zusammenhang die die Bewegung der Körper beschreibenden Verschiebungsvektoren entlang ihrer
Trajektorien, also in ihrer eigenen Richtung, konstant hält. Dies führt auf folgende mathematische Bedingung

\begin{eqnarray}
&&\nabla_\tau \frac{d X^\mu}{d \tau}=0\quad \Leftrightarrow \quad 
\frac{d X^\nu}{d \tau}\nabla_\nu \frac{d X^\mu}{d \tau}=0\quad \Leftrightarrow \quad
\frac{d X^\nu}{d \tau}\left(\frac{d}{d X^\nu}\frac{d X^\mu}{d \tau}
+\Gamma^{\mu}_{\nu\rho}\frac{d X^\rho}{d \tau}\right)=0\nonumber\\
&&\quad \Leftrightarrow \quad \frac{d^2 X^\mu}{d \tau^2}+\Gamma^\mu_{\nu\rho}\frac{d X^\nu}{d \tau}\frac{d X^\rho}{d \tau}=0,
\label{Geodaetengleichung2}
\end{eqnarray}
welche der im letzten Abschnitt beschrieben Bedingung einer Geodäte entspricht ($\ref{Geodaetengleichung1}$), sofern der das 
Gravitationsfeld beschreibende Zusammenhang mit der Geometrie der Raum-Zeit identifiziert wird. Dies geschieht im Sinne
der Riemannschen Geometrie durch die Auszeichnung eines metrischen Feldes, welches Abstandsverhältnisse in der Raum-Zeit 
definiert, und als die eigentlich fundamentale Beschreibung des Gravitationsfeldes postuliert wird. Der durch das metrische 
Feld über die Bedingung des Verschwindens der kovarianten Ableitung bei Anwendung auf dieses induzierte und durch die 
Christoffelsymbole beschriebene Levi-Civita-Zusammenhang ($\ref{Christoffelsymbole}$) definiert dann über die Geodätengleichung 
($\ref{Geodaetengleichung2}$) das Verhältnis der lokalen Trägheitsräume an unterschiedlichen Punkten zueinander und damit das 
Verhalten von Körpern auf der Raum-Zeit. Die Geodätengleichung für den Levi-Civita-Zusammenhang kann aus der folgenden Wirkung 
hergeleitet werden

\begin{equation}
S=\int d \tau \sqrt{g_{\mu\nu}\frac{dx^\mu}{d \tau} \frac{dx^\nu}{d \tau}}.
\label{Wirkung_Geodaetengleichung}
\end{equation}
Da ein metrisches Feld einen Abstandsbegriff konstituiert, beschreibt ($\ref{Wirkung_Geodaetengleichung}$) offensichtlich die 
Länge einer Kurve zwischen den beiden Punkten, zwischen denen die Wirkung bezüglich der Metrik variiert wird. Dies bedeutet, 
dass Kurven, welcher der Geodätengleichung genügen, in der Tat die kürzesten beziehungsweise längsten Wege (dies hängt von der
Vorzeichenkonvention ab) auf einer Riemannschen Raum-Zeit beschreiben, denn die Wirkung wird gemäß dem Hamiltonschen 
Variationsprinzip für die physikalisch realisierte Kurve extremal und es ergibt sich als Bedingung die 
Geodätengleichung ($\ref{Geodaetengleichung2}$).

\subsection{Einsteinsche Feldgleichung}

Die Geodätengleichung beschreibt das Verhalten von Körpern auf einer Raum-Zeit mit vorgegebener Geometrie. Nun stellt sich 
aber die Frage, wodurch die Geometrie (metrische Struktur) der Raum-Zeit selbst bestimmt wird, die ja in der Allgemeinen 
Relativitätstheorie dem Gravitationsfeld als dynamischer Entität entspricht. Da Gravitationsfelder durch die Präsenz von Masse
hervorgerufen werden, welche nach der Speziellen Relativitätstheorie eine bestimmte Form der Energie ist, kann es sich bei der 
physikalischen Größe, welche die Geometrie der Raum-Zeit bestimmt, nur um die Energieverteilung handeln. Die Verteilung der
Energie in der Raum-Zeit wird durch einen Tensor zweiter Ordnung beschrieben, den Energie-Impuls-Tensor $T_{\mu\nu}$. Die 
Krümmungsgrößen der Riemannschen Geometrie müssen also in eine direkte Beziehung zum Energie-Impuls-Tensor gesetzt werden. 
Da der Energie-Impuls-Tensor divergenzfrei ist

\begin{equation}
\nabla^\mu T_{\mu\nu}=0,
\end{equation}
worin sich der Energie- und der Impulserhaltungssatz mathematisch ausdrücken, ist es naheliegend, einen divergenzfreien Tensor
zweiter Ordnung aus den Krümmungsgrößen zu konstruieren und diesen in eine proportionale Beziehung zu $T_{\mu\nu}$ zu setzen.
Der einzige divergenzfreie Tensor zweiter Ordnung $G_{\mu\nu}$, welcher aus den Krümmungsgrößen der Riemannschen Geometrie
gebildet werden kann, hat folgende Gestalt

\begin{equation}
G_{\mu\nu}=R_{\mu\nu}-\frac{1}{2}R g_{\mu\nu}
\end{equation}
und wird als Einstein-Tensor bezeichnet.
\footnote{Natürlich handelt es sich bei dem hier verwendeten Ricci-Tensor beziehungsweise Ricci-Skalar um diejenigen Größen, 
welche aus dem auf den durch die Christoffelsymbole beschriebenen Levi-Civita-Zusammenhang bezogenen Riemann-Tensor gebildet 
werden und damit vom metrischen Feld abhängen.}
Seine Divergenzfreiheit kann wie folgt gezeigt werden: Man geht zunächst von der
Bianchi-Identität ($\ref{Bianchi-Identitaet}$) aus

\begin{equation}
\nabla_\lambda R_{\mu\nu\rho}^{\ \ \ \ \sigma}
+\nabla_\mu R_{\nu\lambda\rho}^{\ \ \ \ \sigma}
+\nabla_\nu R_{\lambda\mu\rho}^{\ \ \ \ \sigma}=0.
\end{equation}
Kontraktion über die Indizes $\nu$ und $\sigma$ liefert

\begin{equation}
\nabla_\lambda R_{\mu\rho}
-\nabla_\mu R_{\lambda\rho}
+\nabla_\nu R_{\lambda\mu\rho}^{\ \ \ \ \nu}=0,
\end{equation} 
wobei die Antisymmetrie des Riemann-Tensors bezüglich der ersten beiden Indizes ($\ref{Antisymmetrie1}$) ausgenutzt wurde.
Eine Kontraktion über $\lambda$ und $\rho$ ergibt weiter

\begin{equation}
\nabla^{\nu} R_{\mu\nu}-\nabla_{\mu} R+\nabla^{\nu} R_{\mu\nu}=0,
\end{equation}
wobei nun neben der Antisymmetrie des Riemann-Tensors bezüglich der ersten beiden Indizes ($\ref{Antisymmetrie1}$) auch die 
Antisymmetrie bezüglich der hinteren beiden Indizes ($\ref{Antisymmetrie2}$) ausgenutzt wurde, welche nur für den speziellen 
aus den Christoffelsymbolen ($\ref{Christoffelsymbole}$) gebildeten Riemann-Tensor gültig ist. Dies führt schließlich auf

\begin{equation}
\nabla^{\nu} \left(R_{\mu\nu}-\frac{1}{2}R g_{\mu\nu}\right)=0,
\end{equation}
womit die Divergenzfreiheit des Einstein-Tensors gezeigt ist. Damit lautet die Gleichung

\begin{equation}
R_{\mu\nu}-\frac{1}{2}R g_{\mu\nu}=-\kappa T_{\mu\nu},
\label{Einsteinsche_Feldgleichung}
\end{equation}
wobei sich im negativen Vorzeichen ausdrückt, dass die Gravitation eine anziehende und keine abstoßende Wechselwirkung ist.
Der Proportionalitätsfaktor kann durch den Übergang zum klassischen Grenzfall der Newtonschen Gravitationstheorie 
zu $\kappa=8\pi G$ ermittelt werden, was auf die Einsteinsche Feldgleichung führt

\begin{equation}
R_{\mu\nu}-\frac{1}{2}R g_{\mu\nu}=-8\pi G T_{\mu\nu}.
\end{equation}
Der spurfreie Teil des Riemann-Tensors $C_{\mu\nu\rho}^{\ \ \ \ \sigma}$ wird durch die Einsteinsche Feldgleichung 
nicht bestimmt und wird als Weyl-Tensor bezeichnet.

\subsection{Einstein-Hilbert-Wirkung}

Ebenso wie die Geodätengleichung kann natürlich auch die Einsteinsche Feldgleichung über das Hamiltonsche Prinzip aus
einer Wirkung gewonnen werden. Da es David Hilbert war, dem es kurz nachdem Einstein die Feldgleichung der Allgemeinen
Relativitätstheorie formuliert hatte gelang, eine der Einsteinschen Feldgleichung entsprechende Wirkung zu finden,
wird sie als Einstein-Hilbert-Wirkung bezeichnet. Die Einstein-Hilbert-Wirkung lautet wie folgt:

\begin{equation}
S_{EH}=\frac{1}{16 \pi G}\int d^4 x \sqrt{-g}R,
\end{equation}
wobei $g$ als Determinante der das metrische Feld $g_{\mu\nu}$ beschreibenden Matrix definiert ist: $g=\det{g_{\mu\nu}}$.
Durch Variation der Einstein-Hilbert-Wirkung nach dem metrischen Tensor $g_{\mu\nu}$ ergibt sich die Einsteinsche Feldgleichung
für das Gravitationsfeld ($\ref{Einsteinsche_Feldgleichung}$). Die Variation der Einstein-Hilbert-Wirkung und damit das
Extrahieren der Einsteinschen Feldgleichung soll nun kurz betrachtet werden. Um die volle Wirkung zu erhalten, muss im Falle
einer Feldtheorie für die Materie auch noch ein Term für die Materiefelder berücksichtigt werden, dessen spezifische Gestalt
aber erst im nächsten Kapitel thematisiert werden wird

\begin{equation}
S=S_G+S_M=S_G+\int_V d^4 x \sqrt{-g} \mathcal{L}_M. 
\label{Wirkung_Gravitation}
\end{equation}
In Bezug auf die Variation der Gravitationswirkung ($\ref{Wirkung_Gravitation}$) ist es zunächst einmal wichtig, dass die 
Variation des Ausdruckes $\sqrt{-g}$ die folgende Gestalt hat

\begin{equation}  
\delta \sqrt{-g}=-\frac{1}{2}\sqrt{-g} g_{\mu\nu} \delta g^{\mu\nu}.
\label{Variation_Determinante_Metrik}
\end{equation}
Mit ($\ref{Variation_Determinante_Metrik}$) ergibt sich für die Variation der Einstein-Hilbert-Wirkung

\begin{eqnarray}
16 \pi G \delta S_{EH}&=&\int_V d^4 x \delta \left(\sqrt{-g}g^{\mu\nu}R_{\mu\nu}\right) \nonumber\\
&=&\int_V d^4 x \left(\delta \sqrt{-g}g^{\mu\nu}R_{\mu\nu}+\sqrt{-g}\delta g^{\mu\nu}R_{\mu\nu}+\sqrt{-g} g^{\mu\nu}\delta
R_{\mu\nu}\right) \nonumber\\
&=&\int_V d^4 x \sqrt{-g} \left(R_{\mu\nu}-\frac{1}{2}R g_{\mu\nu} \right)\delta g^{\mu\nu}+\underbrace{\int_V d^4 x \sqrt{-g}
g^{\mu\nu} \delta R_{\mu\nu}}_{=0}.
\label{Variation_Einstein_Hilbert}
\end{eqnarray}
Der erste Term enthält bereits den Einstein-Tensor, also die linke Seite der Einsteinschen Feldgleichung. Es lässt sich nun mit
Hilfe des Stokesschen Satzes $\int_V A=\int_{\partial V} dA$ 
\footnote{$V$ bezeichnet hier eine Region einer Mannigfaltigkeit, $\partial V$ ihre Oberfläche, $A$ eine Differentialform
auf dieser Mannigfaltigkeit und $dA$ ihre äußere Ableitung.}
zeigen, dass der zweite Term gleich der Variation eines Oberflächentermes ist 

\begin{eqnarray} 
\int_V d^4 x \sqrt{-g} g^{\mu\nu} \delta R_{\mu\nu}&=&\int_V d^4 x \sqrt{-g} g^{\mu\nu}\left(\nabla_\mu \delta
\Gamma_{\lambda\nu}^{\lambda}-\nabla_\lambda \delta \Gamma_{\mu\nu}^{\lambda}\right)\nonumber\\
&=&\int_V d^4 x \nabla_\kappa \left[\sqrt{-g}
\left(g^{\kappa\nu}\delta\Gamma_{\lambda\nu}^{\lambda}-g^{\mu\nu}\delta\Gamma_{\mu\nu}^{\kappa}\right)\right]\nonumber\\
&=&\int_V d^4 x \partial_\kappa \left[\sqrt{-g}
\left(g^{\kappa\nu}\delta\Gamma_{\lambda\nu}^{\lambda}-g^{\mu\nu}\delta\Gamma_{\mu\nu}^{\kappa}\right)\right]\nonumber\\
&=&\oint_{\partial V} d^3 x \left[\sqrt{-g}\left(g^{\kappa\nu}\delta\Gamma_{\lambda\nu}^{\lambda}-g^{\mu\nu}
\delta\Gamma_{\mu\nu}^{\kappa}\right)\right]=0.
\end{eqnarray}
Im Unendlichen soll die Variation des Gravitationsfeldes allerdings verschwinden. Daher verschwindet auch der Oberflächenterm.
Es bleibt also für die Variation der Wirkung für die Gravitation

\begin{equation}
\delta S_G=\int_V \left(R_{\mu\nu}-\frac{1}{2}R g_{\mu\nu} \right)\delta g^{\mu\nu} \sqrt{-g} d^4 x.
\end{equation}
Der Energie-Impuls-Tensor ist definiert als  

\begin{equation}
T_{\mu\nu}=\frac{2}{\sqrt{-g}}\frac{\delta S_M}{\delta g^{\mu\nu}}. 
\label{Energie-Impuls-Tensor}
\end{equation}
Demnach ergibt sich

\begin{equation}
\delta S=\delta(S_G+S_M)=\frac{1}{16 \pi G}\int_V \left(R_{\mu\nu}-\frac{1}{2}R g_{\mu\nu}\right)\delta g^{\mu\nu} 
\sqrt{-g} d^4 x+\frac{1}{2}\int_V T_{\mu\nu}\delta g^{\mu\nu} \sqrt{-g} d^4 x,
\end{equation}
und damit die Einsteinsche Feldgleichung: $R_{\mu\nu}-\frac{1}{2}R g_{\mu\nu}=-8 \pi G T_{\mu\nu}$ 
($\ref{Einsteinsche_Feldgleichung}$). Diese kann mit $T=g^{\mu\nu}T_{\mu\nu}$ auch wie folgt 
geschrieben werden

\begin{equation}
R_{\mu\nu}=-8 \pi G\left(T_{\mu\nu}-\frac{1}{4}Tg_{\mu\nu} \right),
\label{Einsteinsche_Feldgleichung_alternative_Darstellung}
\end{equation}
sodass sich für $T_{\mu\nu}=0$ ergibt, dass der Ricci-Tensor verschwindet:\ $R_{\mu\nu}=0$.

\addtocontents{toc}{\protect\newpage}

\chapter{Beschreibung der Allgemeinen Relativitätstheorie als Eichtheorie}

\section{Feldtheorien auf einer gekrümmten Raum-Zeit und Vierbeinformalismus}

\subsection{Skalarfelder und Vektorfelder}

Die Veränderung der Beschreibung der Raum-Zeit-Geometrie durch die Allgemeine Relativitätstheorie, in der die Metrik selbst
einer inneren Dynamik unterworfen wird, hat natürlich nicht nur einen Einfluss auf die Bewegung von starren Körpern in der 
Raum-Zeit, die gemäß der Allgemeinen Relativitätstheorie der Geodätengleichung genügt. Sie wirkt sich natürlich in der 
gleichen Weise auf die Beschreibung von Materiefeldern
\footnote{Mit Materiefeldern sind in diesem Zusammenhang alle Felder außer dem metrischen Feld gemeint, welche dynamische 
Entitäten beschreiben.} 
auf der Raum-Zeit aus. Das bedeutet, dass Feldtheorien im Sinne der verallgemeinerten Raum-Zeit-Beschreibung der
Allgemeinen Relativitätstheorie umformuliert werden müssen. Bei Skalar- und Vektorfeldern geschieht dies im Wesentlichen
durch die direkte Ersetzung der sich auf eine flache Raum-Zeit beziehenden Minkowski-Metrik der Speziellen Relativitätstheorie
durch die dynamische Metrik der Allgemeinen Relativitätstheorie in den Ausdrücken für die entsprechenden die Dynamik der Felder
beschreibenden Wirkungen. Da die Metrik aber als eine einen Abstandsbegriff konstituierende Größe fungiert, muss auch das sich
hieraus ergebende infinitesimale Volumenelement, das durch die Integration in den Wirkungen für die Felder auftritt, entsprechend
angepasst werden. Dies führt zu einem zusätzlichen Faktor $\sqrt{-g}$ in der Lagrange-Dichte für die Felder. Damit lautet die 
Wirkung eines skalaren Materiefeldes auf einer gekrümmten Raum-Zeit wie folgt

\begin{equation}
S_M=\int d^4 x \sqrt{-g}\left(g^{\mu\nu}\partial_\mu \phi \partial_\nu \phi-m^2 \phi^2 \right),
\end{equation}
was gemäß der Definition des Energie-Impuls-Tensors $(\ref{Energie-Impuls-Tensor})$ auf folgenden Ausdruck für den
entsprechenden Energie-Impuls-Tensor führt

\begin{equation}
T_{\mu\nu}=2 \partial_\mu \phi \partial_\nu \phi-g_{\mu\nu}\left(g^{\rho\sigma}\partial_\rho \phi \partial_\sigma \phi-m^2 \phi^2\right).
\end{equation}
Analog können Wirkungen für Vektorfelder auf einer gekrümmten Raum-Zeit formuliert werden. Um Spinorfelder auf einer gekrümmten
Raum-Zeit zu formulieren, muss allerdings eine alternative mathematische Darstellung des Gravitationsfeldes verwendet werden. Es
handelt sich um die Darstellung durch ein Vierbein.

\subsection{Vierbeinformalismus und Spinorfelder}

Im Vierbeinformalismus wird das Gravitationsfeld nicht durch ein metrisches Feld $g_{\mu\nu}$, sondern durch ein sogenanntes
Vierbeinfeld $e^m_\mu$ beschrieben, das auch als Vierbeinfeld bezeichnet wird. Dieses ist mathematisch ein Kovektorfeld auf der 
die Raum-Zeit beschreibenden Mannigfaltigkeit, wobei jede Komponente selbst einem Vektor in der Minkowski-Raum-Zeit entspricht. 
Es beschreibt die Transformation von beliebigen Koordinaten in dasjenige Koordinatensystem, in welchem die Metrik die 
Gestalt der Minkowski-Metrik besitzt und ist damit über folgende Relation definiert und mit dem metrischen Feld verknüpft

\begin{equation}
g_{\mu\nu}=e^m_\mu e^n_\nu \eta_{mn}.
\end{equation}
Das bedeutet aber auch, dass es aus einem beliebigen Bezugssystem in dasjenige Bezugssystem transformiert, in Bezug auf das
ein Körper eine träge Bewegung ausführt. Aus diesem Grunde stellt es eine noch direktere mathematische Beschreibung des 
Grundgedankens der Allgemeinen Relativitätstheorie dar, der ja gerade in der Äquivalenz zwischen der Wirkung eines 
Gravitationsfeldes und der Transformation in ein diesem Gravitationsfeld entsprechendes Bezugssystem besteht.

Wirkungen für Spinorfelder in einer gekrümmten Raum-Zeit lassen sich nur mit Hilfe des Vierbeinfeldes formulieren. Die
Wirkung eines Spinorfeldes in einer gekrümmten Raum-Zeit lautet wie folgt

\begin{eqnarray}
S_M=\int d^4 x\ e \bar \psi \left[i\gamma^m e^\mu_m\left(\partial_\mu+i\omega_\mu^{ab}\Sigma_{ab}\right)-m\right]\psi
=\int d^4 x\ e \bar \psi \left[i\gamma^m e^\mu_m D_\mu-m\right]\psi, 
\label{Dirac_Materiefeld_Gravitation}
\end{eqnarray}
wobei $e$ als $e=\det{e_\mu^m}=\sqrt{-g}$ definiert ist und die $\Sigma_{ab}=-\frac{i}{4}[\gamma_a,\gamma_b]$ die Generatoren
der Lorentz-Gruppe dargestellt im Dirac-Spinor-Raum sind und damit die folgenden Vertauschungsrelationen erfüllen:
$\left[\Sigma_{ab},\Sigma_{cd}\right]=\eta_{bc}\Sigma_{ad}-\eta_{ac}\Sigma_{bd}+\eta_{bd}\Sigma_{ca}-\eta_{ad}\Sigma_{cb}$. 
$\omega_\mu^{ab}$ beschreibt den Zusammenhang im Dirac-Spinor-Raum und definiert damit die kovariante Ableitung im
Dirac-Spinor-Raum: $D_\mu=\partial_\mu+i\omega_\mu^{ab}\Sigma_{ab}$. Ausgedrückt durch das Vierbein lautet der
Energie-Impuls-Tensor: $T^{\mu}_m=\frac{1}{e}\frac{\delta S_M}{\delta e_\mu^m}$. Der Riemann-Tensor kann nun auch mit Hilfe der
kovarianten Ableitung im Spinorraum beziehungsweise dem entsprechenden Spinzusammenhang beschrieben werden und lautet dann

\begin{equation}
R_{\mu\nu}^{ab}=[D_\mu,D_\nu]=\partial_\mu \omega_\nu^{ab}-\partial_\nu \omega_\mu^{ab}
+\omega_\mu^{ac}\omega_\nu^{cb}-\omega_\nu^{ac}\omega_\mu^{cb},
\label{Riemann_Tensor_Spinzusammenhang}
\end{equation}
wobei er nun zwei Minkowski-Indizes aufweist. Die Einstein-Hilbert-Wirkung wird demgemäß wie folgt ausgedrückt

\begin{equation}
S_{EH}=\int d^4 x\ e e^\mu_a e^\nu_b R_{\mu\nu}^{ab}=\int d^4 x\ e R,
\label{Einstein-Hilbert_Vierbein_Spinzusammenhang}
\end{equation}
was der folgenden Formulierung der Einsteinschen Feldgleichung entspricht

\begin{equation}
R_\mu^a-\frac{1}{2}R e_\mu^a=-8\pi G T_\mu^a.
\end{equation}

\section{Die Allgemeine Relativitätstheorie als Eichtheorie der Translationen}

\subsection{Eichtheorien bezüglich Raum-Zeit-Symmetrien}

Die Idee der Eichtheorien geht ursprünglich auf Hermann Weyl zurück \cite{Weyl:1919fi},\cite{Weyl:1929fm}.
Wie alle anderen bekannten Wechselwirkungen in der Natur lässt sich auch die Allgemeine Relativitätstheorie
als Eichtheorie beschreiben \cite{Hehl:1974cn},\cite{Hehl:1976kj},\cite{Cho:1975dh},\cite{Lyre:2004}.
Im Gegensatz zu den Wechselwirkungen des Standardmodells der Elementarteilchenphysik, wird hier jedoch keine innere Symmetrie
zugrundegelegt, welche sich auf eine Quantenzahl bezieht, sondern eine Symmetrie in Bezug auf die Raum-Zeit selbst. Es wird im
übernächsten Unterabschnitt näher begründet werden, warum es sich gerade um die Eichtheorie der Translationen handelt.
Natürlich ist eine eichtheoretische Beschreibung noch keine quantentheoretische Behandlung. Man muss ja auch im Standardmodell
eine "`klassische"' Eichtheorie erst noch einer Quantisierung unterwerfen, um die entsprechende Quantentheorie zu erhalten.
Dies darf aber dennoch nicht darüber hinwegtäuschen, dass auch eine "`klassische"' Eichtheorie im Rahmen des Standardmodells,
in der die Felder noch keiner "`zweiten"' Quantisierung unterworfen wurden, bereits aus einer quantentheoretischen Beschreibung
hervorgeht. Denn der Spinor, welcher das freie Materiefeld beschreibt, und die Dirac-Gleichung, welcher er genügt, stellen ja
bereits die Quantentheorie eines relativistischen Teilchens dar und der Spin ist von seiner Natur her etwas rein 
Quantentheoretisches. Das gleiche gilt für die Eichsymmetrie, welche die Struktur der Wechselwirkung festlegt und welche sich
auf eine Quantenzahl bezieht. Dies ist auch bereits bei der $U(1)$-Eichtheorie der Fall, welche die Elektrodynamik beschreibt,
denn die Phase einer Wellenfunktion ist ebenfalls ein rein quantentheoretisches Element. Eine eichtheoretische Beschreibung ist
also, zumindest in dem Rahmen, in der sie im Standardmodell realisiert ist, an sich essentiell quantentheoretisch.
Damit stellt aber die Tatsache, dass auch die Allgemeine Relativitätstheorie eichtheoretisch beschrieben werden kann, bereits
ein entscheidendes Bindeglied zu den anderen Wechselwirkungen und damit zur Quantentheorie an sich dar. Bei der
eichtheoretischen Formulierung der Allgemeinen Relativitätstheorie geht man wie bei allen anderen Eichtheorien von einer 
freien Materiefeldgleichung aus, etwa der Dirac-Gleichung mit der folgenden Wirkung

\begin{equation}
S_M=\int d^4 x\ \bar \Psi \left(i\gamma^\mu \partial_\mu-m\right)\Psi.
\label{Dirac_Materiefeld}
\end{equation}
Eine solche Gleichung ist natürlich invariant unter der Poincar\'{e}-Gruppe,
\footnote{Die Dirac-Gleichung und die Poincar\'{e}-Gruppe werden in Kapitel 5 noch einmal thematisiert werden.}
also der Lorentz-Gruppe erweitert um die Raum-Zeit-Translationen, die in infinitesimaler Form folgende Gestalt aufweist

\begin{eqnarray}
x^m &\rightarrow& x^m+\Lambda^m_{\ n} x^n+\xi^m,\nonumber\\
\Psi &\rightarrow& \left(1+\Lambda^{mn}\Sigma_{mn}+\epsilon^m \partial_m \right)\Psi,
\end{eqnarray}
wobei $\epsilon^{m}=\xi^m+\Lambda^m_{\ l}\delta^{l}_n x^{n}$ gilt.
Diesen Symmetrien der freien Materiefeldgleichung entspricht gemäß dem Noetherschen Theorem ein Erhaltungssatz. Bezüglich der
Symmetrie unter Raum-Zeit-Translationen sind es Energie und Impuls, welche erhalten bleiben, was sich in der Divergenzfreiheit
des Energie-Impuls-Tensors ausdrückt. Diese kann aus der Translationssymmetrie wie folgt hergeleitet werden. Die Wirkung 
beziehungsweise Lagrange-Dichte eines Materiefeldes wird im Allgemeinen vom Materiefeld selbst und dessen Ableitungen bezüglich
der verschiedenen Richtungen in der Raum-Zeit abhängen. Daher kann seine Änderung bezüglich einer infinitesimalen 
Raum-Zeit-Translationen in zweierlei Weise ausgedrückt werden

\begin{equation}
\delta \mathcal{L}_M=\frac{\partial{\mathcal{L}_M}}{\partial{x^\mu}}\delta x^\mu\quad,\quad 
\delta \mathcal{L}_M=\frac{\partial{\mathcal{L}_M}}{\partial \psi}\frac{\partial \psi}{\partial x^\mu}\delta x^\mu
+\frac{\partial{\mathcal{L}_M}}{\partial\left[\partial_\mu \psi\right]}
\partial_\mu \frac{\partial \psi}{\partial x_\nu} \delta x^\nu,
\label{Variation_Lagrangedichte_Translationen}
\end{equation}
wobei hier die Tatsache ausgenutzt wurde, dass $\delta \partial_\mu \psi=\partial_\mu \delta \psi$. Gleichsetzen der beiden
Ausdrücke in ($\ref{Variation_Lagrangedichte_Translationen}$), partielle Integration und Ausnutzen der
Euler-Lagrange-Gleichung führen auf

\begin{equation}
\partial_\mu \left[\mathcal{L}_M\delta^\mu_\nu-\frac{\partial \mathcal{L}_M}
{\partial \left[\partial_\mu \psi\right]}\partial_\nu \psi\right]
=\partial_\mu T^{\mu}_{\nu}=0. 
\label{Divergenzfreiheit_Energie-Impuls-Tensor}
\end{equation}
Es kann leicht gezeigt werden, dass der Ausdruck in der Klammer der Definition des Energie-Impuls-Tensors aus 
($\ref{Energie-Impuls-Tensor}$) entspricht, wenn $g_{\mu\nu}=\eta_{\mu\nu}$. Eine analoge Rechnung für 
Lorentz-Transformationen führt auf den folgenden Erhaltungssatz  

\begin{equation}
\partial_\mu\left[\frac{\partial\mathcal{L}_M}{\partial_\mu \psi}\Sigma_{ab}\psi-x_\nu 
\frac{1}{2}\left(\delta^\nu_a \Sigma_b^{\ \mu}-\delta^\nu_b \Sigma_a^{\ \mu}\right)\right]\equiv \partial_\mu \Xi^\mu_{ab}=0,
\label{Erhaltungsgroessen_Lorentzgruppe}
\end{equation}
wobei dies für die räumlichen Rotationen die Erhaltung des Drehimpulses bedeutet. Die Lorentz-Invarianz führt also auf die
Divergenzfreiheit eines Tensors dritter Stufe.

Man kann diese Betrachtung nun auf lokale Eichtransformationen, also Transformationen, die von den
Raum-Zeit-Koordinaten abhängen erweitern

\begin{eqnarray}
x^m &\rightarrow& x^m+\Lambda^m_n(x) x^n+\xi^m(x),\nonumber\\
\psi &\rightarrow& \left(1+\Lambda^{mn}(x)\Sigma_{mn}+\epsilon^m(x)\partial_m \right)\psi,
\label{lokale_Eichtransformation_Poincare}
\end{eqnarray}
wobei $\epsilon^{m}(x)=\xi^m(x)+\Lambda^m_{l}(x)\delta^{l}_n x^{n}$ gilt.
($\ref{Dirac_Materiefeld}$) ist nicht invariant unter lokalen Transformationen der Poincar\'{e}-Gruppe
($\ref{lokale_Eichtransformation_Poincare}$). Um diese Invarianz erneut herzustellen, muss ($\ref{Dirac_Materiefeld}$) durch
die Ersetzung der gewöhnlichen Ableitung mit einer kovarianten Ableitung modifiziert werden

\begin{equation}
\partial_m \rightarrow \nabla_m=e^\mu_m \left(\partial_\mu+i\omega_\mu^{ab}\Sigma_{ab}\right)=e^\mu_m D_\mu,
\label{kovariante_Ableitung_Poincare}
\end{equation}
was auf die Wirkung ($\ref{Dirac_Materiefeld_Gravitation}$) führt. ($\ref{kovariante_Ableitung_Poincare}$) kann auch als

\begin{equation}
\nabla_m=\left(\partial_m+i\omega_m^{ab}\Sigma_{ab}\right)+h^\mu_m \left(\partial_\mu+i\omega_\mu^{ab}\Sigma_{ab}\right),
\label{kovariante_Ableitung_Poincare_2}
\end{equation}
geschrieben werden, wobei $h^\mu_m$ als nicht-trivialer Teil des Vierbeinfeldes definiert ist: $e^\mu_m=\delta^\mu_m+h^\mu_m$, 
sodass $h^\mu_m$ als durch das Postulat der Invarianz unter lokalen Translationen induzierter Zusammenhang explizit in der 
kovarianten Ableitung erscheint.
Mit Hilfe von ($\ref{kovariante_Ableitung_Poincare}$) kann nun der Feldstärketensor des Gravitationsfeldes konstruiert werden

\begin{eqnarray}
F_{mn}&=&[\nabla_m,\nabla_n]=[e_m^\mu D_\mu, e_n^\nu D_\nu]\nonumber\\
&=&e_m^\mu \left(D_\mu e_n^\nu\right) D_\nu-e_n^\mu \left(D_\mu e_m^\nu \right)D_\nu
+e_m^\mu e_n^\nu D_\mu D_\nu-e_n^\mu e_m^\nu D_\mu D_\nu\nonumber\\
&=&e_m^\mu e_n^\nu T_{\mu\nu}^{\ \ \rho}D_\rho+e^\mu_m e^\nu_n R_{\mu\nu}^{ab}\Sigma_{ab},
\label{Feldstaerke_Poincare}
\end{eqnarray}
wobei $T_{\mu\nu}^{\ \ \rho}$ die bereits in ($\ref{Torsion}$) eingeführte Torsion ist, die hier allerdings in Abhängigkeit
des Vierbeins ausgedrückt wird, und $R_{\mu\nu}^{ab}$ der Riemann-Tensor in der bereits in
($\ref{Riemann_Tensor_Spinzusammenhang}$) eingeführten Formulierung über den Spinzusammenhang ist. Aus den Größen in
($\ref{Feldstaerke_Poincare}$) kann nun eine Wirkung für das Gravitationsfeld konstruiert werden, etwa die mit Hilfe
des Vierbeins und ($\ref{Riemann_Tensor_Spinzusammenhang}$) formulierte Einstein-Hilbert-Wirkung
($\ref{Einstein-Hilbert_Vierbein_Spinzusammenhang}$). Die Analogie zu den Yang-Mills-Eichtheorien des Standardmodells
legt eine Wirkung nahe, die quadratisch in den Feldstärken ist. Mit Hilfe der Torsion kann ein quadratische Wirkung definiert
werden, welche der Einstein-Hilbert-Wirkung physikalisch äquivalent ist. Im nächsten Unterabschnitt geht es nun speziell um die
Eichtheorie der lokalen Translationen, zu welcher die Torsion als Feldstärke gehört, und aus der sich bei Voraussetzung einer
in der Feldstärke quadratischen Wirkung in der Tat speziell die zur Einstein-Hilbert-Wirkung äquivalente Wirkung ergibt. 

\subsection{Eichtheorie der Translationen}

Wenn man nun speziell die Gruppe lokaler Translationen betrachtet \cite{Cho:1975dh},\cite{Lyre:2004}, so bedeutet dies, dass
man Invarianz der Wirkung ($\ref{Dirac_Materiefeld}$) unter der folgenden Transformation fordert

\begin{eqnarray}
x^m &\rightarrow& x^m+\xi^m(x),\nonumber\\
\Psi &\rightarrow& [1+i\epsilon^m(x)\partial_m]\Psi,
\end{eqnarray}
wobei nun im Gegensatz zu ($\ref{lokale_Eichtransformation_Poincare}$) $\epsilon^m(x)=\xi^m(x)$.
Um die ($\ref{Dirac_Materiefeld}$) entsprechende Lagrange-Dichte zu erhalten, welche Invarianz unter lokalen
Eichtransformationen aufweist, muss die Ableitung $\partial_m$ nun durch die entsprechende kovariante Ableitung $\nabla_m$
ersetzt werden, die einen Spezialfall von ($\ref{kovariante_Ableitung_Poincare}$) darstellt und ein Eichpotential $h^\mu_m$
einführen, wie es in der Formulierung ($\ref{kovariante_Ableitung_Poincare_2}$) der allgemeineren kovarianten Ableitung der
Poincar\'{e}-Gruppe auftaucht

\begin{equation}
D_m=e_m^\mu \partial_\mu=\left(\delta_m^\mu+h_m^\mu\right)\partial_\mu.    
\end{equation}
Die führt zu folgender Wirkung für das Materiefeld 

\begin{equation}
S_M=\int d^4 x\ e\bar \Psi i\gamma^m \left(\delta_m^\mu+h_m^\mu\right) \partial_\mu\Psi,
\label{Materiewirkung_Translationseichtheorie}
\end{equation}
welche invariant unter lokalen Translationen ist und einen Spezialfall von ($\ref{Dirac_Materiefeld_Gravitation}$)
darstellt. Das bedeutet, dass ($\ref{Materiewirkung_Translationseichtheorie}$) in nicht-infinitesimaler Ausdrucksweise
eine Invarianz unter der folgenden lokalen Transformation aufweist

\begin{eqnarray}
x^m &\rightarrow& x^m+\xi^m(x),\nonumber\\
\Psi &\rightarrow& T \Psi,\nonumber\\
e^m_\mu &\rightarrow& T e^m_\mu T^{-1}-T^{-1}\partial_\mu T,
\end{eqnarray}
wobei der Translationsoperator $T$ wie folgt ausgedrückt werden kann: $T=\exp\left[-ia^\mu(x)\partial_\mu \right]$.
Als Feldstärke ergibt sich der Torsionstensor

\begin{equation}
T_{mn}^{\ \ \ k}=\left(\partial_m e_n^\mu-\partial_n e_m^\mu\right)e_\mu^k.
\end{equation}
Dieser spielt also im Rahmen dieser Beschreibung die entscheidende Rolle und nicht der die Krümmung beschreibende
Riemann-Tensor. Genau genommen muss auch in die kovariante Ableitung in ($\ref{Materiewirkung_Translationseichtheorie}$)
ein Spinzusammenhang eingebunden werden \cite{deAndrade:2001vx},\cite{Maluf:2003fs}, welcher nun vom Torsionstensor abhängt,
da er proportional zum sogenannten Kontorsionstensor $K_{\mu\nu}^{\ \ \rho}$ ist: $\omega_\mu^{ab}=\frac{1}{2}K_{\mu}^{ab}
=\frac{1}{2}e_a^\rho e_b^\nu \left(T_{\rho\mu\nu}+T_{\nu\rho\mu}-T_{\mu\nu\rho}\right)$, was auf die folgende Wirkung führt

\begin{equation}
S_M=\int d^4 x\ e\bar \Psi i\gamma^m  \left(\delta_m^\mu+h_m^\mu\right)
\left(\partial_\mu+\frac{i}{2}K_\mu^{ab}\Sigma_{ab}\right)\Psi.
\end{equation}
Gemäß den Yang-Mills-Eichtheorien ist es wie im letzten Unterabschnitt bereits erwähnt naheliegend,
von einer in der Feldstärke quadratischen Wirkung für das Gravitationsfeld auszugehen. Dies führt
auf eine Wirkung der folgenden Form

\begin{equation}
S_G=\frac{1}{\varkappa^2}\int d^4 x\ e\left(aT_{mn}^{\ \ \ k} T_{\ \ \ \ k}^{mn}
+bT_{mn}^{\ \ \ k} T^{m\ n}_{\ \ k}+cT_{mn}^{\ \ \ n}T^{mk}_{\ \ \ \ k}\right).
\label{Wirkung_Torsion_Quadratisch}
\end{equation} 
Es kann nun gezeigt werden $\cite{Cho:1975dh}$, dass unter der Forderung der Invarianz dieser Wirkung unter einer
Transformation der Orthonormalbasis innerhalb des Minkowski-Raumes, in den hinein das Vierbeinfeld von der 
Raum-Zeit-Mannigfaltigkeit aus eine Abbildung darstellt, die Koeffizienten der quadratischen Terme in der Wirkung
in dem Verhältnis $a=\frac{1}{2}b=-\frac{1}{4}c$ stehen müssen. Dies führt auf folgende Gestalt der Wirkung
($\ref{Wirkung_Torsion_Quadratisch}$)

\begin{equation}
S_G=\frac{1}{\varkappa^2}\int d^4 x\ e\left(\frac{1}{4}T_{mn}^{\ \ \ k} T_{\ \ \ \ k}^{mn}
+\frac{1}{2}T_{mn}^{\ \ \ k} T^{m\ n}_{\ \ k}-T_{mn}^{\ \ \ n}T^{mk}_{\ \ \ \ k}\right).
\label{Wirkung_Torsion_Einstein-Hilbert}
\end{equation} 
Für $\varkappa^2=2 \kappa=16\pi G$ liefert diese Wirkung ($\ref{Wirkung_Torsion_Einstein-Hilbert}$) die Einsteinsche
Feldgleichung, sodass sich die Eichtheorie der Translationen in der Tat als zur Allgemeinen Relativitätstheorie
äquivalent erweist.

\subsection{Argumente für die Eichtheorie der Translationen als Beschreibung der Allgemeinen Relativitätstheorie}

Dass sich aus der Eichtheorie der lokalen Raum-Zeit-Translationen die der Einstein-Hilbert-Wirkung entsprechende
Wirkung ausgedrückt durch die Torsion ergibt, ist bereits ein starkes Indiz dafür, dass die Allgemeine Relativitätstheorie
in der Tat als die diesbezügliche Eichtheorie anzusehen ist. Aber auch unabhängig davon existieren noch diverse konzeptionelle
Gründe, welche die Gültigkeit der Beschreibung der Allgemeinen Relativitätstheorie als Eichtheorie der lokalen
Translationsgruppe rechtfertigen \cite{Lyre:2004}:\\
\\
\textbf{1.} Gemäß dem Noetherschen Theorem entspricht jeder Invarianz unter einer bestimmten Symmetriegruppe eine dazugehörige
Erhaltungsgröße, die man als Noetherstrom bezeichnet. Im Rahmen der eichtheoretischen Beschreibung innerhalb der
Yang-Mills-Eichtheorien koppeln die Eichfeldstärken an den entsprechenden Noetherstrom. Die zu der Translationsgruppe gehörige
Erhaltungsgröße ist der Energie-Impuls-Tensor, was bereits im vorletzten Unterabschnitt thematisiert wurde
($\ref{Variation_Lagrangedichte_Translationen}$), ($\ref{Divergenzfreiheit_Energie-Impuls-Tensor}$),
und in der Einsteinschen Feldgleichung koppeln die im Einsteinschen Tensor enthaltenen Krümmungsgrößen tatsächlich an den
Energie-Impuls-Tensor. Die der Lorentz-Gruppe $SO(3,1)$ entsprechenden Erhaltungsgrößen sind Tensoren dritter Stufe
($\ref{Erhaltungsgroessen_Lorentzgruppe}$).\\
\\
\textbf{2.} Die Feldstärke der Translationseichtheorie ist die Torsion und eine Wirkung, welche quadratisch in den Feldstärken
ist führt zur Dynamik der Allgemeinen Relativitätstheorie ($\ref{Wirkung_Torsion_Einstein-Hilbert}$), was in Einklang mit 
den Yang-Mills-Eichtheorien ist, deren Wirkungen für das Eichfeld auch quadratisch in den Feldstärken sind.\\
\\
\textbf{3.} Die Gruppe beliebiger Diffeomorphismen der Raum-Zeit stellt die allgemeinste Symmetriegruppe der Allgemeinen
Relativitätstheorie dar. Die Diffeomorphismengruppe entspricht der Gruppe beliebiger lokaler Translationen. Damit ist die
Diffeomorphismengruppe als Symmetriegruppe der Allgemeinen Relativitätstheorie zugleich die Eichgruppe, was der
eichtheoretischen Beschreibung im Rahmen der Yang-Mills-Eichtheorien entspricht.\\
\\
\textbf{4.} Das Eichfeld ist selbst das Vierbeinfeld, welches eine noch angemessenere Beschreibung des Gravitationsfeldes darstellt
als das metrische Feld. Dies hat weitere drei Gründe \cite{Rovelli:2004}.\\
a) Die Formulierung einer Wirkung für fermionische Felder auf einer gekrümmten Raum-Zeit kann nur mit Hilfe eines
Vierbeinfeldes gegeben werden.\\
b) Die Vierbeinformulierung wird in modernen Versuchen einer Quantisierung der Gravitation verwendet.\\
c) Das Vierbeinfeld reflektiert die wirkliche Natur des Gravitationsfeldes in einer konsequenteren Weise, da es direkt eine Art
Beziehung zwischen Koordinaten ausdrückt.\\
\\
\newpage
Deshalb sollte die Translationsgruppe als Eichgruppe der Allgemeinen Relativitätstheorie im Vergleich etwa zur
Lorentz-Gruppe $SO(3,1)$ vorgezogen werden. Die Gründe 1-3 für die Formulierung der Gravitation als Translationseichtheorie
können in \cite{Lyre:2004} gefunden werden. Desweiteren sei noch darauf hingewiesen, dass die Anwendung des Eichprinzips
konzeptionell keineswegs unproblematisch ist, weshalb in \cite{Lyre:2004},\cite{Lyre:2000mq} für die Formulierung der
Eichtheorien in der Elementarteilchenphysik ein verallgemeinertes Äquivalenzprinzip vorgeschlagen wird.

\chapter{Diffeomorphismeninvarianz und Hintergrundunabhängigkeit}

\section{Diffeomorphismeninvarianz in der Allgemeinen Relativitätstheorie}

\subsection{Diffeomorphismen}

\textbf{Definition:} Ein Diffeomorphismus $\mathcal{D}$ ist eine stetig differenzierbare bijektive Abbildung zwischen zwei 
Mannigfaltigkeiten $M$ und $N$

\begin{equation}
\mathcal{D}: M \rightarrow N,
\label{Diffeomorphismus_a}
\end{equation}
deren Umkehrabbildung auch stetig differenzierbar ist. Zwei Mannigfaltigkeiten $M$ und $N$ heißen diffeomorph, falls ein
Diffeomorphismus zwischen diesen beiden Mannigfaltigkeiten existiert.\\
Da nun auf jeder Mannigfaltigkeit $M$ jede Teilmenge $U \in M$ mit einer Karte $\varphi$ versehen ist, also einem
Homöomorphismus, der Punkte $p \in U$ in den $\mathbb{R}^d$ abbildet, induziert jeder Diffeomorphismus eine
Kartentransformation. Dieser Zusammenhang lässt sich graphisch wie folgt darstellen:

\begin{eqnarray}
p \in U \subseteq M &\xrightarrow{\quad\quad\quad\quad \mathcal{D}\quad\quad\quad\quad}& \mathcal{D}(p)=q \in V \subseteq N\nonumber\\
\downarrow&&\downarrow\nonumber\\
\varphi\ \ \downarrow&&\downarrow\ \ \chi\nonumber\\
\downarrow&&\downarrow\nonumber\\
\left\{x^i(p)\right\}\in \mathbb{R}^m&\xrightarrow{\ \ \quad\quad \mathbb{D}=\varphi^{-1}\mathcal{D}\chi\quad\quad\ \ }
&\mathbb{D}\left\{x^i(p)\right\}
=\left\{y^j(q)\right\}\in \mathbb{R}^n.
\label{Diffeomorphismus_b}
\end{eqnarray}
Die in ($\ref{Diffeomorphismus_b}$) dargestellte direkte Hintereinanderausführung der Umkehrung $\varphi^{-1}$ der Abbildung
$\varphi$, welche eine Karte auf $M$ darstellt, des Diffeomorphismus $\mathcal{D}$ zwischen den beiden Mannigfaltigkeiten und
der Abbildung $\chi$, welche eine korrespondierende Karte auf $N$ darstellt, liefert also eine Koordinatentransformation
$\mathbb{D}=\varphi^{-1}\mathcal{D}\chi$, welche den Übergang zwischen einer Karte auf $M$ und einer Karte auf $N$
repräsentiert. Nun kann man den Spezialfall eines Automorphismus \footnote{Ein Automorphismus ist ein Isomorphismus einer
mathematischen Struktur auf sich selbst, wobei ein Isomorphismus allgemein eine bijektive strukturerhaltende Abbildung ist.}
einer Mannigfaltigkeit betrachten $\mathcal{D}: M \rightarrow M$.
Da ein solcher Automorphismus einen Punkt einer Mannigfaltigkeit auf einen anderen Punkt der gleichen Mannigfaltigkeit abbildet,
handelt es sich um eine Translation der Punkte einer Mannigfaltigkeit. Dies entspricht gemäß ($\ref{Diffeomorphismus_b}$) dem
Übergang zu einem anderen Koordinatensystem auf der Mannigfaltigkeit.
Die Menge aller Diffeomorphismen einer Mannigfaltigkeit $M$ auf sich selbst wird als $\mathcal{D}\textit{iff}(M)$ bezeichnet.

Es ist nun weiter die Unterscheidung zwischen aktiven und passiven Diffeomorphismen wichtig. Ein aktiver Diffeomorphismus
entspricht einer realen Verschiebung von Punkten auf einer Mannigfaltigkeit. Ein passiver Diffeomorphismus entspricht dem
Übergang in ein neues Koordinatensystem, bei dem die Punkte nur durch neue Karten beschrieben selbst aber nicht aufeinander
abgebildet werden.
Um diese Unterscheidung zu verdeutlichen, sei eine skalare Funktion $f$ auf $M$ definiert $f: p \in M \rightarrow \mathbb{R}$.
Ein aktiver Diffeomorphismus bewirkt nun in Bezug auf $f$ den Übergang zu einer neuen Funktion: $f(p)\rightarrow
f[\mathcal{D}(p)]=f(p^{\prime})=f^{\prime}(p)$, die den Wert, den $f$ vorher am Punkt $p$ hatte, nun am Punkt $p^{\prime}$ hat.
Ein passiver Diffeomorphismus lässt jedoch $f$ gleich, da es die Punkte auf $M$ nicht verschiebt, sondern nur durch neue
Koordinaten darstellt: $f[x^i(p)] \rightarrow f[y^{i}(p)]$.
Nun kann aber in ($\ref{Diffeomorphismus_b}$) gesehen werden, dass jedem aktiven Diffeomorphismus ein passiver
Diffeomorphismus entspricht, da bei zwei gegebenen Karten $\varphi$ und $\chi$ jedem $\mathcal{D}$ ein
$\mathbb{D}=\varphi^{-1}\mathcal{D}\chi$ zugeordnet werden kann: $f[x^{i}(p)] \rightarrow
f\{x^{i}[\mathcal{D}(p)]\}=f[y^{i}(p)]$. Diese Tatsache ist für die Interpretation der
Allgemeinen Relativitätstheorie von großer Bedeutung.

\subsection{Diffeomorphismeninvarianz der Einsteinschen Feldgleichung}

Die Einsteinsche Feldgleichung ($\ref{Einsteinsche_Feldgleichung}$), welche ja bezüglich ihrer mathematischen Gestalt
Tensorfelder beziehungsweise Ableitungen solcher Tensorfelder zueinander in Beziehung setzt, weist nun die besondere
Eigenschaft der Diffeomorphismeninvarianz auf \cite{Einstein:1914bw}, also der Invarianz zwischen der Darstellung durch
beliebige Koordinatensysteme, die durch Diffeomorphismen ineinander überführt werden können, wie sie im letzten Unterabschnitt
besprochen wurden ($\ref{Diffeomorphismus_a}$),($\ref{Diffeomorphismus_b}$). Das bedeutet, dass eine Lösung der Einsteinschen
Feldgleichung für das metrische Feld $g_{\mu\nu}(x)$ bei einem durch einen Diffeomorphismus beschriebenen Übergang in ein
anderes Koordinatensystem $x^i \rightarrow y^j(x^i)$ in eine Lösung der Einsteinschen Feldgleichung
$g_{\mu\nu}^{\prime}[y(x)]$ überführt wird.

Nun kann aber gemäß der im letzten Unterabschnitt beschriebenen Entsprechung zwischen einem aktiven und einem passiven
Diffeomorphismus der Übergang zu der neuen Lösung in zweierlei Weise gedeutet werden. Entweder man sieht 
$g_{\mu\nu}^{\prime}[y(x)]$ als die gleiche Lösung ausgedrückt durch neue Koordinaten an oder man sieht 
$g_{\mu\nu}^{\prime}[y(x)]$ als eine neue Lösung ausgedrückt im alten Koordinatensystem an. Dies bedeutet, dass eine 
beliebige Koordinatentransformation gleichbedeutend ist mit dem Übergang zu einer neuen Lösung für das metrische Feld.
Das allerdings hat nun eine zunächst irritierende Konsequenz, die jedoch letztlich zu einer entscheidenden Einsicht 
bezüglich der Natur des Raumes innerhalb der Allgemeinen Relativitätstheorie führt. Es waren die Eigenschaft der
Diffeomorphismeninvarianz und die mit ihr verbundenen begrifflichen Schwierigkeiten, die Einstein drei Jahre lang
beschäftigten. Bereits im Jahre 1912 hatte Einstein seine Feldgleichung für das Gravitationsfeld gefunden. Es dauerte 
jedoch bis zum Jahre 1915, bis Einstein bezüglich der Eigenschaft der Diffeomorphismeninvarianz zur vollen Klarheit 
gelangt war, in welcher aber wohl tatsächlich die wichtigste begriffliche Errungenschaft der Allgemeinen Relativitätstheorie
besteht, welche von entscheidender Bedeutung ist.
Die Schwierigkeit, welche sich in Bezug auf die Entsprechung zwischen einem Übergang zu einem neuen Koordinatensystem
und dem Übergang zu einer neuen Lösung ergibt, kann anhand der folgenden Überlegung verdeutlicht werden. Eine Lösung der
Einsteinschen Feldgleichung $g_{\mu\nu}(x)$ sei so beschaffen, dass die Raum-Zeit in einer Region $X$ eine große Krümmung
aufweist, während sie in einer anderen Region $Y$ nahezu flach ist. Nun könnte im Prinzip eine Koordinatentransformation
durchgeführt werden, welche die Transformation $g_{\mu\nu}(x) \rightarrow g_{\mu\nu}[y(x)]$ zur Folge hat, die
wiederum einem Übergang zu einer neuen Lösung $g_{\mu\nu}^{\prime}(x)$ entspricht, in der die Raum-Zeit in der Region $Y$ eine
große Krümmung aufweist, während sie in der Region $X$ nahezu flach ist. Das allerdings scheint zunächst zu bedeuten, dass die
Einsteinsche Feldgleichung bei gleichen Rahmenbedingungen zwei völlig unterschiedliche physikalische Szenarien als
gleichberechtigte Lösungen enthält. Um dieses Paradoxon aufzuklären muss zunächst der Raumbegriff näher analysiert werden.

\section{Interpretation der Diffeomorphismeninvarianz}

\subsection{Die naturphilosophische Frage nach der Natur des Raumes}

Die naturphilosophische Analyse des ontologischen Status des Raumes entspringt der Frage, ob der Raum an sich in einem
absoluten Sinne für sich selbst besteht oder ob er nur Relationen zwischen Körpern beschreibt und seine Realität damit nur
indirekt über die Materie erhält. Die beiden diesen verschiedenen Auffassungen des Raumes entsprechenden naturphilosophischen
Positionen bezeichnet man als Substantialismus beziehungsweise Relationalismus.
Zwei bedeutende Vertreter der jeweiligen Position in der Geschichte sind Isaac Newton und Gottfried Wilhelm Leibniz.
Newton glaubte an einen absoluten Raum, in Bezug auf welchen absolute Bewegung von Körpern definiert ist. Diese
naturphilosophische Position legte er der Klassischen Mechanik zu Grunde. Leibniz hingegen war der Auffassung, dass der
Raum nur Relationen zwischen Körpern beschreibt und es sinnlos sei, von einem Raum zu reden, ohne sich auf Abstände zwischen
Körpern zu beziehen. Leibniz hatte vor allem das Argument, dass es bei Annahme eines absoluten Raumes dennoch unmöglich wäre,
eine Verschiebung oder Rotation festzustellen, wenn die genaue gegenseitige Lage aller Körper exakt beibehalten würde. Daher
waren für ihn nur die Lagebeziehungen der Körper untereinander real.

\textbf{Eimerversuch:} Newton hatte ein entscheidendes Argument für die Existenz eines absoluten Raumes, nämlich den Eimerversuch. 
Dieser besteht darin, dass man einen mit Wasser gefüllten Eimer in Rotation versetzt und das Verhalten der Wasseroberfläche
beobachtet. Was man dabei beobachtet, ist das folgende: Während der Eimer in Rotation versetzt wird, steht das Wasser zunächst
still. Der Eimer führt also eine Relativbewegung gegenüber dem sich in ihm befindenden Wasser aus. Die Wasseroberfläche jedoch
bleibt vollkommen flach. Nur allmählich wird das sich in ihm befindliche Wasser durch die Reibung mit in Rotation versetzt.
Irgendwann hat das Wasser die Rotationsgeschwindigkeit des Eimers erreicht und es findet keine Relativbewegung des Wassers
in Bezug auf den Eimer mehr statt. Dennoch ist die Wasseroberfläche nun gewölbt. Da in der ersten Situation eine Relativbewegung
zwischen Eimer und Wasser stattfindet, die Wasseroberfläche jedoch flach bleibt, während in der zweiten Situation keine
Relativbewegung zwischen Eimer und Wasser stattfindet, sich die Wasseroberfläche jedoch wölbt, scheint die Relativbewegung
zwischen Eimer und Wasser also nicht dasjenige zu sein, was über das Verhalten der Wasseroberfläche entscheidet. Es kann also
nur die Bewegung, also die Beschleunigung, in Bezug auf einen absoluten Raum sein. Damit stellt die unbestreitbare Existenz
absoluter Beschleunigungen, die konstitutiver Bestandteil der Klassischen Mechanik sind, ein Indiz für die Existenz eines
absoluten Raumes dar.

\subsection{Diffeomorphismeninvarianz und Relationalismus}

Die Auflösung des Problems der durch die Diffeomorphismeninvarianz bedingten Gleichberechtigung der verschiedenen Lösungen
der Einsteinschen Feldgleichung ist nun direkt mit der Frage nach der Natur des Raumes verbunden. Einstein hatte im Jahre
1915 nämlich erkannt, dass die Idee einer absoluten Position in der Raum-Zeit im Rahmen der Allgemeinen Relativitätstheorie
physikalisch sinnlos ist. Physikalisch relevant sind nur raum-zeitliche Koinzidenzen zwischen dynamischen Entitäten. Dies
bedeutet in Bezug auf die Überlegung am Ende des letzten Abschnittes, dass die Unterscheidung der Region $X$ von der Region
$Y$ für sich genommen nicht sinnvoll ist. Unter der Annahme, dass nur die räumliche Beziehung dynamischer Entitäten zueinander
physikalisch relevant ist, zu denen natürlich auch das lokal den Trägheitsbegriff bestimmende Gravitationsfeld gehört, sind die
beiden zu den beiden Lösungen $g_{\mu\nu}(x)$ und $g_{\mu\nu}^{\prime}(x)$ gehörigen Szenarien nämlich äquivalent, da auch alle
anderen dynamischen Größen, die sich auf die Region $X$ beziehen sich nach der Transformation in der Region $Y$ befinden. Dies
kann man sich beispielsweise anhand eines Körpers klar machen, dessen Trajektorie durch die Region $X$ läuft, in der die
Raum-Zeit eine große Krümmung aufweist, während sie nicht durch die Region $Y$ läuft. Bei einer Koordinatentransformation,
welche einem Übergang zu einer Lösung entspricht, dergemäß nun die Region $Y$ eine große Krümmung aufweist, wird natürlich auch
die Trajektorie des Körpers von der Region $X$ in die Region $Y$ übertragen, sodass sich an der physikalischen Situation, dass
der Körper mit einer starken Raumkrümmung zusammentrifft, nichts ändert. Das gleiche gilt selbstverständlich auch für die
Schnitte von Weltlinien von Körpern. Real ist nur die Relation dynamischer Entitäten zueinander und nicht eine absolute Position
im Raum. Das bedeutet also, dass die Diffeomorphismeninvarianz in Wirklichkeit Ausdruck eines der Allgemeinen Relativitätstheorie
inhärierenden Relationalismus bezüglich der Raum-Zeit ist. Die Raum-Zeit ist nur eine Darstellung der Verhältnisse dynamischer
Entitäten zueinander. Hierzu sein Einstein selbst zitiert:

\begin{quote}
{\small Das physikalisch Reale an dem Weltgeschehen (im Gegensatz zu dem von der Wahl des Bezugssystems Abhängigen) besteht in
raum-zeitlichen Koinzidenzen. Real sind z.B. die Schnittpunkte zweier verschiedener Weltlinien, bezw. die Aussage, dass sie
einander nicht schneiden. Diejenigen Aussagen, welche sich auf das physikalisch Reale beziehen, gehen daher durch keine
(eindeutige) Koordinatentransformation verloren. Wenn zwei Systeme der $g_{\mu\nu}$ (bezw. allg. der zur Beschreibung 
der Welt verwandten Variablen) so beschaffen sind, dass man das zweite aus dem ersten durch bloße Raum-Zeit-Transformation
erhalten kann, so sind sie völlig gleichbedeutend. Denn sie haben alle zeiträumlichen Punktkoinzidenzen gemeinsam, 
d.h. alles Beobachtbare. 

Albert Einstein, in einem Brief an Paul Ehrenfest, Berlin 26.12. 1915}
\end{quote}
Damit scheint die Allgemeine Relativitätstheorie also deutlich auf eine relationalistische Interpretation der Raum-Zeit im
Sinne von Leibniz hinzudeuten. Wie kann aber dann der Eimerversuch, den Newton als Beweis der Existenz absoluter
Beschleunigungen in Bezug auf einen absoluten Raum heranzog, im Rahmen der Allgemeinen Relativitätstheorie gedeutet werden ?
Die Antwort auf diese Frage besteht darin, dass Beschleunigungen in der Allgemeinen Relativitätstheorie auf das Gravitationsfeld
bezogen sind, welches selbst eine dynamische Entität darstellt. Die Wölbung der Wasseroberfläche im Eimer wird also durch eine
Wechselwirkung mit dem Gravitationsfeld hervorgerufen. Das bedeutet, dass Beschleunigung zwar ein wirkliches Phänomen ist, das
unabhängig von der Wahl eines Bezugssystems objektiv existiert. Das Phänomen der Beschleunigung ist aber in Bezug auf das die
metrische Struktur der Raum-Zeit beschreibende Gravitationsfeld definiert, welches selbst eine dynamische Entität darstellt.
Damit kann die Existenz des vom Bezugssystem unabhängigen Phänomens von Beschleunigungen nicht zur Rechtfertigung der Annahme
der Existenz eines absoluten Raumes (im Sinne eines Trägheitsraumes) herangezogen werden. 
Die Allgemeine Relativitätstheorie führt also bezüglich der Frage nach dem ontologischen Status des Raumes zu einer eindeutigen
Entscheidung zugunsten des Relationalismus. In einer noch expliziteren Weise drückt Einstein diese radikalen Konsequenz der
Diffeomorphismeninvarianz in einem weiteren Brief aus:

\begin{quote}
{\small Man kann es scherzhaft so ausdrücken. Wenn ich alle Dinge aus der Welt verschwinden lasse, so bleibt nach Newton der
Galileische Trägheitsraum, nach meiner Auffassung aber nichts übrig.

Albert Einstein, in einem Brief an Karl Schwarzschild, Berlin 9.1.1916}
\end{quote}
Es muss aber noch auf einen weiteren Aspekt eingegangen werden. Man könnte nämlich im Prinzip einwenden, dass die Eigenschaft
der Diffeomorphismeninvarianz eigentlich trivial sei, denn die Wahl der Koordinaten in einer beliebigen Theorie ist immer
willkürlich, da es sich nur um eine Konvention handelt. Dies gilt natürlich auch bereits für die Klassische Mechanik und die
Spezielle Relativitätstheorie, welche von der Existenz eines absoluten Trägheitsraumes ausgehen. Es stellt sich also die Frage,
worin nun der besondere Charakter der Diffeomorphismeninvarianz in der Allgemeinen Relativitätstheorie besteht. Aber das
entscheidende in der Allgemeinen Relativitätstheorie ist eben gerade, dass das den Trägheitsbegriff lokal definierende
metrische Feld als dynamischer Größe selbst den Koordinatentransformationen unterliegt.
Um dies zu präzisieren, muss man den Begriff der absoluten Größe einführen. Eine absolute Größe ist eine Größe, welche in
die Dynamik einer bestimmten physikalischen Theorie nicht miteinbezogen ist. Die Minkowski-Metrik in der Speziellen
Relativitätstheorie ist also eine absolute Größe, da sie nicht vom Koordinatensystem abhängt. Nun gibt es in jeder Theorie eine
größte Symmetriegruppe, welche alle absoluten Größen invariant lässt. Diese sind immer Untergruppen der allgemeinsten
Raum-Zeit-Transformationen, nämlich der Diffeomorphismen. In der Speziellen Relativitätstheorie ist es die Poincar\'{e}-Gruppe,
welche ja gerade so definiert ist, dass sie die Minkowski-Metrik konstant hält. Nun gibt es im Falle der Allgemeinen
Relativitätstheorie aber überhaupt keine absoluten Größen mehr, da die Metrik, welche lokal den Trägheitsraum beschreibt, selbst
eine dynamische Größe ist. Deshalb ist hier die größte Symmetriegruppe, welche alle absoluten Größen invariant lässt, die
Diffeomorphismengruppe selbst. Das entscheidende ist also, dass das metrische Feld $g_{\mu\nu}(x)$, welches lokal Inertialsysteme
auszeichnet, selbst den Koordinatentransformationen unterliegt, da es keine absolute sondern eine dynamische Größe darstellt.
Obwohl also in allen Theorien beliebige Koordinaten verwendet werden können, kommt dieser Tatsache nur in der Allgemeinen
Relativitätstheorie physikalische Signifikanz zu. Hierzu sei Holger Lyre zitiert \cite{Lyre:2004}:

\begin{quote}
{\small Dynamische Objekte können durch die theoriespezifischen Wechselwirkungen verän-dert werden und sind durch sie bestimmt,
absolute Objekte bleiben hingegen invariant. Beispielsweise stellt die Minkowski-Metrik $\eta_{\mu\nu}$ in der SRT eine 
absolute, die Metrik $g_{\mu\nu}$ der ART jedoch eine dynamische Größe dar. Mit einer Raumzeit-Theorie läßt sich nun 
jeweils genau eine Symmetriegruppe assoziieren, deren Definition es ist, diejenige größte Untergruppe der Kovarianzgruppe
zu sein, die die absoluten Objekte der Theorie unverändert lässt. In der Newtonschen Theorie ist dies die Galilei-, in der
SRT die Poincar\'{e}gruppe. Enthält aber eine Raumzeit-Theorie überhaupt keine absoluten Objekte, so degeneriert die Symmetrie
der Theorie zur Kovarianzgruppe. Diese Situation ist in der ART gegeben. Nach Anderson sind mit ihren drei Symmetriegruppen
auch die Relativitätsprinzipien der drei Raumzeit-Theorien verbunden: In der Newtonschen Theorie ist das Galilei-, in der
SRT das spezielle Relativitätsprinzip erfüllt - und die ART erfüllt das allgemeine Relativitätsprinzip in Bezug auf die
Kovarianzgruppe als Symmetriegruppe. Dies scheint soweit analog zu den obigen Ausführungen Einsteins. 
Das Auftreten der Kovarianzgruppe als Symmetriegruppe der ART ist aber wohl zu unterscheiden von ihrem Auftreten im Rahmen
der trivialen Koordinatenkovarianz jeglicher Raumzeit-Theorien ! Demnach besitzt das Kovarianzprinzip also sowohl eine
triviale als auch eine nicht-triviale Komponente.

Holger Lyre, Lokale Symmetrien und Wirklichkeit, 2004 (Seiten 131/132)}
\end{quote}
Die in diesem Unterabschnitt thematisierte entscheidende Konsequenz der Eigenschaft der Diffeomorphismeninvarianz der Allgemeinen
Relativitätstheorie in Bezug auf die Frage nach der Natur des Raumes, die damit definitiv zu Gunsten des Relationalismus
entscheidet, kann auch noch einmal durch das folgende Zitat von Carlo Rovelli verdeutlicht werden \cite{Rovelli:2004}:

\begin{quote}
{\small Concretely, this radically novel understanding of spatial and temporal relations is implemented in the theory by
the invariance of the field equations under diffeomorphisms. Because of background independence-that is, since there are no
nondynamical objects that break this invariance in the theory-diffeomorphism invariance is formally equivalent to general
covariance, namely the invariance of the field equations under arbitrary changes of the spacetime coordinates $\vec x$
and $t$.\\
...\\
A physical theory should not describe the location in space and the evolution in time of dynamical objects. It describes
\textbf{relative} location and \textbf{relative} evolution of dynamical objects. Newton introduced the notion of background spacetime
because he needed the acceleration of a particle to be well defined (so that $\vec F=m \vec a$ could make sense). In the
Newtonian theory and in special relativity, a particle accelerates when it does so with respect to a fixed spacetime in which
the particle moves. In general relativity, a particle (a dynamical object) accelerates when it does so with respect to the
local values of the gravitational field (another dynamical object). There is no meaning for the location of the gravitational
field, or the location of the particle with respect to the gravitational field has physical meaning.

Carlo Rovelli, Quantum Gravity, 2004 (Seiten 74/75)}
\end{quote}
Schließlich sollte noch ein letzter Aspekt thematisiert werden. Es ist in Zusammenhang mit der Allgemeinen Relativitätstheorie
oft davon die Rede, dass Raum und Zeit hier selbst dynamisch werden und damit in das physikalische Geschehen miteinbezogen sind,
während sie in früheren Theorien sozusagen nur den Hintergrund darstellten, auf dem sich das physikalische Geschehen abspielt,
sie aber selbst nicht miteinbezogen sind. Damit scheinen Raum und Zeit innerhalb der Allgemeinen Relativitätstheorie doch
eigentlich in dem Sinne einen höheren Realitätsstatus zu erhalten, als sie nun gerade ein nicht direkt auf Objekte auf der
Raum-Zeit bezogenes Eigenleben führen. Dies scheint aber den obigen Ausführungen entgegenzulaufen, denen gemäß der Raum aufgrund der
Diffeomorphismeninvarianz in der Allgemeinen Relativitätstheorie nur eine Beziehungsstruktur zwischen Objekten beschreibt.
Hierbei ist aber die Unterscheidung zwischen der Struktur der Raum-Zeit als topologischem Raum
beziehungsweise als differenzierbarer Mannigfaltigkeit einerseits und ihrer metrischen Struktur andererseits von Bedeutung.
Die metrische Struktur als Beschreibung des Gravitationsfeldes bezieht sich in der Tat auf etwas für sich selbst
existierendes physikalisch Reales. Was in der Allgemeinen Relativitätstheorie jedoch in einem relationalistischen Sinne gedeutet
werden muss, das sind die topologische Struktur und die Struktur als differenzierbarer Mannigfaltigkeit, von der ausgehend
sowohl das die Gravitation beschreibende metrische Feld als zusätzlicher Struktur als auch andere Felder im Sinne wirklicher
dynamischer Entitäten definiert werden. Durch Koordinaten beschriebene Raum-Zeit-Punkte und -Umgebungen sind an sich nichts
Reales, sondern sie beschreiben tatsächlich nur Beziehungen zwischen dynamischen Entitäten. 
Dieser Beschreibung durch Koordinaten auf einer differenzierbaren Mannigfaltigkeit entspricht also eine bestimmten Darstellung
dieser Beziehungen dynamischer Entitäten, wodurch letztere als Felder in Erscheinung treten.
Die Tatsache, dass in der Allgemeinen Relativitätstheorie keine metrische Hintergrundstruktur existiert, welche einen
nicht-dynamischen Trägheitsraum auszeichnet, bezeichnet man als Hintergrundunabhängigkeit.

\part{Die Quantentheorie}

\chapter{Grundbegriffe der Quantentheorie}

\section{Einleitung}

Der Beginn der Quantentheorie geht auf das Jahr 1900 zurück, in dem Max Planck zur Herleitung des Strahlungsgesetzes Schwarzer
Körper die Quantenhypothese aufstellte und das nach ihm benannte Wirkungsquantum einführte \cite{Planck:1900}. Im Jahre 1905
übertrug Albert Einstein die Quantenhypothese auf das elektromagnetische Feld, indem er die Existenz von Lichtquanten
postulierte und damit den photoelektrischen Effekt einer theoretischen Deutung zugänglich machte \cite{Einstein:1905cc}.
Niels Bohr wandte die Quantenhypothese im Rahmen seines 1913 aufgestellten Atommodells auf die Atomphysik an \cite{Bohr:1913}.
Es ist sein großer Verdienst, erkannt zu haben, dass die adäquate physikalische Beschreibung der Atome nicht allein durch die
Einführung neuer Modelle erreicht werden konnte, sondern die Entwicklung gänzlich neuer Begriffe und einer neuen Art des
Zugangs zur Realität unentbehrlich machte. Er kann wohl als die wichtigste Größe bei der Entstehung der Quantentheorie
angesehen werden. Im Jahre 1925 entdeckte Werner Heisenberg die endgültige Gestalt der Quantentheorie in Form der
Matrizenmechanik \cite{Heisenberg:1925}, die er gemeinsam mit Max Born und Pascual Jordan zu einer vollständigen mathematischen
Theorie ausbaute \cite{Born:1925},\cite{Born:1926}. Bald darauf gelang es Wolfgang Pauli, diese Theorie erfolgreich auf das
Wasserstoffatom anzuwenden \cite{Pauli:1926}. Ein Jahr später, 1926, entdeckte Erwin Schrödinger von Louis de Broglies
Hypothese der Existenz von Materiewellen ausgehend eine andere Formulierung der Quantentheorie in Form der Wellenmechanik
\cite{Schroedinger:1926qe}, deren Äquivalenz zur Heisenbergschen Fassung er bald darauf mathematisch aufzeigen konnte
\cite{Schroedinger:1926vq}. Den entscheidenden Schritt zu einem physikalischen und philosophischen Verständnis der
Quantentheorie taten Heisenberg und Bohr im Jahre 1927 mit der Kopenhagener Interpretation, welche wohl trotz vieler
alternativer Deutungsversuche in ihrem Kern als gültig angesehen werden muss. Sie basiert auf der Heisenbergschen
Unbestimmtheitsrelation \cite{Heisenberg:1927}, welche als direkte Konsequenz der Quantentheorie die entscheidende
physikalische Neuerung der Quantentheorie gegenüber der Klassischen Physik enthält, und dem Bohrschen Komplementaritätsprinzip,
welches den Zugang zu einer philosophischen Deutung der Quantentheorie eröffnet. Paul Adrien Maurice Dirac und Johann von
Neumann lieferten schließlich eine allgemeine von einer speziellen Darstellung abstrahierende Formulierung der Quantentheorie
als abstrakter Theorie des Hilbert-Raumes \cite{Dirac:1927we},\cite{Dirac:1958},\cite{Neumann:1932}. Nur in dieser
verallgemeinerten Fassung kann wohl die Universalität der Quantentheorie vollständig erkannt werden. Denn seit ihrer Entdeckung
in den zwanziger Jahren basiert quasi die gesamte Physik mit Ausnahme der durch die Allgemeine Relativitätstheorie
beschriebenen Erscheinungen auf astronomischer und kosmologischer Ebene, von denen man auf fundamentaler Ebene auch eine
quantentheoretische Beschreibungsweise erwartet, auf den grundlegenden Gesetzten der Quantentheorie.

\section{Allgemeine Quantentheorie}

\subsection{Hilbert-Räume und Operatoren}

Die mathematische Formulierung der Quantentheorie basiert auf der Theorie des Hilbert-Raumes. Diese soll zu Beginn kurz
erläutert werden.\\
\textbf{Definition:} Ein Hilbert-Raum $\mathcal{H}$ ist ein unitärer (komplexer) Vektorraum, also ein Vektorraum $V$, der mit
einem inneren Produkt $\langle\ \cdot\ |\ \cdot\ \rangle$ versehen ist.\\
Die Bedingung der Linearität charakterisiert die Struktur des Hilbert-Raumes $\mathcal{H}$ als Vektorraum und ist wie folgt
definiert:
Wenn zwei beliebige Vektoren $|\varphi\rangle, | \chi \rangle \in \mathcal{H}$ mit komplexen Zahlen $a,b \in \mathbb{C}$
multipliziert und anschließend addiert werden

\begin{equation}
|\psi \rangle=a| \chi \rangle+b| \varphi \rangle,
\label{Linearkombination}
\end{equation}
so ist der sich ergebende Vektor erneut ein Element des Hilbert-Raumes $|\psi \rangle \in \mathcal{H}$.
Die Bedingung der Unitarität bezieht sich auf die zusätzliche durch das innere Produkt $\langle\ \cdot\ |\ \cdot\ \rangle$
konstituierte Struktur, das eine Abbildung $\mathcal{H} \otimes \mathcal{H}\rightarrow \mathbb{C}$ ist, welche die
folgenden Eigenschaften erfüllen muss

\begin{eqnarray}
\langle \chi | \varphi \rangle = \langle \varphi | \chi \rangle^*\quad,\quad
\langle \chi | \varphi_1+\varphi_2 \rangle =\langle \chi | \varphi_1 \rangle+\langle \chi | \varphi_2 \rangle\quad,\quad
\langle \chi |c \varphi \rangle = c\langle \chi | \varphi \rangle\quad c \in \mathbb{C}.
\label{Eigenschaften_Inneres-Produkt}
\end{eqnarray}
Durch das innere Produkt wird jedem Element $|\psi\rangle$ eines Hilbert-Raumes ein dualer Vektor $\langle \psi|$ zugeordnet,
der ein Element des Dualraumes $\mathcal{H}^{*}$ ist, also der Menge der Linearformen auf $\mathcal{H}$.
Innerhalb eines solchen Hilbert-Raumes können nun Operatoren definiert werden.\\
\textbf{Definition:} Ein linearer Operator $A$ in einem Hilbert-Raum ist eine Abbildung des Hilbert-Raumes $\mathcal{H}$ auf
sich selbst $A:\mathcal{H} \rightarrow \mathcal{H}$, welcher die Relation: $A\left(a|\chi\rangle+b|\varphi\rangle\right)
=a A|\chi\rangle+b A|\varphi\rangle$ erfüllt.\\
Als Eigenzustand $|a\rangle$ eines Operators bezeichnet man einen Zustand, für den gilt

\begin{equation}
A|a\rangle=a|a\rangle.
\end{equation}
$a$ ist der zum Eigenzustand $|a\rangle$ gehörige Eigenwert. Der hermitesch-adjungierte Operator $A^{\dagger}$ zu einem 
Operator $A$ ist derjenige Operator, für welchen gilt

\begin{equation}
\langle \chi|A|\varphi\rangle=\langle \varphi|A^{\dagger}|\chi\rangle^{*}.
\end{equation}  
Damit wirkt $A^{\dagger}$ im Dualraum $\mathcal{H}^{*}$ so wie $A$ in $\mathcal{H}$ wirkt. Ein hermitescher Operator $A$ ist
ein Operator für den $A=A^{\dagger}$ gilt. Für einen hermiteschen Operator gilt, dass seine Eigenzustände orthogonal und seine
Eigenwerte reell sind. Ein unitärer Operator $U$ ist ein Operator, für den der hermitesch-adjungierte Operator $U^{\dagger}$
gleich dem inversen Operator $U^{-1}$ ist: $U^{\dagger}=U^{-1}$, wobei $U^{-1}U=\mathbf{1}$. 
Aufbauend auf dem Begriff des Hilbert-Raumes und dem des Operators innerhalb eines Hilbert-Raumes können die Postulate der 
Quantentheorie in ihrer allgemeinen Form formuliert werden.

\subsection{Postulate der Quantentheorie}

\textbf{Postulat 1}
\newline
Ein quantentheoretisches System wird beschrieben durch einen Zustand $| \psi \rangle$, welcher ein Element eines
Vektorraumes darstellt, der mit einem inneren Produkt $\langle\ \cdot\ |\ \cdot\ \rangle$ versehen ist, also
eines Hilbert-Raumes $\mathcal{H}$.
\newline\newline
\textbf{Postulat 2}
\newline
Eine beobachtbare Größe, eine Observable, wird durch einen linearen hermiteschen Operator $A$ innerhalb des Hilbert-Raumes
$\mathcal{H}$ beschrieben. Das bedeutet, dass die Eigenwerte des Operators $A$ den möglichen Messwerten entsprechen, die man
erhalten kann, wenn man eine Messung durchführt. Weil der Operator hermitesch ist, was bedeutet, dass $A=A^{\dagger}$, sind
die Eigenwerte und damit die ihnen entsprechenden Messwerte reell.\newline\newline
\textbf{Postulat 3}
\newline
Es ist grundsätzlich ein spezieller zu einem System gehöriger Operator $H$ ausgezeichnet, der die Zeitentwicklung des Systems
determiniert. Dieser wird als Hamilton-Operator bezeichnet. Solange keine Messung (Wechselwirkung mit einem makroskopischen
System) erfolgt, entwickelt sich der Zustand mit der Zeit kontinuierlich gemäß der zeitabhängigen Schrödinger-Gleichung
$i\hbar \partial_t | \psi \rangle = H | \psi \rangle$, welche ausdrückt, dass die Anwendung des allgemeinen
Zeittranslationsoperators auf einen Zustand der Anwendung des Hamilton-Operators auf diesen Zustand entspricht. Diese
dynamische Zeitentwicklung kann in einer äquivalenten Weise innerhalb des Heisenberg-Bildes formuliert werden, in welchem die
beobachtbaren Größen gemäß dem Kommutator mit dem Hamilton-Operator $\frac{\partial}{\partial t}A=\frac{1}{i\hbar}[A,H]_-$ von
der Zeit abhängen. 
\newline\newline
\textbf{Postulat 4}
\newline
Wenn eine Messung einer bestimmten Größe vorgenommen wird, dann geht der das System beschreibende Vektor $|\psi \rangle$ im
Hilbert-Raum in einen Eigenzustand des entsprechenden Operators über und der gemessene Wert entspricht dem dazugehörigen
Eigenwert. Es ist nicht determiniert, in welchen Eigenzustand der Eigenvektor übergeht.
Aber die Wahrscheinlichkeit $p(| \psi \rangle \rightarrow |a \rangle)$ des Übergangs in einen bestimmten Eigenzustand
entspricht dem Wert des Betragsquadrates des inneren Produktes (das die Hilbert-Raum-Struktur gemäß Postulat 1 bestimmt)
zwischen dem Zustand unmittelbar vor der Messung und dem entsprechenden Eigenzustand $p(| \psi \rangle \rightarrow
|a \rangle)=|\langle a | \psi \rangle|^2$.
\newline\newline
\textbf{Postulat 5}
\newline
Wenn man es mit zwei Systemen zu tun hat, welche durch zwei Vektoren $|\varphi \rangle$ und $| \chi \rangle$ in zwei
unterschiedlichen Hilbert-Räumen beschrieben werden, dann können diese beiden Systeme kombiniert werden und das entstehende
Gesamtsystem wird durch das Tensorprodukt dieser beiden Zustände beschrieben $|\varphi \rangle \otimes | \chi \rangle $.\\
\\
\fbox{\parbox{145 mm}{Es sei darauf hingewiesen, dass die Quantentheorie in dieser allgemeinen Formulierung keinen Ortsraum
und keine Raum-Zeit voraussetzt, sondern lediglich einen mit der Zeit zu identifizierenden reellen Parameter $t$.
Auch sonst wird hier auf keine konkreten physikalischen Begriffe wie Masse, Teilchen oder Kraft Bezug genommen.}}

\subsection{Heisenbergsche Unbestimmtheitsrelation}

Eine entscheidende Eigenschaft der Quantentheorie ist nun, dass in einem System nicht mehr alle prinzipiell beobachtbaren
Größen gleichzeitig mit beliebiger Genauigkeit bestimmt sein können. Diese Eigenschaft ergibt sich als direkte Konsequenz der
allgemeinen Postulate der Quantentheorie. Eine physikalische Größe ist in einem Zustand dann bestimmt, wenn dieser sich in einem
Eigenzustand des diese Größe beschreibenden Operators befindet. Zwei Größen, die durch zwei Operatoren $A$ und $B$ beschrieben
werden, können dann gleichzeitig beliebig genau bestimmt sein, wenn es für die beiden Operatoren einen gemeinsamen Satz von
Eigenzuständen $|ab\rangle$ gibt

\begin{equation}
A|ab\rangle=a|ab\rangle\quad,\quad B|ab\rangle=b|ab\rangle.
\end{equation}
Dies ist dann der Fall, wenn die beiden Operatoren miteinander kommutieren $[A,B]=AB-BA=0$. Wenn dies nicht der Fall
ist und der Kommutator der beiden Operatoren $A$ und $B$ nicht verschwindet $[A,B] \neq 0$, so können die beiden ihnen 
entsprechenden Größen nicht beliebig genau bestimmt sein. Die Streuungen $\Delta A$ und $\Delta B$ der zu den beiden 
Operatoren $A$ und $B$ gehörigen Größen stehen mit dem Kommutator der Operatoren in folgender als verallgemeinerte
Heisenbergsche Unbestimmtheitsrelation bezeichneten Relation
  
\begin{equation}
\Delta A \Delta B \geq \frac{1}{2}|\langle \psi|[A,B]_- |\psi \rangle|.
\label{Allgemeine_Unbestimmtheitsrelation}
\end{equation}
Hierbei sei erwähnt, dass der Ausdruck $\langle \psi|A|\psi\rangle$ allgemein den Erwartungswert eines Operators 
$A$ im Zustand $|\psi\rangle$ bezeichnet.

\subsection{Dynamik der Quantentheorie}

Es existieren grundsätzlich zwei Beschreibungsweisen der quantentheoretischen Dynamik, wie sie in Postulat 3 in der obigen
Klassifizierung vorausgesetzt wird, nämlich die Schrödingersche und die Heisenbergsche Beschreibung.
Die Heisenbergsche Darstellung hat ihren Ursprung in der ersten vollständigen Fassung der Quantenmechanik, nämlich der 1925
durch Heisenberg entdeckten \cite{Heisenberg:1925} und von ihm gemeinsam mit Born und Jordan zu einer vollen Theorie
ausgearbeiteten Matrizenmechanik \cite{Born:1925},\cite{Born:1926}.
Die Schrödingersche Zeitentwicklung der Zustände hat ihren historischen Ursprung in ihrer Manifestation innerhalb der 1926
entdeckten Schrödingerschen Wellenmechanik \cite{Schroedinger:1926qe}, deren Äquivalenz zur Heisenbergschen Version Erwin
Schrödinger selbst zeigte \cite{Schroedinger:1926vq}.
Innerhalb der allgemeinen Dirac-von Neumannschen Darstellung der Quantentheorie als Theorie im Hilbert-Raum, unterscheiden
sich die Heisenbergsche und die Schrödingersche Darstellung der quantentheoretischen Dynamik nur durch eine unitäre
Transformation \cite{Dirac:1927we}. Die Schrödingersche Darstellung der quantentheoretischen Dynamik wird durch die in der
Schrödinger-Gleichung

\begin{equation}
i\hbar \frac{\partial}{\partial t}|\psi \rangle=H |\psi \rangle
\label{Schroedinger-Gleichung}
\end{equation}
ausgedrückte Zeitentwicklung der Zustände beschrieben, in welcher die Observablen konstant sind. Aus der Schrödinger-Gleichung
($\ref{Schroedinger-Gleichung}$) ergibt sich der Zeitentwicklungsoperator $U(t,t_0)$ welcher einen zu einer Zeit $t_0$ gegebenen
Zustand $|\psi(t_0) \rangle$ in den entsprechenden zu einer späteren Zeit $t$ vorliegenden Zustand $|\psi(t) \rangle$ überführt:
$|\psi(t) \rangle=U(t,t_0)|\psi(t_0) \rangle$.
In der Heisenbergschen Darstellung sind die Observablen zeitabhängig und ihre Zeitabhängigkeit wird durch 
die Heisenbergsche Gleichung

\begin{equation}
\frac{\partial}{\partial t}A=\frac{1}{i\hbar}[A,H]
\label{Heisenbergsche_Zeitentwicklung}
\end{equation}
beschrieben, also durch den Kommutator des die Observable beschreibenden Operators mit dem Hamilton-Operator, während die
Zustände konstant bleiben. Diese beiden Beschreibungsweisen sind über die durch den Zeitentwicklungsoperator $U(t,t_0)$
bestimmte unitäre Transformation miteinander verknüpft

\begin{equation}
|\psi_S \rangle=U(t,t_0)|\psi_H \rangle.
\end{equation}
wobei $|\psi_S \rangle$ den Zustand in der Schrödingerschen Beschreibung und $|\psi_H \rangle$ den Zustand in der
Heisenbergschen Beschreibung bezeichnen soll. Die Operatoren $A_S$ und $A_H$ sind entsprechend durch folgende 
Transformation miteinander verknüpft

\begin{equation}
A_S=U(t,t_0) A_H U^{\dagger}(t,t_0),
\end{equation}
wobei die physikalisch relevanten Erwartungswerte in beiden Beschreibungen gleich sind

\begin{equation}
\langle \psi_S|A_S|\psi_S \rangle=\langle \psi_H|U^{\dagger}(t,t_0) U(t,t_0) A_H U^{\dagger}(t,t_0) U(t,t_0)|\psi_H \rangle
=\langle \psi_H|A_H|\psi_H \rangle.
\end{equation}
Aus der Schrödinger-Gleichung ($\ref{Schroedinger-Gleichung}$) ergibt sich, dass der Zeitentwicklungsoperator $U(t,t_0)$
bei bekanntem Hamilton-Operator folgende Gestalt aufweist

\begin{equation}
U(t,t_0)=T\exp\left(-\frac{i}{\hbar}\int_{t_0}^t dt' H(t')\right),
\end{equation}
wobei $T$ den Zeitordnungsoperator bezeichnet, der wenn $A(t_1)$ und $B(t_2)$ zwei Operatoren zu zwei Zeiten $t_1$ und
$t_2$ beschreiben, wie folgt definiert ist

\begin{eqnarray}
T[A(t_1)B(t_2)]=\begin{cases}A(t_1)B(t_2)\ \text{für}\ t_1 > t_2\\ B(t_2)A(t_1)\ \text{für}\ t_2 > t_1\end{cases}.
\label{Zeitordnungsoperator}
\end{eqnarray}

\section{Quantenmechanik}

\subsection{Heisenberg-Algebra und Orts- beziehungsweise Impulsdarstellung}

Die konkrete Gestalt, in welcher die Quantentheorie sich als allgemeine Theorie der Beschreibung von 
Teilchen manifestiert, ist die Quantenmechanik. Neben den allgemeinen Postulaten der 
Quantentheorie ist sie durch die Heisenberg-Algebra bestimmt, welche durch die folgende Vertauschungsrelation 
zwischen den hermiteschen Operatoren $X_i$ und $P_j$ definiert ist

\begin{equation}
[X_i,P_j]=i\hbar\delta_{ij}\quad i,j=1...3,
\label{Heisenberg_Algebra}
\end{equation}
welche mit dem Ort und dem Impuls des beschriebenen Teilchens identifiziert werden. Wenn man nun den Zustand
eines durch ein quantentheoretisch beschriebenes Teilchen charakterisierten Systems auf die Eigenzustände des Orts-
beziehungsweise des Impulsoperators projiziert, so erhält man die Orts- beziehungsweise Impulsdarstellung des Zustandes, also
eine Wellenfunktion im Orts- beziehungsweise Impulsraum

\begin{equation}            
\psi(x_i)=\langle x_i|\psi\rangle\quad,\quad \psi(p_j)=\langle p_j|\psi\rangle.
\end{equation}
Man kann nun die Operatoren $X_i$ und $P_j$, welche die Algebra ($\ref{Heisenberg_Algebra}$) erfüllen, und die entsprechenden
Eigenzustände im Raum der quadratintegrablen Funktionen auf dem Raum der Orts- beziehungsweise Impulseigenwerte darstellen

\begin{eqnarray}
X_i=x_i\quad,\quad P_j=-i\frac{\partial}{\partial x^j}\quad\quad
|x_i\rangle=\delta(x-x^{\prime})\quad,\quad |p_j\rangle=\exp(ip_j x^j)\nonumber\\
X_i=i\frac{\partial}{\partial p^i}\quad,\quad P_j=p_j\quad\quad
|x_i\rangle=\exp(-ip^i x_i)\quad,\quad |p_j\rangle=\delta(p-p^{\prime}).
\end{eqnarray}
Das innere Produkt zwischen zwei Zuständen $\phi(x_i)$,$\chi(x_i)$ beziehungsweise $\phi(p_i)$,$\chi(p_i)$ ist in der Orts-
beziehungsweise Impulsdarstellung gegeben durch 

\begin{equation}
\langle\chi|\phi\rangle=\frac{1}{(2\pi)^3}\int d^3 x\ \chi^{*}(x_i)\phi(x_i)\quad,\quad
\langle\chi|\phi\rangle=\frac{1}{(2\pi)^3}\int d^3 x\ \chi^{*}(p_i)\phi(p_i).
\end{equation}
Damit lautet das innere Produkt zwischen einem Orts- und Impuls-Eigenzustand: $\langle p_i|x_i\rangle=\exp(-ip_i x^i)$. 
Die Menge aller Funktionen $\psi(x)$ mit $\int d^3 x |\psi|^2 < \infty$ bilden den Hilbert-Raum der quadratintegrablen
Funktionen $\mathcal{H}_{L^2}$. Wenn man nun von der Orts- in die Impulsdarstellung eines Zustandes und umgekehrt 
übergehen will, so geschieht das in der folgenden Weise

\begin{eqnarray}
\psi(p_i)&=&\langle p_i|\psi\rangle=\int d^3 x \langle p_i|x_i\rangle \langle x_i|\psi\rangle
=\frac{1}{(2\pi)^3}\int d^3 x \exp(-ip_i x^i)\psi(x_i),\nonumber\\
\psi(x_i)&=&\langle x_i|\psi\rangle=\int d^3 p \langle x_i|p_i\rangle \langle p_i|\psi\rangle
=\frac{1}{(2\pi)^3}\int d^3 p \exp(ip_i x^i)\psi(p_i),
\label{Uebergang_Darstellungen}
\end{eqnarray}
also durch eine Fourier-Transformation, wobei in ($\ref{Uebergang_Darstellungen}$) die Tatsache ausgenutzt wurde, dass die
Summe der Projektion eines Zustandes auf alle Zustände einer Basis multipliziert mit dem jeweiligen Basisvektor gleich dem
Einsoperator ist und diese Summe im kontinuierlichen Fall durch ein Integral ersetzt werden muss:
$\frac{1}{(2\pi)^3}\int d^3 x |x_i\rangle \langle x_i|=\frac{1}{(2\pi)^3}\int d^3 p |p_i\rangle \langle p_i|=\mathbf{1}$.

Wenn man nun ($\ref{Heisenberg_Algebra}$) in der allgemeinen Form der Unbestimmtheitsrelation verwendet
($\ref{Allgemeine_Unbestimmtheitsrelation}$), so ergibt sich

\begin{equation}
\Delta X_i \Delta P_j\geq\frac{1}{2}|\langle \psi|[X_i,P_j]_- |\psi \rangle|\quad
\Leftrightarrow\quad \Delta X_i \Delta P_j\geq\frac{\hbar}{2},
\label{Unbestimmtheitsrelation_Ort-Impuls}
\end{equation}
also die konkrete Form der Unbestimmtheitsrelation, wie sie sich in der Quantenmechanik manifestiert und 1927 von Heisenberg
entdeckt wurde \cite{Heisenberg:1927}.

Um nun zur konkreten Gestalt der Schrödinger-Gleichung innerhalb der Orts- beziehungsweise Impulsdarstellung der Quantenmechanik
zu gelangen, geht man von der klassischen Definition der Energie aus: $E=\frac{\mathbf{p}^2}{2m}+V$. Wenn man in dieser Gleichung
die klassischen Größen durch die entsprechenden Operatoren etwa in der Ortsdarstellung ersetzt, wobei der Energieoperator
als $E=i\hbar\frac{\partial}{\partial t}$ definiert ist, so ergibt sich die konkrete Gestalt der Schrödinger-Gleichung
in der Ortsdarstellung

\begin{equation}
i\hbar\frac{\partial\psi(x,t)}{\partial t}=\left(-\frac{\hbar^2}{2m}\Delta+V \right)\psi(x,t),
\label{Schroedinger-Gleichung_Ortsdarstellung}
\end{equation}
wie sie von Schrödinger 1926 formuliert wurde \cite{Schroedinger:1926qe}, wobei 
$\Delta=\frac{\partial}{\partial x^2}+\frac{\partial}{\partial y^2}+\frac{\partial}{\partial z^2}$. Die spezifische
Definition der Energie eines Systems entspricht also der Auszeichnung eines speziellen zu diesem System gehörigen
Hamilton-Operators. Damit wird die Bezeichnung Hamilton-Operator gerechtfertigt, da der Hamilton-Operator demgemäß
durch die Ersetzung der klassischen Variablen in der Hamilton-Funktion durch Operatoren entsteht.

\subsection{Relativistische Quantenmechanik}

Die relativistische Quantenmechanik stellt eine Realisierung der Quantentheorie in Bezug auf die Relativistische Formulierung
der Mechanik dar, verbindet damit also Quantentheorie und Spezielle Relativitätstheorie
\cite{Dirac:1928hu},\cite{Bjorken:RQM},\cite{Weinberg:1995}.
\footnote{In den folgenden Ausführungen wird die Konvention $\hbar=c=1$ verwendet werden.}
Um zur relativistischen Formulierung der Schrödinger-Gleichung zu kommen, muss man zunächst von der relativistischen Definition
der Energie ausgehen $E^2=\mathbf{p}^2+m^2$. Wenn man in diesem Ausdruck die Größen durch die entsprechenden Operatoren ersetzt, 
so erhält man das relativistische Analogon zu ($\ref{Schroedinger-Gleichung_Ortsdarstellung}$)

\begin{equation}
\left(\partial_\mu \partial^\mu+m^2\right)\psi(x)=0,
\label{Klein-Gordon-Gleichung}
\end{equation}
die Klein-Gordon-Gleichung. Diese Gleichung ist jedoch nicht linear und hat nicht die Gestalt der allgemeinen
Schrödinger-Gleichung ($\ref{Schroedinger-Gleichung}$). Eine linearisierte Version der Klein-Gordon-Gleichung
($\ref{Klein-Gordon-Gleichung}$) ergibt sich durch Einführung der Dirac-Algebra, welche durch die Dirac-Matrizen
$\gamma^\mu$ beschrieben wird, welche den folgenden Antivertauschungsrelationen genügen

\begin{equation}
\left\{\gamma^\mu,\gamma^\nu\right\}=\gamma^\mu \gamma^\nu+\gamma^\nu \gamma^\mu=2\eta^{\mu\nu}.
\label{Relation_Dirac-Matrizen}
\end{equation}
Mit Hilfe von ($\ref{Relation_Dirac-Matrizen}$) kann ($\ref{Klein-Gordon-Gleichung}$) wie folgt umgeschrieben werden

\begin{equation}
\left(\gamma^\mu \gamma^\nu \partial_\mu \partial_\nu+m^2\right)\Psi(x) \Leftrightarrow \left(i\gamma^\mu \partial_\mu+m\right)
\left(i\gamma^\nu \partial_\nu-m\right)\Psi(x)=0
\label{Klein-Gordon-Gleichung_Dirac-Algebra}
\end{equation}
und aus ($\ref{Klein-Gordon-Gleichung_Dirac-Algebra}$) kann die 1928 von Dirac aufgestellte Dirac-Gleichung \cite{Dirac:1928hu}
als linearisierte Form der Klein-Gordon-Gleichung abgelesen werden

\begin{equation}
\left(i\gamma^\mu \partial_\mu-m\right)\Psi(x)=0. 
\label{Dirac-Gleichung}
\end{equation}
Wenn man ($\ref{Dirac-Gleichung}$) wie folgt umschreibt

\begin{equation}
i\frac{\partial}{\partial t}\Psi(x)=\left(-i\gamma^0 \gamma^i \partial_i+\gamma^0 m \right)\Psi(x)\equiv H_D \Psi(x),
\end{equation}
sieht man, dass ($\ref{Dirac-Gleichung}$) eine spezielle Manifestation der allgemeinen Schrödinger-Gleichung
mit dem Hamilton-Operator $H_D$ ($\ref{Schroedinger-Gleichung}$) darstellt.
Die Dirac-Matrizen $\gamma^\mu$ können durch $4\times 4$-Matrizen dargestellt werden. In der chiralen Darstellung haben sie
folgende Gestalt

\begin{equation}
\gamma^0=\left(\begin{matrix}0&\mathbf{1}\\\mathbf{1}&0\end{matrix}\right)\quad,\quad \gamma^i=\left(\begin{matrix}0&\sigma^i
\\-\sigma^i&0\end{matrix}\right)\quad i=1...3,
\label{Dirac-Matrizen}
\end{equation}
wobei \textbf{1} die Einheitsmatrix in zwei Dimensionen ist, die auch als $\sigma^0$ bezeichnet wird, und die $\sigma^i$ die
Pauli-Matrizen sind

\begin{equation}
\mathbf{1}=\sigma^0=\left(\begin{matrix}1&0\\0&1\end{matrix}\right)\quad,\quad \sigma^1=\left(\begin{matrix}0&1\\1&0\end{matrix}\right)\quad,\quad
\sigma^2=\left(\begin{matrix}0&-i\\i&0\end{matrix}\right)\quad,\quad \sigma^3=\left(\begin{matrix}1&0\\0&-1\end{matrix}\right).
\label{Pauli-Matrizen}
\end{equation}
Die Dirac-Matrizen wirken als Operatoren in einem vierdimensionalen komplexen Hilbert-Raum $\mathcal{H}_{S_D}$. Die Elemente
dieses Raumes sind die vierkomponentigen Dirac-Spinoren. Dementsprechend beschreiben die Zustände $\Psi(x)$ nun vierkomponentige
Spinorwellenfunktionen. Ihr Zustandsraum ergibt sich aus dem Tensorprodukt des Hilbert-Raumes der quadratintegrablen
Funktionen und dem Hilbert-Raum der Dirac-Spinoren: $\mathcal{H}_{L^2}\otimes \mathcal{H}_{S_D}$.
Eine Dirac-Spinor-Wellenfunktion enthält zwei zweikomponentige Spinorwellenfunktionen, deren Komponenten sich jeweils auf
die Spinkomponenten beziehen. Zweikomponentige Spinoren werden als Weyl-Spinoren bezeichnet. Die zweite 
Weyl-Spinor-Wellenfunktion beschreibt das zum durch die erste Weyl-Spinor-Wellenfunktion beschrieben Teilchen gehörige
Antiteilchen. In der chiralen Darstellung entspricht dem die Aufspaltung in eine linkshändige und eine rechtshändige Komponente

\begin{equation}
\Psi(x)=\left(\begin{matrix}\Psi_L(x)\\ \Psi_R(x)\end{matrix}\right) \in \mathcal{H}_{L^2}\otimes \mathcal{H}_{S_D}.
\label{Dirac-Spinor_Weyl-Spinoren}
\end{equation}
Die Zustände der relativistischen Quantenmechanik sind natürlich invariant unter der inhomogenen Lorentz-Gruppe beziehungsweise
Poincar\'{e}-Gruppe. Die dieser Gruppe entsprechenden Transformationen können demnach im Hilbert-Raum der Zustände
relativistischer Teilchen unitär
\footnote{Es muss sich um unitäre Transformationen handeln, da unitäre Transformationen die Norm eines Zustandes gleich lassen,
was die Voraussetzung für eine Symmetrietransformation ist, da aufgrund der Wahrscheinlichkeitsbedeutung der Norm eines
Zustandes diese immer eins betragen muss.}
dargestellt werden. Der zu einem Element der Poincar\'{e}-Gruppe gehörige unitäre Transformationsoperator
$U(\Lambda,a)$ wirkt auf einen Zustand $\Psi(x)$ gemäß

\begin{equation}
U(\Lambda,a)\Psi(x)=\Psi(\Lambda x+a).
\end{equation}
Für die Beschreibung von Transformationsgruppen ist der mathematische Begriff der Lie-Algebra von großer Bedeutung.\\
\textbf{Definition:} Eine Lie-Algebra ist eine Abbildung $[\ \cdot\ ,\ \cdot\ ]: V \times V \rightarrow V$, welche die 
Jacobi-Identität erfüllt, was bedeutet, dass für drei Elemente des Vektorraumes $X,Y,Z \in V$ gilt:
[[X,Y],Z]+[[Y,Z],X]+[[Z,X],Y]=0.
\footnote{Eine Lie-Algebra ist nicht zu verwechseln mit einer Lie-Gruppe. Eine Lie-Gruppe ist eine Gruppe, die zugleich die
Struktur einer differenzierbaren Mannigfaltigkeit trägt.}\\
Als Generatoren einer Transformationsgruppe bezeichnet man diejenigen Größen, welche infinitesimale Transformationen
beschreiben. Sie erfüllen eine für die Gruppe charakteristische Lie-Algebra. Die zu einer unitären Gruppe gehörigen
Generatoren sind hermitesche Operatoren. Die Generatoren der inhomogenen Lorentz-Gruppe gehorchen der folgenden Lie-Algebra

\begin{eqnarray} 
i[J^{\mu\nu},J^{\rho\sigma}]&=&\eta^{\nu\rho}J^{\mu\sigma}-\eta^{\mu\rho}J^{\nu\sigma}-\eta^{\sigma\mu}J^{\rho\nu}+\eta^{\sigma\nu}J^{\rho\mu},
\nonumber\\
i[P^\mu,J^{\rho\sigma}]&=&\eta^{\mu\rho}P^\sigma-\eta^{\rho\sigma}P^\rho,\nonumber\\
i[P^\mu,P^\sigma]&=&0.
\label{Algebra_Poincare}
\end{eqnarray}
Hierbei bezeichnen die $J^{\mu\nu}$ die Generatoren der homogenen Lorentz-Gruppe, welche antisymmetrisch sind, wobei diejenigen
Komponenten, bei denen einer der Indizes gleich 0 ist, die eigentlichen Lorentz-Transformationen beschreiben, also
Lorentz-Boosts in die drei Raumrichtungen, und die anderen Komponenten die drei Rotationen im räumlichen Anteil der Raum-Zeit.
Die $P^\mu$ bezeichnen die Generatoren der Translationsgruppe, also den Hamilton-Operator $P^0$ und die drei Impulsoperatoren
$P^k$. Die inhomogene Lorentz-Gruppe oder Poincar\'{e}-Gruppe weist also zehn unabhängige Parameter auf und ergibt sich als
Vereinigung der homogenen Lorentz-Gruppe mit der Translationsgruppe.
Eine beliebige unitäre Darstellung einer inhomogenen Lorentz-Transformation kann demgemäß durch ein Exponential einer 
Linearkombination der Generatoren in ($\ref{Algebra_Poincare}$) beschrieben werden

\begin{equation}
U(\Lambda,a)=\exp\left(i\Lambda_{\mu\nu}J^{\mu\nu}+ia_\mu P^\mu\right).
\label{Unitaerer_Operator_Poincare-Transformation}
\end{equation}

\chapter{Quantenfeldtheorie und Elementarteilchen}

\section{Relativistische Quantenfeldtheorien}

Eine relativistische Quantenfeldtheorie entspricht der Verbindung der in den Postulaten enthaltenen Prinzipien der
allgemeinen Quantentheorie und den Prinzipien einer klassischen relativistischen Feldtheorie
\cite{Weinberg:1995},\cite{Bjorken:RQF}. Die Übertragung der Prinzipien der allgemeinen Quantentheorie auf
eine bestimmte Theorie bezeichnet man allgemein als Quantisierung. Eine Quantisierung wird in Analogie zur Quantenmechanik
durchgeführt, indem man klassische Größen zu Operatoren in einem Hilbert-Raum macht. Dies geschieht in Analogie zur
Quantenmechanik, also der Quantentheorie der Punktteilchenmechanik, durch Forderung von Vertauschungsrelationen zwischen den der
Theorie eigenen kanonischen Variablen, welche in Analogie zur Heisenberg-Algebra ($\ref{Heisenberg_Algebra}$) formuliert werden.
Grundsätzlich gibt es zwei Wege, auf denen man durch Quantisierung zu einer Quantenfeldtheorie geführt wird. Man kann
einerseits von einer klassischen Vielteilchentheorie ausgehen und diese quantisieren oder man geht von einer
quantenmechanischen Gleichung aus, die man als Gleichung für ein klassisches Feld uminterpretiert und einer erneuten
Quantisierung, einer Quantisierung der Feldgrößen unterwirft, was von Heisenberg und Pauli entwickelt wurde \cite{Heisenberg:1929}.
Eine relativistische Gleichung für ein klassisches Feld $\varphi$ entspricht einer Wirkung, also dem Integral über den Raum
und die Zeit zwischen zwei Zeitpunkten über eine für die Dynamik dieses Feldes spezifische Lagrange-Dichte: $S=\int d^4 x
\mathcal{L}$, deren durch Variation nach dem Feld $\varphi$ erhaltene Euler-Lagrange-Gleichung

\begin{equation}
\frac{\partial\mathcal{L}}{\partial \varphi}-\partial_\mu\frac{\partial \mathcal{L}}{\partial \left[\partial_\mu
\varphi\right]}=0
\end{equation}
eben dieser Feldgleichung entspricht. Der zum Feld $\varphi$ gehörige kanonisch konjugierte Impuls $\pi$ ist als die Ableitung
der Lagrange-Dichte nach der Zeitableitung $\partial_t \varphi$ definiert   

\begin{equation}
\pi=\frac{\partial \mathcal{L}}{\partial\left[\partial_t \varphi\right]}.
\label{kanonisch-konjugierter_Impuls_Quantenfeldtheorie}
\end{equation}
Die dieser Feldtheorie entsprechende Quantenfeldtheorie ergibt sich nun durch die erwähnte Übertragung der
Vertauschungsrelationen der Quantenmechanik auf die kanonischen Größen der relativistischen Feldtheorie

\begin{equation}
[\varphi(\mathbf{x},t),\pi(\mathbf{x^{'}},t)]=i\delta(\mathbf{x}-\mathbf{x^{'}})\quad,\quad
[\varphi(\mathbf{x},t),\varphi(\mathbf{x^{'}},t)]=0\quad,\quad
[\pi(\mathbf{ x},t),\pi(\mathbf{ x^{'}},t)]=0.
\label{Vertauschungsrelationen_Quantenfeldtheorie}
\end{equation}
Das bedeutet also, dass sich die den kanonischen Variablen entsprechenden durch
($\ref{Vertauschungsrelationen_Quantenfeldtheorie}$) definierten Operatoren an jedem Raum-Zeit-Punkt unabhängig wie Ort und
Impuls in der Quantenmechanik verhalten. Sie wirken im durch die Quantisierung definierten Hilbert-Raum der möglichen das
Quantenfeld $\varphi$ beschreibenden Zustände $|\Phi\rangle$.
Wenn man gemäß der kanonischen Formulierung eine der Lagrange-Dichte $\mathcal{L}$ entsprechende Hamilton-Dichte
$\mathcal{H}$ durch eine Legendre-Transformation definiert

\begin{equation}
\mathcal{H}=\pi \partial_t \varphi-\mathcal{L},
\label{Hamiltondichte}
\end{equation}
die nach Quantisierung einem Hamilton-Operator für die Quantenfeldtheorie entspricht, so kann man die Dynamik
der Feldgrößen $\varphi$ und $\pi$ im Heisenberg-Bild beschreiben

\begin{equation}
\partial_t \varphi=i\left[\mathcal{H},\varphi\right]\quad,\quad \partial_t \pi=i\left[\mathcal{H},\pi\right].
\end{equation}

\subsection{Quantisierung eines Skalarfeldes}

Im Fall eines skalaren Feldes, dass der Klein-Gordon-Gleichung ($\ref{Klein-Gordon-Gleichung}$) genügt
und dessen Dynamik demnach durch die folgende Lagrange-Dichte beschrieben wird

\begin{equation}
\mathcal{L}=\frac{1}{2}\partial_\mu \varphi \partial^\mu \varphi-\frac{1}{2}m^2 \varphi^2,
\end{equation}
kann man das Feld durch eine Entwicklung nach ebenen Wellen beschreiben

\begin{equation}
\varphi\left(\mathbf{p},t\right)=\int \frac{d^3 p}{\left(2\pi\right)^3\sqrt{2 p_0}}
\left[a\left(\mathbf{p}\right)e^{ip_\mu x^\mu}+a^{\dagger}\left(\mathbf{p}\right)
e^{-ip_\mu x^\mu}\right],\quad p_0=\sqrt{\mathbf{p}^2+m^2}.
\label{Entwicklung_Ebene-Wellen}
\end{equation}
Der kanonisch konjugierte Impuls entspricht in diesem Fall des skalaren Feldes der Zeitableitung des Feldes:
$\pi=\partial_t \varphi$. Die Quantisierung ($\ref{Vertauschungsrelationen_Quantenfeldtheorie}$) führt
bei Zugrundelegung der Entwicklung des skalaren Feldes nach ebenen Wellen ($\ref{Entwicklung_Ebene-Wellen}$)
auf Vertauschungsrelationen für die Koeffizienten $a(\mathbf{p})$ und $a^{\dagger}(\mathbf{p})$

\begin{equation}
[a(\mathbf{ p}),a^{\dagger}(\mathbf{ p^{\prime}})]=\delta^3(\mathbf{ p}-\mathbf{ p^{\prime}})\quad,\quad
[a(\mathbf{ p}),a(\mathbf{ p^{\prime}})]=0\quad,\quad
[a^{\dagger}(\mathbf{ p}),a^{\dagger}(\mathbf{ p^{\prime}})]=0,
\label{Vertauschungsrelationen_Erzeuger_Vernichter}
\end{equation}
die damit also zu Operatoren werden. Die Operatoren $a(\mathbf{p})$ und $a^{\dagger}(\mathbf{p})$ können aufgrund ihrer
Vertauschungsrelationen ($\ref{Vertauschungsrelationen_Erzeuger_Vernichter}$) als Erzeugungs- und Vernichtungsoperatoren für
Teilchen interpretiert werden, welche einem Zustand $|\Phi\rangle$ ein Teilchen im Impulseigenzustand $|p\rangle$ hinzufügen
oder wegnehmen. Damit kann ein Quantenfeld als aus einzelnen Teilchen aufgebaut gedacht werden. Die Teilchenbesetzungszahlen
$n_k$ für die jeweiligen Impulse definieren damit eine Basis von Zuständen $|\Phi_{n_p}\rangle$ im Hilbert-Raum des
Quantenfeldes $\varphi$, welche Eigenzustände des Besetzungszahloperators $N_p=a(p)^{\dagger}a(p)$ darstellen. Der Zustand des
Feldes, in dem kein Teilchen enthalten ist wird als Vakuumzustand $|0\rangle$ bezeichnet, für den gilt: $a(p)|0\rangle$=0.
Demgemäß kann ein beliebiger Basiszustand mit definierter Besetzungszahl für die jeweiligen Impulse $p$ aus dem Vakuumzustand
mit Hilfe der Erzeugungsoperatoren $a^{\dagger}_p$ erzeugt werden

\begin{equation}
|\Phi_{n_p}\rangle=\prod_p \left[a^{\dagger}(p)\right]^{n_p}|0\rangle.
\end{equation}    

\subsection{Quantisierung eines Spinorfeldes}

Bei einem Spinorfeld $\Psi(x)$, dass der Dirac-Gleichung ($\ref{Dirac-Gleichung}$) gehorcht und durch die Lagrange-Dichte

\begin{equation}
\mathcal{L}=\bar \Psi\left(i\gamma^\mu \partial_\mu-m\right)\Psi
\label{Dirac-Lagrangedichte}
\end{equation}
beschrieben wird, wobei $\bar \Psi=\Psi^{\dagger}\gamma^0$, sieht die Entwicklung nach ebenen Wellen wie folgt aus

\begin{eqnarray}
\Psi\left(\mathbf{x},t\right)&=&\sum_{\pm s} \int\frac{d^3 p}{\left(2\pi\right)^{\frac{3}{2}}}\sqrt{\frac{m}{p_0}}
\left[b\left(\mathbf{p},s\right) u\left(\mathbf{p},s\right)e^{ip_\mu x^\mu}+d^{\dagger}
\left(\mathbf{p},s\right) v\left(\mathbf{p},s\right) e^{-ip_\mu x^\mu}\right],\nonumber\\
\Psi^{\dagger}\left(\mathbf{x},t\right)&=&\sum_{\pm s}\int\frac{d^3 p}{\left(2\pi\right)^{\frac{3}{2}}}\sqrt{\frac{m}{p_0}}
\left[b^{\dagger}\left(\mathbf{p},s\right) u^{\dagger}\left(\mathbf{p},s\right)e^{-ip_\mu x^\mu}
+d\left(\mathbf{p},s\right) v^{\dagger}\left(\mathbf{p},s\right) e^{ip_\mu x^\mu}\right],\nonumber\\
\label{Spinorfeld_Entwicklung_Ebene-Wellen}
\end{eqnarray}
wobei die Operatoren $b(\mathbf{p},s)$ und $b^{\dagger}(\mathbf{p},s)$ ein Teilchen mit dem Impuls $p$ und der Spineinstellung $s$
erzeugen beziehungsweise vernichten, während die Operatoren $d(\mathbf{p},s)$ und $d^{\dagger}(\mathbf{p},s)$ ein Antiteilchen mit
dem Impuls $p$ und der Spineinstellung $s$ vernichten.
Für ein Spinorfeld $\Psi$ müssen statt der Vertauschungsrelationen ($\ref{Vertauschungsrelationen_Quantenfeldtheorie}$)
Antivertauschungsrelationen postuliert werden, da Teilchen mit Spin $1/2$ dem Paulischen Ausschließungsprinzip gehorchen
\cite{Pauli:1925},\cite{Pauli:1940}, demgemäß sich nicht zwei Teilchen im selben Quantenzustand aufhalten dürfen. Der kanonisch
konjugierte Impuls zu einem Spinorfeld ist $\pi=i\Psi^{\dagger}$. Damit lautet die Quantisierungsbedingung für ein Spinorfeld

\begin{equation}
\left\{\Psi(\mathbf{x},t),\Psi^{\dagger}(\mathbf{x}^{\prime},t)\right\}=\delta^3(\mathbf{x}-\mathbf{x}^{\prime})\quad,\quad
\left\{\Psi(\mathbf{x},t),\Psi(\mathbf{x}^{\prime},t)\right\}=0\quad,\quad
\left\{\Psi^{\dagger}(\mathbf{x},t),\Psi^{\dagger}(\mathbf{x}^{\prime},t)\right\}=0,
\label{Anti-Vertauschungsrelationen_Quantenfeldtheorie}
\end{equation}
was folgenden Vertauschungsrelationen für die Erzeugungs- und Vernichtungsoperatoren entspricht

\begin{eqnarray}
\{b(\mathbf{ p},s),b^{\dagger}(\mathbf{ p^{\prime}},s^{\prime})\}&=&\delta^3(\mathbf{ p}-\mathbf{ p^{\prime}})\delta_{s s^{\prime}},\quad 
\{b(\mathbf{ p},s),b(\mathbf{ p^{\prime}},s^{\prime})\}=0,\quad
\{b^{\dagger}(\mathbf{ p},s),b^{\dagger}(\mathbf{ p^{\prime}},s^{\prime})\}=0,\nonumber\\
\{d(\mathbf{ p},s),d^{\dagger}(\mathbf{ p^{\prime}},s^{\prime})\}&=&\delta^3(\mathbf{ p}-\mathbf{ p^{\prime}})\delta_{s s^{\prime}},\quad 
\{d(\mathbf{ p},s),d(\mathbf{ p^{\prime}},s^{\prime})\}=0,\quad
\{d^{\dagger}(\mathbf{ p},s),d^{\dagger}(\mathbf{ p^{\prime}},s^{\prime})\}=0\nonumber\\
\{b(\mathbf{ p},s),d(\mathbf{ p^{\prime}},s^{\prime})\}&=&
\{b(\mathbf{ p},s),d^{\dagger}(\mathbf{ p^{\prime}},s^{\prime})\}=
\{b^{\dagger}(\mathbf{ p},s),d(\mathbf{ p^{\prime}},s^{\prime})\}=
\{b^{\dagger}(\mathbf{ p},s),d^{\dagger}(\mathbf{ p^{\prime}},s^{\prime})\}=0.\nonumber\\
\label{Anti-Vertauschungsrelationen_Erzeuger_Vernichter}
\end{eqnarray}
Die Operatoren in ($\ref{Anti-Vertauschungsrelationen_Erzeuger_Vernichter}$) beschreiben also die Erzeugung und Vernichtung von
Teilchen mit Spin $1/2$, welche gemäß ($\ref{Anti-Vertauschungsrelationen_Erzeuger_Vernichter}$) dem Paulischen 
Ausschließungsprinzip gehorchen, wie man leicht zeigen kann.

\subsection{Quantenfeldtheorie und Vielteilchentheorie}

Eine Quantenfeldtheorie entspricht also einer quantenmechanischen Beschreibung vieler Teilchen, weshalb man, wie bereits
erwähnt, auch über die Einführung von Produktzuständen, welche ein System vieler Teilchen beschreiben, zu einer
Quantenfeldtheorie gelangt.
Hierbei geht man von $N$ Zuständen $\psi_{n_i}(x_i), i=1...N$ aus, welche alle einer Schrödinger-Gleichung für ein Teilchen
genügen $i\partial_t \psi_{n_i}=H \psi_{n_i}$. Ein Zustand des Gesamtsystems wird durch ein Produkt der Zustände $\psi_{n_i}(x)$
beschrieben und liegt gemäß Postulat 5 der allgemeinen Quantentheorie in demjenigen Hilbert-Raum $\mathcal{H}$, welcher sich
aus dem Tensorprodukt der Hilbert-Räume $\mathcal{H}_i$ der einzelnen Teilchen ergibt: $\mathcal{H}=\mathcal{H}_1
\otimes...\otimes \mathcal{H}_N$.
Da die einzelnen Teilchen in der Quantenmechanik als ununterscheidbar vorausgesetzt werden, muss ein symmetrisiertes (Bosonen) 
beziehungsweise antisymmetrisiertes (Fermionen) Produkt der Zustände gebildet werden

\begin{equation}
\psi=\sum_{P\pm} \psi_{n_1}(x_1)...\psi_{n_N}(x_N),
\end{equation}
wobei $\sum_{P\pm}$ die Summe über alle Permutationen der $n_i$ andeutet, bei der das Vorzeichen bei Vertauschung zweier Indizes
im symmetrischen Fall gleich bleibt und sich im antisymmetrischen Fall umkehrt. Ein Zustand ist also damit durch die Anzahl
der Teilchen in den verschiedenen Einteilchenzuständen charakterisiert. Erzeugungs- und Vernichtungsoperatoren sind dann dadurch
definiert, dass sie verschiedene Zustände ineinander überführen und genügen den Relationen
($\ref{Vertauschungsrelationen_Erzeuger_Vernichter}$) beziehungsweise ($\ref{Anti-Vertauschungsrelationen_Erzeuger_Vernichter}$),
abhängig davon, ob es sich um totalsymmetrische oder totalantisymmetrische Zustände handelt. Mit Hilfe der Erzeugungs- und
Vernichtungsoperatoren können dann Feldoperatoren konstruiert werden, welche den Quantisierungsbedingungen
($\ref{Vertauschungsrelationen_Quantenfeldtheorie}$) beziehungsweise ($\ref{Anti-Vertauschungsrelationen_Quantenfeldtheorie}$)
genügen. Da im Rahmen der Feldquantisierung ein quantenmechanischer Zustand, der einer Schrödinger-Gleichung für ein Teilchen
genügt, als klassisches Feld uminterpretiert wird, um anschließend durch die Quantisierung selbst zu einem Operator zu werden,
kann man die Feldquantisierung auch als eine iterierte Quantisierung betrachten. Die Quantisierung der klassischen 
Punktteilchenmechanik, deren Zustand im Phasenraum durch die Variablen $x_i$ und $p_i$ charakterisiert, führt nach der
Quantisierung zu einer von einer dieser Variablen anhängigen Wellenfunktion, welche den Zustand eines einzelnen
Teilchens beschreibt. In der Quantenfeldtheorie wird diese Wellenfunktion jedoch als klassische Variable in Bezug 
auf einen weiteren Quantisierungsschritt gedeutet, und selbst zu einem Operator, welcher der Schrödinger-Gleichung 
für ein einzelnes Teilchen genügt und auf einen dadurch neu definierten quantenfeldtheoretischen Zustand wirkt, welcher 
durch die Zahl der Teilchen in den einzelnen Einteilchenzuständen charakterisiert ist.

\subsection{Propagatoren}

Eine entscheidende Größe in der Quantenfeldtheorie ist der Feynman-Propagator $i\Delta_F$. Dieser beschreibt die
Übergangsamplitude zwischen zwei Zuständen zu verschiedenen Zeiten, damit also die Wahrscheinlichkeit, ein Teilchen zum
Zeitpunkt $t^{\prime}$ im Zustand $\psi(x,t^{\prime})$ zu finden, wenn es sich zum Zeitpunkt $t$ im Zustand $\psi(x,t)$
befunden hat. Er entspricht demnach dem Erwartungswert des zeitgeordneten Produktes zweier Feldoperatoren im
Vakuumzustand $|0\rangle$

\begin{equation}
i\Delta(x^{\prime}-x)=\langle 0|T\left[\psi(x^{\prime})\psi(x)\right]|0\rangle,
\label{Feynman_Propagator}
\end{equation}
wobei $T$ der Zeitordnungsoperator ($\ref{Zeitordnungsoperator}$) ist. Verwendet man ($\ref{Entwicklung_Ebene-Wellen}$)
in ($\ref{Feynman_Propagator}$), so ergibt sich

\begin{equation}
i\Delta(x^{\prime}-x)=i \int \frac{d^4 p}{(2\pi)^4}\frac{1}{p^2-m^2+i\epsilon}e^{-ip(x^{\prime}-x)}\quad,\quad
\Delta(p)=\frac{1}{p^2-m^2+i\epsilon}
\end{equation}
Für ein Spinorfeld ergibt sich unter Verwendung von ($\ref{Spinorfeld_Entwicklung_Ebene-Wellen}$) in ($\ref{Feynman_Propagator}$)
in analoger Weise

\begin{equation}
i\Delta(x^{\prime}-x)=i \int \frac{d^4 p}{(2\pi)^4}\frac{1}{p\!\!\!/-m+i\epsilon}e^{-ip(x^{\prime}-x)}\quad,\quad
\Delta(p)=\frac{1}{p\!\!\!/-m+i\epsilon},
\end{equation}
wobei gilt: $p\!\!\!/=\gamma^\mu \partial_\mu$ und $\sum_{\pm s}u(\mathbf{p},s)\bar u(\mathbf{p},s)=\frac{p\!\!\!/+m}{2m}$
beziehungsweise $\sum_{\pm s}v(\mathbf{p},s)\bar v(\mathbf{p},s)=\frac{p\!\!\!/-m}{2m}$ zur Herleitung verwendet werden muss.

\subsection{Pfadintegrale}

Eine alternative Formulierung der Quantentheorie besteht in der von Richard P. Feynman entwickelten Pfadintegralmethode 
\cite{Feynman:1942us}, mit Hilfe derer man den Propagator eines Teilchens dadurch erhält, dass man über alle möglichen Bahnen
zwischen zwei Punkten summiert, auf denen ein Teilchen überhaupt laufen kann. Auf die Pfadintegralformulierung gelangt man wie
folgt: Man geht zunächst von der Übergangsamplitude zwischen einem Ortseigenzustand $|x\rangle$ zur Zeit $t$ und einem
Ortseigenzustand $|x^{\prime}\rangle$ zur Zeit $t^{\prime}$ aus

\begin{equation}
\langle x^{\prime}(t^{\prime})|x(t)\rangle=\langle x^{\prime}|U(t^{\prime},t)|x\rangle
=\langle x^{\prime}|\exp\left[-i\hat H(t^{\prime}-t)(\hat x,\hat p)\right]|x\rangle.
\label{Inneres_Produkt_Ortseigenzustaende}
\end{equation}
Durch Einfügen des Einsoperators $\int d^3 p |p\rangle \langle p|$ und Ausnutzen von $\langle x_i|p_i \rangle=
\exp\left(ip_ix^i\right)$ sowie $\langle x|\hat H(\hat x,\hat p)|p \rangle=\langle x|H(x,p)|p \rangle$
in ($\ref{Inneres_Produkt_Ortseigenzustaende}$) ergibt sich

\begin{eqnarray}
\langle x^{\prime}(t^{\prime})|x(t)\rangle
&=&\int d^3 p\ \langle x^{\prime}|\exp\left[-i\hat H\left(\hat x,\hat p\right)\left(t^{\prime}-t\right)\right]|p\rangle \langle p|x\rangle\nonumber\\
&=&\int d^3 p\ \langle x^{\prime}|\exp\left[-iH\left(x^{\prime},p\right)\left(t^{\prime}-t\right)\right]|p\rangle \langle p|x\rangle\nonumber\\
&=&\int d^3 p\ \exp\left[-iH\left(x^{\prime},p\right)\left(t^{\prime}-t\right)+ip_i\left(x_i^{\prime}-x_i\right)\right].
\label{Inneres_Produkt_Ortseigenzustaende_2}
\end{eqnarray}
Man kann nun das Zeitintervall zwischen $t$ und $t^{\prime}$ in $N$ gleiche Teilintervalle mit der Länge
$\Delta t=\frac{t^{\prime}-t}{N}$ zerlegen und zwischen jedem dieser Zeitintervalle einen Einsoperator
$\int d^3 x |x(t_i)\rangle\langle x(t_i)|,\ i=1...N-1$ einschieben

\begin{equation}
\langle x^{\prime}(t^{\prime})|x(t)\rangle=\int dx_{N-1}...d x_{1}\langle x^{\prime}(t^{\prime})|x_{N-1}(t_{N-1})\rangle
\langle x_{N-1}(t_{N-1})|...|x_1(t_1)\rangle \langle x_1(t_1)|x(t)\rangle.
\label{Zeitintervalle}
\end{equation}
Verwendung von ($\ref{Inneres_Produkt_Ortseigenzustaende}$) in ($\ref{Zeitintervalle}$) führt auf

\begin{equation}
\langle x^{\prime}(t^{\prime})|x(t)\rangle=\int \left[\prod_{n=1}^{N-1}dx_n\right]\left[\prod_{n=0}^{N-1}dp_n\right]
\exp\left[-i\sum_{n=0}^{N}\left\{H\left(x_{k},p_{k-1}\right)\left(t_{k}-t_{k-1}\right)+ip_{k-1}\left(x_{k}-x_{k-1}\right)\right\}\right].
\end{equation}
Wenn man nun zum Grenzfall $N \rightarrow \infty$ übergeht, so erhält man

\begin{equation}
\langle x^{\prime}(t^{\prime})|x(t)\rangle=\int \mathcal{D}x \mathcal{D}p
\exp\left[-i\int_{t}^{t^{\prime}} dt\left\{H\left(x(t),p(t)\right)+ip(t)\frac{dx(t)}{dt}\right\}\right], 
\end{equation}
wobei $\mathcal{D}x=\lim_{N \to \infty}\prod_{n=1}^{N-1}dx_n$ und $\mathcal{D}p=\lim_{N \to \infty}\prod_{n=0}^{N-1}dp_n$.
Bei Hamilton-Operatoren, welche quadratisch in den Impulsen sind, kann man das Integral über die Impulse $\int \mathcal{D}p$
direkt lösen, sodass sich ergibt

\begin{equation}
\langle x^{\prime}(t^{\prime})|x(t)\rangle=\mathcal{N}\int \mathcal{D}x \exp\left[-i\int_{t}^{t^{\prime}} dtL(t)\right]
=\mathcal{N}\int \mathcal{D}x \exp\left[-iS\right],
\label{Feynmansches_Pfadintegral}
\end{equation}
wobei $\mathcal{N}$ eine Konstante ist.
\footnote{An dieser Stelle muss daran erinnert werden, dass von der Konvention $\hbar=1$ ausgegangen wurde. Obwohl $\hbar=1$
nicht explizit in Erscheinung tritt, steht in ($\ref{Feynmansches_Pfadintegral}$) im Exponenten eigentlich ein $\hbar$ im 
Nenner. Dies ist für die Argumentation bezüglich des Übergangs zum klassischen Grenzfall des Hamiltonschen Prinzips 
bedeutsam.}

Überträgt man dies auf die Übergangsamplitude für Felder, so lautet diese ausgedrückt durch ein Pfadintegral

\begin{equation}
\langle 0| T\left[\varphi(x^{\prime})\varphi(x)\right]|0\rangle
=\mathcal{N}\int \mathcal{D}\varphi\exp\left[i\int d^4 x \mathcal{L}[\varphi(x)]\right]
\label{Feynmansches_Pfadintegral_Felder}.
\end{equation}
In der Pfadintegralformulierung kann man auf sehr direkte Weise den Übergang zum klassischen Grenzfall deutlich machen. Die
Übergangsamplitude zwischen zwei Zuständen zu zwei Zeiten ergibt sich als Überlagerung der Amplituden aller formal möglichen
klassischen dynamischen Entwicklungen eines Systems, welche zwischen den beiden Zeitpunkten im Prinzip von dem einen auf den
anderen Zustand führen. Nun werden sich aber die zu klassischen dynamische Entwicklungen gehörigen Amplituden, die sich
bezüglich ihrer Wirkung stark unterscheiden, gegenseitig auslöschen, da sie nicht kohärent sind. Daher tragen nur diejenigen
Entwicklungen signifikant bei, die in der Nähe des Extremums der Wirkung des Systems liegen, denn die Wirkungen für die 
verschiedenen Entwicklungen unterscheiden sich dort nur wenig und die Amplituden werden dort deshalb nahezu kohärent, sodass
sie sich gegenseitig verstärken. 
Im klassischen Grenzfall $\hbar \rightarrow 0$ liefern nur noch diejenigen Entwicklungen einen Beitrag, deren Wirkungen
beliebig nahe am Maximum liegen, sodass ($\ref{Feynmansches_Pfadintegral}$) beziehungsweise
($\ref{Feynmansches_Pfadintegral_Felder}$) in das Hamiltonsche Wirkungsprinzip übergeht.

\section{Elementarteilchen, Eichprinzip und Wechselwirkung}

\subsection{Innere Symmetrien}

Relativistische Quantenfeldtheorien stellen eine Erweiterung der Relativistischen Quantenmechanik dar. Sie weisen daher
eine Symmetrie bezüglich der inhomogenen Lorentz-Gruppe auf ($\ref{Algebra_Poincare}$). Nun ist es aber entscheidend, dass
es noch eine weitere Gruppe von Symmetrien in der Elementarteilchenphysik gibt. Diese Symmetrien beziehen sich auf
Quantenzahlen. Eine Quantenzahl ist eine Charakterisierung eines Teilchens, welche eine Eigenschaft eines Teilchens mit einem
Zustand in einem zusätzlichen Hilbert-Raum identifiziert. Der Spin ist also eine Quantenzahl. Sein Zustandsraum $\mathcal{H}_S$
ist kein Teilraum des Hilbert-Raumes der quadratintegrablen Funktionen $\mathcal{H}_L^2$ über dem Orts- oder dem Impulsraum,
sondern er wird durch das Tensorprodukt mit diesem verbunden: $\mathcal{H}_L^2 \otimes \mathcal{H}_S$, wie es im Unterabschnitt
über Relativistische Quantenmechanik bereits thematisiert wurde. Der Spin bezeichnet aber dennoch eine auf den Ortsraum bezogene
Ausrichtung. Daher unterliegt er auch den Lorentz-Transformationen. Insofern muss
($\ref{Unitaerer_Operator_Poincare-Transformation}$) noch genauer erläutert werden, da die Lorentz-Gruppe auch auf den Spinorraum
wirkt, im Falle von Teilchen mit Spin $1/2$ auf den Dirac-Spinor-Raum. Deshalb muss die Lorentz-Gruppe mit ihren Generatoren in
diesem Raum dargestellt werden. Das bedeutet, dass es neben den im Ortsraum wirkenden Generatoren
$M^{\mu\nu}=i\left(x^\mu\partial^\nu-x^\nu \partial^\mu\right)$ der homogenen Lorentz-Gruppe zusätzliche Generatoren
$S^{\mu\nu}$ gibt, welche zwar untereinander die gleiche Algebra wie $M^{\mu\nu}$ erfüllen, aber mit den $P^\mu$ kommutieren,
da sie nicht im auf den Ortsraum bezogenen Hilbert-Raum sondern im Hilbert-Raum des Spin wirken. Das bedeutet, dass die in
($\ref{Algebra_Poincare}$) und ($\ref{Unitaerer_Operator_Poincare-Transformation}$) auftauchenden Generatoren $J^{\mu\nu}$ der
homogenen Lorentz-Gruppe in beiden Räumen wirken: $J^{\mu\nu}=M^{\mu\nu}+S^{\mu\nu}$, was für die Wirkung auf einen Zustand
$\Psi_\alpha(x)$ bedeutet, wenn $\alpha$ die Spinkomponente bezeichnet

\begin{equation}
U(\Lambda,a)\Psi_\alpha(x)=\exp\left[i\Lambda_{\mu\nu}\left(M^{\mu\nu}+iS^{\mu\nu}_{\alpha\beta}\right)
+ia_\mu P^\mu\right]\Psi_\beta(x).
\end{equation}
\textbf{Definition:} Eine Darstellung einer Transformationsgruppe $G$ auf einem Vektorraum $V$ heißt irreduzibel, wenn es keine
Zerlegung von $V$ in lineare Teilräume $V=V_1\oplus...\oplus V_n$ gibt, sodass für alle Abbildungen $A \in G$ gilt:
$v \in V_n \rightarrow Av \in V_n$, Elemente der Gruppe also in der Weise wirken, dass ein Element eines Teilraumes der
Zerlegung in ein Element ebendieses Teilraumes überführt wird.\\
Gemäß Eugene Wigner werden Elementarteilchen durch Zustandsräume beschrieben, in welchen die Poincar\'{e}-Gruppe irreduzibel
dargestellt werden kann \cite{Wigner:1939}. Diese Räume bestehen eben aus dem Hilbert-Raum der Wellenfunktionen über dem Orts-
beziehungsweise Impulsraum und dem Raum, in welchem der Spin beschrieben wird. Die Poincar\'{e}-Gruppe hat den Operator
$P_\mu P^\mu S^2$ mit den Eigenwerten $m^2 s(s+1)$ als Casimir-Operator,
\footnote{Ein Casimir-Operator ist ein Operator, der mit allen Generatoren einer Gruppe kommutiert.}
wobei $S$ der Spinoperator ist und $s$ den Spin des Teilchens beschreibt.
Damit ist ein Teilchen durch seine Masse $m$ und seinen Spin $s$ charakterisiert, welche Invarianten bezüglich der
Poincar\'{e}-Gruppe darstellen.

Die möglichen Darstellungen der homogenen Lorentz-Gruppe kann man nun wie folgt finden. Zunächst definiert man

\begin{eqnarray}
J_{12}&=&J_1\quad,\quad J_{13}=J_2\quad,\quad J_{23}=J_3,\nonumber\\
J_{01}&=&K_1\quad,\quad J_{02}=K_2\quad,\quad J_{03}=K_3.
\label{Operatoren_JK}
\end{eqnarray}
Mit Hilfe dieser Operatoren kann man nun die Operatoren $\mathcal{A}$ und $\mathcal{B}$ definieren

\begin{equation}
\mathcal{A}_i=J_i+iK_i,\quad i=1...3\quad,\quad
\mathcal{B}_i=J_i-iK_i,\quad i=1...3.
\label{Operatoren_AB}
\end{equation}
Diese Operatoren erfüllen die Vertauschungsrelationen

\begin{equation}
\left[\mathcal{A}_i,\mathcal{A}_j\right]=i\epsilon_{ijk}\mathcal{A}_k\quad,\quad\left[\mathcal{B}_i,\mathcal{B}_j\right]
=i\epsilon_{ijk}\mathcal{B}_k\quad,\quad\left[\mathcal{A}_i,\mathcal{B}_j\right]=0,
\label{Algebra_AB}
\end{equation}
wobei $\epsilon_{ijk}$ den totalantisymmetrischen Tensor dritter Stufe bezeichnet.
Die Operatoren $\mathcal{A}$ und $\mathcal{B}$ entsprechen der Darstellung der Lorentz-Gruppe auf dem Raum der links-
beziehungsweise rechtshändigen Spinoren, wie sie in ($\ref{Dirac-Spinor_Weyl-Spinoren}$) erscheinen.
Diese Gruppe weist nun zwei Casimir-Operatoren auf, nämlich $\mathcal{A}^2$ und $\mathcal{B}^2$. Diese Operatoren haben die
Eigenwerte $a(a+1)$ beziehungsweise $b(b+1)$ für $a,b=0,\frac{1}{2},1,\frac{3}{2},2, ...$\ . Damit ist eine Darstellung der
inhomogenen Lorentz-Gruppe durch ein Paar $\left(a,b\right)$ gekennzeichnet. Den folgenden Paaren $\left(a,b\right)$, welche
als die einfachsten Beispiele aufgeführt werden, entsprechen die folgenden Darstellungen 

\begin{eqnarray}
&&\left(\frac{1}{2},0\right)\ \widehat{=}\ \text{linkshändiger Weyl-Spinor}\nonumber\\
&&\left(0,\frac{1}{2}\right)\ \widehat{=}\ \text{rechtshändiger Weyl-Spinor}\nonumber\\
&&\left(\frac{1}{2},0\right)\oplus \left(0,\frac{1}{2}\right)\ \widehat{=}\ \text{Dirac-Spinor}\nonumber\\
&&\left(\frac{1}{2},0\right)\otimes\left(0,\frac{1}{2}\right)=\left(\frac{1}{2},\frac{1}{2}\right)\ \widehat{=}\ \text{Minkowski-Vektor}.
\label{Darstellungen_Lorentz-Gruppe}
\end{eqnarray}
Neben dem Spin gibt es nun aber auch noch Quantenzahlen, welche überhaupt keinen Bezug zum Ortsraum haben, auf deren Raum die 
Poincar\'{e}-Gruppe also nicht wirkt. Solche Quantenzahlen unterscheiden verschiedene Typen von Elementarteilchen, welche den
gleichen Spin aufweisen, aber sich bezüglich ihrer Masse oder ihrem Wechselwirkungsverhalten voneinander unterscheiden.
Wie im Falle des Spin wird der Zustandsraum des Teilchens durch das Tensorprodukt um den zur Quantenzahl gehörigen
Zustandsraum $\mathcal{H}_Q$ erweitert, sodass sich der gesamte Hilbert-Raum eines Teilchens als 
$\mathcal{H}=\mathcal{H}_{L^2}\otimes \mathcal{H}_{S_D}\otimes \mathcal{H}_Q$ ergibt. 
Das historisch erste Beispiel einer solchen Quantenzahl ist der von Heisenberg eingeführte Isospin, welcher ein Neutron und
ein Proton voneinander unterscheidet, welche sich in Bezug auf die starke Wechselwirkung vollkommen gleich verhalten, aber
eine unterschiedliche elektrische Ladung und eine geringfügig unterschiedliche Masse aufweisen. Es gibt aber auch noch 
andere Quantenzahlen. Im Standardmodell der Elementarteilchenphysik gibt es insgesamt zwölf elementare Teilchen, aus denen 
alle anderen Teilchen aufgebaut sind. Es handelt sich um die Quarks und die Leptonen.
Diese sind durch die Quantenzahlen der Ladung, des schwachen Isospins, des Flavour und der Farbe gekennzeichnet.
Entspreche nun einer bestimmten Quantenzahl ein Zustand in einem $N$-dimensionalen Hilbert-Raum $\mathcal{H}_Q$, so ist die
entsprechende Symmetrie-Gruppe die $SU(N)$, also die Gruppe aller unitären Matrizen in $N$ komplexen Dimensionen mit $\det U=1$.
Die $N^2-1$ Generatoren der $SU(N)$, sie seien als $T^a, a=1...N^2-1$ bezeichnet, erfüllen die folgende Lie-Algebra

\begin{equation}
[T^a,T^b]=f^{abc}T^c,
\label{Lie-Algebra_Generatoren_SU(N)}
\end{equation}
wobei die $f^{abc}$ die Strukturkonstanten sind.
Man bezeichnet diese zu den Quantenzahlen gehörigen Symmetrien als innere Symmetrien in Abgrenzung zu den auf die Raum-Zeit
bezogenen Symmetrien, welche als äußere Symmetrien bezeichnet werden. Damit ist ein Elementarteilchen also als irreduzible
Darstellung der Poincar\'{e}-Gruppe gekennzeichnet, weist aber darüber hinaus auch noch die durch die Quantenzahlen begründeten
inneren Symmetrien auf.

\subsection{Lokale Eichtheorien}

Es ist nun eine sehr interessante Tatsache, dass sich alle bekannten Wechselwirkungen in der Natur durch Eichtheorien
beschreiben lassen.
\footnote{Die Selbstwechselwirkung des Higgs-Feldes wurde hierbei außer acht gelassen.}
Dies gilt auch für die im Rahmen der Allgemeinen Relativitätstheorie beschriebene Gravitation, deren
eichtheoretische Beschreibung in einem separaten Kapitel (siehe \textbf{[3.2]}) bereits behandelt wurde. In einer Eichtheorie
geht man von einer freien als Materiefeldgleichung interpretierten relativistischen quantenmechanischen Gleichung aus,
\footnote{Mit Materiefeldern beziehungsweise Teilchen sind hier Felder beziehungsweise Teilchen mit halbzahligem Spin gemeint,
welche dem Paulischen Ausschließungsprinzip gehorchen. Es wird später noch die Frage behandelt werden, ob Spin $1/2$ Felder nicht
grundsätzlich fundamentaler sind als andere Felder, was mit Heisenbergs einheitlicher Spinorfeldtheorie in Zusammenhang steht.
}
wie etwa der Dirac-Gleichung ($\ref{Dirac-Gleichung}$).
Diese weist bestimmte innere und äußere Symmetrien auf, die durch einen unitären Operator $U$ in dem
Hilbert-Raum des entsprechenden Zustandes dargestellt werden: $\Psi \rightarrow U\Psi$ und als Eichsymmetrien bezeichnet werden
Dies liegt einfach daran, dass in der Quantentheorie innere Produkte messbaren Größen entsprechen und diese ändern sich 
durch unitäre Transformation nicht.
\footnote{Zu diesen Symmetrien gehören auch die Phasentransformationen der $U(1)$, welche zur Quantenelektrodynamik führen.
Das bedeutet jedoch nicht, dass nicht Phasenbeziehungen ganz entscheidende ontologische Bedeutung zu käme. Aber auch
in diesem Falle geht es ja um die Phasenbeziehung als eine Relation zweier Zustände und in einen solchen Fall, dass zwei
Systeme in einer Wechselbeziehung stehen, besteht die Symmetrie ja zumindest für die Einzelobjekte auch nicht mehr.}
Allerdings gelten diese Eichsymmetrien nur für eine globale Transformation, also für eine Transformation, bei der die
Parameter der speziellen Transformation nicht von Raum-Zeit-Punkt abhängen. 
Bezüglich Transformationen, die vom Raum-Zeit-Punkt abhängen, denen ein unitärer Operator $U=U(x)$
entspricht, ist die freie Materiefeldgleichung nicht symmetrisch, da durch die Wirkung des Ableitungsoperators auf den unitären
Operator ein zusätzlicher Term entsteht. Indem man nun Invarianz der Gleichung unter der zu einer bestimmten globalen Symmetrie
gehörigen lokalen Symmetrie postuliert, muss die gewöhnliche Ableitung durch eine kovariante Ableitung ersetzt werden

\begin{equation}
\partial_\mu \rightarrow D_\mu=\partial_\mu+iA_\mu,
\label{kovariante_Ableitung}
\end{equation}
die sich unter einer Eichtransformation genau so transformiert, dass der zusätzlich entstehende Term exakt kompensiert wird,
sodass die Gleichung insgesamt wieder Eichinvarianz aufweist, wobei $A_\mu$ das Eichfeld ist, das einen Zusammenhang in dem
Raum darstellt, auf den sich die Symmetriegruppe bezieht. Wenn sich die kovariante Ableitung wie folgt transformiert

\begin{equation}
D_\mu \rightarrow U(x)D_\mu U^{\dagger}(x),
\label{Transformation_kovariante_Ableitung}
\end{equation}
so weist die etwa aus der Dirac-Gleichung ($\ref{Dirac-Gleichung}$) entstehende Gleichung beziehungsweise die aus der
entsprechenden Lagrange-Dichte ($\ref{Dirac-Lagrangedichte}$) entstehende erweiterte Lagrange-Dichte 

\begin{equation}
\mathcal{L}=\bar \Psi\left(i\gamma^\mu D_\mu-m\right)\Psi.
\end{equation}
Invarianz unter der zum unitären Operator $U(x)$ gehörigen lokalen Transformation auf

\begin{equation}
\bar \Psi\left(i\gamma^\mu D_\mu-m\right)\Psi \rightarrow 
\bar \Psi U^{\dagger}(x)\left(i\gamma^\mu U(x)D_\mu U^{\dagger}(x)-m\right)U(x)\Psi
=\bar \Psi\left(i\gamma^\mu D_\mu-m\right)\Psi,
\end{equation}
wobei hier die Tatsache ausgenutzt wurde, dass $U^{\dagger}(x)U(x)=\mathbf{1}$ und die Dirac-Matrizen $\gamma^\mu$ entweder mit dem
Transformationsoperator $U(x)$ kommutieren, bei einer inneren Transformation, oder sich selbst gemäß $\gamma^\mu \rightarrow
U(x)\gamma^\mu U^{\dagger}(x)$ transformieren, bei einer äußeren Transformation.
Der Transformation ($\ref{Transformation_kovariante_Ableitung}$) entspricht eine Transformation des Feldes $A_\mu$ der
folgenden Form: $A_\mu \rightarrow U(x)A_\mu U^{\dagger}(x)-U^{\dagger}\partial_\mu U(x)$, sodass die durch die Ersetzung
($\ref{kovariante_Ableitung}$) entstehende Gleichung Invarianz unter der folgenden Eichtransformation aufweist

\begin{equation}
\Psi \rightarrow U(x)\Psi \quad,\quad A_\mu \rightarrow U(x)A_\mu U^{\dagger}(x)-U^{\dagger}\partial_\mu U(x).
\label{Eichtransformation}
\end{equation}
Man kann nun mit Hilfe der kovarianten Ableitung ($\ref{kovariante_Ableitung}$) einen Feldstärketensor über den Kommutator
der kovarianten Ableitungen definieren

\begin{equation}
F_{\mu\nu}=-i[D_\mu,D_\nu]
\end{equation}
und mit Hilfe dessen eine Lagrange-Dichte für das Eichfeld $A_\mu$ konstruieren, deren Gestalt durch die Forderung nach
Eichinvarianz im Wesentlichen bestimmt ist und wie folgt lautet: $\mathcal{L}_E=\frac{1}{4}F_{\mu\nu}F^{\mu\nu}$,
sodass sich als Gesamt-Lagrange-Dichte der Eichtheorie ergibt

\begin{equation}
\mathcal{L}=\bar \Psi \left(i\gamma^\mu D_\mu-m\right)\Psi+\frac{1}{4}F_{\mu\nu}F^{\mu\nu}.
\end{equation}
Damit ist also durch das Postulat lokaler Eichinvarianz unter einer bestimmten Symmetriegruppe die Existenz eines Eichfeldes
als neuem physikalischen Freiheitsgrad, die Wechselwirkungsstruktur des Materiefeldes mit diesem Eichfeld und die Dynamik
des Eichfeldes begründet. Im Standardmodell der Elementarteilchenphysik ist das Eichprinzip nun in Bezug auf die Gruppe $U(1)$
der Phasentransformationen und in Bezug auf spezifische Quantenzahlen realisiert. Im ersten Fall, welcher der Formulierung der
Quantenelektrodynamik als Eichtheorie entspricht, lauten die dazugehörigen unitären Operatoren: $U=\exp\left[i\omega(x)\right]$.
Im Falle der Quantenzahlen bedeutet dies, dass eine Eichtransformation ($\ref{Eichtransformation}$) sich auf den internen Raum
der entsprechenden Quantenzahl bezieht. Ein dazugehöriger unitärer Operator hat damit folgende Gestalt:
$U(x)=\exp\left[i\omega^a(x) T^a\right]$, wobei die $T^a$ die als Generatoren der $SU(N)$ die Lie-Algebra 
($\ref{Lie-Algebra_Generatoren_SU(N)}$) erfüllen, und das entsprechende Feld wird durch eine Linearkombination
der Generatoren ausgedrückt

\begin{equation}
A_\mu=A_\mu^a T^a
\end{equation}
und ist damit Lie-Algebra-wertig. Für die konkrete Formulierung der Wirkung für das Eichfeld bedeutet das, dass die Spur über
die entsprechende Matrix gebildet werden muss: $\mathcal{L}_E=\text{tr}\left[F_{\mu\nu}^a F^{\mu\nu a}\right]$. Solche auf 
interne Symmetrien bezogene Eichtheorien wurden zuerst von Yang und Mills formuliert und werden daher als Yang-Mills-Theorien
bezeichnet \cite{Yang:1954}. Im Rahmen der Eichtheorie der elektroschwachen Wechselwirkung \cite{Weinberg:1967tq}
wird die sich auf den schwachen Isospin beziehenden $SU(2)$ mit einer $U(1)$ gemischt, welche sich auf die sogenannte
Hyperladung bezieht. Die Ladung, auf die sich die $U(1)$ der Elektrodynamik bezieht, entspricht dort einer bestimmten
Linearkombination der dritten Komponente des schwachen Isospins mit der Hyperladung.

\chapter{Die Deutung der Quantentheorie}

\section{Vorbemerkung zur Deutung der Quantentheorie}

Seit ihrer Entstehung im ersten Viertel des zwanzigsten Jahrhunderts ist die Quantentheorie die grundlegende Naturtheorie,
dies gilt zumindest im mikroskopischen Bereich, aber man vermutet das auch die im makroskopischen Bereich gültige Allgemeine
Relativitätstheorie im Rahmen einer einheitlichen Naturtheorie quantentheoretisch beschrieben werden muss. Sie ist in den 
unterschiedlichsten Bereichen erfolgreich angewandt und empirisch bestätigt worden. Die Deutung der Quantentheorie,
also die Frage, was sie eigentlich über die physikalische Realität aussagt, ist jedoch noch immer nicht vollständig aufgeklärt.
Man kann wohl mit Recht behaupten, dass die Kopenhagener Interpretation der Quantentheorie im Allgemeinen als die etablierte
Interpretation der Quantentheorie angesehen wird. Es sind zwar auch einige alternative Deutungen angeboten worden. Aber meinem
Empfinden nach kann keine dieser alternativen Deutungen die Strenge der Analyse und philosophische Tiefe der Argumente
erreichen, wie sie die Kopenhagener Deutung der Quantentheorie aufweist. Selbst dann, wenn auch sie nicht alle sich in
Zusammenhang mit der Quantentheorie ergebenden Interpretationsfragen vollständig aufgeklärt haben sollte, so hat
sie wohl doch einen wesentlichen Kern des neuartigen Verständnisses der Realität und ihrer Beschreibung, wie sie durch die
Quantentheorie nahegelegt wird, in seiner Tiefe berührt. Es wird hier daher davon ausgegangen, dass ihr grundlegender Gehalt
für ein wirkliches Verständnis der Quantentheorie unentbehrlich ist.
Dies aber impliziert, dass alle weiteren Überlegungen zur Interpretation der Quantentheorie, diesen grundlegenden Gehalt der
Kopenhagener Deutung in sich aufnehmen müssen, wenn sie der Quantentheorie gerecht werden wollen. Da sie wohl in der
Vergangenheit selbst nicht immer völlig einheitlich verstanden wurde, scheint die Notwendigkeit zu bestehen, sie selbst zu
interpretieren. Hierbei kommt einem hohen Maß an philosophischer Präzision bei der Auseinandersetzung mit den Anschauungen
Bohrs und Heisenbergs eine entscheidende Bedeutung zu, wenn sie in ihrem genauen Sinn erfasst werden soll. Vielfach
ist sie in einer zu subjektivistischen oder positivistischen Weise interpretiert worden. Daher muss in diesem Zusammenhang vor
allem in Bezug auf den Unterschied des epistemologischen und des ontologischen Sinnes eines Argumentes sehr behutsam vorgegangen
werden. Um die Kopenhagener Deutung in ihrem Grundgehalt zu verstehen und die Wichtigkeit der Anschauungen Heisenbergs und
Bohrs zu erkennen, ist ihre Beziehung zur Erkenntnistheorie Immanuel Kants von entscheidender Bedeutung. Carl Friedrich von
Weizsäcker hat auf diese sehr enge Beziehung zwischen dem Denken Kants und Bohrs hingewiesen \cite{Weizsaecker:1971}.

\section{Die Kopenhagener Deutung der Quantentheorie}

Bezüglich der Darstellung des Inhalts der Kopenhagener Deutung soll zunächst mit einem Zitat von Heisenberg selbst
begonnen werden, dessen eigene Beschreibung der Kopenhagener Deutung ihrerseits mit den folgenden Sätzen beginnt
\cite{Heisenberg:1958}:

\begin{quote}
{\small Die Kopenhagener Deutung der Quantentheorie beginnt mit einem Paradox. Jedes physikalische Experiment, gleichgültig,
ob es sich auf Erscheinungen des täglichen Lebens oder auf Atomphysik bezieht, muss in den Begriffen der Klassischen Physik
beschrieben werden. Diese Begriffe der Klassischen Physik bilden die Sprache, in der wir die Anordnung unserer Versuche angeben
und die Ergebnisse festlegen. Wir können sie nicht durch andere ersetzen. Trotzdem ist die Anwendbarkeit dieser Begriffe
begrenzt durch die Unbestimmtheitsrelationen.

Werner Heisenberg, Physik und Philosophie, 1958 (Seite 67)} 
\end{quote}
Wir haben es an dieser Stelle bereits mit der entscheidenden Aussage der Kopenhagener Deutung zu tun, welche sich auf die
Notwendigkeit der Verwendung klassischer Begriffe in der konkreten Beschreibung empirischer Phänomene bezieht. 
Aber diesbezüglich ist es nun andererseits ebenso bedeutsam, dass die Anwendbarkeit dieser Begriffe gemäß den
Unbestimmtheitsrelationen, welche doch die entscheidende Eigenschaft der Quantentheorie repräsentieren,
in einer grundsätzlichen Weise beschränkt ist.
Aufgrund dieses Spannungsverhältnisses der Notwendigkeit der Verwendung klassischer Begriffe bei der Beschreibung eines
Experimentes einerseits und der begrenzten Gültigkeit innerhalb der Physik der Quantentheorie an sich, gewinnt die Beziehung
zwischen erkenntnistheoretischen und ontologischen Argumenten fundamentale Bedeutung. 
In Bezug auf diese Betrachtung ist es weiter entscheidend, dass die Unbestimmtheitsrelation an sich eine ontologische
von der menschlichen Art der Beschreibung der Natur völlig unabhängige Bedeutung aufweist. Die positivistische Behauptung etwa,
dass ein exakter Ort und ein exakter Impuls deshalb nicht gleichzeitig existieren könnten, weil es unmöglich sei, sie zu messen,
gehört also dezidiert nicht zur Kopenhagener Deutung. Heisenberg entdeckte die Unbestimmtheitsrelation, indem er sich an eine
Unterhaltung mit Einstein erinnerte, in der Einstein gerade die entgegengesetzte Position vertrat, dass nämlich eine Theorie
notwendig sei, um überhaupt erst entscheiden zu können, welche Größen messbar seien \cite{Heisenberg:1969}. Die Argumentation
Heisenbergs war demgemäß vielmehr die, dass innerhalb der Naturbeschreibung durch die Quantentheorie keine Zustände existieren,
innerhalb derer ein exakter Ort und ein exakter Impuls zur gleichen Zeit definiert sind und deshalb muss es unmöglich sein, sie
zu messen, wenn man die Quantentheorie als wahr voraussetzt.
Sein berühmtes Gedankenexperiment mit dem Photon, das zur Beobachtung eines Elektrons verwendet wird, und welches die
Unmöglichkeit einer gleichzeitigen beliebig genauen Messung beider Größen zeigt, diente lediglich dazu, dem möglichen Einwand
zu begegnen, dass es doch bis dahin als möglich erschien, eine entsprechende Messung durchzuführen und dies hätte einen
Widerspruch der Quantentheorie mit der Erfahrung impliziert \cite{Heisenberg:1927}.

Die Unbestimmtheitsrelation kann weiter nicht als eine Unvollständigkeit der Quantentheorie interpretiert werden, weil viele
Experimente existieren, wie etwa das Doppelspaltexperiment, innerhalb derer bereits die Annahme einer exakt definierten
Teilchenbahn Widersprüche mit den Ergebnissen des Experimentes zur Folge hätte. Das Interferenzmuster des
Doppelspaltexperimentes mit einzelnen Photonen zum Beispiel kann nur adäquat interpretiert werden, wenn an sich unbestimmt ist,
durch welchen Spalt die Bahn eines Photons läuft. Insbesondere den Phasenrelationen, welche nicht direkt messbar sind, kommt
konstitutive Bedeutung bezüglich des Verständnisses dieses und ähnlicher Phänomene zu. Aus diesem Grunde erscheint eine Theorie
versteckter Variablen, wie sie in \cite{Bohm:1951} vorgeschlagen und in \cite{Bell:1964fg} diskutiert wird, von vorneherein als
unplausibel. Phänomene, wie sie in solchen Experimenten beobachtet werden, scheinen vielmehr nahezulegen, dass die Natur
tatsächlich grundsätzlich keiner Beschreibung durch klassische Begriffe wie etwa einer Beschreibung durch eine Teilchenbahn
gehorcht, wie es die aus der mathematischen Beschreibung der Quantentheorie hervorgehende Unbestimmtheitsrelation nahelegt.
Und dennoch behauptet die Kopenhagener Deutung, dass diese Begriffe auch dann unentbehrlich für eine Beschreibung von
Experimenten sind, wenn sie sich auf die Untersuchung spezifisch quantentheoretischer Phänomene beziehen.
Warum ist es unmöglich, sie durch andere Begriffe zu ersetzen ? Die Antwort auf diese Frage kann nicht in der
Natur selbst gefunden werden. Sie kann nur in der Art und Weise menschlicher Wahrnehmung gefunden werden.\\
Der Grund, warum wir die Natur durch klassische Begriffe beschreiben müssen, hat seinen Ursprung in der Art und Weise, wie wir
Erfahrung machen und das bedeutet, dass er in der Beziehung des menschlichen Geistes zur Natur und nicht in der Natur selbst
liegt. Wie bereits im ersten Kapitel dieser Dissertation thematisiert, sind Raum und Zeit gemäß der Kantischen Philosophie
grundlegende Konstituenten menschlicher Erfahrung \cite{Kant:1781},\cite{Kant:1783}.
Kant bezeichnet sie als Grundformen der Anschauung. Sie stellen Vorvoraussetzungen des menschlichen Geistes dar, um mit der Welt
in Berührung gelangen zu können und sind damit zwangsläufig Bestandteil der Erfahrung, also erfahrungskonstitutiv (zumindest in
Bezug auf menschliche Erfahrung).
Aus dieser grundlegenden erkenntnistheoretischen Einsicht, dass Raum und Zeit fundamentale Eigenschaften jeder
Wahrnehmung darstellen und unentbehrliche Konstituenten für den menschlichen Geist sind, sofern jener Zugang zur wirklichen Welt
erhalten soll, schloss Kant, dass Raum und Zeit a priori vor aller Erfahrung gegeben sind. Die Kausalität wurde als
grundlegender Begriff menschlicher Erkenntnis, welche Kant als Kategorien bezeichnet, in einer ähnlichen Weise gedeutet.
Die Begriffe der Klassischen Mechanik stehen in Übereinstimmung mit diesen seitens Kants als für den menschlichen Geist
konstitutiven Strukturen, welche jenem a priori gegeben sind. Sie entsprechen also den Begriffen unseres Verstandes. Die
Quantentheorie beschreibt die Realität jedoch in einer viel allgemeineren Art und Weise. Die Unbestimmtheitsrelation begrenzt
die Gültigkeit der klassischen Begriffe in der wirklichen Welt. Aber dennoch bleiben sie für menschliche Erfahrung konstitutiv.
Somit erweist sich Bohrs Behauptung bezüglich der Notwendigkeit klassischer Begriffe als in enger Beziehung zur Anschauung
Kants stehend. Kant konnte die Entwicklung der modernen Physik nicht vorausahnen,
\footnote{Er hätte sie sogar für unmöglich gehalten. Daher wird die Kantische Philosophie hier auch nicht in ihrer streng
idealistischen Ausprägung als zur Philosophie Bohrs korrespondierend dargestellt, sondern in einer Uminterpretation, welche
bestimmten transzendentalen Bedingungen der Erfahrung zwar ihre rein geistige Natur belässt, welche a priori gegeben ist, ihr
jedoch eine zu ihr korrespondierende reale Struktur gegenüberstellt, der sie gewissermaßen, zumindest bis zu einem gewissen
Grade, isomorph ist.}
welche Bohr zu der Einsicht führte, dass in der realen Welt Phänomene existieren, welche nicht in Übereinstimmung mit den
Strukturen menschlicher Wahrnehmung stehen, aber welche sich dennoch innerhalb dieser Strukturen widerspiegeln. Bohrs Konzept
der Komplementarität erwächst aus dieser Beziehung zwischen der Begrenztheit der Anwendbarkeit klassischer Begriffe und der
Unumgänglichkeit, sie im Rahmen konkreter Erfahrung zu gebrauchen. Klassische Begriffe sind eigentlich nicht geeignet, um diese
Phänomene zu beschreiben. Aber indem wir mehrere Begriffe zur Beschreibung eines Phänomens verwenden, welche innerhalb unseres
klassischen Denkens inkommensurabel erscheinen, können wir dennoch ein wirkliches Verständnis der Realität hinter diesen
Begriffen erlangen. Dies vermag uns jedoch dennoch nur eine Idee jener größeren Realität zu vermitteln, welche durch Verwendung
des jeweiligen klassischen Begriffes nur inadäquat beschrieben werden kann.
Und in diesem Sinne, dass sich verschiedene Begriffe widersprechen und dennoch zur Erfassung des den Rahmen des durch sie
Beschreibbaren Übersteigenden einander ergänzend wirken, sind sie als komplementär zu bezeichnen. Ein Elementarteilchen wie
das Elektron etwa kann in Bezug auf seine wirkliche Natur grundsätzlich nicht Objekt menschlicher Erfahrung sein.
Es ist lediglich möglich, seine Darstellung innerhalb der Strukturen menschlicher Erfahrung und menschlichen Denkens
wahrzunehmen, worin Elektronen als Teilchen und Wellen erscheinen können, was von der Situation des Beobachters abhängig ist,
aber nicht in ihrer wahren Natur, welche das menschliche Wahrnehmungsvermögen und das menschliche Vorstellungsvermögen
übersteigt. Was diese Betrachtungen und ihren Bezug zur Erkenntnistheorie Kants betrifft ist ein Zitat von Weizsäckers sehr
aufschlussreich \cite{Weizsaecker:1971}:

\begin{quote}
{\small Dies kommt der Meinung Kants sehr Nahe, dass der Objektbegriff eine Bedingung der Möglichkeit von Erfahrung ist;
Bohrs Dichotomie der Raum-Zeit-Beschreibung und der Kausalität entspricht der Dichotomie Kants der Anschauungsformen und der
Kategorien (und Grundsätze) des Verstandes, die nur durch ihr Zusammenwirken die Erfahrung möglich machen. Die Parallelität
beider Auffassungen ist umso bemerkenswerter, als Bohr nie viel Kant gelesen zu haben scheint. Im Unterschied zu Kant hat Bohr
aus der modernen Atomphysik die Lehre gezogen, dass es Wissenschaft jenseits des Bereichs gibt, in dem man Vorgänge sinnvoll
durch Eigenschaften beschreiben kann, die von der Situation des Beobachters unabhängig wären; das drückt sein Gedanke der
Komplementarität aus.

Carl Friedrich von Weizsäcker, Die Einheit der Natur, 1971 (Seite 228)} 
\end{quote}
Durch Einbeziehung des Gedankens der biologischen Evolution in die Kantische Erkenntnistheorie erscheinen klassische Begriffe,
insbesondere Raum und Zeit, welche in der Kantischen Erkenntnistheorie als Grundformen der Anschauung rein epistemischen
Charakter haben, auch wieder als reale Entitäten, die den Anschauungsformen korrespondieren. Zumindest scheinen die den
Anschauungsformen inhärierenden Strukturen demnach reale Strukturen widerzuspiegeln, denn die Strukturen des menschlichen
Wahrnehmungsapparates haben sich gemäß der Evolutionären Erkenntnistheorie an die Strukturen der realen Welt angepasst, um so
das Überleben zu ermöglichen. Aber diese Idee der Evolutionären Erkenntnistheorie impliziert nur, dass die Erkenntnisstrukturen
des Menschen (und im Prinzip jedes anderen Lebewesens) der realen Welt in einer Weise korrespondieren, welche adäquat genug
war, um das Überleben in der Stammesgeschichtlichen Entwicklung zu ermöglichen, aber nicht um den Menschen in die Lage zu
versetzen, objektive Wahrheit in einem absoluten Sinne zu erkennen. Insofern erweist sich die Naturbeschreibung innerhalb der
Physik durch klassische Begriffe, wie etwa Teilchen im Ortsraum, als wahr in dem Sinne, dass sie eine Annäherung an die
Strukturen der realen Welt darstellt, aber sie muss eben nicht in einem absoluten Sinne wahr sein. Und die Quantentheorie, wie
auch schon die Relativitätstheorie, scheinen deutlich zu zeigen, dass dies auch tatsächlich nicht der Fall ist. Dies ist der
Grund, warum wir einerseits klassische Begriffe benutzen müssen, zumindest in Bezug auf die Beschreibung von Experimenten, und
warum diese andererseits nicht in Übereinstimmung mit den Strukturen der realen Welt stehen.

Die erkenntnistheoretische Anschauung der Kopenhagener Deutung erwächst aus dieser Dualität zwischen der realen Welt und
unserem Verstand. Solange die Begriffe unseres Denkens den untersuchten Phänomenen sehr eng entsprechen, wie dies im Rahmen der
Klassischen Physik der Fall ist, können wir den Einfluss unserer angeborenen Erkenntnisstrukturen auf die Wahrnehmung 
vernachlässigen. Aber in der Quantentheorie beginnen diese Strukturen zu versagen und deshalb müssen wir bei der Beschreibung
der Natur auf das Verhältnis unserer spezifisch menschlichen Art des Denkens und Wahrnehmens, oder des menschlichen Geistes an
sich, zur Natur Bezug nehmen. Dies bedeutet nicht, dass das Verhalten der Natur von der Art und Weise abhinge, wie wir sie
beschreiben oder dass unser Bewusstsein einen Einfluss auf das Resultat einer Messung hätte. Das wäre eine vollkommene
Fehlinterpretation der Kopenhagener Deutung. Aber es bedeutet, dass wenn wir die Natur beschreiben, wir dies als Menschen tun
müssen und deshalb ist unsere Beschreibung der Natur (nicht die Natur an sich) auch bestimmt durch und bezogen auf die
inhärenten Strukturen menschlicher Wahrnehmung. Diese Schilderung der Kopenhagener Deutung, sowie die im nächsten Abschnitt
vorzufindende Behandlung des Messproblems sind auch in \cite{Kober:2009} zu finden.

\section{Das Messproblem in der Quantentheorie}

\subsection{Das Messproblem als grundlegendes Problem der Deutung}

Ein entscheidendes konzeptionelles Problem der Quantentheorie steht in Zusammenhang mit den beiden unterschiedlichen Weisen,
nach denen sich ein Quantenzustand gemäß den Postulaten der Quantentheorie ändern kann. Solange keine Messung durchgeführt wird
(keine Wechselwirkung mit einem makroskopischen Objekt stattfindet), entwickelt sich ein Quantenzustand nach der zeitabhängigen
Schrödinger-Gleichung ($\ref{Schroedinger-Gleichung}$). Mit Schrödinger-Gleichung ist natürlich in diesem Zusammenhang nicht
der Spezialfall der nicht-relativistischen Quantenmechanik gemeint, sondern die Schrödinger-Gleichung als abstrakte Gleichung,
welche auch nicht auf eine bestimmte Darstellung in Orts-oder Impulsraum, sondern auf jeden beliebigen abstrakten Quantenzustand
bezogen ist. Diese Art der Zustandsänderung wird durch Postulat 3 der oben gegebenen Klassifikation der Postulate zum Ausdruck
gebracht. In der Heisenbergschen Darstellung entwickeln sich statt der Zustände die Observablen gemäß dem Kommutator mit dem
Hamilton-Operator ($\ref{Heisenbergsche_Zeitentwicklung}$). Diese Beschreibung der Zeitentwicklung stellt eine vollkommen
deterministische Beschreibungsweise dar. Findet jedoch eine Messung statt, so ändert sich der Zustand instantan und dies
geschieht in einer völlig indeterministischen Art und Weise, wobei der Zustand nur die Wahrscheinlichkeit eines Übergangs
in einen Eigenzustand der gemessenen Variable bestimmt, was durch Postulat 4 zum Ausdruck gebracht wird. Diese beiden Arten und
Weisen, nach denen sich ein Zustand gemäß der Quantentheorie entwickeln kann, wie sie Postulat 3 und 4 der obigen Klassifikation
enthalten, scheinen nicht kommensurabel zu sein, wenn man zusätzlich die Gültigkeit des Postulats 5 voraussetzt, der sogenannten
Kompositionsregel. Der Grund dieser Unvereinbarkeit kann durch die folgende Argumentation eingesehen werden: Man betrachte
zunächst ein System $S$, dessen Zustand durch einen Vektor $|\psi_S(t)\rangle$ beschrieben wird. Wenn man an diesem System zur
Zeit $t$ eine Messung einer Größe vornimmt, welche durch den Operator $A$ beschrieben wird, so geht der Zustand instantan in
einen Eigenzustand $|a\rangle$ des Operators über, und dies geschieht mit einer Wahrscheinlichkeit, welche durch das
Betragsquadrat des inneren Produkts zwischen dem Zustand unmittelbar vor der Wechselwirkung mit dem Messapparat $M$ und dem
Eigenzustand gegeben ist $|\langle a |\psi_S(t)\rangle|^2$. Diese Beschreibung ist im Gegensatz zur Zeitentwicklung gemäß der
Schrödinger-Gleichung nicht deterministisch, denn dem Zufall kommt im Rahmen dieser Beschreibung eine prinzipielle Bedeutung zu.
Aber nun muss man sich im Hinblick hierauf vergegenwärtigen, dass eine Messung nichts anderes als eine Wechselwirkung mit einem
makroskopischen System darstellt. Und wenn man weiter annimmt, dass der Quantentheorie fundamentale Gültigkeit zukommt, oder
dass sie zumindest fundamentaler ist als jede klassische Theorie wie etwa die Klassische Mechanik, welche auf makroskopische
Systeme bezogen ist, so muss $M$ auch den Gesetzen der Quantentheorie unterliegen. Es mag durchaus sein, dass eine rein
klassische Beschreibung ausreicht, aber eine solche Beschreibungsweise muss selbst als eine Annäherung an die Quantentheorie
angesehen werden, welche nur für makroskopische Grenzfälle gültig ist, sofern man davon ausgeht, dass die Quantentheorie im oben
genannten Sinne fundamental ist. Das bedeutet nun aber, dass $M$ selbst als quantentheoretisches System behandelt
und sein Zustand deshalb als Vektor $|\psi_M(t)\rangle$ in einem Hilbert-Raum beschrieben werden kann. Und dies impliziert gemäß
Postulat 5, dass man das System des beobachteten Objekts $S$, dass ursprünglich betrachtet wurde und welches durch den Zustand
$|\psi_S(t)\rangle$ beschrieben wird und das System des Messinstrumentes $M$, welche durch den Zustand $|\psi_M(t)\rangle$
repräsentiert wird, zu einem Gesamtsystem $G$ zusammenfassen kann, dessen Zustand durch das Tensorprodukt der beiden die
Systeme $S$ und $M$ beschreibenden Zustände beschrieben wird $|\psi_G(t)\rangle=|\psi_S(t)\rangle \otimes |\psi_M(t)\rangle$.
Die Dynamik dieses zusammengefassten Systems wird natürlich ebenfalls durch einen Hamilton-Operator $H_G$ beschrieben, welcher
durch die Eigenschaften der beiden Systeme bestimmt ist, aus denen es zusammengesetzt ist, und das bedeutet, dass die
Zeitentwicklung dieses Gesamtsystems sich deterministisch verhalten muss, solange keine Messung in Bezug auf einen weiteren
Messapparat stattfindet, welche sich auf das zusammengesetzte System $G$ bezieht. Der Hamilton-Operator des zusammengesetzten
Systems $H_G$ kann als die Summe des freien Hamilton-Operators des gemessenen Systems $H_S$, des freien Hamilton-Operators
des Messinstruments $H_M$ und eines Wechselwirkungs-Hamilton-Operators $H_W$ geschrieben werden 

\begin{equation}
H_G=H_S+H_M+H_W
\end{equation}
und determiniert die Dynamik des zusammengesetzten Systems gemäß der entsprechenden Schrödinger-Gleichung

\begin{equation}
i \partial_t \left[|\psi_S(t) \rangle \otimes |\psi_M(t) \rangle \right]
=\left[H_S+H_M+H_W\right]\left[|\psi_S(t)\rangle \otimes |\psi_M(t)\rangle\right]
\Leftrightarrow i \partial_t |\psi_G(t)\rangle=H_G |\psi_G(t)\rangle.
\end{equation}
Da der Messprozess eine Wechselwirkung zwischen $S$ und $M$ darstellt, impliziert dies eine dynamische und somit
deterministische Beschreibung des Messprozesses. Das bedeutet, dass im Rahmen der ersten Betrachtung eine indeterministische
Beschreibung des Systems $G$, welches aus $S$ und $M$ besteht, vorzuliegen scheint, da der Zustand des gemessenen
Objektes sich spontan ändert, und im Rahmen der zweiten Betrachtung eine deterministische Beschreibung der selben Situation
vorzuliegen scheint, innerhalb derer $S$ und $M$ als ein großes System $G$ in Bezug auf ein neues Messinstrument angesehen
werden, und dies impliziert, dass die Quantentheorie sich in Bezug auf diese Dichotomie der Beschreibung selbst zu
widersprechen scheint.

Das bedeutet aber, dass die Tatsache, dass die Beschreibung der Natur gemäß der Quantentheorie nicht kausal oder nicht lokal
ist, nicht das eigentliche konzeptionelle Problem innerhalb der Quantentheorie darstellt. Diese Aspekte der Quantentheorie
stehen nicht in Übereinstimmung mit den Strukturen menschlichen Wahrnehmens und Denkens, aber dies stellt nicht im Geringsten
einen Hinweis auf eine inhärente Problematik einer Theorie dar. Das Gegenteil ist der Fall: Die Tatsache, dass die basalen
Begriffe einer Theorie nicht in Übereinstimmung mit unserer menschlichen Vorstellung von Zusammenhängen in der Welt sind,
stellt ein großes Indiz dafür dar, dass in dieser Theorie eine fundamentalere Beschreibung der Natur entdeckt wurde. 
Es ist nämlich nur zu erwarten, dass das Verhalten der Natur auf einer sehr basalen Ebene von den Strukturen menschlichen
Erkennens abweicht.
Dies kann man eben dann einsehen, wenn man bei den Überlegungen zur Interpretation der Quantentheorie die Erkenntnistheorie
unter Einbeziehung der biologischen Entwicklungsgeschichte gemäß den epistemologischen Betrachtungen beachtet, welche innerhalb
der obigen Beschreibung der Kopenhagener Deutung der Quantentheorie und im ersten Kapitel dieser Arbeit gegeben wurde. Dann
sollte man in der Tat nicht sonderlich erstaunt darüber sein, dass die Strukturen der realen Welt sich sehr von den Strukturen 
menschlichen Denkens und menschlicher Erfahrung unterscheiden, denn letztere haben sich nach dem Kriterium des Überlebensvorteil
entwickelt, den sie dem Menschen eingebracht haben, und nicht etwa gemäß dem Kriterium des Grades an objektivem Wissen, dass
der Mensch mit ihrer Hilfe über die Welt erhalten kann.
Somit stellt das tatsächliche konzeptionelle Problem der Quantentheorie eine inhärente Schwierigkeit dar, welche in dem
Paradoxon zu bestehen scheint, dass sie im Hinblick auf die beiden Wege, nach denen sich ein Zustand gemäß Postulat 3 und 4
ändern kann, in sich bei logischer Strenge nicht vollständig konsistent zu sein scheint. Diesbezüglich hat man es mit einer aus
der Theorie selbst erwachsenden Schwierigkeit zu tun, nicht mit einer Unvereinbarkeit mit unseren angeborenen Vorstellungen über
die Welt, sondern einer Schwierigkeit ihrer Postulate selbst in Bezug aufeinander.
Somit erscheint es als eine zentrale Aufgabe bezüglich eines weitergehenden Verständnisses der Quantentheorie, Postulat 3 und 4
unter Berücksichtigung von Postulat 5 miteinander zu versöhnen. Das scheint zumindest nahezulegen, dass jene Postulate in Bezug
auf ihre Behauptung über das Verhalten der Natur auf ganz fundamentaler Ebene uminterpretiert werden müssen.

\subsection{Versuch einer Auflösung des Konsistenzproblems} 

Wenn man den Widerspruch zwischen Postulat 3 auf der einen Seite und Postulat 4 unter Einbeziehung von Postulat 5 auf der
anderen Seite auf dieser Stufe des Verständnisgrades akzeptiert, so ergeben sich im Prinzip zunächst mehrere Möglichkeiten,
wie man versuchen könnte, diese scheinbare innere Problematik der Quantentheorie zu beheben. Eine Möglichkeit besteht darin,
dass eines der beiden Postulate, welche sich auf eine Änderung von Zuständen in der Quantentheorie beziehen, als falsch
angenommen wird. Aber diese Annahme der schlichten Ungültigkeit eines der Postulate scheint nicht sehr plausibel, da die
Postulate empirisch sehr gut bestätigt sind. Darüber hinaus gehören die verschiedenen Postulate der Quantentheorie zusammen,
denn sie bilden einen zusammengehörigen einheitlichen Rahmen zur Beschreibung der Natur.
Falls eines der Postulate nicht wahr sein sollte, dann würde man also erwarten, dass der gesamte theoretische Rahmen der
Quantentheorie durch eine völlig neue Theorie relativiert werden müsste, welche noch fundamentaler ist und die Quantentheorie
als Grenzfall enthält, aber es erscheint nicht sehr plausibel anzunehmen, dass die Quantentheorie, welche in großen Bereichen
der Erfahrung universelle Gültigkeit zu haben scheint, an sich in dem Sinne abgeändert werden müsste, dass neue Postulate
aufgestellt würden, ohne diese mit einer völlig neuen Theorie und einem weiteren Erfahrungsbereich zu identifizieren.
Es ergibt sich also die Aufgabe eines tieferen Verständnisses der Bedeutung der Postulate im Hinblick auf ihre Beziehung
zueinander, was eine andere Möglichkeit eröffnet, welche in der Annahme besteht, dass eines der Postulate in einer bestimmten
Weise uminterpretiert beziehungsweise in Bezug auf seinen genauen Sinn präzisiert werden muss. Carlo Rovelli hat die sogenannte
relationale Interpretation der Quantentheorie vorgeschlagen \cite{Rovelli:2004},\cite{Rovelli:1995fv},\cite{Smerlak:2006gi}.
Gemäß dieser Interpretation ist ein Zustand nur in Bezug auf einen bestimmten Beobachter definiert. Der Zustand des Systems in
Bezug auf einen Beobachter, welcher eine Messung an einem System $S$ mit einem Messinstrument $M_1$ durchführt, muss von
demjenigen Zustand unterschieden werden, welcher in Bezug auf einen weiteren Beobachter definiert ist, welcher an dem aus $S$
und $M_1$ zusammengesetzten System eine Messung mit einem weiteren Messinstrument $M_2$ durchführt. Somit ist das Messproblem
in einer Weise gelöst, welche dem Geist der Speziellen Relativitätstheorie sehr ähnlich zu sein scheint. Nach meiner Auffassung
stellt dieser Ansatz einen sehr interessanten Versuch dar, das oben beschriebene Problem der Postulate in Bezug aufeinander
aufzulösen. Diesem Ansatz wird hier aber nicht weiter nachgegangen. Es existiert nämlich noch eine weitere Möglichkeit, das
Problem zu lösen. Diese Deutung wird in dieser Arbeit vorgezogen, weil sie konservativer ist. Das bedeutet, dass sie keine
zusätzlichen Annahmen in Bezug auf die Postulate der Quantentheorie benötigt. Diese würde in der Annahme bestehen, dass eines
der beiden Postulate keine fundamentale Beschreibung der Natur darstellt, sondern als effektive Beschreibung in Bezug auf das
andere Postulat dient, aus welchem es im Prinzip hergeleitet werden kann. In dieser Arbeit wird genau diese Anschauung
vertreten. Und sie scheint in der Tat sehr plausibel zu sein, wenn man sich die zentrale Aussage der Kopenhagener Deutung der
Quantentheorie in Erinnerung ruft, nämlich dass Experimente in klassischen Begriffen beschrieben werden müssen, was impliziert,
dass ein Messinstrument ein makroskopisches Gebilde darstellen muss. Ein solches System gehorcht natürlich auch den Gesetzen
der Quantentheorie, aber es kann effektiv auch in einer klassischen Weise beschrieben werden. Um eine näherungsweise klassische
Beschreibung zu ermöglichen, muss das Messinstrument aus einer sehr großen Zahl von Teilchen bestehen und diese Zahl muss so
groß sein, dass es völlig unmöglich wird, seinen exakten Zustand zu kennen, der aus all den Zuständen seiner Bestandteile und
deren Verschränkungen besteht. Aber wenn der exakte Zustand des Messinstruments nicht exakt bekannt sein kann, so ist es
natürlich ebenso unmöglich, die Wechselwirkung zwischen dem gemessenen Objekt und dem Messinstrument in einer exakten Art und
Weise zu beschreiben, weil die Reaktion des Systems, an dem die Messung durchgeführt wird, das vielleicht nur aus einem
einzelnen Teilchen besteht, sehr sensibel auf kleine Differenzen der exakten Zustände der Teilchen reagieren wird, aus denen
das Messinstrument besteht und mit denen das Quantensystem während des Messprozesses verschränkt ist.
Dies steht in Übereinstimmung mit dem Phänomen der Dekohärenz, welche als Konsequenz der Korrelation von Quantensystemen mit
der Umgebung erscheint, welche eine Aufhebung der Phasenrelationen und somit der Interferenzeffekte zur Folge hat und zu einem
Übergang zu klassischen Eigenschaften führt. Betrachtungen der Dekohärenz in Bezug auf die Interpretationen der Quantentheorie
sind in \cite{Schlosshauer:2003zy},\cite{Schlosshauer:2007},\cite{Schlosshauer:2008} behandelt worden. Es scheint also die
Möglichkeit eröffnet zu werden, dass Postulat 4 in der Tat aus Postulat 3 hergeleitet werden könnte. Im Rahmen einer
fundamentalen Beschreibung würde der Messprozess demgemäß einen deterministischen Prozess darstellen, aber es würde nur eine
effektive Beschreibung zugänglich sein, da das Messinstrument als klassisches System beschrieben werden muss, um menschlicher
Wahrnehmung zugänglich zu sein. Wenn diese Annahme, dass Postulat 4 im Prinzip aus Postulat 3 hergeleitet werden kann, korrekt
wäre, dann wäre die Quantentheorie in einem grundsätzlichen ontologischen Sinne deterministisch. Das Element der
Indeterminiertheit in der Quantentheorie würde also eine ähnliche Rolle spielen wie in der Statistischen Mechanik, wo der
Zufall aus einem mangelnden Wissen über die einzelnen Bestandteile eines komplexen Systems erwächst. Dennoch wäre die Situation
in der Quantentheorie ein wenig anders, denn in der Quantentheorie ist der Indeterminismus eine direkte Konsequenz der Art und
Weise, wie Menschen Erfahrung machen, und der Unvereinbarkeit dessen mit der Realität an sich. Um den ontologischen Status des
Indeterminismus zu beschreiben, der dieser Interpretation der Quantentheorie entspricht, ist es sinnvoll, zunächst verschiedene
Stufen zu klassifizieren, auf denen ein Element der Indeterminiertheit in einer physikalischen Theorie auftreten kann. Im
Hinblick hierauf soll zwischen verschiedenen Bedeutungen der Indeterminiertheit in einer physikalischen Theorie unterschieden
werden. Es handelt sich dabei um die folgenden vier Bedeutungen:
\newline\newline
\noindent
\textbf{1 kein Indeterminismus (beispielsweise Klassische Mechanik, Klassische Elektrodynamik):}
\newline
In einer solchen Theorie existiert kein Element des Indeterminismus. Die dynamische Entwicklung aller Konstituenten, welche
notwendig sind, um eine vollständige Beschreibung des Systems zu liefern, ist durch die Theorie determiniert.
\newline\newline
\textbf{2 Indeterminismus in einem epistemologischen Sinne der praktische Gründe hat (beispielsweise Statistische Mechanik):}
\newline
In einer solchen Theorie hat der auftretende Indeterminismus mit einem Mangel an Wissen aus praktischen Gründen zu tun. Es ist
aufgrund der Komplexität der beschriebenen Systeme nicht möglich, alle die Dynamik des Systems bestimmenden Parameter zu
kennen. Aber im Prinzip wäre es möglich, alle Konstituenten zu messen, welche die dynamische Entwicklung bestimmen und eine
deterministische Beschreibung ist daher prinzipiell zumindest denkbar.
\newline\newline
\textbf{3 Indeterminismus in einem epistemologischen Sinne der prinzipielle Gründe hat (beispielsweise die in dieser Arbeit vertretene
Interpretation der Quantentheorie um Postulat 3 und 4 miteinander in Einklang zu bringen):}
\newline
In einer solchen Theorie erwächst der Indeterminismus ebenfalls aus einem Mangel an Wissen. Aber im Gegensatz zum obigen Fall,
ist es (für Menschen) prinzipiell unmöglich, die Information zu erhalten, welche für eine deterministische Beschreibung
notwendig wäre. Die Natur menschlicher Wahrnehmung schließt eine deterministische Beschreibung prinzipiell aus. (Man muss
makroskopische Systeme verwenden, um beispielsweise den Zustand eines Teilchens zu messen.)
\footnote{Dies hat absolut nichts mit der Unbestimmtheitsrelation der Quantentheorie zu tun, die eine wirkliche Unbestimmtheit
unabhängig von der spezifischen Art und Weise menschlicher Erfahrung beschreibt, also im vollen Sinne ontologisch zu verstehen
ist, was weiter unten noch einmal explizit thematisiert werden wird.}
\newline\newline
\textbf{4 Indeterminismus in einem ontologischen Sinne (beispielsweise Quantentheorie, wobei Postulat 4 als fundamental
interpretiert wird):}
\newline
In einer solchen Theorie existiert ein Element realer Indeterminiertheit. Dieses ist unabhängig von unserer menschlichen Art
der Beschreibung der Welt und der Unmöglichkeit, ein exaktes Wissen aller Konstituenten zu erhalten, die das System
beschreiben, sondern es hat ontologischen Status.
\newline\newline
Wie bereits innerhalb der Klammern angedeutet, entspricht das Element der Indeterminiertheit innerhalb der Interpretation des
quantentheoretischen Messproblems, welche in dieser Arbeit vertreten wird, einer Indeterminiertheit im Sinne 3 der obigen
Klassifikation. Der Indeterminismus hat keinen ontologischen Status, zumindest nicht gemäß der Beschreibung der Natur, wie sie
die Quantentheorie liefert. Nichtsdestotrotz existiert mit Bezug auf die Beschreibung der Natur durch Menschen ein
fundamentales Element des Indeterminismus, welches unumgänglich ist. Es hat seinen Ursprung in der für Menschen bestehenden
Notwendigkeit, physikalische Objekte als in einem Ortsraum befindlich zu beschreiben.
Alle Messungen, sogar Messungen interner Variablen, werden aus Messungen des Ortes hergeleitet. Wenn der Spin eines Teilchens
zum Beispiel in einem Stern-Gerlach-Experiment gemessen wird, so handelt es sich bei derjenigen Größe, welche direkt gemessen
wird, um den Ort, wobei diese Messung durchgeführt wird, indem man feststellt, an welcher Stelle eine Schwärzung auftritt.
Hieraus wird der entsprechende Impuls hergeleitet, welcher in direkter Beziehung zum Spin des Teilchens steht.
Die fundamentalsten in der Natur bekannten Objekte werden durch die Quantentheorie beschrieben, welche Zustände impliziert,
welche gemäß der Unbestimmtheitsrelation nicht Trajektorien im kinematischen Sinne der Klassischen Mechanik beschreibt und die
keinen scharfen Ort und Impuls zur selben Zeit haben. Sie sind noch nicht einmal auf eine kleine Region lokalisiert,
wenn sie als freie Objekte in Erscheinung treten. Nur im Rahmen gebundener Zustände innerhalb größerer Systeme,
welche andere Objekte mitumfassen, kann eine schärfere Lokalisation erreicht werden (innerhalb der Grenzen, welche durch die
Unbestimmtheitsrelation definiert werden).
Um eine Beschreibung als klassisches System zu ermöglichen, muss ein Objekt aus einer so großen Zahl an Teilchen bestehen, dass
es unmöglich ist, seinen Zustand exakt zu kennen. Raum ist eine Vorbedingung menschlicher Erfahrung. Aber nur makroskopische
Objekte erfüllen die Voraussetzung, solch eine Beschreibung zuzulassen. Aus diesem Grunde ist es notwendig, eine Wechselwirkung
eines Elementarteilchens mit einem makroskopischen Objekt zu betrachten, um Wissen über es zu erlangen. Und dies impliziert,
dass nur eine indeterministische Beschreibung möglich ist. Gemäß der obigen Argumentation kann dieses Element der 
Indeterminiertheit prinzipiell nicht umgangen werden. In Bezug auf die statistische Mechanik ist es zumindest denkbar, dass der
exakte Zustand jedes Konstituenten eines komplexen Systems gemessen wird und somit ist eine deterministische Beschreibung nicht
prinzipiell sondern aus praktischen Gründen ausgeschlossen. In der Quantentheorie ist solch eine Beschreibung noch nicht einmal
denkbar, da man um alle Zustände der Teilchen zu messen, aus denen das Messinstrument besteht, ein weiteres makroskopisches
Messinstrument bräuchte, für welches die gleiche Argumentation gültig wäre. Eine sehr ähnliche Anschauung wird explizit seitens
Peter Mittelstaedt beschrieben \cite{Mittelstaedt:1963}:

\begin{quote}
{\small Zu diesen Voraussetzungen gehört, dass ein Messgerät von makroskopischen Dimensionen verwendet wird. Denn bei einer
Wechselwirkung dieses Geräts mit dem Objekt verteilt sich die ursprünglich in kompakter Form vorliegende Information auf die
etwa $10^{20}$ Freiheitsgrade des des Messgeräts und wird dadurch - jedenfalls zu einem gewissen Teil - praktisch unzugänglich.
Denn auch die rein rechnerische Verfolgung der komplizierten Gleichungssysteme, die in dem hochdimensionalen Zustandsraum von
S+M gegeben sind, liegt vollkommen außerhalb der realen Möglichkeiten.

Peter Mittelstaedt, Philosophische Probleme der modernen Physik, 1963 (Seite 68)}
\end{quote}
Die Tatsache, dass die Theorie gemäß der hier vertretenen Möglichkeit einer Versöhnung von Postulat 3 und 4 in dem Sinne nicht
deterministisch ist, dass der Zufall nicht in einem ontologischen sondern in einem prinzipiell epistemologischen Sinne
erscheint, ist logisch völlig unabhängig von der Interpretation der Unbestimmtheitsrelation, insbesondere der Tatsache, dass es
sinnlos ist, über einen exakten Ort und einen exakten Impuls eines Teilchens zur gleichen Zeit zu sprechen, welche die
Genauigkeit übersteigen, welche durch die Unbestimmtheitsrelation als direkter Konsequenz der Postulate der Quantentheorie
determiniert wird. Oft wird die neuartige Beschreibung der Natur durch die Quantentheorie im Gegensatz zur Klassischen
Mechanik in der Weise dargestellt, dass man es in der Quantentheorie mit Wahrscheinlichkeiten für reale Fakten zu tun hat, man
aber erst dann etwas physikalisch Reales erhält, wenn man Messungen durchführt, also Observablen wie Ort und Impuls betrachtet.
Im Rahmen einer solchen Anschauung kann der Quantenzustand an sich keine wirkliche physikalische Realität beanspruchen, sondern
nur in Bezug auf die Observablen, für deren mögliche Messwerte er Wahrscheinlichkeiten determiniert. Es ist aber vermutlich in
Bezug auf die Art und Weise, wie in dieser Arbeit Postulat 3 und Postulat 4 miteinander versöhnt werden, wesentlich plausibler,
den Zustand an sich als die fundamentale Beschreibung einer realen physikalischen Gegebenheit zu interpretieren.
In Bezug auf diese Interpretation wird die Struktur des abstrakten Hilbert-Raumes als fundamentaler angesehen und die
Darstellung innerhalb eines Orts- oder Impulsraumes wird als eine abgeleitete Darstellung betrachtet. Was diesen Aspekt angeht,
ist es hilfreich, sich in Erinnerung zu rufen, dass die Postulate der Quantentheorie keinen Orts- und Impulsraum voraussetzen,
und dies noch nicht einmal im Sinne eines Darstellungsraumes für die Zustände, welche in einer solchen Darstellung als
Wellenfunktionen erscheinen. Dies liefert eine starke Rechtfertigung der Meinung, dass die Zustände an sich einen höheren
Realitätsgrad aufweisen als Darstellungen durch klassische Größen wie Ort und Impuls, welche als Observablen direkt messbar
sind. Bei einem allgemeinen Zustand, welcher in Bezug auf eine bestimmte Observable entartet ist, ist es demgemäß sinnlos
anzunehmen, dass nur ein definierter Wert dieser beobachtbaren Größe eine wirkliche Beschreibung des Systems darstellen
würde und der Zustand nur als Wahrscheinlichkeitsverteilung wirkliche Realität beansprucht. Der Zustand selbst muss als die
eigentliche fundamentale Beschreibung der physikalische Realität angesehen werden, während die Observablen nur unvollkommenen
makroskopischen Begriffen entsprechen, denen lediglich in Bezug auf menschliche Erfahrung grundsätzliche Bedeutung zukommt.
Hierzu sei erneut Mittelstadt zitiert \cite{Mittelstaedt:1963}:

\begin{quote}
{\small Die Tatsache, dass $|\varphi \rangle$ der Zustand des Systems ist, ist eine objektive Eigenschaft von $S$ in dem Sinne, 
dass der Operator $P_\varphi=|\varphi \rangle \langle \varphi|$ den Meßwert 1 besitzt und nicht 0. Die Koeffizienten $\langle
A_i|\varphi \rangle=\varphi(A_i)$ sind nichts anderes als eine spezielle Darstellung des Zustandes $|\varphi \rangle$ und
besagen in ihrer Gesamtheit nichts weiter, als dass das System $S$ sich eben in diesem Zustand $|\varphi \rangle$ 
befindet. $|\varphi \rangle$ beziehungsweise $P_\varphi$ ist daher als eine objektive Beschreibung des Systems aufzufassen.

Peter Mittelstaedt, Philosophische Probleme der modernen Physik, 1963 (Seite 70)}
\end{quote}
Diese Anschauung, welche davon ausgeht, dass Zustände einen höheren ontologischen Status als direkt messbare Größen haben, 
wird durch das Faktum gestützt, dass die Phase einer quantentheoretischen Wellenfunktion im Ortsraum noch nicht einmal
Wahrscheinlichkeiten direkt messbarer Größen korrespondiert. Dennoch sind Phasenbeziehungen für die Interpretation
bestimmter Interferenzeffekte wie etwa des Aharanov-Bohm-Effektes \cite{Aharanov:1959} unentbehrlich und müssen daher
als ontische Realität angesehen werden.

\section{Die Nichtlokalität der Quantentheorie}

Die für die Kopenhagener Deutung und die obige Behandlung des Messproblems konstitutive Eigenschaft der Nichtlokalität der
Quantentheorie soll nun in systematischer Weise durch eine Liste von Argumenten aufgezeigt werden. Sie manifestiert sich
sowohl in konkreten Experimenten als auch im allgemeinen mathematischen Formalismus.

\subsection{Manifestation der Nichtlokalität in konkreten Phänomenen}

\noindent
\textbf{Doppelspaltexperiment:}\\
\\
Beim Doppelspaltexperiment wird monochromatisches kohärentes Licht auf eine Blende
mit zwei Spalten gestrahlt. Ein sich hinter der Blende befindender Schirm fängt das 
Licht auf. Man kann nun die folgende Beobachtung machen:\\
Wenn beide Spalten der Blende geöffnet sind, so ergibt sich ein anderes Interferenzbild als wenn nacheinander
jeweils nur ein Spalt geöffnet wird und der andere verschlossen bleibt. Diese Situation liegt auch dann vor,
wenn die Intensität der Strahlung so gering ist, dass die Einschläge einzelner Photonen auf dem Schirm
nacheinander registriert werden können. In Anlehnung an diese Beobachtung kann man nun wie folgt
argumentieren:\\ 
1) Wenn das Photon im Falle, dass beide Spalten geöffnet sind, nur durch einen der beiden Spalten geht,
so muss sein Verhalten unabhängig davon sein, ob der andere Spalt geöffnet ist.\\
2) Daraus ergibt sich, dass sich unter der Annahme, dass das Photon nur durch einen der beiden Spalten geht,
im Falle, dass beide Spalten geöffnet sind, das gleiche Interferenzmuster ergeben müsste, wie im Falle, dass
nacheinander jeweils nur ein Spalt geöffnet wird.\\
3) Unter dieser Annahme ergibt sich aber offenkundig ein Widerspruch zur Beobachtung, die zwei verschiedene
Interferenzmuster zeigt.\\
4) Die Annahme, dass das Lichtquant nur durch einen der beiden Spalten geht, muss also falsch sein.\\
Weiter tritt das Lichtquant aber als ein einheitliches Objekt auf, dessen Wechselwirkungsverhalten nicht weiter unterteilbar
ist. Dies zeigt sich an der Wechselwirkung mit dem Schirm. Das Interferenzmuster besteht aus diskreten Punkten, die einzelne
diskret voneinander unterscheidbare Prozesse darstellen. Das Lichtquant lässt sich also nicht wie ein Feld im Sinne einer
lokalen Feldtheorie in verschiedene Sektionen räumlich aufteilen, die miteinander in Wechselwirkung stehen, und dennoch läuft
es sozusagen durch beide Spalten gleichzeitig. Dies scheint nun zwangsläufig zu der Deutung der Beobachtung im Rahmen des
Doppelspaltexperimentes und der sich daran anschließenden Argumentation zu führen, dass es tatsächlich nicht sinnvoll ist,
davon auszugehen, dass sich das Photon auf einer klassischen Teilchenbahn  bewegt. Es handelt sich beim im
Doppelspaltexperiment beobachteten Phänomen vielmehr um einen Vorgang, der durch eine räumliche Beschreibungsweise im Sinne
eines lokalisierten Objektes auf einer Bahn nicht adäquat verstanden werden kann.\\
\\
\textbf{Einstein-Podolsky-Rosen-Paradoxon:}\\
\\
Das Einstein-Podolsky-Rosen-Paradoxon wurde von Einstein und seinen Assistenten Boris Podolsky und Nathan Rosen im Rahmen
eines Gedankenexperimentes formuliert, mit dem Einstein die Unvollständigkeit der Quantenmechanik aufzuzeigen glaubte
\cite{Einstein:1935rr}. Bohr reagierte auf die Arbeit, in der Einstein, Podolsky und Rosen dieses Gedankenexperiment
darstellten, mit einer eigenen den gleichen Titel tragenden Arbeit, indem er seine Einwände gegenüber der Einsteinschen
Deutung des Gedankenexperimentes im Sinne einer Unvollständigkeit der Quantenmechanik formulierte \cite{Bohr:1935af}.
Mittlerweile wurde dieses Experiment in sehr vielen unterschiedlichen Varianten durchgeführt.
Das Einstein-Podolsky-Rosen-Paradoxon bezieht sich auf ein Phänomen, dass gemäß der Quantenmechanik in folgenden experimentellen
Situation in Erscheinung tritt und mittlerweile auch beobachtet wurde: Zwei Teilchen wechselwirken miteinander. Nach der
Wechselwirkung sei eine bestimmte Observable des aus beiden Teilchen bestehenden quantenmechanischen Systems bekannt,
beispielsweise der Gesamtspin des Systems in Bezug auf eine bestimmte Richtung. Die diesbezügliche Observable jedes einzelnen
Teilchens sei jedoch entartet, also in diesem Falle die Spineinstellung in Bezug auf eine bestimmte Raumrichtung der beiden
Teilchen für sich genommen. Weiter soll die Situation so gewählt sein, dass sich die beiden Teilchen nach der Wechselwirkung
voneinander entfernen. Wenn nun, nachdem die Teilchen eine bestimmte Distanz zueinander erreicht haben, die Spineinstellung in
Bezug auf die entsprechende Raumrichtung eines der beiden Teilchen gemessen wird, so stellt sich heraus, dass bei einer
anschließenden Messung der Spineinstellung des anderen Teilchens derjenige Wert herauskommt, der zu dem vorher bereits bekannten
Gesamtspin des Systems führt.\\
Dies bedeutet, dass die Messung an dem einen Teilchen eine instantane Wirkung auf den Zustand des anderen Teilchens hervorruft.
Dies scheint zur Konsequenz zu haben, dass die beiden Teilchen ein System bilden, das auf einer nicht-räumlichen Ebene ein
individuelles Objekt darstellt. Genau auf dieses Phänomen bezieht sich auch der Bohrsche Begriff der Individualität.

\subsection{Manifestation der Nichtlokalität im mathematischen Formalismus}

Natürlich ist die in den oben thematisierten Phänomenen beobachtete für die Quantentheorie charakteristische Eigenschaft der
Nichtlokalität in dazu korrespondierender Weise auch im mathematischen Formalismus der Quantentheorie enthalten, der ja ein
allgemeines Schema zur Beschreibung dieser Phänomene darstellt.\\
\\
\newpage\noindent
\textbf{Heisenbergsche Unbestimmtheitsrelation:}\\
\\
Die Nichtlokalität der Quantentheorie zeigt sich natürlich in der Heisenbergschen Unbestimmtheitsrelation
($\ref{Unbestimmtheitsrelation_Ort-Impuls}$),
welche die Möglichkeit einer gleichzeitigen beliebig genauen Bestimmung von Ort
und Impuls begrenzt. Sie ist nicht Ausdruck einer praktischen Unmöglichkeit der Durchführung einer solchen Messung, sondern
hat in dem Sinne prinzipiellen Charakter, als sie eine Konsequenz der Quantenmechanik ist, in welcher dementsprechend keine
Zustände existieren, in denen Ort und Impuls in einer über durch die Unbestimmtheitsrelation definierte Grenze hinausgehenden
Genauigkeit bestimmt wären. Damit existiert aber auch keine exakt definierte Bahn für die Bewegung eines Teilchens, den diese
würde die Existenz eines exakten Ortes und eines exakten Impulses voraussetzen. Und eben die Nichtexistenz einer Teilchenbahn
war ja auch das, was aus dem im Doppelspaltexperiment beobachteten Phänomen geschlossen wurde. Aus diesem Grunde hat Heisenberg
davon gesprochen, dass quantenmechanische Vorgänge keine Vorgänge in Raum und Zeit darstellen \cite{Heisenberg:1958}.\\
\\
\textbf{Charakterisierung von Elementarteilchen:}\\
\\
Durch die Unbestimmtheitsrelation wird auch der Teilchenbegriff an sich problematisch. Eigentlich ist es nicht möglich, in
anschaulichen Begriffen zu sagen, was ein Elementarteilchen eigentlich ist. Es existiert allerdings eine bestimmte Art und
Weise, wie ein Teilchen im Rahmen relativistischer Quantenfeldtheorien beschrieben wird. Es wird im Rahmen dieser Beschreibung
durch einen Zustand in einem Hilbert-Raum charakterisiert, in dem eine irreduzible Darstellung der Poincar\'{e}-Gruppe
möglich ist und weist zudem bestimmte durch Quantenzahlen bezeichnete Eigenschaften mit dazugehörigen inneren Symmetriegruppen
auf (siehe \textbf{[6.2.1]}). Die mit den inneren Symmetrien verknüpften Eigenschaften haben endgültig überhaupt keinen Bezug zum
physikalischen Ortsraum und zur Anschauungswelt mehr. Es ist vielleicht die entscheidende Entdeckung der Elementarteilchenphysik,
dass auf fundamentaler Ebene zur Charakterisierung dessen, was wir als Materie bezeichnen, nicht konkrete Begriffe aus der
Anschauungswelt, sondern abstrakte mathematische Begriffe verwendet werden müssen. Am Beginn steht nicht der Begriff eines
ausgedehnten Teilchens, sondern der einer mathematischen Größe, nämlich einer Symmetrie. Die Elementarteilchenphysik ist also
damit tatsächlich zu jener Art der Naturbeschreibung vorgedrungen, wie sie, zumindest bezüglich des eigentlichen philosophischen
Kernes, gerade bereits Platon im Dialog Timaios dargestellt hat (siehe \textbf{[1.3.2]}). Dort waren es zwar regelmäßige Dreiecke,
die auf fundamentaler Ebene die Materie charakterisierten und damit war also in der konkreten Ausführung der Platonischen
Philosophie noch eine spezielle näherungsweise auch in der Anschauungswelt auftauchende Gestalt dominierend. In dieser Hinsicht
kann die moderne Physik Platon nicht folgen. Was aber die eigentliche philosophische Anschauung anbelangt, die darin besteht,
die Natur auf fundamentaler Ebene in reine mathematische Struktur oder Form aufzulösen, so scheint sich die Entwicklung der
modernen Physik in völligem Einklang mit dem philosophischen Geist Platons zu ereignen, worauf Heisenberg immer wieder hinwies,
wie etwa im folgenden Zitat \cite{Heisenberg:1969}:

\begin{quote}
{\small "`Am Anfang war die Symmetrie"', das ist sicher richtiger als die Demokritsche Behauptung "`Am Anfang war
das Teilchen"'. Die Elementarteilchen verkörpern die Symmetrien, sie sind ihre einfachsten Darstellungen, aber sie sind
erst eine Folge der Symmetrien.

Werner Heisenberg, Der Teil und das Ganze, 1969 (Seite 280)}
\end{quote}

\noindent
\textbf{Allgemeiner mathematischer Formalismus im Sinne Diracs und von Neumanns:}\\
\\
In der allgemeinen mathematischen Fassung der Quantentheorie (siehe \textbf{[5.2]}), wie sie durch Dirac und von Neumann
formuliert wurde, tauchen daher konkrete auf die Anschauungswelt verweisende Begriffe auch überhaupt nicht mehr auf, worauf
bereits in \textbf{[5.2.2]} hingewiesen wurde. Nur der Zeitbegriff als basalster der physikalischen Begriffe erscheint in dieser
Beschreibung. Die allgemeine Quantentheorie kann damit völlig ohne Bezug auf den physikalischen Anschauungsraum formuliert
werden. Dies ist letztlich die Zusammenfassung der beiden oben beschriebenen Aspekte der Heisenbergschen
Unbestimmtheitsrelation und der Charakterisierung der Elementarteilchen durch abstrakte Symmetrien. In der Quantentheorie
erscheinen der Orts- und der Impulsraum als Darstellungsräume für abstrakte Zustände (siehe \textbf{[5.3.1]}). Die
Unbestimmtheitsrelation ist letztlich Ausdruck der Grenze der Möglichkeit einer Erfassung der Realität mit diesen Begriffen,
die an sich nicht-räumlich ist. Die Existenz der inneren Symmetrien, die einer räumlichen Beschreibung grundsätzlich nicht mehr
zugänglich sind, zeigt dies in gleicher Weise. Und die diesen Beschreibungsweisen zugrundeliegende allgemeine Theorie ist eben
die Quantentheorie in ihrer abstrakten Form. Die Tatsache, dass sich diese allgemeine Formulierung ohne Verwendung des
Raumbegriffes ergibt, zeigt sehr deutlich, dass die Räumlichkeit gemäß der Quantentheorie keine fundamentale Eigenschaft
der Natur darzustellen scheint.

\part{Die Quantentheorie der Ur-Alternativen}

\chapter{Die abstrakte Quantentheorie als fundamentale Naturtheorie}

\section{Einleitung}

\begin{quote}
{\small
Wir haben es mit einer unvollendeten, aber, wie mir scheint, aussichtsreichen Theorie zu tun. Ihr Fortschritt wäre rascher
gewesen, wenn es mir gelungen wäre, mehr von den heute lebenden theoretischen Physikern für Ihre Fragestellung zu
interessieren. Sie versucht nicht, wie es häufig vor einer Kuhnschen Revolution und in der Elementarteilchenphysik auch
heute der Fall ist, die ungelösten Probleme im Rahmen der alten Begriffe durch Modelle von zunehmender Kompliziertheit zu
lösen. Sie versucht auch nicht, ein neues Paradigma durch Phantasie zu erraten. Sie versucht vielmehr, die prinzipiellen
Probleme der Quantentheorie so konsequent wie möglich zu behandeln und die Lösungsansätze für spezielle Probleme
hieraus zu gewinnen. "`Nur der wahre Konservative kann ein wahrer Revolutionär sein."'(Heisenberg)

Carl Friedrich von Weizsäcker, Zeit und Wissen, 1992 (Seite 318)}
\end{quote}

\noindent
Im Rahmen sehr grundsätzlicher Überlegungen über die begrifflichen Grundlagen der Physik versuchte Carl Friedrich von Weizsäcker
in der zweiten Hälfte des zwanzigsten Jahrhunderts in Anlehnung an bestimmte für die Kantische Philosophie konstitutive Ideen
die bekannte empirisch gefundene Physik philosophisch zu begründen \cite{Weizsaecker:1971},\cite{Weizsaecker:1955},
\cite{Weizsaecker:1985},\cite{Weizsaecker:1992}. Dieses von Weizsäckersche Programm der Begründung der Physik, auf das sich das
oben gegebene Zitat bezieht \cite{Weizsaecker:1992}, setzt sich prinzipiell aus zwei Teilen zusammen, wobei der zweite Teil aus
dem ersten Teil in natürlicher Weise erwächst.

Der erste Teil des Programmes besteht in der Rekonstruktion der abstrakten Quantentheorie. Im Rahmen dieses Unternehmens
versucht von Weizsäcker zunächst zu verstehen, warum gerade die Quantentheorie in ihrer allgemeinen Form (siehe \textbf{[5.2]})
so universell gültig zu sein scheint. Hierbei wird davon ausgegangen, dass die Quantentheorie als
allgemeine Theorie menschlichen Wissens über die Natur anzusehen sei. Dies ist im Sinne der Kopenhagener Deutung so zu
verstehen, dass die Quantentheorie von einer objektiven Realität in der Natur handelt, diese aber in einer für den menschlichen
Geist spezifischen Weise darstellt. Die Rekonstruktion geschieht durch ein philosophisches Programm, welches durch die
Erkenntnistheorie Kants, wie sie in der "`Kritik der reinen Vernunft"' dargestellt ist, inspiriert ist. Hierbei wird versucht,
die allgemeine Quantentheorie aus den Bedingungen der Möglichkeit von Erfahrung herzuleiten. Wenn man nämlich zeigen könnte,
dass die allgemeinen Naturgesetze, die wir ausgehend von der Erfahrung gewonnen haben, sich als Konsequenz der Bedingungen der
Möglichkeit von (menschlicher) Erfahrung herausstellen würden, so wären sie deshalb notwendigerweise wahr, da wir ansonsten
eben überhaupt keine Erfahrung über die Natur haben könnten. Wesentlich hierfür sind der Begriff der Zeit in ihren drei Modi
Vergangenheit, Gegenwart und Zukunft, welche von Weizsäcker als die grundlegendste Bedingung der Möglichkeit von Erfahrung sieht
und die allgemeine Schematisierung des Inhalts einer Erfahrung durch Alternativen.

Der zweite Teil des von Weizsäckerschen Programmes besteht darin, anschließend an die Begründung der universellen Gültigkeit
der Quantentheorie im Rahmen dieses Programmes den Versuch zu unternehmen, die reale Physik konkreter Objekte, wie sie etwa im
Rahmen relativistischer Quantenfeldtheorien beschrieben wird, aus der abstrakten Quantentheorie zu begründen, die ja zunächst
von speziellen Objekten abstrahiert. Konstitutiv für dieses Unternehmen ist die Möglichkeit der logischen Zerlegung einer
Alternative in Subalternativen, was schließlich auf binäre Alternativen als den basalsten Konstituenten menschlicher Erfahrung
führt, welche quantentheoretisch betrachtet durch zweidimensionale komplexe Vektorräume beschrieben werden. Da eine Zerlegung
in binäre Alternativen eine grundsätzliche Grenze für eine weitere logische Zerlegung darstellt, bezeichnet von Weizsäcker
diese Alternativen als Ur-Alternativen, um ihren grundsätzlichen Charakter zum Ausdruck zu bringen.
Dementsprechend wird dieser zweite Teil aufgrund der hierfür konstitutiven Annahme der Zerlegbarkeit aller Zustände in
Ur-Alternativen beziehungsweise den dieser Annahme korrespondierenden Ur-Objekten als "`Quantentheorie der Ur-Alternativen"' oder
als "`Ur-Theorie"' bezeichnet. Eine wesentliche Errungenschaft dieser Betrachtung ist die Möglichkeit, die Existenz eines
dreidimensionalen Ortsraumes, der mit der Zeit im Sinne der Speziellen Relativitätstheorie zur Minkowski-Raum-Zeit verbunden
ist, in welcher alle Objekte dargestellt werden können, aus der Quantentheorie ohne weitere spezielle Annahmen über die in der
Natur vorkommenden Entitäten zu begründen. Dies steht mathematisch gesehen in Zusammenhang mit der Tatsache, dass die
$SU(2)$ und die $SL(2,\mathbb{C})$ inhärent im Begriff der quantentheoretischen Fassung einer Ur-Alternative enthalten
sind, sofern diese mathematisch beschrieben wird.
Die $U(2)=SU(2)\times U(1)$ ergibt sich als universelle Symmetriegruppe der Quantentheorie der Ur-Alternativen, da die
Ur-Alternativen eben diese Symmetriegruppe konstituieren, was auf einen dreidimensionalen kompakten Raum als Ortsanteil des
Kosmos führt. Weiter kann man unter Betrachtung des Zustandsraumes vieler Ur-Alternativen auf direktem Wege zur konformen Gruppe
der Speziellen Relativitätstheorie als fundamentaler Symmetriegruppe der Natur gelangen. Damit wäre auch die Beschreibung von
Elementarteilchen begründet, deren mögliche Zustände im Sinne Eugene Wigners quantentheoretische Zustandsräume bilden,
welche irreduzible Darstellungen der Poincar\'{e}-Gruppe ermöglichen. Zur Begründung der Systematik der Elementarteilchen,
ihrer inneren Symmetrien und vor allem der Wechselwirkungen, insbesondere der Gravitation als allgemeinster Wechselwirkung
beziehungsweise dem Bezug zur Allgemeinen Relativitätstheorie als Rahmen für die Beschreibung der Gravitation, existieren bisher
nur aussichtsreiche Ansätze, die als programmatisch verstanden werden können.

Der Sinn der Kapitel 8,9 und 10 besteht darin, die wesentlichen Errungenschaften der von Weizsäckerschen Quantentheorie der
Ur-Alternativen herauszustellen, wie sie in den Schriften \cite{Weizsaecker:1971},\cite{Weizsaecker:1985} und \cite{Weizsaecker:1992}
dargestellt ist. Bestimmte Aspekte der Theorie sind auch in
\cite{Lyre:1994eg},\cite{Lyre:1995gm},\cite{Lyre:2001},\cite{Lyre:2003tr},\cite{Drieschner:2002} widergegeben.

\section{Rekonstruktion der abstrakten Quantentheorie und Ur-Alternativen}

\subsection{Allgemeines Programm}

Das von Weizsäckersche Programm der Rekonstruktion der Physik, welche die Quantentheorie als fundamentale Naturtheorie
begründen soll, ist durch die Kantische Transzendentalphilosophie inspiriert. Wesentlich ist, wie bereits erwähnt, die Kantische
Anschauung, dass es Bedingungen der Möglichkeit von (menschlicher) Erfahrung gibt (siehe $\textbf{[1.1.1]}$), welche bereits
a priori bestimmen, in welche grundlegende Form die speziellen Erfahrungen (des Menschen) eingebettet sein werden.
Ähnlich wie bei Kant sollen die grundlegenden Naturgesetze, welche aus der Erfahrung gewonnen werden können, Bedingungen der
Möglichkeit von Erfahrung darstellen. Wenn es gelingen würde, die grundlegenden Gesetze der Physik aus den Bedingungen der
Möglichkeit von Erfahrung herzuleiten, müssten sie zwangsläufig in einer beliebigen Erfahrung gültig sein, da sie ja eben
Bedingungen dieser Erfahrung darstellen, ohne ihre Gültigkeit diese Erfahrung also überhaupt nicht möglich wäre.
Das von Weizsäckersche Programm ist aber bezüglich dessen noch konsequenter als Kant in seiner Erkenntnistheorie. Denn Kant
ging nur davon aus, dass die allgemeine Form unserer Erfahrung festgelegt sei, nicht aber die konkreten Inhalte. Hingegen
basiert das von Weizsäckersche Programm auf der Hoffnung, nicht nur die allgemeine Form unserer Erfahrung, sondern auch deren
konkrete Inhalte selbst herzuleiten, also auch die speziellen Naturgesetze. Da diese nach bisheriger Kenntnis auf der
Quantentheorie basieren, versucht von Weizsäcker also die allgemeine Quantentheorie aus den Bedingungen der Möglichkeit von
Erfahrung herzuleiten beziehungsweise ihre Identität mit letzteren zu zeigen. Dieses Idealziel ist bisher nicht erreicht worden.
\footnote{Ob es überhaupt möglich ist, es zu erreichen, darüber soll hier nicht diskutiert werden. Es ist allerdings für die
weitere Argumentation entscheidend, dass die Rekonstruktion der Quantentheorie in jedem Falle zu einer prinzipiellen
Interpretation der Quantentheorie als Theorie der Information führt, aus der konkrete physikalische Begriffe erst begründet
werden. Dies bedeutet, dass es in Bezug auf die aus dieser Rekonstruktion erwachsende Quantentheorie der Ur-Alternativen wichtig
ist, dass die abstrakte Quantentheorie als fundamentale Naturtheorie vorausgesetzt und als allgemeine Theorie menschlichen
Wissens über die Natur interpretiert wird.}
Es ist für die von Weizsäckersche Begründung der allgemeinen Quantentheorie neben epistemischen Postulaten, welche sich auf
die Bedingungen der Möglichkeit von Erfahrung beziehen, zusätzlich die Einführung realistischer Postulate notwendig, die
ontische Annahmen über die Welt darstellen, ohne dass (bisher) hätte gezeigt werden können, dass es sich bei Ihnen um 
Bedingungen der Möglichkeit von Erfahrung handelt.

Als grundlegende Bedingung der Möglichkeit von Erfahrung sieht von Weizsäcker die Struktur der Zeit in ihren drei Modi
Vergangenheit, Gegenwart und Zukunft: Erfahrung zu machen bedeutet aus der Vergangenheit für die Zukunft zu lernen. In der Physik
werden mit Hilfe von Erfahrung über vergangene Ereignisse, Vorhersagen für zukünftige Ereignisse gemacht. Die Physik stellt
also basierend auf Dokumenten über die Vergangenheit Aussagen über die Zukunft auf. Für Aussagen, welche sich auf die Zukunft
beziehen, gilt aber das "`Tertium non Datur"', der Satz vom ausgeschlossenen Dritten, als grundlegender Bestandteil der
Klassischen Logik nicht mehr, der besagt, dass einer Aussage als Wahrheitswert entweder "`wahr"' oder "`falsch"' zukommt.
Diese Aufhebung der Gültigkeit des "`Tertium non Datur"' in Bezug auf zeitliche Aussagen, rührt von der Unbestimmtheit der
Zukunft als Wesensmerkmal der Struktur der Zeit.
Damit wird man in Bezug auf die Beschreibung von zeitlichen Aussagen zum Begriff der Wahrscheinlichkeit geführt, die ja in der
Quantentheorie eine wichtige Rolle spielt. Wahrscheinlichkeiten beschreiben in Bezug auf eine Alternative von Aussagen
allgemeinere Wahrheitswerte als "`wahr"' und "`falsch"'. Dies definiert eine sogenannte Metaalternative, welche nämlich durch die
Menge der Wahrscheinlichkeitsaussagen über die Ursprungsalternative konstituiert wird. Auf diese Metaalternative muss aber
erneut die zeitliche Logik angewandt werden, was zu einer Iteration führt.
Die grundlegende Rolle, welche die Zeit im Rahmen des Programmes der Rekonstruktion der abstrakten Quantentheorie spielt,
impliziert nun, dass der Zeit eine fundamentalere Bedeutung als dem Raum zukommt, der im Gegensatz zur Zeit weder in den
Postulaten der allgemeinen Quantentheorie vorkommt noch in der von Weizsäckerschen Rekonstruktion als epistemisches oder
ontisches Postulat zu ihrer Rekonstruktion vorausgesetzt wird.
\footnote{Der genaue Zusammenhang der Kantischen Vorstellung der Natur des Raumes zu der durch die Quantentheorie der
Ur-Alternativen implizierten Vorstellung der Natur des Raumes wird in \textbf{[9.4]} noch genauer erörtert werden und ist von
zentraler Bedeutung für die Quantentheorie der Ur-Alternativen und vermutlich auch für das Verhältnis von Quantentheorie und
Allgemeiner Relativitätstheorie.}
Dies widerspricht der Speziellen Relativitätstheorie allerdings in keiner Weise, denn diese hat den Unterschied von Raum und
Zeit nicht aufgehoben, sondern diese beiden physikalischen Realitäten nur in eine neue Beziehung zueinander gesetzt. Auch kann
die Zeit nicht aus der Thermodynamik begründet werden, da man zur Begründung des zweiten Hauptsatzes den
Wahrscheinlichkeitsbegriff voraussetzen muss, der aber seinerseits ein auf die Struktur der Zeit bezogener Begriff ist.
Generell ist es wohl völlig ausgeschlossen, die Zeit in irgendeiner Weise auf andere physikalische Begriffe zurückzuführen,
da ihr ein viel grundsätzlicherer, über die Physik weit hinausgehender ontologischer Status zukommt.
\footnote{Die Zeit selbst wird deshalb in diesem Zusammenhang nicht näher philosophisch erörtert werden. Es sei hier nur darauf
hingewiesen, dass auch Kant die Zeit als noch grundsätzlicher als den Raum ansah, denn sie liegt nach Kant unserer inneren
Anschauung zu Grunde und ist damit auch aller Anschauung der Außenwelt übergeordnet, während der Raum nur unserer Anschauung
der äußeren Gegebenheiten zu Grunde liegt. Dies könnte man auch so ausdrücken, dass die Zeit im Gegensatz zum Raum nicht nur ein
auf die Wahrnehmung physikalischer Geschehnisse bezogenes, sondern auch ein unser Bewusstsein selbst bestimmendes Phänomen ist.}

Die von Weizsäckersche Rekonstruktion der Quantentheorie basiert nun weiter auf dem Begriff der empirisch entscheidbaren
Alternative. Dieser Begriff stellt eine allgemeine Schematisierung der Information dar, die man im Rahmen einer beliebigen
Beobachtungssituation über eine bestimmte durch diese Beobachtung untersuchte Gegebenheit gewinnen kann.\\
\\
\fbox{\parbox{145 mm}{\textbf{Definition:} Eine n-fache empirisch entscheidbare Alternative ist eine Menge von n
Aussagen/Zuständen, von denen sich genau eine als wahr/gegenwärtig erweisen wird, wenn eine empirische Prüfung
gemacht wird.}}\\
\\
Aufgrund der Struktur der Zeit ist es nun im Allgemeinen nicht möglich, eine definitive Vorhersage darüber zu machen,
welche Aussage einer solchen empirisch entscheidbaren Alternative sich als gegenwärtig erweisen wird, sondern es sind nur
Wahrscheinlichkeitsaussagen möglich. $n$ beschreibt eine natürliche Zahl, wobei $n \geq 2$, da eine binäre Alternative die 
einfachste Alternative darstellt, deren Entscheidung zu einem Informationsgewinn führt.

In seinem Buch "`Aufbau der Physik"'\cite{Weizsaecker:1985} schildert Carl Friedrich von Weizsäcker hiervon ausgehend vier
verschiedene Wege, um die Quantentheorie zu rekonstruieren. Alle vier Wege setzen jedoch auch noch Annahmen über die Realität
voraus, die keine Bedingungen der Möglichkeiten von Erfahrung darstellen oder von denen dies bisher zumindest noch nicht gezeigt
werden konnte. Im Rahmen dieses Kapitels soll nun der Weg "`über Wahrscheinlichkeiten direkt zum Vektorraum"'
(im Buch "`Aufbau der Physik"' als "`zweiter Weg"' aufgeführt) geschildert werden, der laut von Weizsäcker das Problem der
Rekonstruktion bisher am besten löst. Der "`vierte Weg"' geht bereits von Ur-Alternativen aus, welche zur konkreten Beschreibung
der realen Welt führen sollen, wie sie sich uns in der Erfahrung darstellt.

\subsection{Rekonstruktion auf dem Weg über Wahrscheinlichkeiten direkt zum Vektorraum}

Ausgehend von der Struktur der Zeit als grundlegender Bedingung der Möglichkeit von Erfahrung und dem Begriff der empirisch
entscheidbaren Alternative werden nun auf dem "`zweiten Weg"' der Rekonstruktion der Quantentheorie \cite{Weizsaecker:1985}
die folgenden drei grundlegenden Postulate formuliert:\\
\\
\textbf{1.Trennbarkeit der Alternativen:} Es gibt trennbare Alternativen, also Alternativen, deren Ergebnis nicht von der
Entscheidung anderer Alternativen abhängt.\\
\\
\textbf{2.Erweiterung der Menge der Zustände:} Zu jedem Paar $x$ und $y$ von Zuständen gibt es (wenigstens) einen von ihnen
untrennbaren Zustand $z$ (Trennbarkeit ist hier im Sinne des ersten Postulats zu verstehen), der keinen der Zustände
ausschließt, sondern bedingte Wahrscheinlichkeiten $p(z,x)$ und $p(z,y)$ bestimmt, die von null und eins verschieden sind.\\
\\
\textbf{3.Kinematik:} Zustände ändern sich mit der Zeit. Dabei bleiben Wahrscheinlichkeitsrelationen zur selben Alternative
gehöriger Zustände unverändert:\\ $p[x(t),y(t)]=p[x(t_0),y(t_0)]$.\\
\\
\textbf{1.} Das Postulat der Trennbarkeit der Alternativen ist epistemisch, denn nur dann wenn eine Alternative unabhängig für sich
entschieden werden kann, ist es möglich, Information zu gewinnen. Wenn Alternativen grundsätzlich von anderen Alternativen
abhängen, so kann man nur Information gewinnen, indem man alle überhaupt denkbaren Alternativen entscheidet, was der Kenntnis
der in diesem Sinne verfügbaren Information über den gesamten Kosmos entsprechen würde, was natürlich unmöglich ist.
Allerdings kommen in der realen Welt keine völlig isolierten Objekte vor, da jedes Objekt grundsätzlich mit (allen)
anderen Objekten in Wechselwirkung steht. Damit formuliert dieses Postulat eine notwendige Näherung, die der Annahme
der Vernachlässigbarkeit äußerer Einflüsse durch Wechselwirkung entspricht, welche in der Physik grundsätzlich gemacht wird.
Dies bedeutet auch, dass die Quantentheorie an sich nur in der Näherung gültig ist, in der man von freien Objekten sprechen kann.
Man kann nun umgekehrt in der Weise argumentieren, dass man die Existenz des Phänomenes der Wechselwirkung als eine Folge der
Beschreibung der Welt durch als näherungsweise isoliert existierende Objekte annimmt. Die Wechselwirkung würde dann einer
Korrektur dieser Näherung entsprechen.
\footnote{Diese Thematik wird in \textbf{[10.2.3]} noch ausführlicher thematisiert werden und steht mit der Frage nach
der Natur des Raumes und der Natur von Objekten in Zusammenhang.}

\textbf{2.} Das Postulat der Erweiterung der Menge der Zustände ist ontisch, da es keine Bedingung der Möglichkeit von Erfahrung
darzustellen scheint. $z$ ist natürlich nicht von $x$ und $y$ trennbar, da $z$ bedingte Wahrscheinlichkeiten $p(z,x)$ und
$p(z,y)$ definiert, also Wahrscheinlichkeiten, $x$ beziehungsweise $y$ zu finden, wenn $z$ wahr ist. $z$ ist aber auch kein
Element der Alternative, der $x$ und $y$ angehören, denn eine Alternative ist eine Menge einander ausschließender Aussagen,
was bedeutet, dass $p(x,y)=0$ für alle Elemente der Alternative gilt. Deshalb wird das Verhältnis, indem $z$ zur Alternative
steht, in der Weise ausgedrückt, dass $z$ zur Alternative gehöre. Die Elemente einer Alternative sollen in diesem Sinne als
eine Teilmenge aller zu einer Alternative gehörigen Elemente verstanden werden.

\textbf{3.} Das Postulat der Kinematik stellt natürlich eine Annahme über die Wirklichkeit dar. Dass sich Zustände zwangsläufig
kontinuierlich mit der Zeit entwickeln, scheint nicht zwangsläufig eine Bedingung der Möglichkeit von Erfahrung zu sein.
\footnote{Von Weizsäcker fasst dieses Postulat als epistemisch auf. Der Zusammenhang der kontinuierlichen Zeitentwicklung
zur Offenheit des Ausganges einer Entscheidung einer Alternative ist eine grundsätzliche Schwierigkeit, die mit dem
quantentheoretischen Messproblem in Zusammenhang steht.}
Die Tatsache, dass die Wahrscheinlichkeitsrelationen der zu einer Alternative gehörigen Zustände sich mit der Zeit nicht ändern,
ist aber dennoch epistemisch zu verstehen, da ein zu einer Alternative gehöriger Zustand nur durch seine
Wahrscheinlichkeitsrelationen zu anderen Zuständen identifiziert werden kann. Dies können im Prinzip auch zu anderen
Alternativen gehörige Zustände sein, da eine Alternative von außen im Prinzip identifizierbar sein muss. Aber eine zweite
Alternative, mit Hilfe derer die Alternative von außen identifiziert werden kann, lässt sich durch das Cartesische Produkt zu
einer einzigen Alternative zusammenfassen, was bedeutet, dass ein Zustand in dieser zusammengefassten Beschreibungsweise wieder
durch Wahrscheinlichkeitsrelationen zu den anderen Alternativen definiert ist. Wenn aber eine Alternative in der Zeit bestehen
soll, so müssen die Zustände identifizierbar bleiben, was gemäß der obigen Argumentation nur durch gleichbleibende
Wahrscheinlichkeitsrelationen gewährleistet werden kann.\\
\\  
Diese drei Postulate führen zu den folgenden direkten Konsequenzen:\\
\\
\textbf{1.Zustandsraum:} Die Menge aller Zustände, welche zu einer Alternative gehören, also entweder Element der Alternative sind
oder gemäß dem Postulat der Erweiterung durch bedingte Wahrscheinlichkeiten bezüglich dieser Zustände definiert sind, bilden
den zur Alternative gehörigen Zustandsraum $S(n)$.\\
\\
\textbf{2.Symmetrie:} Alle Zustände des $S(n)$ sind äquivalent, lassen also keine innere Auszeichnung voreinander zu, sondern nur
in Bezug auf eine andere Alternative (Symmetrie ist also eine Folge der Trennbarkeit der Alternativen).\\
\\
\textbf{3.Dynamik:} Die Entwicklung aller Zustände muss durch eine einparametrige Untergruppe der Symmetriegruppe beschrieben
werden, deren Parameter die Zeit ist.\\
\\
\textbf{1.} Die Konsequenz des Zustandsraumes ergibt sich aus dem Postulat der Erweiterung.
Es muss immer genau $n$ Zustände der Alternative geben, für welche $p(x_i,x_j)=\delta_{ij},\ i,j=1...n$, da eine $n$-fache
Alternative aus $n$ einander ausschließenden Elementen besteht und die weiteren zur Alternative gehörigen Zustände durch
bedingte Wahrscheinlichkeiten dieser Elemente gekennzeichnet sind. Dies führt zur Struktur eines Vektorraumes $\mathbb{R}^n$,
wobei die Wahrscheinlichkeitsrelationen zwischen den Zuständen $p(x,y)$ eine Metrik definieren, da sie sich als Funktion
einer Bilinearform in $\mathbb{R}^n$ darstellen lassen.

\textbf{2.} Die Konsequenz der Symmetrie ergibt sich aus dem Postulat der Trennbarkeit der Alternativen. Wenn eine Alternative von
allen anderen Alternativen getrennt ist, so ist vor der Entscheidung der Alternative unbekannt, welches Element $x_i$ der
Alternative gefunden werden wird. Da zu einer Alternative gehörige Zustände durch ihre Wahrscheinlichkeitsrelationen zu anderen
Zuständen gekennzeichnet sind und die Elemente einer Alternative $p(x_i,x_j)=\delta_{ij},\ i,j=1...n$ erfüllen, sind die
Zustände isomorph zueinander. Ohne Bezug zu einer anderen Alternative ist auch nicht bestimmt, welche Zustände aus $S(n)$
Elemente einer Alternative sind, die entschieden werden wird. Das bedeutet, dass jeder Zustand $z_i$ im Zustandsraum $S(n)$ als
Element einer Alternative angesehen werden kann, deren Elemente eben diejenigen Zustände $z_j$ sind, welche bezüglich $z_i$ die
Wahrscheinlichkeiten $p(z_i,z_j)=\delta_{ij},\ i,j=1...n$ erfüllen. Die Symmetrie des Zustandsraumes entspricht also der
Tatsache, dass in der Alternative selbst ohne Bezug zu anderen Alternativen nur die Wahrscheinlichkeitsrelationen zwischen den
Zuständen absolut sind. Eine Zuordnung $x \rightarrow x^{\prime}$, bei der jedem Zustand $x$ aus $S(n)$ ein Zustand $x^{\prime}$ aus $S(n)$ zugeordnet wird, wobei für zwei beliebige Zustände $x$ und $y$ gilt: $p(x^{\prime},y^{\prime})=p(x,y)$, ist also eine Symmetrietransformation innerhalb von $S(n)$. Da sich die Wahrscheinlichkeitsrelationen gemäß der Charakterisierung des Zustandsraumes $S(n)$ als Vektorraum über Bilinearformen definieren lassen, führt dies zur $SO(n)$ als Symmetriegruppe.

\textbf{3.} Die Konsequenz der Dynamik ergibt sich aus dem Postulat der Kinematik zusammen mit der Konsequenz der Symmetrie, denn
gemäß dem Postulat der Kinematik müssen Wahrscheinlichkeitsrelationen zwischen Zuständen in der Zeit erhalten bleiben und der
Zustandsraum einer Alternative weist eben gerade eine Symmetrie unter der Menge aller wahrscheinlichkeitserhaltenden
Zuordnungen auf. Damit muss die Zeittransformation durch eine Untergruppe der Gruppe der Symmetrietransformationen beschrieben
werden, die durch einen Parameter beschrieben wird, der mit der Zeit identifiziert wird. Einer solchen Symmetrietransformation
entspricht ein Operator $H$, was auf die Gleichung $\frac{\partial x}{\partial t}=Hx$ führt.
Eine einparametrige Untergruppe der $SO(n)$ ist die $SO(2)$. Das bedeutet, dass die durch $H$ ausgedrückte Zeitentwicklung
Rotationen in jeweils zweidimensionalen Untervektorräumen von $S(n)$ entspricht. Daher ist es sinnvoll, einem Vektor $x$
aus $S(n)$ einen komplexen Vektor $c$ zuzuordnen: $\varphi_{j}=x_{2j-1}+ix_{2j},\ j=1...n$, wobei die Zustände $x_{2j-1}$ und
$x_{2j}$ jeweils eine Basis im entsprechenden Unterraum bilden, auf den die $SO(2)$ wirkt.
\footnote{Hierzu muss natürlich angenommen werden, dass Alternativen grundsätzlich eine gerade Zahl von Elementen aufweisen.}
$H$ wirkt daher auf den komplexen Vektor $\varphi$ in der Weise, dass er für jedes $j$ jeweils die komplexe Komponente in die reelle überführt und umgekehrt, was einer $U(1)$-Transformation entspricht: $\varphi_{j}(t)=e^{-i\omega_j t}\varphi_j^0$. Das bedeutet, dass sich als Gleichung für die Zeitentwicklung $i\frac{\partial \varphi_{j}}{\partial t}=H_{jk}\varphi_k$ mit 
$H_{jk}=\omega_{j}\delta_{jk}$ oder einfacher geschrieben $i\frac{\partial \varphi}{\partial t}=H\varphi$ ergibt. Dies kann als Schrödinger-Gleichung einer freien Alternative aufgefasst werden, wobei $H$ den freien Hamilton-Operator beschreibt.
\footnote{Es sei darauf hingewiesen, dass hier die Zeit beziehungsweise die Zeitentwicklung wie in den Postulaten der
Quantentheorie in einer zweifachen Weise auftritt, einerseits im Sinne der drei Modi, welche zur Unbestimmtheit der Entscheidung der Alternativen führen und andererseits im Sinne der kontinuierlichen Zeitentwicklung der Zustände, solange die Alternative nicht entschieden wird. Im Rahmen der Besprechung des Messproblems der dort einfach vorausgesetzten Quantentheorie in Kapitel 7 wurde diese Thematik bereits besprochen. Ich vermute ganz im Sinne meiner Ausführungen in Kapitel 7, dass die kontinuierliche Zeitentwicklung einen rein ontologischen Sinn hat und die Offenheit der Entscheidung einer Alternative eben mit jenem Mangel an Wissen zu tun hat, der durch das Postulat der Trennbarkeit der Alternativen zwangsläufig in die Beschreibung eingeführt wird. Dementsprechend hätte die Unbestimmtheit der Entscheidung einer Alternative eine rein epistemologische Bedeutung, was genau den Ausführungen in Kapitel 7 entspricht.}

Damit ist die Quantentheorie in ihrer allgemeinen Gestalt zumindest für den Fall endlicher Alternativen rekonstruiert, denen
endlichdimensionale Vektorräume entsprechen. Der Annahme, dass alle Alternativen im Prinzip endlich sind und die potentielle
Information in der Welt endlich ist, es aber keine obere Schranke für $n$ gibt, bezeichnet von Weizsäcker als offenen
Finitismus. Durch die Diskretheit der Basiszustände ergibt sich vor allem die Aussicht darauf, die in der üblichen
Quantenfeldtheorie auftretenden Unendlichkeiten von vorneherein in natürlicher Weise zu vermeiden.

\subsection{Ur-Alternativen}

Es ist nun im Prinzip möglich, jede endliche Alternative als das Cartesische Produkt von $n$ zweidimensionalen 
Alternativen beziehungsweise jeden endlichdimensionalen Hilbert-Raum als das Tensorprodukt von $n$ zweidimensionalen 
Zustandsräumen darzustellen

\begin{equation}
\mathcal{H}^m \subseteq T^n=\bigotimes_n \mathbb{C}^2,\quad m<2^n.
\end{equation}
Dies führt zum Begriff der Ur-Alternative beziehungsweise des durch diese quantentheoretisch formulierte Alternative
beschriebenen Ur-Objektes. Hiermit ist im Rahmen des Begriffes der Alternative eine grundsätzliche nämlich logische
Grenze der Teilbarkeit erreicht. Wenn der Begriff der Alternative wirklich als prinzipiell zur Beschreibung der Natur 
vorausgesetzt wird, so ist mit der Ur-Alternative wirklich eine fundamentale Ebene schlechthinniger Unteilbarkeit erreicht.
Insbesondere wird dem Problem der zweiten Kantischen Antinomie (siehe $\textbf{[1.1.2]}$) entgangen, denn auf der Ebene einer
Ur-Alternative existiert der Raumbegriff noch nicht, der sich ja gemäß der zweiten Kantischen Antinomie in Bezug auf die Annahme
der Existenz kleinster Objekte als so problematisch erwies. Wenn man nun von der prinzipiell verstandenen und mit Hilfe des
Begriffes der Alternative rekonstruierten allgemeinen Quantentheorie ausgeht und diese mit Hilfe von Ur-Alternativen darstellt,
so ergibt sich, dass alle in Bezug auf die Erfahrung beschriebenen konkreten physikalischen Entitäten sich grundsätzlich aus
Ur-Alternativen zusammensetzen und ihre Eigenschaften zumindest im Prinzip aus Ihnen begründet werden können. Der Versuch,
eine solche Beschreibung zu erreichen, ist der Inhalt der als zweitem Teil des von Weizsäckerschen Rekonstruktionsprogrammes
gekennzeichneten Theorie der Ur-Alternativen.\\
\\ 
\fbox{\parbox{145 mm}{Die Quantentheorie der Ur-Alternativen versucht die konkreten in der Natur existierenden Entitäten,
so wie die Gesetze, denen sie gehorchen, aus der allgemeinen Quantentheorie zu begründen, welche als eine allgemeine Theorie
der Information angesehen wird. Das bedeutet, dass auf der basalen Ebene nicht mehr von konkreten klassischen Begriffen
ausgegangen wird, auf welche dann nachträglich die Quantentheorie übertragen wird. Vielmehr wird die Quantentheorie mit
ihren sehr grundsätzlichen und abstrakten Begriffen als die eigentliche Naturbeschreibung vorausgesetzt und der Versuch
unternommen, die konkreten in der Physik auftauchenden Begriffe aus ihr zu begründen. Das bedeutet, dass man es hier mit
einer Beschreibung zu tun hat, die von ihrer Natur her rein quantentheoretisch ist, und daher insbesondere den Raumbegriff
(im Sinne des physikalischen Raumes) nicht voraussetzt.}}

\subsection{Die ontologische Bedeutung der Ur-Alternativen}

\begin{quote}
{\small Die Welt ist alles, was der Fall ist. Die Welt ist die Gesamtheit der Tatsachen, nicht der Dinge.

Ludwig Wittgenstein, Tractatus Logico-Philosophicus, 1918}
\end{quote}
Die Quantentheorie der Ur-Alternativen setzt also den Begriff der empirisch entscheidbaren Alternative beziehungsweise eben den
Begriff der Ur-Alternative als einfachster Alternative, in die sich alle Alternativen aus rein logischen Gründen im Prinzip
auflösen lassen, als die fundamentale Größe in der Naturbeschreibung voraus, die natürlich ihrerseits auf dem Begriff der Zeit
basiert, welche die Voraussetzung für eine empirische Entscheidung und überhaupt für jedes reale Geschehen darstellt, und dies,
wie bereits erwähnt, in einem viel allgemeineren als nur dem physikalischen Zusammenhang.
Es stellt sich aber nun die Frage nach dem ontologischen Status einer Ur-Alternative. Ist eine Ur-Alternative etwas objektiv
Existierendes oder handelt es sich nur um eine rein subjektive Beschreibungsweise, die sich auf etwas anderes objektiv
Existierendes bezieht ? In gewissem Sinne ist sie eigentlich beides. In diesem Zusammenhang ist natürlich wieder die Anschauung
wichtig, in welcher sich die Kantische Philosophie mit der Kopenhagener Deutung verbindet, nämlich dass die Realität an sich
nicht erkennbar ist, sondern nur ihre Darstellung innerhalb unserer Wahrnehmung. Diese Anschauung wird in der Theorie der
Ur-Alternativen selbst bis auf den Begriff der Ur-Alternative als basalem logischen Begriff zurückgeführt. Eine
quantentheoretische Ur-Alternative beschreibt nun in diesem Sinne eine elementare objektive Realität in einer für unseren Geist
erfassbaren Weise. Diese objektive Realität hat aber nicht materiellen, gegenständlichen Charakter in dem Sinne, dass die
Ur-Alternative die Information über die Eigenschaft eines materiellen Objektes enthalten würde. Vielmehr soll ja aus der
Kombination von Ur-Alternativen die Existenz konkreter Objekte erst begründet werden. Daher ist es auch gänzlich ausgeschlossen,
eine Ur-Alternative in irgendeiner Weise zu charakterisieren, die unabhängig von ihrer formalen Struktur wäre.
Eine Ur-Alternative ist das basalste Element der Darstellung einer an sich nicht erkennbaren Realität. In diesem
Sinne kann man auch von einem der Ur-Alternative entsprechenden Ur-Objekt reden, sofern man sich der Tatsache bewusst
bleibt, dass für dieses prinzipiell keine über die mit der Ur-Alternative gegebene Beschreibung hinausgehende konkrete 
Beschreibung auch nur denkbar ist.

Das bedeutet, dass die Quantentheorie der Ur-Alternativen in einem gewissen Sinne platonisch ist. Sie ist es insofern, als die
im Platonischen Dialog Timaios \cite{Platon:Timaios} dargelegte Vorstellung von den Grundkonstituenten der Materie davon
ausgeht, dass auf der basalen Ebene mathematische Formen anzutreffen sind. Diese stellen aber nicht im Sinne eines Dualismus
von Stoff und Form nur eine Art Akzidenz zu einer materiellen Substanz dar, sondern sie sind selbst Substanz. Dadurch wird die
Physik auf der fundamentalen Ebene auf Mathematik zurückgeführt. Dies wurde bereits in \textbf{[1.3.2]} behandelt. Im Unterschied
zu Platon, der diese mathematischen Grundgestalten als geometrische Größen ansieht, geht die Quantentheorie der
Ur-Alternativen von einer rein logischen Gestalt aus, aus der auch die Geometrie erst abgeleitet werden soll.

Deshalb zeigt sich hier auch eine deutliche Verwandtschaft zur Philosophie Wittgensteins, wie sie im
Tractatus Logico-Philosophicus \cite{Wittgenstein:1921} dargestellt ist, aus dem das obige Zitat stammt. Denn auch in der
Wittgensteinschen Philosophie basiert die Beschreibung der Welt auf Sachverhalten, die durch Sätze ausgedrückt werden, die
entweder wahr oder falsch sein können, und nicht auf Gegenständen. Diese Sachverhalte können in elementare Sachverhalte
aufgelöst werden. Sie beziehen sich aber nicht auf etwas Materielles, was unabhängig von ihnen bestehen würde, sondern sie
haben selbst ontologische Bedeutung.

\section{Zusammenfassende Darstellung der Indizien, welche die Quantentheorie der Ur-Alternativen als fundamentale Naturtheorie nahelegen}

In diesem Anschnitt sollen nun die grundlegenden Argumente, welche die von Weizsäckersche Quantentheorie der Ur-Alternativen als
Ansatz zu einer fundamentalen Naturtheorie nahelegen, in einer zusammenfassenden Weise aufgelistet werden.\\
\\
\textbf{1. Der ontologische Status des Raumes in der Quantentheorie und der Allgemeinen Relativitätstheorie}\\
\\
Beide Fundamentaltheorien, deren Zusammenführung in eine begriffliche Einheit ja als das wichtigste Problem bezüglich der Suche
nach einer einheitlichen Naturbeschreibung angesehen wird, sowohl die Quantentheorie als auch die Allgemeine Relativitätstheorie,
legen eine relationalistische Raumanschauung nahe, also eine Raumanschauung, welche den Raum nur als eine Beziehungsstruktur
zwischen Objekten und nicht als eigenständige ontologische Entität ansieht. In der Quantentheorie drückt sich dies in konkreten
Phänomenen wie dem Interferenzmuster beim Doppelspaltexperiment, sowie im allgemeinen mathematischen Formalismus aus, der keinen
physikalischen Raum voraussetzt (siehe \textbf{[7.4]}). In der Allgemeinen Relativitätstheorie ist es die Diffeomorphismeninvarianz
beziehungsweise die Tatsache, dass diese Theorie keine absoluten auf die Raum-Zeit bezogenen Größen aufweist, welche auf eine
relationalistische Raumanschauung führt (siehe \textbf{[4.2]}). In beiden Theorien geht man aber bei der konkreten Formulierung von
einer gegebenen Raum-Zeit aus. Der Relationalismus ist daher in der tatsächlichen Beschreibung nur implizit sichtbar. Auf einer
begrifflichen Ebene, auf der diese beiden Theorien zu einer Einheit zusammengefasst sind, aus der die beiden Theorien
hervorgehen, ist daher zu erwarten, dass der Raumbegriff überhaupt nicht mehr auftaucht, sich die Möglichkeit der räumliche
Beschreibungsweise allerdings als Konsequenz ergeben muss, da eine solche Beschreibungsweise ja zumindest näherungsweise
möglich ist.

Die Quantentheorie der Ur-Alternativen trägt dieser entscheidenden Erkenntnis bezüg-lich der Natur des Raumes wie sie
Quantentheorie und Allgemeine Relativitätstheorie nahelegen in grundsätzlicher Weise Rechnung, indem sie auf der fundamentalen
Ebene Objekte annimmt, die selbst noch nicht im Raum sind. Auf dieser Ebene gibt es gemäß der Theorie noch gar keinen
Raumbegriff. Hingegen wird die Existenz des physikalischen Raumes/der Minkowski-Raum-Zeit durch die Ur-Alternativen erst
konstituiert (siehe \textbf{[9.2]},\textbf{[9.3]} und \textbf{[9.4]}).\\
\\
\textbf{2. Raum und Materie innerhalb der Kantischen Erkenntnislehre}\\
\\
a) Raum und Zeit als Grundformen der Anschauung in der transzendentalen Ästhetik\\
\\
Laut der Kantischen Erkenntnistheorie, wie sie in der "`Kritik der reinen Vernunft"' dargestellt ist, sind Raum und Zeit an sich
keine ontischen Realitäten, sondern Arten und Weisen, wie der menschliche Geist objektive Inhalte einer an sich nicht
erkennbaren ontischen Realität darstellt (siehe \textbf{[1.1.1]}). Auch nach Umdeutung durch die Evolutionäre
Erkenntnistheorie kann nur der abstrakten Struktur des Raumes eine (näherungsweise) gültige Realität zugeschrieben werden
(siehe \textbf{[1.2.3]}). Es ist allerdings plausibel, den Raum als Anschauungsform hingegen weiterhin als eine Veranschaulichung
dieser Struktur anzusehen. Damit liefert Kant das erkenntnistheoretische Analogon zum Relationalismus der Quantentheorie und
der Allgemeinen Relativitätstheorie, welches sich darin ausdrückt, dass der physikalische Raum als Anschauungsform konstitutiver
Bestandteil menschlicher Wahrnehmung ist, obwohl ihm in der Realität lediglich eine Struktur entspricht.
In Bezug hierauf ist nun nicht nur entscheidend, wie bereits unter 1 erwähnt, dass der Quantentheorie der Ur-Alternativen die
Annahme einer nicht-räumlichen Realität auf der fundamentalen Ebene entspricht, sondern auch, dass durch sie plausibel gemacht
werden kann, warum es für den menschlichen Geist höchst zweckmäßig ist, die Welt im Rahmen der Anschauungsform des Raumes
darzustellen. Denn die eigentliche Realität ist hier eine abstrakte logische Struktur des menschlichen Wissens über die Natur.
Deren Darstellung durch Ur-Alternativen führt zu einer Struktur, welche der Raum-Zeit isomorph ist (siehe \textbf{[9.2]},
\textbf{[9.3]} und \textbf{[9.4]}). Diese Struktur ist aber nur eine Darstellung logischer Verhältnisse zwischen Informationseinheiten
über die Welt. Es ist jedoch möglich, sie ihrerseits in der (näherungsweise) isomorphen Anschauungsform des Raumes darzustellen.\\
\\
b) Zweite Kantische Antinomie\\
\\
Nach der Argumentation innerhalb der zweiten Kantischen Antinomie kann es keine kleinsten
räumlichen Objekte geben, da diese zumindest begrifflich immer noch weiter teilbar sind und sich die
Frage nach der Beziehung ihrer Teile stellt (siehe \textbf{[1.2.2]}).
Die Quantentheorie der Ur-Alternativen trägt dieser Erkenntnis Rechnung, indem sie keine kleinsten
räumlichen, sondern fundamentale einfachste Objekte in einem logischen Sinne annimmt. Diese
sind wirklich unteilbar, in einem schlechthinnigen Sinne. Denn eine weitere Teilbarkeit ist begrifflich
ausgeschlossen.\\
\\
c) Materie als "`Ding an sich"'\\
\\
In der Kantischen Philosophie ist das "`Ding an sich"' als reiner ontischer Realität unerkennbar (siehe \textbf{[1.1.1]}).
Es gibt Bedingungen der Möglichkeit von Erfahrung, welche über seine Erscheinung im menschlichen
Geist präjudizieren, und das sind eben vor allem Raum und Zeit als Grundformen der Anschauung. Dies ist im
Grunde eine Tatsache, die direkt mit 1a und 1b in Zusammenhang steht. 
In der Quantentheorie der Ur-Alternativen manifestiert sich diese Tatsache dadurch, dass sie keine speziellen
Annahmen über ein an sich existierendes "`Ding an sich"' macht, welche unabhängig von der Art und Weise wären,
wie dessen Eigenschaften im Rahmen menschlicher Erfahrung beschrieben werden können. Die Ur-Alternativen sind
ja gerade keine für sich existierenden Objekte, sondern vielmehr bereits eine Schematisierung einer
fundamentalen ontischen Realität in unserem erfahrenden Geist. Sie haben demnach sozusagen halb
ontischen und halb epistemischen Charakter.\\
\\
\textbf{3. Die grundsätzliche Rolle der Ur-Alternativen in einer beliebigen Quantentheorie}\\
\\
Die Ur-Alternativen werden durch die einfachsten überhaupt denkbaren Zustandsräume beschrieben, die in
in einer beliebigen Quantentheorie überhaupt vorkommen können. Die ihnen entsprechende ontische Realität
ist also die fundamentalste, deren Beschreibung der mathematische Formalismus der Quantentheorie zulässt.
Selbst eine Quantentheorie spezieller als fundamental angenommener konkreter Objekte würde grundsätzlich
auf Zustände in Hilbert-Räumen führen, die man prinzipiell in Ur-Alternativen zerlegen könnte.\\
\\
\newpage

\noindent
\textbf{4. Beziehung der mathematischen Struktur der Ur-Alternativen und ihres Tensorraumes zu den empirisch
gefundenen Symmetrien der realen Welt}\\
\\
Die Eigenschaften des Zustandsraumes einer Ur-Alternative beziehungsweise vieler Ur-Alternativen führen auf direktem Wege
zu den empirisch gefundenen Raum-Zeit-Symmet-rien der realen Welt, wie im nächsten Kapitel gezeigt werden wird.
Die Betrachtung der Symmetriegruppe einer einzelnen Ur-Alternative führt für den Ortsanteil der räumlichen Struktur 
des Kosmos zu einem kompakten dreidimensionalen Raum (siehe \textbf{[9.3]}), der mit einem Zeitparameter zu einer Raum-Zeit
verbunden ist. Die Betrachtung des Tensorraumes vieler Ur-Alternativen führt zur sich auf den (3+1)-dimensionalen
Minkowski-Raum beziehenden 15-parametrigen konformen Gruppe, welche die Poincar\'{e}-Gruppe als Untergruppe
aufweist (siehe\textbf{[9.2]}).\\
\\
\textbf{5. Aussicht auf Vermeidung der gewöhnlich auftretenden Divergenzen in natürlicher Weise}\\
\\
Im Rahmen der gewöhnlichen Formulierung relativistischer Quantenfeldtheorien auf einem vorgegebenen
kontinuierlichen Raum-Zeit-Hintergrund treten bei der Berechnung messbarer Größen Divergenzen auf,
die im Rahmen des Verfahrens der Renormalisierung auf ziemlich künstliche Weise im nachhinein entfernt
werden müssen, was bei bestimmten Versuchen der Quantisierung der Gravitation zu Problemen führt. 
Da im Rahmen der Quantentheorie der Ur-Alternativen die Zahl der Basiszustände grundsätzlich unbeschränkt
aber endlich ist, ist zu erwarten, dass sich divergente Ausdrücke ohne weitere Annahmen von vorneherein
nicht ergeben.\\
\\
\textbf{6. Bezug zum bisherigen physikalischen Wissen}\\
\\
Der Ansatz der Quantentheorie der Ur-Alternativen ist ontologisch höchst sparsam. Er führt keinerlei neue Entitäten
ein. Nur die bereits bekannte abstrakte Quantentheorie wird in einer völlig neuen Weise interpretiert
beziehungsweise aus noch grundsätzlicheren Postulaten über die Bedingungen der Möglichkeit von Erfahrung hergeleitet.
Die Annahme der Existenz der Ur-Alternativen entspricht einer bestimmten Darstellung der mit Hilfe des Begriffes der Alternative
konstruierten abstrakten Quantentheorie, ergibt sich also in natürlicher Weise und stellt damit keine von der abstrakten
Quantentheorie getrennte Annahme dar. Diese ontologische Sparsamkeit und das Hervorgehen aus einer konsequenten
Analyse des bisher schon Bekannten stellen aus wissenschaftstheoretischer Sicht ein Gütesiegel dar.
(siehe auch Zitat am Anfang des Kapitels)\\
\\
\textbf{7. Die Natur der Materie in der Platonischen Philosophie}\\
\\
In der Platonischen Philosophie wird die Materie auf basaler Ebene auf grundlegende geometrische Formen
zurückgeführt. Diese können im Sinne der Ideenlehre als Urbilder der Materie angesehen werden,
welche die Existenz der Materie erst konstituieren, selbst aber noch nicht Materie sind (siehe \textbf{[1.3.2]}).
Im Rahmen der modernen Physik kann dies so gedeutet werden, dass die Symmetrien, denen die fundamentalen
Gleichungen unterliegen, als das wirklich in einem fundamentalen Sinne existierende angenommen werden
(siehe \textbf{[7.4.2]}). Es ist auch völlig unmöglich in anschaulicher Sprache zu sagen, was ein Elementarteilchen ist.
Man kann nur sagen, dass seine möglichen Zustände einen Raum bilden, welcher irreduzible Darstellung
der Poincar\'{e}-Gruppe ermöglicht und weiter innere Symmetrien aufweist (siehe \textbf{[6.2.1]}).
Die Quantentheorie der Ur-Alternativen nimmt dementsprechend als grundlegende Entität nicht-räumliche
und daher nicht-materielle Objekte an, aus denen sich die Eigenschaften der Anschauungswelt erst ergeben.\\
\\
\textbf{8. Bezug zur Sprachphilosophie Wittgensteins}\\
\\
In der Philosophie (Sprachphilosophie) Ludwig Wittgensteins stehen am Anfang unserer Erkenntnis nicht wahrgenommene Dinge,
sondern sprachlich formulierte Tatsachen über die Welt (siehe \textbf{[8.2.4]}). Diese Tatsachen können auf einfachere und
schließlich auf elementare Tatsachen zurückgeführt werden, denen zugleich eine fundamentale ontische Realität
entspricht. Die Quantentheorie der Ur-Alternativen erscheint beinahe wie eine Konkretisierung dieser Anschauung im Rahmen der
philosophischen Analyse einer grundsätzlichen physikalischen Theorie. Die Ur-Alternativen stellen genau solche von Wittgenstein
beschriebenen elementaren Tatsachen (Möglichkeiten) dar, welche quantentheoretisch formuliert werden. Ihnen und denen aus Ihnen
konstruierten logischen Strukturen (Alternativen) kommt als logischer Form wie bei Wittgenstein auch an sich ontologische
Bedeutung zu.

\chapter{Die Struktur der Raum-Zeit als Konsequenz der Quantentheorie}

Als entscheidende Konsequenz der Quantentheorie der Ur-Alternativen ergibt sich die Existenz eines dreidimensionalen Ortsraumes,
der mit der Zeit zu einer Minkowski-Raum-Zeit verknüpft ist. Hierzu sei von Weizsäcker zitiert \cite{Weizsaecker:1985}:

\begin{quote}
{\small Im Sinne von Felix Kleins Erlanger Programm betrachten wir eine Geometrie als bestimmt durch ihre Symmetriegruppe. In
der Rekonstruktion der Quantentheorie erscheint uns die Symmetrie des Zustandsraums als ein Ausdruck der Trennbarkeit der
Alternativen. Dies liefert aber zunächst zu einer n-fachen Alternative die Gruppe $U(n)$, also die komplexe metrische Geometrie
des Hilbertraums. Wenn aber alle real vorkommenden Wechselwirkungen von einer dreidimensionalen reellen Geometrie abhängen, so
muss das bedeuten, dass alle real vorkommenden dynamischen Gesetze eine gemeinsame Symmetriegruppe haben, die sehr viel kleiner
ist als $U(n)$ für größere $n$. Dies betrachten wir als das zentrale Phänomen der konkreten Quantentheorie. Wir werden es durch
das Postulat der Uralternativen zu erklären versuchen. Der Ortsraum wird dann als ein homogener Raum der universalen
Symmetriegruppe der Dynamik erklärt.

Carl Friedrich von Weizsäcker, Aufbau der Physik, 1985 (Seite 382)}
\end{quote}
Zunächst wird der Tensorraum der Ur-Alternativen eingeführt, innerhalb dessen sich die konforme Gruppe darstellen lässt.
Anschließend daran wird in Bezug auf die Symmetrieeigenschaften einzelner Ur-Alternativen die topologische Struktur
des Kosmos begründet werden.  

\section{Der Tensorraum der Ur-Alternativen}

Da natürlich jedes reale Objekt aus vielen Ur-Alternativen besteht, müssen Produktzustände aus diesen Ur-Alternativen konstruiert
werden. Dies führt zu einem Tensorraum für die Ur-Alternativen. Eine einzelne Ur-Alternative ist die quantentheoretische Version
einer binären Alternative $(1,2)$, entsteht also durch die Quantisierung einer einfachen ja-nein-Entscheidung und wird demnach
durch einen zweidimensionalen komplexen Vektor dargestellt, also durch einen Weyl-Spinor 

\begin{equation}
u=\left(\begin{matrix}u_1\\u_2\end{matrix}\right)=\left(\begin{matrix}w+ix\\y+iz\end{matrix}\right).
\label{Ur-Alternative}
\end{equation}
Die Menge der einzelne Ur-Alternativen beschreibenden Weyl-Spinoren bilden einen Hilbert-Raum sofern man das innere Produkt
zweier solcher Ur-Alternativen beschreibender Weyl-Spinoren $u$ und $v$ als $|\langle u|v \rangle|=u_1^{*}v_1+u_2^{*}v_2$
definiert.
 
Zum Tensorraum vieler Ur-Alternativen gelangt man nun durch eine weitere Quantisierung, die Quantisierung einer
ihrerseits bereits einen Quantenzustand beschreibenden Ur-Alternative, welche durchgeführt wird, indem für sie die folgende
Vertauschungsrelation postuliert wird

\begin{equation}
[u^{\dagger}_r,u_s]=\delta_{rs}\quad\quad r,s=1...2,
\label{Quantisierung_Ur-Alternative}
\end{equation}
wobei $r$ und $s$ den Basiszustand der Ur-Alternative beschreiben.
Der die Ur-Alternative beschreibende Weyl-Spinor wird durch diese Quantisierungsprozedur also selbst zu einem Operator

\begin{equation}
u \rightarrow \hat u,
\end{equation}
der auf einen Zustand in einem weiteren Hilbert-Raum wirkt.
Dadurch, dass die beiden Einstellungen der Ur-Alternative gemäß der Quantentheorie durch komplexe Zahlen beschrieben werden,
gibt es insgesamt vier verschiedene voneinander unabhängige Zustände (Basis-Zustände), in denen sich eine einzelne Ur-Alternative
befinden kann. Dem entsprechen nach der Quantisierung ($\ref{Quantisierung_Ur-Alternative}$) vier Operatoren und die
dazugehörigen hermitesch-adjungierten Operatoren. Sie seien gemäß der Notation in ($\ref{Ur-Alternative}$) mit $a_w$, $a_x$,
$a_y$ und $a_z$ beziehungsweise $a_w^{\dagger}$, $a_x^{\dagger}$, $a_y^{\dagger}$ und $a_z^{\dagger}$ bezeichnet und erfüllen die
folgenden Vertauschungsrelationen

\begin{equation}
[a_w,a_w^{\dagger}]=1\quad,\quad [a_x,a_x^{\dagger}]=1\quad,\quad [a_y,a_y^{\dagger}]=1\quad,\quad [a_z,a_z^{\dagger}]=1.
\label{Operatoren_Ur-Alternative}
\end{equation}
Da die durch ($\ref{Quantisierung_Ur-Alternative}$) aus ($\ref{Ur-Alternative}$) entstandenen Operatoren die
Vertauschungsrelationen von Erzeugungs- und Vernichtungsoperatoren erfüllen ($\ref{Operatoren_Ur-Alternative}$), kann man sie
demgemäß als Erzeugungs- und Vernichtungsoperatoren für Ur-Alternativen interpretieren.
Die Operatoren $a_w$, $a_x$, $a_y$ und $a_z$ definieren damit einen Tensorraum $T_u$ vieler Ur-Alternativen über dem Hilbert-Raum
der Ur-Alternativen, genauer den Unterraum der bezüglich Vertauschung der in einem Zustand enthaltenen Ur-Alternativen
symmetrischen Tensoren. Dies bedeutet natürlich, dass hier für die Ur-Alternativen, die als ununterscheidbar angesehen werden,
\footnote{Die Ununterscheidbarkeit der Ur-Alternativen ergibt sich daraus, dass sie gemäß der Quantentheorie der Ur-Alternativen
als die fundamentale Entität in der Natur angesehen werden, sich also jede Information in der Natur zumindest im Prinzip in
Ur-Alternativen auflösen lässt. Eine Unterscheidung zwischen Ur-Alternativen würde selbst wieder nur durch Ur-Alternativen
beschrieben werden können, weshalb sich eine einzelne Ur-Alternative nicht von einer anderen unterscheiden lässt.}
zunächst Bose-Statistik vorausgesetzt wird. Bei Fermi-Statistik könnte es aufgrund des Ausschließungsprinzips insgesamt nur
vier Ur-Alternativen geben, nämlich jeweils eine in den vier sich auf eine einzelne Ur-Alternative beziehenden Basiszuständen.
Zur Beschreibung der Zustände von Teilchen und der Wechselwirkung verschiedener Teilchen untereinander ist es anschließend
notwendig, zu allgemeineren Symmetrieklassen im Tensorraum überzugehen.
Eine Basis im Tensorraum der Ur-Alternativen ist durch alle möglichen Kombinationen von Anzahlen von Ur-Alternativen in den vier
voneinander unabhängigen Zuständen einer einzelnen Ur-Alternative gegeben. Diese seien mit $N_w$, $N_x$, $N_y$ und $N_z$
bezeichnet und damit können die Basiszustände als $|N_w,N_x,N_y,N_z\rangle$ bezeichnet werden.
Die Operatoren $a_w$, $a_x$, $a_y$ und $a_z$ beziehungsweise $a_w^{\dagger}$, $a_x^{\dagger}$,$a_y^{\dagger}$ und
$a_z^{\dagger}$ repräsentieren demnach Erzeugungs- und Vernichtungsoperatoren im Tensorraum der Ur-Alternativen, durch deren
Anwendung die sich auf unterschiedliche Besetzungszahlen beziehenden Basiszustände ineinander überführt werden
\footnote{Die Normierung der Zustände wird hier zunächst außer acht gelassen.}

\begin{eqnarray}
a_w |N_w,N_x,N_y,N_z\rangle=|N_w-1,N_x,N_y,N_z\rangle\quad,\quad a_w^{\dagger}
|N_w,N_x,N_y,N_z\rangle=|N_w+1,N_x,N_y,N_z\rangle,\nonumber\\
a_x |N_w,N_x,N_y,N_z\rangle=|N_w,N_x-1,N_y,N_z\rangle\quad,\quad a_x^{\dagger}
|N_w,N_x,N_y,N_z\rangle=|N_w,N_x+1,N_y,N_z\rangle,\nonumber\\
a_y |N_w,N_x,N_y,N_z\rangle=|N_w,N_x,N_y-1,N_z\rangle\quad,\quad a_y^{\dagger}
|N_w,N_x,N_y,N_z\rangle=|N_w,N_x,N_y+1,N_z\rangle,\nonumber\\
a_z |N_w,N_x,N_y,N_z\rangle=|N_w,N_x,N_y,N_z-1\rangle\quad,\quad a_z^{\dagger}
|N_w,N_x,N_y,N_z\rangle=|N_w,N_x,N_y,N_z+1\rangle.\nonumber\\
\label{Erzeuger-Vernichter-Operatoren_Basiszustaende_Tensorraum}
\end{eqnarray}
Der Vakuumzustand $|0\rangle$ entspricht den Besetzungszahlen $N_w=N_x=N_y=N_z=0$ und verschwindet bei
Anwendung der Vernichtungsoperatoren auf ihn

\begin{equation}
a_w|0\rangle=a_x|0\rangle=a_y|0\rangle=a_z|0\rangle=0.
\end{equation}
Die Operatoren $a_w^{\dagger}a_w$, $a_x^{\dagger}a_x$, $a_y^{\dagger}a_y$ und $a_z^{\dagger}a_z$ haben als
Besetzungszahloperatoren angewandt auf einen Basiszustand $|N_w,N_x,N_y,N_z\rangle$ im Tensorraum die Anzahl der Ur-Alternativen
als Eigenwerte

\begin{eqnarray}
a_w^{\dagger}a_w|N_w,N_x,N_y,N_z\rangle=N_w|N_w,N_x,N_y,N_z\rangle\quad,\quad
a_x^{\dagger}a_x|N_w,N_x,N_y,N_z\rangle=N_x|N_w,N_x,N_y,N_z\rangle,\nonumber\\
a_y^{\dagger}a_y|N_w,N_x,N_y,N_z\rangle=N_y|N_w,N_x,N_y,N_z\rangle\quad,\quad
a_z^{\dagger}a_z|N_w,N_x,N_y,N_z\rangle=N_z|N_w,N_x,N_y,N_z\rangle.\nonumber\\
\label{Besetzungszahl-Operator_Basiszustaende_Tensorraum}
\end{eqnarray}
Ein Zustand $|N_w,N_x,N_y,N_z\rangle$ kann auch als Produkt von vier Teilräumen dargestellt werden, auf die sich Paare von
Erzeugungs- und Vernichtungsoperatoren jeweils beziehen

\begin{equation}
|N_w,N_x,N_y,N_z\rangle=|N_w\rangle \otimes |N_x\rangle \otimes |N_y\rangle \otimes |N_z\rangle.
\label{Basiszustaende_Tensorraum_Teilraeume}
\end{equation} 
Das bedeutet natürlich auch, dass nicht nur die Anwendung eines Vernichtungsoperators auf den Vakuumzustand, sondern auch
auf jeden Basiszustand verschwindet, bei dem nur die zu ihm gehörige Besetzungszahl gleich null ist.
Ein beliebiger Zustand $|\Psi\rangle$ im Tensorraum der Ur-Alternativen kann also durch eine Linearkombination der
in ($\ref{Erzeuger-Vernichter-Operatoren_Basiszustaende_Tensorraum}$),($\ref{Besetzungszahl-Operator_Basiszustaende_Tensorraum}$)
und ($\ref{Basiszustaende_Tensorraum_Teilraeume}$) beschriebenen Basiszustände dargestellt werden

\begin{equation}
|\psi\rangle=\sum_{N_w}\sum_{N_x}\sum_{N_y}\sum_{N_z} c(N_w,N_x,N_y,N_z)|N_w,N_x,N_y,N_z\rangle.
\label{Allgemeiner_Zustand_Symmetrischer_Tensorraum}
\end{equation}
Ein inneres Produkt durch das der Tensorraum $T_u$ zu einem Hilbert-Raum wird, ist durch die Bedingung definiert, dass zwei
Basiszustände dann orthogonal zueinander sein sollen, wenn sie sich in mindestens einer Besetzungszahl voneinander
unterscheidenden

\begin{equation}
\langle N_w^{\prime},N_x^{\prime},N_y^{\prime},N_z^{\prime}|N_w,N_x,N_y,N_z\rangle
=\delta_{N_w N_w^{\prime}}\delta_{N_x N_x^{\prime}}\delta_{N_y N_y^{\prime}}\delta_{N_z N_z^{\prime}},
\end{equation}
wodurch sich für das innere Produkt zweier beliebiger Zustände $|\psi\rangle$ und  $|\psi^{\prime}\rangle$ die folgende
Definition ergibt

\begin{eqnarray}
\langle \psi^{\prime}|\psi \rangle&=&\sum_{N_w,N_w^{\prime}}\sum_{N_x,N_x^{\prime}}\sum_{N_y,N_y^{\prime}}\sum_{N_z,N_z^{\prime}}
c^{\prime}(N_w^{\prime},N_x^{\prime},N_y^{\prime},N_z^{\prime})c(N_w,N_x,N_y,N_z)\nonumber\\
&&\quad\quad\quad\quad\quad\quad\quad\quad\quad\quad\quad\quad\cdot\langle N_w^{\prime}
,N_x^{\prime},N_y^{\prime},N_z^{\prime}|N_w,N_x,N_y,N_z\rangle\nonumber\\
&=&\sum_{N_w,N_w^{\prime}}\sum_{N_x,N_x^{\prime}}\sum_{N_y,N_y^{\prime}}\sum_{N_z,N_z^{\prime}}
c^{\prime}(N_w^{\prime},N_x^{\prime},N_y^{\prime},N_z^{\prime})c(N_w,N_x,N_y,N_z)
\nonumber\\&&\quad\quad\quad\quad\quad\quad\quad\quad\quad\quad\quad\quad\cdot
\delta_{N_w N_w^{\prime}}\delta_{N_x N_x^{\prime}}\delta_{N_y N_y^{\prime}}\delta_{N_z N_z^{\prime}}\nonumber\\
&=&\sum_{N_w}\sum_{N_x}\sum_{N_y}\sum_{N_z} c^{\prime}(N_w,N_x,N_y,N_z)c(N_w,N_x,N_y,N_z).
\end{eqnarray}

\section{Die Quantentheorie der Ur-Alternativen und die Struktur der Raum-Zeit im Sinne der Speziellen Relativitätstheorie}

In diesem Abschnitt soll die Struktur der Raum-Zeit im Sinne der Speziellen Relativitätstheorie als Konsequenz der
Quantentheorie begründet werden. Hierzu ist es zunächst wichtig, dass es möglich ist, eine zu einer Ur-Alternative gehörige
Anti-Ur-Alternative durch folgende Transformation    

\begin{equation}
u \rightarrow i\sigma^2 u^{*}=\left(\begin{matrix}0&1\\-1&0\end{matrix}\right)\left(\begin{matrix}u_1^{*}\\u_2^{*}
\end{matrix}\right)=\left(\begin{matrix}u_2^{*}\\-u_1^{*}\end{matrix}\right)
=\left(\begin{matrix}y-iz\\-w+ix\end{matrix}\right)
\end{equation}
zu erhalten. Damit ist es weiter möglich, einen entsprechenden Dirac-Spinor zu definieren, bei dem es sich genauer um einen
Majorana-Spinor handelt, da die untere Komponenten dieses Dirac-Spinors nur eine andere Darstellung des Weyl-Spinors ist und
sie daher gegenüber diesem keine zusätzlichen Freiheitsgrade aufweist

\begin{equation}
\psi=\left(\begin{matrix}\psi_a\\\psi_b\\\psi_c\\\psi_d\end{matrix}\right)
=\left(\begin{matrix}u_1\\u_2\\u_2^{*}\\-u_1^{*}\end{matrix}\right)
=\left(\begin{matrix}w+ix\\y+iz\\y-iz\\-w+ix\end{matrix}\right).
\label{Majorana-Spinor_Ur-Alternative}
\end{equation}
Quantisierung der Ur-Alternative ($\ref{Quantisierung_Ur-Alternative}$) führt natürlich auch zu einer Quantisierung
des Spinors ($\ref{Majorana-Spinor_Ur-Alternative}$)

\begin{equation}
\left[\hat \psi_r,\hat \psi_s^{\dagger}\right]=\delta_{rs},
\label{Quantisierung_Majorana-Spinor_Ur-Alternative}
\end{equation}
und dies bedeutet, dass die Komponenten von ($\ref{Majorana-Spinor_Ur-Alternative}$) analog zu den Operatoren
($\ref{Operatoren_Ur-Alternative}$) zu Erzeugungs- und Vernichtungsoperatoren für Ur-Alternativen werden

\begin{equation}
\psi_a \rightarrow \hat \psi_a \equiv a\quad,\quad \psi_b \rightarrow \hat \psi_b \equiv b\quad,\quad
\psi_c \rightarrow \hat \psi_c \equiv c\quad,\quad \psi_d \rightarrow \hat \psi_d \equiv d,
\end{equation}
und in Bezug auf die Operatoren in ($\ref{Operatoren_Ur-Alternative}$) wie folgt definiert sind

\begin{eqnarray}
&&a=\frac{1}{\sqrt{2}}\left(a_w+ia_x\right),\quad
b=\frac{1}{\sqrt{2}}\left(a_y+ia_z\right),\quad
c=\frac{1}{\sqrt{2}}\left(a_y-ia_z\right),\quad
d=\frac{1}{\sqrt{2}}\left(-a_w+ia_x\right),\\
&&a^{\dagger}=\frac{1}{\sqrt{2}}\left(a_w^{\dagger}-ia_x^{\dagger}\right),\quad
b^{\dagger}=\frac{1}{\sqrt{2}}\left(a_y^{\dagger}-ia_z^{\dagger}\right),\quad
c^{\dagger}=\frac{1}{\sqrt{2}}\left(a_y^{\dagger}+ia_z^{\dagger}\right),\quad
d^{\dagger}=\frac{1}{\sqrt{2}}\left(-a_w^{\dagger}-ia_x^{\dagger}\right).\nonumber
\label{Definition_abcd}
\end{eqnarray}
Diese Operatoren erfüllen die folgenden Vertauschungsrelationen

\begin{equation}
[a,a^{\dagger}]=1\quad,\quad [b,b^{\dagger}]=1\quad,\quad [c,c^{\dagger}]=1\quad,\quad [d,d^{\dagger}]=1.
\label{Vertauschungsrelationen_abcd}
\end{equation}
Das bedeutet, dass man nun mit Hilfe dieser Konstruktion und einer Übertragung der Quantisierung der Ur-Alternative ($\ref{Quantisierung_Ur-Alternative}$) auf ($\ref{Majorana-Spinor_Ur-Alternative}$) eine neue Darstellung des Tensorraumes
der Ur-Alternativen erhalten hat. Basiszustände im Tensorraum der Ur-Alternativen in Bezug auf die Operatoren
($\ref{Definition_abcd}$) sind analog zu den zu den Operatoren $a_w$,$a_x$,$a_y$,$a_z$ gehörigen in ($\ref{Erzeuger-Vernichter-Operatoren_Basiszustaende_Tensorraum}$), ($\ref{Besetzungszahl-Operator_Basiszustaende_Tensorraum}$)
und ($\ref{Basiszustaende_Tensorraum_Teilraeume}$) beschriebenen Zuständen durch die Besetzungszahlen $N_a$,$N_b$,$N_c$,$N_d$
als $|N_a,N_b,N_c,N_d\rangle$ definiert. Aus dem durch den Spinor $\psi$ beschriebenen Operator können mit Hilfe
bilinearer Ausdrücke im Dirac-Spinor-Raum die folgenden Operatoren konstruiert werden

\begin{eqnarray}
&&\Sigma^{\mu\nu}=\frac{i}{4}\psi^{\dagger}\left[\gamma^\mu,\gamma^\nu\right]\psi\quad,\quad
P^\mu=\frac{1}{2}\psi^{\dagger}\left(\gamma^\mu-\gamma^5 \gamma^\mu\right)\psi\quad,\quad
K^\mu=\frac{1}{2}\psi^{\dagger}\left(\gamma^\mu+\gamma^5 \gamma^\mu\right)\psi,\nonumber\\
&&D=i\psi^{\dagger}\gamma^5\psi\quad,\quad
N=\psi^{\dagger}\psi\quad\quad\quad\quad \text{mit}\quad\quad\quad\quad \mu={0...3},\label{Groessen_Dirac-Spinor}
\end{eqnarray}
wobei die $\gamma^\mu$ die in ($\ref{Dirac-Matrizen}$) und mit Hilfe der Pauli-Matrizen ($\ref{Pauli-Matrizen}$)
definierten Gamma-Matrizen sind und $\gamma^5=i\gamma^0 \gamma^1 \gamma^2 \gamma^3$. Diese Operatoren
($\ref{Groessen_Dirac-Spinor}$) wirken natürlich im Tensorraum der Ur-Alternativen, da sie aus den
Operatoren ($\ref{Definition_abcd}$) bestehen. Außer dem Operator $N$ sind sie nicht hermitesch, sondern
sind selbstadjungiert in einem verallgemeinerten Sinne. Die Operatoren in ($\ref{Groessen_Dirac-Spinor}$)
haben die allgemeine Form $\psi^{\dagger}A\psi$, wobei $A$ Matrizen in Bezug auf den Dirac-Spinor-Raum
beschreibt. Man kann nun eine adjungierte Matrix in der folgenden Weise definieren: $A \rightarrow \gamma^0 A^{\dagger} \gamma^0$. 
Und wenn man die Matrizen $A$ in dieser Weise adjungiert, dann sind sie selbstadjungiert und das bedeutet, dass gilt

\begin{eqnarray}
\psi^{\dagger} \gamma^0 A^{\dagger} \gamma^0 \psi=\psi^{\dagger} A \psi.
\label{Groessen_Dirac-Spinor_selbstadjungiert}
\end{eqnarray}
Und in diesem Sinne ($\ref{Groessen_Dirac-Spinor_selbstadjungiert}$) sind die Opertatoren in ($\ref{Groessen_Dirac-Spinor}$)
selbstadjungiert. Aufgrund der Tatsache, dass mit ($\ref{Quantisierung_Majorana-Spinor_Ur-Alternative}$) gilt

\begin{equation}
\left[\psi^{\dagger}A \psi,\psi^{\dagger}B\psi\right]=\psi^{\dagger}\left[A,B\right]\psi,
\end{equation}
können die Vertauschungsrelationen zwischen den Operatoren ($\ref{Groessen_Dirac-Spinor}$) leicht auf die
Vertauschungsrelationen der Matrizen $A$ im Dirac-Spinor-Raum zurückgeführt werden und es ergibt sich, dass
die Operatoren ($\ref{Groessen_Dirac-Spinor}$) folgende Vertauschungsrelationen erfüllen

\begin{eqnarray}
&&\left[\Sigma_{\mu\nu},\Sigma_{\rho\sigma}\right]=
\eta_{\nu\rho}\Sigma_{\mu\sigma}-\eta_{\mu\rho}\Sigma_{\nu\sigma}+\eta_{\nu\sigma}\Sigma_{\rho\mu}
-\eta_{\mu\sigma}\Sigma_{\rho\nu},\nonumber\\
&&\left[\Sigma_{\mu\nu},P_{\rho}\right]=\eta_{\nu\rho}P_{\mu}-\eta_{\mu\rho}P_{\nu}\quad,\quad
\left[\Sigma_{\mu\nu},K_{\rho}\right]=\eta_{\nu\rho}K_{\mu}-\eta_{\mu\rho}K_{\nu},\nonumber\\
&&\left[\Sigma_{\mu\nu},D\right]=0\quad,\quad
\left[D,P_{\mu}\right]=-P_{\mu}\quad,\quad
\left[D,K_{\mu}\right]=K_{\mu},\nonumber\\
&&\left[P_{\mu},K_{\nu}\right]=-2\Sigma_{\mu\nu}+2\eta_{\mu\nu}D \quad,\quad
\left[K_{\mu},K_{\nu}\right]=0\quad,\quad
\left[P_{\mu},P_{\nu}\right]=0,\nonumber\\
&&\left[\Sigma_{\mu\nu},N\right]=\left[P_\mu,N\right]=\left[K_\mu,N\right]=\left[D,N\right]=0.
\label{Algebra_Konforme-Gruppe}
\end{eqnarray}
Das bedeutet, dass es sich bei den Operatoren $\Sigma^{\mu\nu}$,$P^\mu$,$K^\mu$ und $D$ um die Generatoren der auf den
$(3+1)$-dimensionalen Raum der Speziellen Relativitätstheorie bezogenen konformen Gruppe handelt, die damit im
Tensorraum der Ur-Alternativen dargestellt wird. Daher erscheint es sinnvoll, zu postulieren, dass die den Größen
aus ($\ref{Groessen_Dirac-Spinor}$) entsprechenden Operatoren mit den auf Zustände in der realen Minkowski-Raum-Zeit bezogenen 
isomorphen Operatoren identisch sind und dementsprechend auch die durch diese Operatoren beschriebenen physikalischen Größen
darstellen. Dies impliziert die folgende Zuordnung der den Größen aus ($\ref{Groessen_Dirac-Spinor}$) entsprechenden
Operatoren

\begin{eqnarray}
M_{ik}&\widehat{=}& Drehimpulse\nonumber\\
M_{i0}&\widehat{=}& Eigentliche\ Lorentz-Transformationen\nonumber\\
P_{i}&\widehat{=}& Impulse\nonumber\\
P_{0}&\widehat{=}& Energie\nonumber\\
K_{i}&\widehat{=}& Spezielle\ konforme\ Transformationen\ (Raumkomponenten)\nonumber\\
K_{0}&\widehat{=}& Spezielle\ konforme\ Transformationen\ (Zeitkomponente)\nonumber\\
D&\widehat{=}& Dilatation.\nonumber\\
\end{eqnarray}
$N$ beschreibt die Summe der Besetzungszahloperatoren für die vier Basis-Zustände der einzelnen Alternative, also die Gesamtzahl
der Ur-Alternativen. Da $N$ mit allen die konforme Gruppe generierenden Operatoren kommutiert, ändert eine Transformation des
Zustandes mit einem Element der konformen Gruppe die Zahl der Ur-Alternativen nicht.
Daher scheint es plausibel, da alle Transformationen, welche die Zahl der Ur-Alternativen unverändert lassen, sich mit Hilfe
dieser Generatoren beschreiben lassen, davon auszugehen, dass diese Gruppe damit als fundamentale Symmetriegruppe konstituiert
ist. Wenn man diese Argumentation nun umkehrt, bedeutet dies, dass die reale Raum-Zeit, welche die Struktur eines
Minkowski-Raumes aufweist, in Wirklichkeit die Darstellung einer dahinterliegenden Informationsstruktur ist, welche ihrerseits
bereits durch die Aufspaltung in binäre Alternativen auf eine bestimmte Art und Weise dargestellt ist.
Dass sich hier die volle konforme Gruppe als Symmetriegruppe ergibt und nicht nur die Poincar\'{e}-Gruppe als Untergruppe der
konformen Gruppe, erscheint insofern sehr plausibel, als sie die allgemeinste Gruppe ist, welche metrische Verhältnisse
beziehungsweise Skalen unverändert lässt. Begriffe wie Metrik, Ruhemasse und Wechselwirkung sind aber auf dieser Ebene
noch nicht definiert, denn eine Raum-Zeit-Metrik entspricht einem Gravitationsfeld und dieses muss als dynamische Entität
natürlich selbst aus Ur-Alternativen aufgebaut werden. Weiter ist die Existenz von Massen sowohl im Standardmodell der
Elementarteilchenphysik als auch in Heisenbergs nichtlinearer Spinorfeldtheorie durch Wechselwirkung begründet. Bisher wurden
allerdings nur freie Alternativen betrachtet. Eine Wechselbeziehung zwischen Alternativen entsteht erst, indem eine Alternative
mit allgemeiner Permutationssymmetrie in totalsymmetrische Alternativen aufgespalten wird. Dies wird in \textbf{[10.2]} näher
thematisiert werden.

\section{Die Quantentheorie der Ur-Alternativen und die Struktur des Kosmos}

Im letzten Abschnitt wurde gezeigt, dass sich die konforme Gruppe der Speziellen Relativitätstheorie in natürlicher Weise 
als Symmetrie-Gruppe von Zuständen im Tensorraum der Ur-Alternativen ergibt. Hieraus wurde der Schluss gezogen, dass damit
die Existenz der Raum-Zeit im Sinne der Speziellen Relativitätstheorie aus der Quantentheorie der Ur-Alternativen begründet
wurde, die damit als eine Art Darstellung dahinterliegender dynamischer Verhältnisse von Objekten angesehen würde, die
ihrerseits aus Ur-Alternativen bestehen.

Wenn man nun die Frage nach der globalen Struktur der Raum-Zeit, also die Frage nach der Struktur des Kosmos 
als Ganzem stellt, so liefert die Quantentheorie der Ur-Alternativen auch hierauf eine Antwort. Man muss hierzu 
zunächst wieder die Struktur einer einzelnen Ur-Alternative ($\ref{Ur-Alternative}$) betrachten

\begin{equation}
u=\left(\begin{matrix}u_1\\u_2 \end{matrix}\right)=\left(\begin{matrix}w+ix\\y+iz \end{matrix}\right).
\end{equation}
Da eine einzelne Ur-Alternative durch ein Element eines zweidimensionalen komplexen Vektorraumes, also durch einen Weyl-Spinor,
beschrieben wird, hat es als allgemeine Symmetriegruppe die $U(2)$, also die unitäre Gruppe in zwei komplexen Dimensionen.
Diese kann aufgespalten werden in die $SU(2)$ und die $U(1)$

\begin{equation}
U(2)=SU(2)\otimes U(1).
\end{equation}
Die $U(1)$ beschreibt gemäß der in \textbf{[8.2.2]} dargestellten Rekonstruktion der Quantentheorie die Dynamik einer
beliebigen Alternative, also auch einer Ur-Alternative. Damit kann ihr Parameter also mit der Zeit identifiziert werden.
Wenn man nun jede einzelne in der Welt existierende Ur-Alternative mit dem selben Element der $SU(2)$ transformiert,
so ändert dies natürlich am Zustand der Welt nichts, da Alternativen nur relativ zueinander definiert sind, was sich im
in \textbf{[8.2.2]} aufgeführten grundlegenden Postulat der Symmetrie ausdrückt. In Bezug auf die nun folgende Argumentation ist
der mathematische Begriff des homogenen Raumes von Bedeutung.\\ 
\textbf{Definition:} Ein homogener Raum $(M,G)$ ist eine Kombination aus einer differenzierbaren Mannigfaltigkeit $M$ und einer
Lieschen Gruppe $G$, wobei $G$ auf $M$ transitiv und glatt operiert, was bedeutet, dass für zwei beliebige Punkte
$x,y \in M$ ein Element $g \in G$ existiert, sodass $g(x)=y$.\\
Gemäß dem zu Beginn dieses Kapitels angeführten Zitat von Weizsäckers kann der Raum als homogener Raum
der universalen Symmetriegruppe angesehen werden, also als ein homogener Raum der $SU(2)$. Der naheliegendste
Raum hierfür ist die $SU(2)$ selbst, die als Lie-Gruppe natürlich selbst eine Mannigfaltigkeit darstellt und damit sowohl
als Gruppe als auch als Mannigfaltigkeit fungiert, auf der die Gruppe wirkt. Da die $SU(2)$ der $SO(3)$
isomorph ist, also der Gruppe der Drehungen in einem reellen dreidimensionalen Raum, ergibt sich von vorneherein eine direkte
Beziehung zur mathematischen Struktur des realen physikalischen Raumes. Das allgemeine Element der $SU(2)$ hat folgende Gestalt

\begin{equation}
U=\left(\begin{matrix}a+bi&-c+di\\c+di&a-bi\end{matrix}\right)\quad,\quad \det U=a^2+b^2+c^2+d^2=1.
\end{equation}
Damit entspricht der $SU(2)$ topologisch eine $\mathbb{S}^3$. Es wird nun postuliert, dass diese $\mathbb{S}^3$ in der Tat die
reale Struktur des Kosmos widerspiegelt. Dieses Postulat entspricht der Annahme, dass uns die realen physikalischen Objekte
deshalb als in einem dreidimensionalen Ortsraum existierend erscheinen, weil dessen Struktur, wie aufgrund der
Symmetrieeigenschaften gezeigt wurde, da sich die universalen Symmetrieeigenschaften aus Ur-Alternativen bestehender Zustände
in ihm in natürlicher Weise darstellen lassen, dem Raum der Zustände der Ur-Alternativen isomorph ist. Da die $U(1)$ mit der
$SU(2)$ zur $U(2)$ verbunden ist, ergibt sich also gemäß den obigen Argumentationen auf kosmologischer Ebene, dass der Raum
mit der Zeit zu einer Raum-Zeit verbunden ist, deren lokale Minkowski-Struktur aus dem Tensorraum der Ur-Alternativen begründet
wurde, da sich dort die konforme Gruppe als Symmetriegruppe ergab, welche die Poincar\'{e}-Gruppe als Untergruppe enthält.
Nun kann man Zuständen im Tensorraum der Ur-Alternativen in der folgenden Weise Punkte auf der $\mathbb{S}^3$ zuordnen.
Es sei der Operator $X$ wie folgt definiert

\begin{eqnarray}
X&=&a_w^{\dagger}a_w u_w+a_x^{\dagger}a_x u_x+a_y^{\dagger}a_y u_y+a_z^{\dagger}a_z u_z
=\left(\begin{matrix}a_w^{\dagger}a_w+ia_x^{\dagger}a_x\\a_y^{\dagger}a_y+ia_z^{\dagger}a_z\end{matrix}\right).
\label{X-Operator}
\end{eqnarray}
Für den Operator ($\ref{X-Operator}$) ergibt sich bei Anwendung auf einen Basiszustand $|N_w,N_x,N_y,N_z \rangle$

\begin{eqnarray}
&&\left(a_w^{\dagger}a_w u_w+a_x^{\dagger}a_x u_x+a_y^{\dagger}a_y u_y+a_z^{\dagger}a_z u_z\right)|N_w,N_x,N_y,N_z \rangle\nonumber\\
&&=\left(N_w u_w+N_x u_x+N_y u_y+N_z u_z\right)|N_w,N_x,N_y,N_z \rangle
=\left(\begin{matrix}N_w+iN_x\\N_y+iN_z\end{matrix}\right)|N_w,N_x,N_y,N_z\rangle.\nonumber\\
\label{Eigenwerte_X-Operator}
\end{eqnarray}
Wenn man nun den entsprechenden Eigenwert des Operators ($\ref{X-Operator}$) in ($\ref{Eigenwerte_X-Operator}$) normiert

\begin{equation}
\left(\begin{matrix}\bar N_w+i\bar N_x\\\bar N_y+i\bar N_z\end{matrix}\right)=\frac{1}{\sqrt{N_w^2+N_x^2+N_y^2+N_z^2}}
\left(\begin{matrix}N_w+iN_x\\N_y+iN_z\end{matrix}\right),
\end{equation}
so kann man, da $\bar N_w^2+\bar N_x^2+\bar N_y^2+\bar N_z^2=1$, jedem Eigenwert einen Punkt auf der den Kosmos beschreibenden
$\mathbb{S}^3$ zuordnen. Ein allgemeiner Zustand im Tensorraum ($\ref{Allgemeiner_Zustand_Symmetrischer_Tensorraum}$) entspricht
damit einer Wellenfunktion auf dieser $\mathbb{S}^3$. Die Punkte auf dieser $\mathbb{S}^3$ sind allerdings diskret.
Je höher die Gesamtzahl an Ur-Alternativen in einem Zustand ist, desto mehr Kombinationen von Besetzungszahlen für die einzelnen
Basiszustände gibt es, welche zu dieser Gesamtanzahl an Ur-Alternativen führen, und demnach wird die räumliche Bestimmung auf der 
$\mathbb{S}^3$ für einen Basiszustand immer genauer. Das bedeutet, dass eine große Anzahl von Ur-Alternativen nötig ist, um eine
genaue örtliche Bestimmung zu erreichen. Dies ist plausibel, da eine genaue örtliche Bestimmung einer großen Menge an Information
entspricht.

\section{Die Natur des Raumes in der Quantentheorie der Ur-Alternativen}

\begin{figure}[ht]
\centering
\epsfig{figure=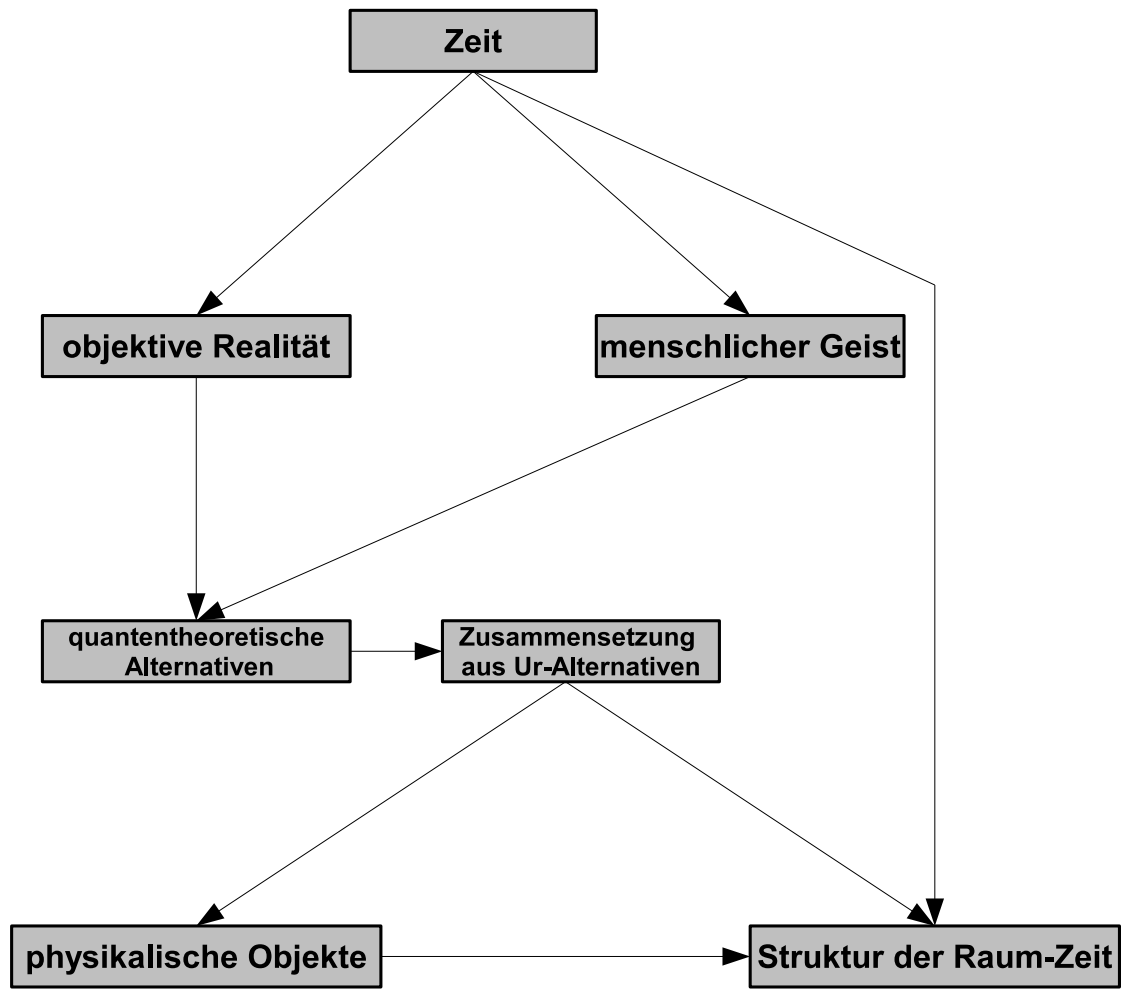,width=17cm}
\caption{\label{Ur-Alternativen_Raum-Zeit} Die Natur der Raum-Zeit: Als sowohl den menschlichen Geist als auch die objektive
Realität umfassende Größe wird die Zeit angenommen. Die Darstellung der objektiven Realität im menschlichen Geist geschieht
durch quantentheoretisch beschriebene Alternativen. Deren Darstellung durch die Kombination von Ur-Alternativen
führt schließlich zu konkreten physikalischen Objekten, wie quantentheoretisch beschriebenen Teilchen und Feldern,
die als in einem gemeinsamen physikalischen Raum existierend dargestellt werden können, der seinerseits mit der
Zeit zu einer Raum-Zeit im Einsteinschen Sinne verbunden ist.}
\end{figure}

Der Raum ist in der Quantentheorie der Ur-Alternativen also eine hergeleitete Entität. Ur-Alternativen sind an sich nicht
räumlich. Sie begründen die Existenz eines mit der Zeit zu einer Minkowski-Raum-Zeit verbundenen Ortsraumes. Die Frage ist nun,
inwiefern man den Raum im Rahmen der Quantentheorie der Ur-Alternativen als objektiv existierend bezeichnen kann. Er ist in dem
Sinne als objektiv existierend anzusehen, als er eine wirkliche in den Ur-Alternativen und ihrer Kombination zu komplexeren
Objekten vorhandene Struktur beschreibt. Diese Struktur ist jedoch von dem Raum als Anschauungsform in unserem Geist zu
unterscheiden. Wie bereits in \textbf{[1.2.3]} erwähnt, kann der Raum als Anschauungsform in unserem Geist
als zum wirklichen Raum näherungsweise isomorph, aber nicht wesensgleich angesehen werden. In der Quantentheorie der
Ur-Alternativen ergibt sich nun, dass der reale Raum eigentlich selbst nichts anderes als eine abstrakte Struktur ist, welche
sich aus den Ur-Alternativen und ihren wechselseitigen Beziehungen konstituiert. Hierbei ist aber wichtig, dass die
Ur-Alternativen selbst bereits einer formalen Darstellung der Realität auf basaler Ebene entsprechen, die an sich unerkennbar
ist. Damit ergibt sich die Kantische Erkenntnis, dass der Raum als Grundform der Anschauung eine Bedingung der Möglichkeit von
Erfahrung darstellt, gewissermaßen als Konsequenz der Quantentheorie der Ur-Alternativen, da sich die Existenz des Ortsraumes
wiederum als Darstellung der Ur-Alternativen ergibt, deren Struktur gemäß der Rekonstruktion der Quantentheorie ihrerseits eine
Bedingung der Möglichkeit von Erfahrung darstellt, da sie eine basale Darstellung der Realität liefern. Die Quantentheorie der
Ur-Alternativen basiert also auf der Anschauung, dass nicht etwa die Ortskoordinaten der Klassischen Physik die eigentlich
richtige Naturbeschreibung darstellen, welche die Quantentheorie nun speziellen neuen Gesetzen unterwirft. Vielmehr stellt die
Quantentheorie eine für sich selbst bestehende unabhängige Theorie dar, die an sich keine klassischen Begriffe benötigt. Die
Möglichkeit der Beschreibung durch klassische Begriffe wie Raum-Zeit-Koordinaten ist vielmehr der Grenzfall einer bestimmten
Weise, die Quantentheorie darzustellen. Dieser Zusammenhang soll durch die Abbildung ($\ref{Ur-Alternativen_Raum-Zeit}$)
verdeutlicht werden.

\chapter{Programm für die Beschreibung der Elementarteilchen und ihrer Wechselwirkungen in der Quantentheorie der Ur-Alternativen}

Natürlich muss sich aus der Quantentheorie der Ur-Alternativen nicht nur eine Beschreibung der Struktur der Raum-Zeit, sondern
schließlich auch eine Beschreibung für die Elementarteilchen und ihre spezifischen Wechselwirkungen ergeben, welche als
aus Ur-Alternativen zusammengesetzt beschrieben werden müssen. Hierzu sei von Weizsäcker zitiert \cite{Weizsaecker:1985}:

\begin{quote}
{\small Die Elementarteilchentheorie müßte dann aus der Theorie der einfachsten nach der Quantenmechanik überhaupt möglichen
Objekte aufgebaut werden können; diese wären zugleich die einzigen Atome im ursprünglichen philosophischen Sinn schlechthinniger
Unteilbarkeit. Solche Objekte wären durch eine einzige einfache Meß-Alter-native, eine Ja-Nein Entscheidung definiert. Ihr
quantenmechanischer Zustandsraum ist ein zweidimensionaler komplexer Vektorraum, der in bekannter Weise auf einen
dreidimensionalen reellen Raum abgebildet werden kann. In diesem mathematischen Faktum möchte ich den physikalischen 
Grund der Dreidimensionalität des Weltraums vermuten. Die sogenannten Elementarteilchen müssen als Komplexe solcher
"`Urobjekte"' und eben darum als ineinander unwandelbar erscheinen. Die Symmetriegruppen der Elementarteilchenphysik
ebenso wie die Topologie des Weltraums sollten auf diese Weise aus der Struktur des quantenmechanischen Zustandsraums der
"`Urobjekte"' folgen. Ich nenne diese Hypothesen nicht, um sie schon als richtig anzukündigen, sondern um zu zeigen, dass
wir keinen Grund haben, eine Herleitung der Elementarteilchenphysik und der Kosmologie aus der Quantentheorie für unmöglich
zu halten.

Carl Friedrich von Weizsäcker, Die Einheit der Natur, 1971 (Seite 222)}
\end{quote}

\section{Elementarteilchen und Feldgleichungen}

\subsection{Konkrete Gestalt der Impulsoperatoren}

Wenn man die Komponenten des Impulsvierervektors, der in ($\ref{Groessen_Dirac-Spinor}$) definiert wurde, explizit in Abhängigkeit der Erzeugungs-
und Vernichtungsoperatoren ($\ref{Definition_abcd}$) ausdrückt, so haben sie folgende Gestalt

\begin{eqnarray}
P_0=c^{\dagger}a+d^{\dagger}b\quad,\quad
P_1=c^{\dagger}b+d^{\dagger}a\quad,\quad
P_2=-ic^{\dagger}b+id^{\dagger}a\quad,\quad
P_3=c^{\dagger}a-d^{\dagger}b.
\label{Impulsoperatoren}
\end{eqnarray}
Es ist nicht möglich, Eigenzustände zu den Impulsoperatoren ($\ref{Impulsoperatoren}$) gemäß der
entsprechenden Eigenwertgleichung

\begin{equation}
\hat P^\mu |P^\mu \rangle =P^\mu |P^\mu \rangle
\end{equation}
zu konstruieren.  

\subsection{Feldgleichungen}

Die in ($\ref{Groessen_Dirac-Spinor}$) und ($\ref{Quantisierung_Majorana-Spinor_Ur-Alternative}$) definierten
Impuls-Operatoren $P_\mu$ erfüllen die folgende Relation 

\begin{equation}
P^\mu P_\mu=0.
\label{Relation_Impuls-Operatoren}
\end{equation}
Angewandt auf einen Zustand im Tensorraum ($\ref{Allgemeiner_Zustand_Symmetrischer_Tensorraum}$)
bedeutet dies

\begin{equation}
P^\mu P_\mu|\Psi\rangle=0.
\label{Klein-Gordon-Gleichung_Tensorraum_Ur-Alternativen}
\end{equation}
Dies entspricht der Klein-Gordon-Gleichung für ein masseloses Teilchen. Es ist nun im Prinzip möglich,
($\ref{Klein-Gordon-Gleichung_Tensorraum_Ur-Alternativen}$) gemäß ($\ref{Klein-Gordon-Gleichung_Dirac-Algebra}$)
umzuschreiben

\begin{equation}
\gamma^\mu \gamma^\nu P_\mu P_\mu|\Psi\rangle=0,
\label{Klein-Gordon-Gleichung_Dirac-Algebra_Tensorraum_Ur-Alternativen_}
\end{equation}
wobei $|\Psi\rangle$ nun als das Tensorprodukt eines Zustandes im symmetrischen Tensorraum der Ur-Alternativen und
eines Dirac-Spinors, also einer weiteren nicht in die Symmetrisierung der Ur-Alternativen einbezogenen Ur-Alternative,
interpretiert werden muss

\begin{equation}
|\Psi\rangle=c(N_w,N_x,N_y,N_z)|N_w,N_x,N_y,N_z\rangle \otimes
\left(\begin{matrix} u_1\\u_2\\u_2^{*}\\-u_1^{*}\end{matrix}\right).
\label{Zustand_Teilchen_Spin}
\end{equation}
Dies führt auf die masselose Version der Dirac-Gleichung ($\ref{Dirac-Gleichung}$) dargestellt
im Tensorraum der Ur-Alternativen

\begin{equation}
\gamma^\mu P_\mu|\Psi\rangle=0,
\end{equation}
welche gemäß ($\ref{Dirac-Spinor_Weyl-Spinoren}$) der Weyl-Gleichung entspricht.
Wie bereits im letzten Abschnitt erwähnt erscheinen Masse und Wechselwirkung erst in einer allgemeineren Beschreibung
in einem Tensorraum mit allgemeiner Permutationssymmetrie. Bevor dieser jedoch behandelt wird, soll zunächst auf die
Beschreibung von Teilchen eingegangen werden. Aus dieser ergibt sich dann in natürlicher Weise die Notwendigkeit,
zu allgemeineren Symmetrieklassen im Tensorraum überzugehen.

\subsection{Zustände für Teilchen und Felder}

Es soll von der Annahme ausgegangen werden, dass ein freies Teilchen durch einen total-symmetrischen Zustand im Tensorraum
der Ur-Alternativen beschrieben wird. Ein dementsprechender Basiszustand eines Teilchens kann auch wie folgt geschrieben werden

\begin{equation}
|N_1,N_2,N_3,N_4\rangle=\prod_{i=1}^4 \left({a_i}^{\dagger}\right)^{N_i}|0\rangle,
\label{Teilchen-Basiszustand_Ur-Alternativen}
\end{equation}
falls die Zahlen in ($\ref{Teilchen-Basiszustand_Ur-Alternativen}$) in folgender Weise mit den die Basiszustände der
Ur-Alternativen ($\ref{Ur-Alternative}$) bezeichnenden Buchstaben identifiziert werden: $w \widehat{=} 1$, $x \widehat{=} 2$,
$y \widehat{=} 3$, $z \widehat{=} 4$. Ein allgemeiner Zustand eines Teilchens $|\Psi\rangle$ entspricht einer beliebigen
Linearkombination solcher Zustände und lautet damit

\begin{equation}
|\Psi \rangle=\sum_{N_1}\sum_{N_2}\sum_{N_3}\sum_{N_4}c(N_1,N_2,N_3,N_4)\prod_{i=1}^4 \left({a_i}^{\dagger}\right)^{N_i}|0\rangle.
\label{allgemeiner_Teilchen-Zustand_Ur-Alternativen}
\end{equation}
Man kann nun dementsprechend einen Teilchenerzeugungsoperator $a_{N_1,N_2,N_3,N_4}^{\dagger}$ beziehungsweise den dazugehörigen
Teilchenvernichtungsoperator $a_{N_1,N_2,N_3,N_4}$ definieren

\begin{equation}
a_{N_1,N_2,N_3,N_4}^{\dagger}=\prod_{i=1}^4 \left({a_i}^{\dagger}\right)^{N_i}\quad,\quad
a_{N_1,N_2,N_3,N_4}=\prod_{i=1}^4 \left({a_i}\right)^{N_i}.
\label{Operatoren_Ur-Alternativen_Teilchen}
\end{equation}
Wenn man nun ein symmetrisches Produkt solcher symmetrisierter Zustände bildet, so entspricht dies einem Vielteilchenzustand
beziehungsweise dem Zustand eines aus Feldquanten, denen Teilchen entsprechen, aufgebauten Feldes. Der Erzeugung beziehungsweise
Vernichtung von Teilchen in solchen symmetrischer Produktzuständen symmetrischer Zustände im Tensorraum der Ur-Alternativen
entsprechen die folgenden Vertauschungsrelationen zwischen den Erzeugungs- und Vernichtungsoperatoren für Teilchen

\begin{equation}
\left[a_{N_1,N_2,N_3,N_4},a_{N_{1}^{\prime},N_{2}^{\prime},N_{3}^{\prime},N_{4}^{\prime}}^{\dagger}\right]
=\delta_{N_{1},N_{1}^{\prime}}\delta_{N_{2},N_{2}^{\prime}}\delta_{N_{3},N_{3}^{\prime}}\delta_{N_{4},N_{4}^{\prime}}.
\end{equation}
Während ein Basiszustand für ein Teilchen (\ref{Teilchen-Basiszustand_Ur-Alternativen}) also durch die Anzahl von
Ur-Alternativen $N_w,N_x,N_y,N_z$ in den ihnen entsprechenden Basiszuständen charakterisiert ist, ist der Zustand eines
Quantenfeldes $|\Phi\rangle$ durch die Anzahl von Teilchen in solchen Basiszuständen charakterisiert: $\bar N=\bar
N(N_1,N_2,N_3,N_4)$, und lautet damit wie folgt

\begin{equation}
|\bar N\rangle=\sum_{N_1}\sum_{N_2}\sum_{N_3}\sum_{N_4}
\left[\prod_{i=1}^4 \left({a_i}^{\dagger}\right)^{N_i}\right]^{\bar N}|0\rangle.
\label{Quantenfeld-Basiszustand_Ur-Alternativen}
\end{equation}
Damit kann ein allgemeiner Zustand eines Quantenfeldes $|\Phi\rangle$ als

\begin{equation}
|\Phi\rangle=\sum_{N_1}\sum_{N_2}\sum_{N_3}\sum_{N_4}\sum_{\bar N}c[\bar N]
\left[\prod_{i=1}^4 \left({a_i}^{\dagger}\right)^{N_i}\right]^{\bar N}|0\rangle
\label{allgemeiner_Quantenfeld-Zustand_Ur-Alternativen}
\end{equation}
geschrieben werden.
Da ein symmetrisiertes Produkt symmetrisierter Zustände nicht wieder ein symmetrischer Zustand ist, kann ein solcher
Vielteilchenzustand also von einem Einteilchenzustand unterschieden werden. Er weist eine allgemeinere Symmetrieklasse auf.
Ein Ansatz zur Konstruktion einer Quantenfeldtheorie im Rahmen der Quantentheorie der Ur-Alternativen ist in \cite{Goernitz:1992}
zu finden.

\section{Allgemeine Symmetrieklassen und Wechselwirkung}

\subsection{Parabose-Statistik}

Es sollen nun Zustände mit allgemeiner Permutationssymmetrie, also nicht nur total-symmetrische oder total-anti-symmetrische
Zustände betrachtet werden. Diese entsprechen nicht der Bose- oder Fermi-Statistik, sondern der Parabose-Statistik.
Solche Zustände lassen sich mit Hilfe von Erzeugungs- und Vernichtungsoperatoren beschreiben, welche
verallgemeinerten Vertauschungsrelationen genügen, die folgende Gestalt aufweisen

\begin{eqnarray}
\left[\frac{1}{2}\{a_r, a_s^{\dagger}\},a_t\right]=-\delta_{st}a_r,\quad
\left[\{a_r, a_s\},a_t\right]=\left[\{a_r^{\dagger}, a_s^{\dagger}\},a_t^{\dagger}\right]=0\quad\quad r,s=1...4,
\label{Algebra_Parabose-Statistik}
\end{eqnarray}
wobei der untere Index den entsprechenden Basiszustand der erzeugten beziehungsweise vernichteten Ur-Alternative bezeichnet.
Die Vertauschungsrelationen ($\ref{Algebra_Parabose-Statistik}$) lassen sich realisieren, indem man neue Operatoren
$b_r^{\alpha},b_r^{\alpha\dagger}$ einführt und die Operatoren $a_r,a_r^{\dagger}$ in der folgenden Weise durch sie definiert

\begin{equation}
a_r=\sum_{\alpha=1}^{p} b_r^{\alpha}\quad,\quad a_r^{\dagger}=\sum_{\alpha=1}^{p} b_r^{\alpha\dagger},
\end{equation}
wobei $p$ die sogenannte Parabose-Ordnung bezeichnet.
Die Operatoren $b_r^{\alpha},b_r^{\alpha\dagger}$ erfüllen folgende Vertauschungsrelationen

\begin{eqnarray}
\left[b_r^{\alpha},b_s^{\alpha\dagger}\right]=\delta_{rs}\quad,\quad
\left[b_r^{\alpha},b_s^{\alpha}\right]=\left[b_r^{\alpha\dagger},b_s^{\alpha\dagger}\right]=0,\nonumber\\
\{b_r^{\alpha},b_s^{\beta\dagger}\}=\{b_r^{\alpha},b_s^{\beta}\}=\{b_r^{\alpha\dagger},b_s^{\beta\dagger}\}=0\quad,\quad \alpha
\neq \beta.
\label{Parabose-Operatoren}
\end{eqnarray}
Sie verhalten sich also bei gleichem oberen Index wie Bose-Operatoren und bei ungleichem oberen Index antikommutieren
alle Operatoren miteinander. $p$ beschreibt die sogenannte Parabose-Ordnung, wobei $p \in \mathbb{N}$. 
Ein Zustand $|\Psi_P\rangle$ mit $n$ Ur-Alternativen, welcher der Parabose-Statistik gehorcht, geht aus dem Vakuum-Zustand
$|0\rangle$ hervor, indem $n$ Parabose-Erzeugungsoperatoren ($\ref{Parabose-Operatoren}$) auf $|0\rangle$ angewandt
werden

\begin{equation}
|\Psi_P\rangle=\prod_{i=1}^{n} b_{r_i}^{\alpha_i}|0\rangle. 
\end{equation}
Ein solcher Zustand entspricht einem Zustand, der dadurch gewonnen wird, dass man von einem
einfachen Produktzustand von $n$ Ur-Alternativen ausgeht

\begin{equation}
|\varphi\rangle=\bigotimes_n  |u_{r_n}^{\alpha_n}\rangle
\end{equation}
und anschließend über alle Permutationen summiert, wobei das Vorzeichen bei einer Transposition zweier
Ur-Alternativen dann gleich bleibt, wenn die oberen Indizes übereinstimmen und sich dann umkehrt, wenn
die oberen Indizes verschieden sind.\\

\subsection{Young-Diagramme für Ur-Alternativen}
 
Die allgemeinen Zustände der Parabose-Statistik für Ur-Alternativen können natürlich wie die Zustände von Teilchen
beschreibenden Quantenzahlen auch durch Young-Diagram-me beschrieben werden. Ein Young-Rahmen des Rangs $n$ ist
allgemein eine Konstruktion, welche aus $l$ Zeilen besteht, deren jede $m_l$ Elemente enthält, wobei gilt:
$i > j \rightarrow m_i \leq m_j,\ i,j=1...l$.
Ein Standard-Tableau entspricht nun einem Young-Rahmen, in den $n$ Zahlen $z_i,\ i=1...n$ als seine Elemente eingesetzt
werden, wobei gilt, dass $z_i \leq d$, wenn $d$ die Dimension des Vektorraumes beschreibt, über dem der entsprechende
Tensorraum definiert ist, und die Zahlen in jeder Zeile nach rechts und in jeder Spalte nach unten zunehmen
müssen. Die verschieden Rahmen einer Ordnung $n$ seien durch den Index $k$ gekennzeichnet. Es gibt nun zu jedem Rahmen
einer Ordnung $n$ $f_k$ verschiedene Standard-Tableaux und jeder dieser Rahmen definiert $f_k$ verschiedene Darstellungen
der symmetrischen Gruppe, wobei gilt: $\sum_k f_k^2=n !$. Außerdem definiert der Rahmen $f_k$ Darstellungen der
$GL(d)$, also der Gruppe der umkehrbaren Matrizen in $d$-Dimensionen. Die Basistensoren jeder dieser Darstellungen können
nun durch Standard-Schemata gewonnen werden. Ein Standard-Schema entspricht einem Young-Rahmen, in den ebenso wie bei einem
Standard-Tableau $n$ Zahlen als seine Elemente eingesetzt werden, wobei gilt, dass $z_i \leq n$ und die Zahlen in jeder Spalte
nach unten zunehmen müssen. Im Gegensatz zu einem Standard-Tableau gilt allerdings jetzt in Bezug auf die Zeilen die schwächere
Bedingung, dass die Zahlen nach links nicht zunehmen dürfen, anstatt nach rechts zunehmen zu müssen. Sowohl ein
Standard-Tableau als auch ein Standard-Schema können nicht mehr als $d$ Zeilen enthalten. Wenn man dies nun auf die
Ur-Alternativen überträgt, so bedeutet dies, dass ein entsprechendes Standard-Tableau oder Standard-Schema nicht mehr als $d$
Zeilen enthalten kann.

\subsection{Wechselwirkung}

Um nun Wechselwirkungen beschreiben zu können, muss eine Beziehung zwischen Zustän-den betrachtet werden.
Hierzu sei zunächst der Begriff des Singletons definiert. Ein Singleton ist ein Zustand im Tensorraum der Ur-Alternativen zu
$p=1$. Damit entspricht er im Grunde dem Zustand eines masselosen Teilchens gemäß
($\ref{allgemeiner_Teilchen-Zustand_Ur-Alternativen}$), wird jedoch jetzt als Element eines Unterraumes des Raumes der
Zustände mit beliebigem $p$ angesehen 

\begin{equation}
|\varphi^{\alpha}(N_1^{\alpha},N_2^{\alpha},N_3^{\alpha},N_4^{\alpha})\rangle
=\prod_{i=1}^4 \left(b_{i}^{\alpha}\right)^{N_i}|0\rangle.
\end{equation}
Man nun das Tensorprodukt von $p$ Zuständen der Ordnung $p=1$ bilden

\begin{equation}
|\phi\rangle=\bigotimes_{\alpha=1}^{p}|\varphi^{\alpha}\rangle
\label{Tensorprodukt_Singletonen}
\end{equation}
und dieses symmetrisieren

\begin{equation}
|\phi_S\rangle=\Sigma_{\pi(\alpha)} \bigotimes_{\alpha=1}^{p}|\varphi^{\alpha}\rangle,
\label{Tensorprodukt_Singletonen_symmetrisiert}
\end{equation}
wobei $\pi(\alpha)$ alle Permutation über $\alpha$ bezeichnen soll. In dem Zustand
($\ref{Tensorprodukt_Singletonen}$) ist die Information enthalten, welche Ur-Alternativen
zu welchem Singleton gehören, während in dem Zustand ($\ref{Tensorprodukt_Singletonen_symmetrisiert}$)
nur die Information steckt, wie viele Ur-Alternativen, die sich in einem bestimmten Basiszustand
befinden, in jedem Singleton $\alpha$ enthalten sind.
Ein Zustand ($\ref{Tensorprodukt_Singletonen_symmetrisiert}$) entspricht in eindeutiger Weise einem Zustand der
Ordnung $p$. Dies ist in umgekehrter Richtung nicht der Fall. Ein Zustand der Ordnung $p$ kann in unterschiedlicher
Weise als symmetrisiertes Produkt von Zuständen der Ordnung $p=1$ dargestellt werden. Diese verschiedenen Darstellungen
sind allerdings nicht unabhängig voneinander, sondern weisen lineare Beziehungen zueinander auf.
Wenn man also $p$ Zustände zu $p=1$, die in einem Zustand der Parabose-Ordnung $p$ enthalten sind, als getrennte Objekte
beschreibt, in dem man ein symmetrisches Produkt ihrer jeweils in sich bereits symmetrischen Zustände betrachtet, so
stellt sich die im Gesamtzustand enthaltene Information über ihre linearen Beziehungen als Wechselwirkung zwischen ihnen dar.
Es ist nun eine Vermutung, dass sich die Gesamtdarstellung der Ordnung $p$ gemäß der Dynamik einer freien
Alternative entwickelt: $i\frac{\partial \varphi}{\partial t}=H \varphi$. Diese freie Dynamik der
Gesamtdarstellung stellt sich aber innerhalb der einzelnen Unterräume als Wechselwirkung dar, da in
den Zuständen der totalsymmetrischen Unterräume auf die im Gesamtzustand enthaltene Information, in welche
diese Zustand in linearen Beziehungen stehen, verzichtet wird.\\
\\
\fbox{\parbox{145 mm}{Wenn ein Zustand der Ordnung $p=k$ als ein symmetrisiertes Produkt von $k$
Zuständen der Ordnung $p=1$ dargestellt wird, so besteht eine Wechselbeziehung zwischen den 
möglichen Produktzuständen. Dies definiert eine Wechselwirkung.}}\\
\\
Damit kann man nun allgemein sagen, dass Wechselwirkung eine Folge der Beschreibung eines Ganzen durch Teile ist.
Wenn ein Ganzes in für sich getrennt beschrieben Teile aufgespalten wird, denen trennbare empirisch entscheidbare
Alternativen entsprechen, so entspricht dem eine Näherung, in der auf Information verzichtet wird. Die Existenz
der Wechselwirkung ergibt sich dann als eine Korrektur dieser Näherung. Hiermit in Zusammenhang stehende Überlegungen
von Weizsäckers in Bezug auf die Platonische Philosophie sind in \cite{Weizsaecker:1971} und auch in \cite{Weizsaecker:1981}
zu finden. 
Da Alternativen in ihrer Darstellung durch Ur-Alternativen als in einer Minkowski-Raum-Zeit existierend 
dargestellt werden können, ist die räumliche Beschreibung eine Folge beziehungsweise eine Darstellung der Trennung
und die Wechselwirkung als Korrektur dieser Näherung hängt demgemäß von räumlichen Verhältnissen ab.
Hierzu sei von Weizsäcker selbst zitiert \cite{Weizsaecker:1985}:
\newpage

\begin{quote}
{\small Da wir nur durch Wechselwirkung beobachten, ist jede Messung zunächst eine Ortsmessung. Diese Abhängigkeit der
Wechselwirkung vom Ort musste man in der korrespondenzmäßigen Auffassung als eine empirische Tatsache schlicht hinnehmen.
In einer abstrakten Auffassung liegt es wieder nahe, die Anordnung der Argumente umzukehren. Wenn es überhaupt einen
Zustandsparameter gibt, von dem alle Wechselwirkungen abhängen, so darf man erwarten, dass die real beobachtbaren Objekte
und ihre Zustände am direktesten in einer Darstellung beschrieben werden können, die diesen Parameter als unabhängige
Variable zugrunde legt.

Carl Friedrich von Weizsäcker, Aufbau der Physik, 1985 (Seite 382)}
\end{quote}
In der zweiten Kantischen Antinomie (siehe \textbf{[1.2.2]}) wird in der Weise argumentiert, dass ein räumliches Objekt immer noch
weiter unterteilbar ist. Die Idee der Separation einzelner Teile eines physikalischen Systems ist mit der Raumanschauung
verbunden, welche in der Quantentheorie der Ur-Alternativen einer Darstellung der Beschreibung durch Alternativen ist, die als
trennbar angesehen werden. Wenn nun die Räumlichkeit der Naturbeschreibung letztlich eine Folge der Trennbarkeit der
Alternativen ist, da sie einer Darstellung von ausgedehnten teilbaren Objekten entspricht, die miteinander in einer
Wechselwirkungsbeziehung stehen, so ist die Räumlichkeit der Wahrnehmung, welche bei Kant eine Bedingung der Möglichkeit von
Erfahrung darstellt, im Rahmen des von Weizsäckerschen Programmes indirekt gewissermaßen in einem hergeleiteten
Sinne Bedingung der Möglichkeit von Erfahrung, also eine spezielle Manifestation der Trennbarkeit der Alternativen als
Bedingung der Möglichkeit von Erfahrung.
Wenn man sich nun daran erinnert, dass sich in \textbf{[8.2.2]} die Symmetrie der Zustandsräume der Alternativen als eine Folge der
Trennbarkeit der Alternativen ergab, so zeigt sich die Konsequenz dessen in der Tatsache, dass ein physikalischer Zustand  
im Allgemeinen nur dann invariant unter einer Poincar\'{e}-Transformation sein wird, wenn er ein isoliertes Objekt beschreibt,
das also näherungsweise als wechselwirkungsfrei angenommen wird.
Sobald andere Objekte existieren, wird die Homogenität und Isotropie des Raumes zerstört und eine Änderung der räumlichen
Lage des betrachteten Objektes ändert die räumliche Beziehung zu den anderen Objekten, mit denen es in Wechselwirkung steht.
Wechselwirkung ist aber eben eine Korrektur der approximativen Annahme der Trennbarkeit.\\
\\
\fbox{\parbox{145 mm}{Die Existenz des Phänomens der Wechselwirkung zwischen verschiedenen Objekten sollte in diesem
Sinne also wohl als eine Folge der Beschreibung der Welt innerhalb der Physik durch getrennte Objekte angesehen werden.}}

\subsection{Masse und innere Symmetrien}

Wenn ein Objekt durch einen Zustand zu einer Parabose-Ordnung $p \geq 1$ beschrieben wird beziehungsweise aus mehreren Singletonen
zusammengesetzt ist, so wird seine Ruhemasse im Allgemeinen nicht mehr gleich null sein. Man kann nämlich nun zu jedem 
$\alpha,\alpha=1...p$ der Darstellung der Ordnung $p$ zu den entsprechenden Parabose-Operatoren ($\ref{Parabose-Operatoren}$)
gehörige Generatoren definieren, welche mit den durch ($\ref{Groessen_Dirac-Spinor}$) und
($\ref{Quantisierung_Majorana-Spinor_Ur-Alternative}$) definierten
Operatoren im total-symmetrischen Tensorraum identisch sind und die konforme Gruppe beziehungsweise die ihr entsprechende Algebra
($\ref{Algebra_Konforme-Gruppe}$) mit der in ihr enthaltenen Poincar\'{e}-Gruppe im entsprechenden Unterraum $\alpha$
darstellen. Die Impuls-Operatoren haben damit folgende Gestalt:

\begin{eqnarray}
P_0^{\alpha}=b_3^{\alpha\dagger}b_1^{\alpha}+b_4^{\alpha\dagger}b_2^{\alpha},\quad
P_1^{\alpha}=b_3^{\alpha\dagger}b_2^{\alpha}+b_4^{\alpha\dagger}b_1^{\alpha},\quad
P_2^{\alpha}=-ib_3^{\alpha\dagger}b_2^{\alpha}+ib_4^{\alpha\dagger}b_1^{\alpha},\quad
P_3^{\alpha}=b_3^{\alpha\dagger}b_1^{\alpha}-b_4^{\alpha\dagger}b_2^{\alpha}.\nonumber\\
\label{Impulsoperatoren_Unterraum_alpha_Parabose-Darstellung}
\end{eqnarray}
Gemäß ($\ref{Relation_Impuls-Operatoren}$) erfüllen die Impuls-Operatoren ($\ref{Impulsoperatoren_Unterraum_alpha_Parabose-Darstellung}$)
im jeweiligen Unterraum $\alpha$ die Bedingung $P^{\alpha\mu} P^{\alpha}_{\mu}=0$. Jeder Zustand eines Unterraumes $\alpha$ ist also masselos.
Man kann aber nun auch einen Gesamtimpulsoperator $P_{G \mu}$ für die Parabose-Darstellung der Ordnung $p$ als die Summe der Impulsoperatoren
der einzelnen Unterräume definieren

\begin{equation}
P_{G \mu}=\sum_{\alpha=1}^{p}P_\mu^{\alpha}.
\end{equation}
Da eine Summe von lichtartigen Minkowski-Vektoren im allgemeinen nicht wieder lichtartig ist, was auch für die entsprechenden
Operatoren gilt, ergibt sich

\begin{equation}
P_{G}^{\mu}P_{G \mu}=\left(\sum_{\alpha=1}^{p}P^{\mu\alpha}\right)
\left(\sum_{\alpha=1}^{p}P_\mu^{\alpha}\right)=m^2,
\end{equation}
wobei $m$ im Allgemeinen von null verschieden ist. Wenn also nun Masse durch eine Kombination mehrerer Unterräume zu $p=1$
in einer Darstellung zu $p \geq 1$ entsteht und Wechselwirkung ein Phänomen ist, dass durch die Beziehungen von Zuständen
solcher Unterräume in einem Zustand im gesamten Raum der Parabose-Darstellung definiert ist, so ist der Begriff der Masse
also mit dem Begriff der Wechselwirkung verknüpft. Massebehaftete Teilchen müssen damit also als aus zwei oder mehr Singletonen
zusammengesetzt angenommen werden.
Ein hiermit im Zusammenhang stehender Aspekt ist nun die Existenz von Quantenzahlen. Wenn die Symmetrieeigenschaften der
Raum-Zeit gemäß der Poincar\'{e}-Gruppe Eigenschaften der Ur-Alternativen und ihres Tensorraumes widerspiegeln 
(siehe \textbf{[9.2]}) beziehungsweise die entsprechenden Zustände in einer Raum-Zeit dargestellt werden können, so stellt sich die
Frage, in welcher Weise sich die inneren Symmetrien und die Quantenzahlen, auf die sie sich beziehen, in der Quantentheorie der
Ur-Alternativen ergeben sollen. Warum gibt es also noch zusätzliche interne Zustandsräume, wenn die Raum-Zeit-Beschreibung von
Zuständen schon eine Darstellung solcher diskreter spezifisch quantentheoretischer Zustände ist ?
Wenn man diesbezüglich zunächst auf den Spin eingeht, so könnte dieser gemäß ($\ref{Zustand_Teilchen_Spin}$) beispielsweise
durch einen Produktzustand einer einzelnen Ur-Alternative mit einem total-symmetrischen Zustand beschrieben werden, wobei die
einzelne den Spin beschreibende Ur-Alternative nicht in die Symmetrisierung miteinbezogen ist. Was ist aber mit den rein
internen Quantenzahlen wie etwa dem schwachen Isospin oder dem Flavour ?
Hierbei ist nun wichtig, dass Quantenzahlen mit Eigenschaften von Teilchen wie etwa der Ladung oder der Ruhemasse verbunden
sind, die rein in Ihnen selbst liegen und nicht in Bezug auf andere Objekte definiert sind, was bei einer räumlichen
Eigenschaft wie dem Ort, der Geschwindigkeit oder der räumlichen Richtungsorientierung sehr wohl der Fall ist, da diese nicht
absolut definiert sind. Daher ist zu vermuten, dass einer Quantenzahl und einer entsprechenden Transformation im ihr
entsprechenden Zustandsraum eine Beziehung von Ur-Alternativen innerhalb eines Teilchens entspricht, während einer auf den Raum
bezogenen Eigenschaft, die Beziehung der Ur-Alternativen eines Teilchens zu allen anderen Ur-Alternativen im Kosmos entspricht.
Das bedeutet, dass etwa einer Translation oder Drehung, eine Transformation aller Ur-Alternativen innerhalb des Zustandes eines
Teilchens entspricht, wodurch diese ihre Beziehung zu allen anderen Ur-Alternativen des Kosmos verändern. Einer internen
Transformation hingegen würde eine Transformation eines Teiles der Ur-Alternativen, oder wie etwa im Falle des schwachen
Isospins, eine auf eine einzige Ur-Alternative bezogene Transformation in Bezug auf die anderen Ur-Alternativen innerhalb
des Teilchens entsprechen, also eine Transformation von Ur-Alternativen innerhalb des Zustandes eines Teilchens in Bezug
aufeinander.

\addtocontents{toc}{\protect\newpage}

\part{Das Verhältnis der Quantentheorie zur Allgemeinen\\ Relativitätstheorie}

\chapter{Das Problem der Vereinheitlichung der beiden Theorien}

Es existieren verschiedene Ansätze zu einer quantentheoretischen Fassung der Allgemeinen Relativitätstheorie. Grundsätzlich
kann man zwischen zwei Gruppen von Ansätzen zur Quantisierung des Gravitationsfeldes unterscheiden. Es gibt zum einen die Gruppe
der kovarianten Quantisierungsansätze und zum anderen diejenigen Ansätze der kanonischen Quantisierung. Im Falle der kanonischen
Ansätze zu einer Quantisierung wird die Raum-Zeit im Unterschied zu den kovarianten Quantisierungsansätzen in eine raumartige
Untermannigfaltigkeit sowie eine Zeitkomponente formal aufgespalten.
Eine Einführung in die verschiedenen Quantisierungsansätze ist beispielsweise in \cite{Kiefer:2004} oder
\cite{Kiefer:2005uk} gegeben.

\section{Kovariante Quantisierung}

Der naheliegendste Ansatz zu einer Quantisierung der Gravitation versucht die Gravitation in Analogie zu den anderen
Wechselwirkungen des Standardmodells als Quantenfeldtheorie auf einer Minkowski-Raum-Zeit zu beschreiben
(siehe beispielsweise \cite{Donoghue:1995cz}).
Hierbei wird von einer Aufspaltung des metrischen Feldes $g_{\mu\nu}$ in die Minkowski-Metrik einer flachen
Minkowski-Hintergrund-Raum-Zeit $\eta_{\mu\nu}$ und einer Störung $h_{\mu\nu}$ ausgegangen

\begin{equation}
g_{\mu\nu}=\eta_{\mu\nu}+h_{\mu\nu},
\label{Aufspaltung_Metrik}
\end{equation}
wobei die Störung dem Gravitationsfeld entspricht, das nach der Quantisierung im Sinne üblicher relativistischer
Quantenfeldtheorien als aus Feldquanten aufgebaut gedacht wird, welche analog als Austauschteilchen der Gravitation
zu interpretieren sind und daher als Gravitonen bezeichnet werden.

Wenn man ($\ref{Aufspaltung_Metrik}$) in ($\ref{Christoffelsymbole}$) einsetzt und ($\ref{Christoffelsymbole}$)
in ($\ref{Riemann-Tensor}$) verwendet, so ergeben sich für ($\ref{Ricci-Tensor}$) und ($\ref{Ricci-Skalar}$) in einer linearen
Näherung, die dann sinnvoll ist, wenn man davon ausgeht, dass die Störung $h_{\mu\nu}$ sehr klein ist, die folgenden Ausdrücke

\begin{eqnarray}
R_{\mu\nu}&=&\frac{1}{2}\left[\partial_\mu \partial_\nu h^{\lambda}_{\ \lambda}-\partial_\mu \partial_\lambda
h^{\lambda}_{\ \nu}-\partial_\nu \partial_\lambda h^{\lambda}_{\ \mu}+\partial_\lambda
\partial^{\lambda}h_{\mu\nu}\right]+\mathcal{O}\left(h^2\right)\nonumber\\
R&=&\left[\partial_\mu \partial^\mu h^{\lambda}_{\ \lambda}-\partial_\lambda \partial^\mu
h^{\lambda}_{\ \mu}\right]+\mathcal{O}\left(h^2\right).
\label{Ricci-Tensor-Skalar_Lineare_Naeherung}
\end{eqnarray}
Die sich aus ($\ref{Ricci-Tensor-Skalar_Lineare_Naeherung}$) ergebende und der Einsteinschen Gleichung
($\ref{Einsteinsche_Feldgleichung}$) ergebende linearisierte Einsteinsche Gleichung ist eichinvariant unter
der folgenden Transformation

\begin{equation}
x^\mu \rightarrow x^\mu+\epsilon^\mu(x)\quad,\quad h_{\mu\nu} \rightarrow h_{\mu\nu}-\partial_\mu \epsilon_\nu(x)
-\partial_\nu \epsilon_\mu(x).  
\end{equation}
Man kann nun die Eichung wie folgt fixieren: $\partial_\lambda h^{\lambda}_\mu=\frac{1}{2}\partial_\mu h^{\lambda}_\lambda$,
sodass sich für die linearisierte Einsteinsche Gleichung in der Formulierung ($\ref{Einsteinsche_Feldgleichung_alternative_Darstellung}$) ergibt

\begin{equation}
\partial_\lambda \partial^\lambda h_{\mu\nu}=-8\pi G\left(T_{\mu\nu}-\frac{1}{4}Tg_{\mu\nu}\right),
\label{linearisierte_Einsteinsche_Gleichung}
\end{equation}
was für $T_{\mu\nu}=0$ bedeutet, dass $\partial_\lambda \partial^{\lambda}h_{\mu\nu}=0$. Dies entspricht einer freien
Klein-Gordon-Gleichung ($\ref{Klein-Gordon-Gleichung}$). Dementsprechend kann das freie Gravitationsfeld $h_{\mu\nu}(x)$
nach ebenen Wellen entwickelt werden

\begin{eqnarray}
h_{\mu\nu}\left(\mathbf{x},t\right)=\sum_{\sigma}\int \frac{d^3 p}{\left(2\pi\right)^3\sqrt{2 p_0}}
\left[a\left(\mathbf{p},\sigma\right)e_{\mu\nu}\left(\mathbf{ p},\sigma\right) e^{ip_\mu x^\mu}
+a^{\dagger}\left(\mathbf{ p},\sigma\right)e_{\mu\nu}^{*}\left(\mathbf{ p},\sigma\right)e^{-ip_\mu x^\mu}\right],\nonumber\\
p_0=\sqrt{\mathbf{ p}^2+m^2}\quad\quad
\end{eqnarray}
und durch Forderung von Vertauschungsrelationen $\left[a(\mathbf{ p},\sigma),a^{\dagger}(\mathbf{p^{\prime}},\sigma^{\prime})\right]
=\delta_{\sigma,\sigma^{\prime}}\delta^3(\mathbf{ p}-\mathbf{ p}^{\prime})$ eine Quantisierung durchgeführt werden.
Die Polarisationsvektoren $e_{\mu\nu}$ beschreiben den Spin der Gravitonen, welche Spin $2$ aufweisen, was einer Darstellung
der Lorentz-Gruppe aus der in ($\ref{Darstellungen_Lorentz-Gruppe}$) begonnenen Reihe entspricht.
Dieses Quantisierungsverfahren führt jedoch auf eine nicht-renormalisierbare Quantenfeldtheorie. Zudem ist sie es nicht
hintergrundunabhängig, da die Minkowski-Metrik als nicht dynamische Hintergrundmetrik vorausgesetzt wird.
 
In einem weiteren vielversprechenden Ansatz geht man von der Methode der Pfadintegralquantisierung
($\ref{Feynmansches_Pfadintegral_Felder}$) aus, um zu einer quantentheoretischen Beschreibung der Gravitation zu gelangen

\begin{equation}
Z=\int \mathcal{D}g_{\mu\nu}\exp\left[iS(g_{\mu\nu})\right].
\end{equation}
Dieser Ansatz wurde in der Euklidischen Formulierung von Hawking ausgearbeitet \cite{Hawking:1978jz}.

\section{Kanonische Quantisierung}

Die zweite Gruppe der Quantisierungsversuche besteht in den kanonischen Quantisierungen. Im Rahmen der kanonischen
Quantisierungsansätze bringt man die Allgemeine Relativitätstheorie zunächst in die Hamiltonsche Form,
um dann eine kanonische Quantisierung im Sinne von ($\ref{Vertauschungsrelationen_Quantenfeldtheorie}$) durchführen zu können
(siehe beispielsweise \cite{Giulini:2006xi}).  
Hierzu muss zunächst eine (3+1)-Zerlegung der Raum-Zeit-Mannigfaltigkeit $M$ durchgeführt werden, also eine Zerlegung in eine
Zeitkoordinate $t$ und eine raumartige Untermannigfaltigkeit, eine Cauchy-Fläche $\Sigma_t$: $M=t \times \Sigma_t$.
Dies entspricht der Einführung eines Vektorfeldes $t^\mu$, das auf der Raum-Zeit-Mannigfaltigkeit lokal die Richtung der
Zeitkoordinate definiert. Dieses Vektorfeld $t^\mu$ stellt an jedem Raum-Zeit-Punkt eine Linearkombination einer Komponente 
innerhalb der Cauchy-Fläche und einer Komponente senkrecht zu dieser Cauchy-Fläche dar: $t^\mu=N n^\mu+N^\mu$, wobei $n^\mu$
den am jeweiligen Punkt zur Cauchy-Fläche orthogonalen Einheitsvektor, $N$ den Betrag der durchlaufenen Zeit und $N^\mu$ die
entsprechende Verschiebung parallel zur Cauchy-Fläche bezeichnet, die mit einer Verschiebung entlang der Zeitkoordinate einhergeht.

\subsection{Quantengeometrodynamik}

Es kann nun der sich auf die Cauchy-Fläche $\Sigma_t$ als dreidimensionaler Untermannigfaltigkeit der Raum-Zeit-Mannigfaltigkeit
beziehende Anteil $h_{ab}$ der Raum-Zeit-Metrik $g_{\mu\nu}$ separat betrachtet werden. Mit Hilfe von $h_{ab}$ kann die 
Einstein-Hilbert-Wirkung beziehungsweise die dazugehörige Lagrange-Dichte $\mathcal{L}_G$ ausgedrückt und somit ein kanonisch
konjugierter Impuls $p^{ab}$ zu $h_{ab}$ definiert werden

\begin{equation}
p^{ab}=\frac{\partial \mathcal{L}_G}{\partial h_{ab}}.
\label{kanonisch-konjugierter_Impuls_Quantengeometrodynamik}
\end{equation}
Mit Hilfe von ($\ref{kanonisch-konjugierter_Impuls_Quantengeometrodynamik}$) kann weiter eine entsprechende Hamilton-Dichte
$\mathcal{H}_g$ gemäß ($\ref{Hamiltondichte}$) definiert werden, welche sich aus einem auf die zur Cauchy-Fläche 
$\Sigma_t$ senkrechten Anteil des Vektorfeldes $t^\mu$ und aus einem in $\Sigma_t$ liegenden Anteil von $t^\mu$
zusammensetzt: $\mathcal{H}_g=N\mathcal{H}_0+N_a \mathcal{H}_a$, was für die Lagrange-Dichte entsprechend bedeutet:
$\mathcal{L}_G=p^{ij}\partial_t h_{ij}-N\mathcal{H}_0-N_a \mathcal{H}_a$. Es gelten die beiden Zwangsbedingungen
$\mathcal{H}_0=0$ und $\mathcal{H}_a=0$.
Die Quantisierung geschieht nun gemäß ($\ref{Vertauschungsrelationen_Quantenfeldtheorie}$), indem für die kanonischen
Variablen $h_{ab}$ und $p^{ab}$ die folgende Vertauschungsrelation postuliert wird

\begin{equation}
\left[h_{ab}(x),p^{cd}(y)\right]=\left(\delta_a^c\delta_b^d-\delta_b^c\delta_a^d\right)\delta\left(x-y\right),
\label{kanonische_Quantisierung_Gravitation_Quantengeometrodynamik}
\end{equation}
wodurch $h_{ab}$ und $p^{ab}$ zu Operatoren werden, welche auf Zustände $\Psi(h_{ab})$ wirken

\begin{equation}
h_{ab}(x)\Psi\left[h_{ab}(x)\right]=h_{ab} \Psi\left[h_{ab}(x)\right]\quad,\quad p^{ab}(x)\Psi\left[h_{ab}(x)\right]
=i\frac{\delta}{\delta h_{ab}}\Psi\left[h_{ab}(x)\right].
\label{Operatoren_Zustaende_Quantengeometrodynamik}
\end{equation}
Dies führt zur Quantengeometrodynamik.

\subsection{Quantisierung mit Zusammenhängen und Holonomien}

Abhay Ashtekar hat neue Variablen eingeführt, auf denen eine alternative Formulierung der kanonischen Quantisierung der
Gravitation basiert \cite{Ashtekar:1986yd},\cite{Ashtekar:1987gu},\cite{Ashtekar:1989ju}. Die von Carlo Rovelli und Lee Smolin
formulierte Schleifenquantengravitation \cite{Rovelli:2004},\cite{Rovelli:1987df},\cite{Rovelli:1989za} verwendet wiederum die
Ashtekar-Variablen, um mit Ihnen Holonomien zu definieren, auf denen dann die quantentheoretische Beschreibung der Gravitation
im Rahmen dieser Theorie basiert. Dieser Ansatz stellt unter den bisherigen mathematisch weit ausgearbeiteten Theorien einen sehr aussichtsreichsten Versuch einer Quantisierung der Gravitation dar. Bei den Ashtekar Variablen handelt es sich um den mit
seiner eigenen Determinante multiplizierten räumlichen Anteil des Vierbeins $E^a_i$ (auch als Dreibein bezeichnet) 

\begin{equation}
E^a_i(x)=\sqrt{h}e^a_i(x),
\label{gewichtetes_Vierbein_Ashtekar}
\end{equation}
wobei $h_{ij}=e^a_i e^b_j \delta_{ab}$ und $\sqrt{h}=\det\left(e^a_i\right)$, und um den Zusammenhang $A_a^i$, dessen kanonisch
konjugierte Variable $E^a_i$ ist und welcher wie folgt definiert ist

\begin{equation}
G A_a^i=\Gamma_a^i(x)+\beta K_a^i(x),
\label{Zusammenhang_Ashtekar}
\end{equation}
wobei $K_a^i$ die extrinsische Krümmung beschreibt $K_{\mu\nu}=h_\mu^\rho \nabla_\rho n_\nu$
und $\Gamma_k^i$ über den Spinzusammenhang $\omega_{ijk}$ in der folgenden Weise definiert ist:
$\Gamma_k^i=-\frac{1}{2}\omega_{ijk}\epsilon^{ijk}$. $\beta$ bezeichnet den sogenannten Immirzi-Parameter.
Da die Variablen $E^a_i$ und $A_a^i$ zueinander kanonisch konjugiert sind, was bedeutet, dass sie
die folgende Relation erfüllen

\begin{equation}
\left\{A_a^i(x),E_j^b(y)\right\}_P=8\pi \beta G\delta_a^b \delta_j^i \delta^3(x-y),
\end{equation}
wobei $\{\ \cdot\ ,\ \cdot \}_P$ in diesem Zusammenhang die Poisson-Klammer beschreibt, und $E^a_i$ daher
auch über die durch $A_a^i$ ausgedrückte Wirkung definiert werden kann: $E_i^a=\frac{\delta S[A]}{\delta A_a^i}$.
Die kovariante Ableitung angewandt auf einen Vektor $v_i$ in der Cauchy-Fläche hat in dieser Formulierung
der Allgemeinen Relativitätstheorie folgende Gestalt

\begin{equation}
D_a v_i=\partial_a v_i-\epsilon_{ijk}A_a^j v^k,
\label{Kovariante_Ableitung_Ashtekar}
\end{equation}
wobei der Zusammenhang $(\ref{Zusammenhang_Ashtekar}$), da er sich auf den dreidimensionalen Raum $\Sigma_t$
bezieht und ein Paralleltransport eines Vektors $v^i$ im Tangentialraum von $\Sigma_t$ einer Rotation
entspricht, als Eichzusammenhang der $SO(3)$ interpretiert werden kann, welche wiederum isomorph
zur $SU(2)$ ist.
\footnote{Dies gilt in dem Sinne, dass jedem Element der $SO(3)$ zwei Elemente der $SU(2)$ entsprechen.}
Das bedeutet, dass der der Variablen ($\ref{Zusammenhang_Ashtekar}$) entsprechende Zusammenhang auch als Element
der $SU(2)$ interpretiert und dementsprechend als Linearkombination ihrer Generatoren, also der Pauli-Matrizen,
ausgedrückt werden kann $A_a^i \sigma^a$. Mit dieser kovarianten Ableitung
($\ref{Kovariante_Ableitung_Ashtekar}$) kann man nun auch eine entsprechende Feldstärke definieren

\begin{equation}
F_{ab}^i=\partial_a A_b^i-\partial_b A_a^i+\epsilon^{i}_{jk}A_a^jA_b^k.
\end{equation}
Die Dynamik der Allgemeinen Relativitätstheorie kann mit Hilfe der Ashtekarschen Variablen
($\ref{gewichtetes_Vierbein_Ashtekar}$) und ($\ref{Zusammenhang_Ashtekar}$) in den folgenden Gleichungen
ausgedrückt werden 

\begin{equation}
D_a E_i^a=0\quad,\quad E_i^a F^i_{ab}=0\quad,\quad F_{ab}^{ij}E_i^a E_j^b=0,
\label{Dynamik_Gravitation_Ashtekar}
\end{equation}
wobei die beiden hinteren Gleichungen den Zwangsbedingungen $\mathcal{H}_0=0$ und $\mathcal{H}_a=0$ entsprechen.
Die Quantisierung erfolgt nun völlig analog zu ($\ref{kanonische_Quantisierung_Gravitation_Quantengeometrodynamik}$)
und ($\ref{Operatoren_Zustaende_Quantengeometrodynamik}$) durch Postulieren einer Vertauschungsrelation zwischen
den kanonisch konjugierten Variablen

\begin{equation}
\left[A_a^i(x),E_j^b(y)\right]=(8\pi \beta i \hbar G)\delta_a^b \delta_j^i \delta^3(x-y),
\label{kanonische_Quantisierung_Gravitation_Ashtekar}
\end{equation}
welche damit auch zu Operatoren werden, welche auf Zustände in einem Hilbert-Raum $\mathcal{H}$ wirken

\begin{equation}
A_a^i(x)\Psi[A]=A_a^i(x)\Psi[A]\quad,\quad E_j^b(y)\Psi[A]=8\pi\beta\frac{\hbar}{i}\frac{\delta}{\delta A_j^b}\Psi[A].
\label{Operatoren_Zustaende_Ashtekar}
\end{equation}
Indem man im Sinne Diracs in den dynamischen Gleichungen die auftretenden Größen durch die durch die Quantisierung entstehenden
Operatoren ($\ref{kanonische_Quantisierung_Gravitation_Ashtekar}$) ersetzt und gemäß ($\ref{Operatoren_Zustaende_Ashtekar}$)
auf den entsprechenden quantentheoretischen Zustand anwendet, ergeben sich aus ($\ref{Dynamik_Gravitation_Ashtekar}$)
die analogen quantentheoretischen dynamischen Grundgleichungen

\begin{equation}
D_a\frac{\delta}{\delta A_a^i}\Psi[A]=0\quad,\quad
F_{ab}^i\frac{\delta}{\delta A_a^i}\Psi[A]=0\quad,\quad
F_{ab}^{ij}\frac{\delta}{\delta A_a^i}\frac{\delta}{\delta A_b^j}\Psi[A]=0.
\label{Dynamik_Gravitation_Ashtekar_Quantentheoretisch}
\end{equation}
Der Hilbert-Raum der physikalischen Zustände $\mathcal{H}_{Phys}$ ergibt sich damit durch die
dynamischen Bedingungen ($\ref{Dynamik_Gravitation_Ashtekar_Quantentheoretisch}$) als Teilraum
von $\mathcal{H}$: $\mathcal{H}_{Phys} \in \mathcal{H}_{N} \in \mathcal{H}$, wobei $\mathcal{H}_N$
bereits den Teilraum der auf eins normierten Zustände beschreibt.
Im Rahmen dieser Quantisierung treten nun Probleme in Bezug auf die Definition eines inneren Produktes zur Auszeichnung
einer Hilbert-Raum-Struktur im Raum der in ($\ref{Operatoren_Zustaende_Ashtekar}$) definierten Zustände auf. Daher wird in der
Schleifenquantengravitation der Zusammenhang durch eine neue über ihn definierte Größe ersetzt, nämlich die Holonomie,
weshalb zunächst der Begriff der Holonomie definiert sei.\\
\textbf{Definition:} Eine Holonomie eines Zusammenhangs ist die Menge linearer Transformationen, welche durch den Paralleltransport
entlang einer Kurve eines Elementes des Vektorraumes, auf den sich der Zusammenhang bezieht, in diesem Vektorraum
induziert wird. Bezeichne $\gamma$ eine Kurve gemäß

\begin{equation}
\gamma:[0,1] \rightarrow M,\quad,\quad \gamma:s \rightarrow x^\mu(s),
\end{equation}
so ist die Holonomie eines Zusammenhangs $A$ gemäß den folgenden Bedingungen an ein Element dieser Holonomie definiert

\begin{eqnarray}
U[A,\gamma](0)=\mathbf{1},\quad
\frac{d}{ds}U[A,\gamma](s)+\frac{d\gamma^\mu(s)}{ds}A_\mu\left[\gamma(s)\right]U[A,\gamma](s)=0,\quad
U[A,\gamma]=U[A,\gamma](1).\nonumber\\
\end{eqnarray}
Ein Element der Holonomie hat damit folgende allgemeine Gestalt

\begin{equation}
U[A,\gamma]=\mathcal{P}\exp\left(\int_0^1 ds \frac{d \gamma^\mu}{ds}A_\mu^i(\gamma(s))\right)
\equiv \mathcal{P}\exp\left(\int_\gamma {A}\right),
\end{equation}
wobei $\mathcal{P}$ das Produkt der Operatoren entlang des Weges ordnet.
Man kann natürlich auch eine Holonomie in Bezug auf Ashtekars Zusammenhang ($\ref{Zusammenhang_Ashtekar}$)
definieren. In der Schleifenquantengravitation werden die Zustände nun durch eine sogenannte Spinnetzwerkbasis
ausgedrückt. Dies ist eine orthonormale Basis, durch welche alle Zustände ausgedrückt werden können. Ein Spinnetzwerk
ist ein System von Knotenpunkten, welche durch Kurven miteinander verbunden sind, denen jeweils eine Holonomie in Bezug auf
den Zusammenhang ($\ref{Zusammenhang_Ashtekar}$) entspricht. Ein Zustand ist demnach durch eine Funktion $f$
charakterisiert, welche eine Abbildung $\left[SU(2)\right]^n \rightarrow \mathbb{C}$ darstellt

\begin{equation}
\Psi[A]=f\left(U_1,...,U_n\right). 
\end{equation}
Dem entspricht ein inneres Produkt der folgenden Form

\begin{equation}
\langle \Psi_{\Gamma,f}|\Psi_{\Gamma,g}\rangle = \int_{[SU(2)]^n} dU_1...dU_L f(U_a,...,U_L)g(U_1,...,U_L).
\end{equation}

\section{Begriffliche Grundfragen der Vereinheitlichung}

Die Formulierung einer Quantentheorie der Gravitation stellt eine Übertragung der Gesetze der Quantentheorie
auf die Allgemeine Relativitätstheorie dar. Eine solche Theorie muss also die beiden Grundprinzipien der beiden
Theorien im Rahmen einer einheitlichen Theorie enthalten.
Diesbezüglich scheint nun eine Tatsache von entscheidender Bedeutung zu sein. Beide Theorien, sowohl die Quantentheorie
als auch die Allgemeine Relativitätstheorie legen eine relationalistische Raumanschauung zugrunde. Sie sind also beide in ihrer
begrifflichen Grundstruktur so beschaffen, dass der Raum ein sich nur in Bezug auf dynamische Entitäten konstituierender
Begriff ist. In beiden Theorien scheint das zwar zunächst nicht explizit erkennbar, kann aber durch eine sorgfältige Analyse der
Grundbegriffe gezeigt werden.
In der Allgemeinen Relativitätstheorie geht man vom Feldbegriff aus und beschreibt die physikalischen Ereignisse mit Hilfe 
des Begriffes einer pseudo-Riemannschen Mannigfaltigkeit, auf der bereits das metrische Feld als Beschreibung des
Gravitationsfeldes ausgezeichnet ist. Diese Riemannsche Mannigfaltigkeit beschreibt die mathematische Struktur der Raum-Zeit,
auf welcher dann weitere Felder definiert werden können. Die Diffeomorphismeninvarianz der Allgemeinen Relativitätstheorie zeigt
aber nun, wie in \textbf{[4.2]} ausführlich dargelegt wurde, dass die Raum-Zeit-Punkte der Mannigfaltigkeit für sich selbst
genommen keinerlei Bedeutung haben. Nur der relativen Lage von Körpern oder Feldern, also raum-zeitlichen Koinzidenzen, kommt
physikalische Relevanz zu. Das bedeutet aber, dass obwohl es im Hinblick auf die Formulierung der Theorie zunächst so
erscheint, als komme der Raum-Zeit im Rahmen der Allgemeinen Relativitätstheorie eine unabhängige Existenz zu und als sei
diese unabhängige Existenz sogar die Voraussetzung für die Existenz dynamischer Entitäten, die als Felder auf der Raum-Zeit
erscheinen, sie in Wahrheit nur eine Art und Weise ist, die dynamischen Relationen dieser Entitäten zueinander darzustellen.
In der Quantentheorie geht man von klassischen Grenzfällen aus, spezifisch von der klassischen Punktteilchenmechanik
beziehungsweise von relativistischen Feldtheorien, auf welche dann durch Quantisierung die Axiomatik der allgemeinen
Quantentheorie übertragen wird.
Das bedeutet, dass sich die Quantentheorie im Rahmen klassischer unter anderem auf dem Raumbegriff basierender Theorien
manifestiert. Die Unbestimmtheitsrelation zeigt aber dann, dass diese klassischen Ausgangsbegriffe in Wirklichkeit auf
fundamentaler Ebene gar keine adäquate Beschreibung der Natur liefern und die Vorstellung genau lokalisierter Objekte
nicht zutreffend ist. In Zusammenhang mit der Kopenhagener Deutung der Quantentheorie (siehe \textbf{[7.2]}) wurde in 
\textbf{[7.4]} in der Weise argumentiert, dass es sinnvoll ist, davon auszugehen, dass der Raum in Wirklichkeit eine Art der
Darstellung einer an sich unanschaulichen Realität ist. 
Wenn man sich dies nun umgekehrt ausgehend von der abstrakten Quantentheorie betrachtet, so erscheint der Ortsraum in der Tat
nicht als fundamental, da diese ihn nicht voraussetzt. In der Theorie der Ur-Alternativen wird nun in konsequenter Weise der
deduktive Versuch unternommen, die Existenz des Raumes als Konsequenz der abstrakten Quantentheorie darzustellen. Es ist also
im Hinblick auf die Vereinheitlichung von Allgemeiner Relativitätstheorie und Quantentheorie zu erwarten, dass der Knotenpunkt,
an dem beide Theorien sich vereinigen, begrifflich gesehen auf einer Ebene liegt, auf welcher der Raumbegriff noch nicht sinnvoll
verwendet werden kann.

In diesem Sinne muss also das Bestreben dahin gehen, bei der Formulierung einer Theorie, welche Quantentheorie und
Allgemeine Relativitätstheorie vereinigen soll, so wenig räumliche Struktur wie möglich vorauszusetzen. Eine jegliche
Quantisierung der Gravitation, welche von einer Feldtheorie mit einer Minkowski-Hintergrund-Metrik ausgeht, macht die
wahrscheinlich wichtigste begriffliche Errungenschaft der Allgemeinen Relativitätstheorie gewissermaßen wieder rückgängig.
Die hintergrundunabhängigen Quantisierungsversuche der Allgemeinen Relativitätstheorie nehmen diese begriffliche Neuerung in
sich auf, gehen aber dennoch von einer feldtheoretischen Formulierung aus. Auch wird die Dualität zwischen dem Gravitationsfeld
auf der einen Seite und den Materiefeldern auf der anderen Seite, also die Dualität von Geometrie und Materie, welche Albert
Einstein in einer einheitlichen Feldtheorie überwinden wollte, in keiner Weise abgeschwächt. Der im folgenden Kapitel
vorgestellte Ansatz, in welchem das Gravitationsfeld sich als Konsequenz des Zusammenhangs eines einheitlichen Spinorfeldes
ergibt, geht bezüglich dieser begrifflichen Fragen schon einen wichtigen Schritt weiter. Hier wird nämlich zu Beginn nur noch
ein einziges Feld vorausgesetzt, namentlich das fundamentale Spinorfeld, aus dem sich ganz gemäß der einheitlichen
Spinorfeldtheorie Werner Heisenbergs alle Wechselwirkungen außer der Gravitation ergeben sollen. Das Gravitationsfeld ist dann
von seiner Natur her ein auf dieses Feld bezogenes Phänomen. Aus ihm wird die metrische Struktur der Raum-Zeit hergeleitet,
welche damit begrifflich mit der Existenz von Materie verbunden ist, was in der Allgemeinen Relativitätstheorie in dieser Weise
noch nicht der Fall ist. Letztendlich muss es aber das Ziel sein, mit einer Beschreibung zu beginnen, welche überhaupt keine
Raum-Zeit mehr voraussetzt.
\footnote{Diese Aussage bezieht sich auf den physikalischen Ortsraum und seine Verbindung mit der Zeit zu einer
Raum-Zeit-Mannigfaltigkeit. Wie in \textbf{[8.2.1]} thematisiert und in Abbildung ($\ref{Ur-Alternativen_Raum-Zeit}$)
aus Kapitel \textbf{[9.4]} vorausgesetzt sollte aber davon ausgegangen werden, dass die Zeit an sich fundamentaleren Charakter
hat, also auf der basalen Ebene vorausgesetzt werden muss.}
Und die Quantentheorie der Ur-Alternativen bietet eben die Aussicht darauf. Daher wird im daran
anschließenden Kapitel das Programm der Beschreibung der Gravitation innerhalb der Quantentheorie der Ur-Alternativen
thematisiert werden. Das bedeutet, dass die Quantentheorie der Ur-Alternativen als der bislang beste begriffliche Rahmen
angesehen wird, um eine Quantentheorie der Gravitation zu formulieren.

\chapter{Die Beziehung einer einheitlichen Quantenfeldtheorie der Spinoren zur Struktur der Allgemeinen Relativitätstheorie}

\section{Einleitung}

Im letzten Kapitel wurden Ansätze zu einer Quantisierung des Gravitationsfeldes vorgestellt. In diesem Kapitel wird nun die
Frage nach dem Verhältnis relativistischer Quantenfeldtheorien, welche auf einem gegebenen Minkowski-Raum-Zeit-Hintergrund
formuliert sind und den konzeptionellen Rahmen des Standardmodells der Elementarteilchenphysik liefern, zur allgemein
kovarianten Beschreibung der Raum-Zeit und der Gravitation gemäß der Allgemeinen Relativitätstheorie gestellt,
in welcher sich das Problem der Vereinheitlichung von Quantentheorie und Allgemeiner Relativitätstheorie manifestiert.

In der hier vorliegenden Betrachtung wird nun versucht werden, zunächst in dem Sinne eine begriffliche Veränderung zu
unternehmen, als die Allgemeine Relativitätstheorie und die in ihrem Rahmen erfolgende Beschreibung der metrischen Struktur
der Raum-Zeit als Konsequenz einer einheitlichen Quantenfeldtheorie der Spinoren dargestellt werden.
Solch eine vereinheitlichte Quantenfeldtheorie der Spinoren mit einem Selbstwechselwirkungsterm des fundamentalen Spinorfeldes,
welcher den Ursprung der Existenz der Massen und Wechselwirkungen der Teilchen beschreiben soll, wurde von Heisenberg
entwickelt \cite{Heisenberg:1957},\cite{Heisenberg:1967},\cite{Heisenberg:1974du}. Im Rahmen dieser Betrachtungen wird diese
Theorie zugrunde gelegt werden, sie wird jedoch in einer Weise formuliert werden, welche in einem strikten Sinne
hintergrundunabhängig ist. In Heisenbergs ursprünglicher Fassung war die Theorie auf einem gegebenen
Minkowski-Raum-Zeit-Hintergrund formuliert. Im Gegensatz zu Heisenberg wird hier nicht eine a priori gegebene
metrische Struktur der Raum-Zeit vorausgesetzt und dies noch nicht einmal im weiteren Sinne der Allgemeinen
Relativitätstheorie. Damit wird also der Begriff der Metrik auf den eines fundamentalen Materiefeldes auf einer Raum-Zeit ohne
weitere Struktur zurückgeführt, wobei das Materiefeld als Spinorfeld beschrieben wird.
Ausgehend von dieser begrifflichen Neuerung, welcher eine weitergehende Einheit der Naturbeschreibung entspricht,
wird ein Ansatz zu einer Quantisierung dargestellt, der auf der gewöhnlichen Quantisierung von Spinorfeldern basiert.
Das Kapitel gibt den Inhalt einer von mir bereits veröffentlichten Arbeit wider \cite{Kober:2008xp}.
Es besteht aus drei Teilen. Im ersten Teil wird zunächst die grundlegende
Idee der vereinheitlichten Quantenfeldtheorie der Spinoren gemäß Heisenberg dargestellt. Danach wird eine kurze Einführung
einer Beschreibung der Raum-Zeit-Metrik durch Spinoren in dem Sinne gegeben, wie sie von Roger Penrose entwickelt wurde
\cite{Penrose:1960eq},\cite{Penrose:1977in},\cite{Penrose:1984},\cite{Stewart:1990}. Mit Hilfe dieser Konzepte wird im
zweiten Teil eine Theorie formuliert, in welcher die metrische Struktur der Raum-Zeit als Konsequenz einer symplektischen
Struktur kombiniert mit einem allgemeinen Zusammenhang innerhalb des abstrakten Raumes des fundamentalen Spinorfeldes erscheint.
Die Wirkung für dieses fundamentale Spinorfeld wird formuliert, indem eine Metrik verwendet wird, welche aus einer Basis von
Spinorfeldern konstruiert wird, welche diesem Zusammenhang entsprechen, aus welchem dann anschließend auch eine Wirkung für
die Gravitation gebildet wird.
Im dritten Teil wird schließlich ein Versuch unternommen, eine Quantisierung des Gravitationsfeldes durchzuführen, welches mit
Hilfe der Basis im Raum des fundamentalen Spinorfeldes beschrieben wird. Somit erscheint die quantentheoretische Beschreibung
des Vierbeinfeldes beziehungsweise des metrischen Feldes als eine Konsequenz einer auf die Spinoren bezogenen Quantisierung,
deren Status in diesem Ansatz als fundamentaler angenommen wird.

\section{Vorbereitungen}

\subsection{Einheitliche Quantenfeldtheorie der Spinoren}

\begin{figure}[ht]
\centering
\epsfig{figure=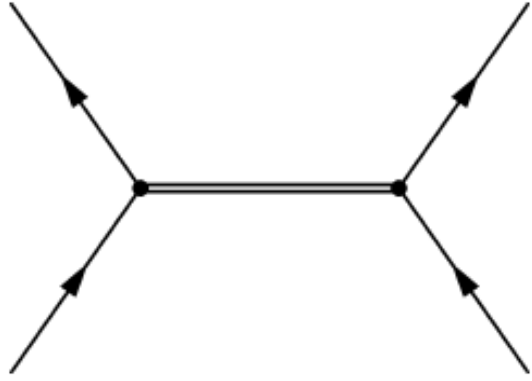,width=7cm}
\caption{\label{Feynman} Ein Feynman-Graph des selbstwechselwirkenden fundamentalen Spinorfeldes: Die horizontale Doppellinie
stellt einen zusammengesetzten Zustand aus einem Spinorzustand, welcher ein Teilchen beschreibt, und einem Spinorzustand,
welcher ein Antiteilchen beschreibt, dar. An den Vertizes treffen ein Zustand eines freien Teilchens und ein Zustand eines
freien Antiteilchens zusammen, welches einem Teilchen entspricht, das in der Zeit rückwärts läuft.}
\end{figure}

Im Rahmen relativistischer Quantenfeldtheorien werden Elementarteilchen, wie bereits in \textbf{[6.2.1]} und \textbf{[7.4.2]}
thematisiert, durch irreduzible Darstellungen der Poincar\'{e}-Gruppe beschrieben \cite{Wigner:1939},\cite{Weinberg:1995}.
Die einfachste Darstellung der Poincar\'{e}-Gruppe ist innerhalb eines Spinorraumes gegeben, innerhalb dessen Teilchen mit Spin
$1/2$ dargestellt werden. Wenn man von möglichen supersymmetrischen Erweiterungen einmal absieht, werden alle elementaren
Materiefelder des Standardmodells der Elementarteilchenphysik durch Spinorfelder beschrieben. Alle anderen Arten von Feldern,
wie etwa Wechselwirkungsfelder mit Spin $1$, können im Prinzip als aus Spinorfeldern aufgebaut gedacht werden
(siehe \textbf{[6.2.1]}). Aus diesem Grunde schlug Heisenberg eine einheitliche Quantenfeldtheorie eines fundamentalen
Spinorfeldes vor, aus welcher die Beschreibung aller Materiefelder und ihrer Wechselwirkungen hervorgehen soll. Die Massen und
Wechselwirkungen der Teilchen stellen in dieser Theorie eine Konsequenz eines Selbstwechselwirkungstermes des elementaren
Spinorfeldes dar. Durch Weyl-Spinoren ausgedrückt lautet die postulierte fundamentale Feldgleichung wie folgt

\begin{equation}
i\sigma^\mu \partial_\mu \psi \pm l^2 \sigma^\mu \psi \bar \psi \sigma_\mu \psi=0,
\label{fieldequation_Weyl}
\end{equation}
wobei $\psi$ einen Weyl-Spinor, $\bar \psi$ den adjungierten Weyl Spinor und die $\sigma^\mu$ die Pauli-Matrizen beschreiben,
wobei die Einheitsmatrix in zwei Dimensionen $\sigma^0$ miteinbezogen ist. Die Größe $l$ stellt eine fundamentale Konstante der
Natur dar, welche die Dimension einer Länge aufweist. Es ist wichtig zu erwähnen, dass in Heisenbergs ursprünglicher Fassung der
Theorie, welche auf das Postulat der Symmetrie unter der Lorentz-Gruppe und der auf den schwachen Isospin bezogenen $SU(2)$
Symmetriegruppe gegründet war, der Bezug zur Allgemeine Relativitätstheorie zunächst nicht explizit thematisiert wurde.
Vektorbosonen mit Spin $1$, welche die Wechselwirkungen im Standardmodell der Elementarteilchenphysik übertragen, können im
Rahmen der Heisenbergschen Theorie als Zustände angesehen werden, welche aus einem Spinorzustand, welcher ein Teilchen und einem
Spinorzustand, welcher ein Antiteilchen beschreibt, zusammengesetzt sind. Dies entspricht der Darstellung
$D(\frac{1}{2},0) \otimes D(0,\frac{1}{2})=D(\frac{1}{2},\frac{1}{2}) \oplus D(0,0)$ der Lorentz-Gruppe. In der ursprünglichen
Fassung Heisenbergs war das elementare Spinorfeld ein Dublett bezüglich der $SU(2)$ Isospin Symmetriegruppe, aber nicht unter
den Symmetriegruppen $SU(3)_{flavour}$ und $SU(3)_{colour}$ der starken Wechselwirkung, welche gemäß Heisenbergs Theorie nur
angenäherte Symmetrien darstellen. Ansätze auch die Symmetrien der starken Wechselwirkung und eine Supersymmetrie in die Theorie
einzubinden, sind in \cite{Durr:1979fi},\cite{Durr:1982fk},\cite{Durr:1982nn} zu finden. Aber dieses Thema ist hier nicht von
Interesse, da es hier nur um die Beziehung zur Allgemeinen Relativitätstheorie geht. Somit kann der Spinor als ein Multiplett
in Bezug auf jeden internen Freiheitsgrad zusätzlich zur Spinstruktur angesehen werden. Das Feld muss gemäß den
Quantisierungsregeln für fermionische Felder quantisiert werden, welche Antivertauschungsregeln für das Feld postulieren,
welche den Ursprung des Ausschließungsprinzips im Rahmen relativistischer Quantenfeldtheorien darstellen. Somit wird man zu den
folgenden Vertauschungsrelationen für das Feld geführt

\begin{equation}
\left\{\psi^\alpha(x,t) ,\bar \psi^\beta(x^\prime,t) \right\}=\delta^3 \left(x-x^\prime \right) \delta^{\alpha \beta},\\
\label{quantization_spinorfield}
\end{equation}
wobei gilt: $\left\{A,B\right\} \equiv AB+BA$.
Ein Wechselwirkungsprozess, bei dem Fermionen Vektorbosonen austauschen, zum Beispiel zwei Elektronen ein Photon austauschen,
kann interpretiert werden als ein Teilchenzustand, welcher mit einem Antiteilchenzustand wechselwirkt, der in der Zeit
rückwärts läuft und mit diesem einen zusammengesetzten Zustand bildet bevor dieser Zustand sich aufspaltet und wieder zwei
separierte Zustände entstehen, welche erneut freie Teilchen beschreiben (siehe Abbildung ($\ref{Feynman}$)).

Die Frage, ob die fundamentale Konstante $l$ in ($\ref{fieldequation_Weyl}$) die Rolle einer kleinsten Länge wie der
Planck-Länge spielt, wird später behandelt werden, wenn eine Lagrange-Dichte für die Gravitation eingeführt sein wird.
Da der Selbstwechselwirkungsterm den Ursprung der Massen und der Wechselwirkungen darstellt, wie sie durch das
Standardmodell beschrieben werden, muss sie in jedem Falle mit der elektroschwachen Skala verknüpft sein.

\subsection{Spinoren und die Raum-Zeit-Metrik}

Es soll nun ein zweidimensionaler abstrakter Vektorraum betrachtet werden, dessen Elemente zweikomponentige Weyl-Spinoren
sind. Der komplex konjugierte Spinor zu einem Spinor $\varphi$ in diesem Raum soll mit $\bar \varphi$ bezeichnet werden. Ein
beliebiger Spinor $\varphi$ kann auf einen Minkowski-Vektor gemäß der Gleichung

\begin{equation}
k^\mu=\bar \varphi \sigma^\mu \varphi
\label{spinortovector}
\end{equation}
abgebildet werden. Weiter soll angenommen werden, dass der Raum mit einer symplektischen Struktur versehen ist, welche durch
ein schiefsymmetrisches Skalarprodukt induziert wird, welche mit $[\ \cdot\ ,\ \cdot\ ]$ bezeichnet werden soll. Die Eigenschaft,
dass $[\ \cdot\ ,\ \cdot\ ]$ schiefsymmetrisch ist, entspricht der Tatsache, dass für zwei beliebige Spinoren $\varphi$ und $\chi$ die
Beziehung $[\varphi,\chi]=-[\chi,\varphi]$ gültig ist. Die Bilinearform $[\ \cdot\ ,\ \cdot\ ]$ kann durch folgende Matrix
ausgedrückt werden

\begin{equation}
\epsilon_{\alpha\beta}=\begin{pmatrix} 0 & 1 \\ -1 & 0 \end{pmatrix},
\label{symplectic_matrix}
\end{equation}
welche impliziert, dass das schiefsymmetrische Skalarprodukt zweier Spinoren $\varphi$ und $\chi$ wie folgt aussieht

\begin{equation}
[\varphi,\chi]=\epsilon_{\alpha\beta} \varphi^\alpha \chi^\beta.
\label{symplectic_innerproduct}
\end{equation}
Somit kann $\epsilon_{\alpha\beta}$ benutzt werden, um Indizes von Spinoren zu erhöhen und zu erniedrigen. Einen Spinor
$\varphi^\alpha$ auf den adjungierten Spinor $\varphi_\alpha$ mit Bezug auf das schiefsymmetrische Skalarprodukt
($\ref{symplectic_innerproduct}$) abzubilden, bedeutet den kontravarianten Vektor im Minkowski-Raum $k^\mu$, den man mit
Gleichung ($\ref{spinortovector}$) erhält, auf den entsprechenden kovarianten Vektor $k_\mu$ abzubilden. Die Gruppe
$Sp(2,\mathcal{C})$, welche das schiefsymmetrische Skalarprodukt ($\ref{symplectic_innerproduct}$) invariant lässt, ist
isomorph zur $SL(2,\mathbb{C})$ und somit zur homogenen Lorentz-Gruppe, welche eine direkte Beziehung zur Minkowski-Raum-Zeit
konstituiert. Aus einer beliebigen Basis innerhalb des Spinorraumes, welcher gemäß den beiden Dimensionen des Raumes aus zwei
Spinoren besteht, sie seien durch $\varphi$ und $\chi$ bezeichnet, kann ein auf die Minkowski-Raum-Zeit bezogenes Vierbeinfeld
in der folgenden Weise konstruiert werden

\begin{eqnarray}
e_0^m=\frac{1}{2}(\bar \varphi \sigma^m \varphi+\bar \chi \sigma^m \chi),\quad e_1^m=\frac{1}{2}(\bar \varphi \sigma^m
\chi+\bar \chi \sigma^m \varphi), \nonumber\\
e_2^m=\frac{1}{2}i(\bar \varphi \sigma^m \chi-\bar \chi \sigma^m \varphi),\quad e_3^m=\frac{1}{2}(\bar \varphi \sigma^m
\varphi-\bar \chi \sigma^m \chi).
\label{tetrad}
\end{eqnarray}
Gemäß der üblichen Relation, welche das Vierbeinfeld und das metrisches Feld zueinander in Beziehung setzt

\begin{equation}
g_{\mu\nu}=e_{\mu}^m e_{\nu m},
\label{metric}
\end{equation}
korrespondiert das Vierbeinfeld ($\ref{tetrad}$) einem metrischen Tensor $g_{\mu\nu}$ mit der Lorentz-Signatur $(+ ,- ,- ,- )$.
In ($\ref{metric}$) wurde die Tatsache ausgenutzt, dass man $e_{\mu m}$ aus $e_{\mu}^m$ gewinnen kann, indem man die dualen
Spinoren von $\varphi$ und $\chi$ gemäß ($\ref{symplectic_innerproduct}$) verwendet. Es ist wichtig anzumerken, dass
lateinische Indizes flache Indizes und griechische Indizes gekrümmte Indizes bezeichnen. Wenn die folgende Wahl für die 
Basisspinoren gemacht wird

\begin{equation}
\varphi=\begin{pmatrix} 1\\0 \end{pmatrix}\quad,\quad \chi=\begin{pmatrix} 0\\1 \end{pmatrix},
\end{equation}
so liefern ($\ref{tetrad}$) und ($\ref{metric}$) die Metrik $g_{\mu\nu}=(1,-1,-1,-1) \equiv \eta_{\mu\nu}$ der flachen
Minkowski-Raum-Zeit. Gemäß diesen Betrachtungen existiert eine natürliche Korrespondenz zwischen einem zweidimensionalen
Spinorraum, der mit einer symplektischen Struktur ausgestattet ist, und einer Minkowski-Raum-Zeit. Eine ausführlichere
Behandlung dieser Fragen kann in \cite{Penrose:1984},\cite{Stewart:1990} vorgefunden werden.

\section{Die Allgemeine Relativitätstheorie und die metrische Struktur der Raum-Zeit als Konsequenz des Zusammenhangs eines 
fundamentalen Spinorfeldes}

\subsection{Zusammenhang des Spinorfeldes und metrische Struktur}

Gemäß Heisenberg wird vorgeschlagen, dass die Elementarteilchen und ihre Wechselwirkungen durch ein fundamentales Spinorfeld
beschrieben werden. Heisenbergs Theorie in der ursprünglichen Fassung ist auf einem gegebenen Minkowski-Raum-Zeit-Hintergrund
formuliert. Ansätze die Gravitation zu integrieren und die Spinorfeldtheorie in einer erweiterten Fassung im Sinne der
Allgemeinen Relativitätstheorie zu formulieren, können in \cite{Treder:1967},\cite{Durr:1982ie},\cite{Durr:1983uw} gefunden
werden. Aber hier wird die metrische Struktur der Raum-Zeit vorausgesetzt und die Theorie ist nur auf einem bereits gegebenen
Raum-Zeit-Hintergrund formuliert.
Eine Formulierung der Allgemeinen Relativitätstheorie mit Spinoren wird in der Twistortheorie gegeben
\cite{Penrose:1960eq},\cite{Penrose:1977in}. Hier erscheinen Raum-Zeit-Vektoren selbst als Ausdruck einer postulierten
dahinterliegenden Spinorstruktur. Im Rahmen der Betrachtungen dieses Kapitels wird ein anderes Ziel angestrebt. 
Es wird der Versuch unternommen, die Eigenschaften des Gravitationsfeldes und somit der metrischen Struktur der Raum-Zeit aus
den Eigenschaften des abstrakten internen Raumes des fundamentalen Spinorfeldes herzuleiten. Zu Beginn wird lediglich die
Annahme gemacht, dass ein selbstwechselwirkendes fundamentales Spinorfeld $\psi(x^\mu)$ auf einer vierdimensionalen
Mannigfaltigkeit existiert, welche die Raum-Zeit vor der Einführung einer metrischen Struktur repräsentiert. Der entsprechende
Spinorraum soll mit einer ($\ref{symplectic_matrix}$) und ($\ref{symplectic_innerproduct}$) entsprechenden symplektischen
Struktur versehen sein. Wenn man zwei Werte des Spinorfeldes an zwei unterschiedlichen Raum-Zeit-Punkten vergleichen will, so
muss man einen Spinzusammenhang definieren. Solch ein Spinzusammenhang, er sei mit $A_{\mu\beta}^{\alpha}$ bezeichnet, liefert
eine Vorschrift, wie dies zu erfolgen hat und er repräsentiert die Eigenschaft, dass es möglich ist, an jedem Raum-Zeit-Punkt
eine andere Basis für die Spinoren zu verwenden, was der Definition eines nicht-trivialen Zusammenhangs äquivalent ist.

Da beliebige Koordinaten gewählt werden können, wird man zu der Gruppe $GL(2,\mathbb{C})$ für den Zusammenhang geführt. Die
$GL(2,\mathbb{C})$ hat die $SL(2,\mathbb{C})$ und somit die Lorentz-Gruppe als Untergruppe. Entsprechend zu diesem Zusammenhang
kann man eine kovariante Ableitung $\nabla_\mu$ in Bezug auf den Spinorraum definieren

\begin{equation}
\nabla_\mu=\partial_\mu \mathbf{1}+i A_{\mu\beta}^{\alpha}.
\label{covariant_derivative}
\end{equation}
Die Definition eines Spinzusammenhangs $A_{\mu\beta}^{\alpha}$ gemäß ($\ref{covariant_derivative}$) ist äquivalent zur
Definition zweier unabhängiger Spinorfelder, sie seien mit $\varphi$ und $\chi$ bezeichnet, welche vom
Raum-Zeit-Punkt abhängig sind, eine Basis von Spinoren bilden und konstant in Bezug auf die kovariante Ableitung
($\ref{covariant_derivative}$) sind, was bedeutet, dass sie durch die folgenden Relationen definiert sind

\begin{eqnarray}
\nabla_\mu \varphi^\alpha=\partial_\mu \varphi^\alpha+iA_{\mu\beta}^{\alpha} \varphi^\beta=0,\nonumber\\
\nabla_\mu \chi^\alpha=\partial_\mu \chi^\alpha+iA_{\mu\beta}^{\alpha} \chi^\beta=0.
\label{spinor_basis}
\end{eqnarray}
Es macht keinen Unterschied, ob der Zusammenhang $A_{\mu\beta}^{\alpha}$ oder die aus $\varphi$ und $\chi$ bestehende Basis von
Spinorfeldern als fundamentaler angenommen wird. Beide Darstellungen enthalten in sich die Information, wie Werte des
fundamentalen Spinorfeldes $\psi(x^\mu)$ an unterschiedlichen Raum-Zeit-Punkten verglichen werden müssen. Deshalb ist es auch
möglich, den Zusammenhang durch ($\ref{spinor_basis}$) über die Basis der Spinoren zu definieren. Wenn $\varphi$ und $\chi$
gegeben sind, dann hat der Zusammenhang $A_{\mu\beta}^{\alpha}$, welcher die Relationen ($\ref{spinor_basis}$) erfüllt, die
folgende Form

\begin{equation}
A_{\mu\beta}^{\alpha}=-i\frac{\partial_\mu \chi^\alpha \varphi_\beta-\partial_\mu \varphi^\alpha\chi_\beta}
{\epsilon_{\gamma\delta}\varphi^{\gamma} \chi^{\delta}},
\label{connection}
\end{equation}
wobei der symplektische Ausdruck $\epsilon_{\alpha\beta}$ benutzt wurde, welcher in ($\ref{symplectic_matrix}$) eingeführt
wurde und die Beziehung zwischen $\varphi^\alpha$ und $\varphi_\alpha$ definiert. In diesem Sinne hängt die kovariante
Ableitung ($\ref{covariant_derivative}$) von $\varphi$ und $\chi$ ab

\begin{equation}
\nabla_\mu=\partial_\mu \mathbf{1}+i A_{\mu\beta}^{\alpha}(\varphi,\chi).
\label{covariant_derivative_basis}
\end{equation}
Wenn ein Spinor von einem Punkt zu einem anderen verschoben wird, hat man es entsprechend ($\ref{covariant_derivative}$) und
($\ref{spinor_basis}$) mit lokalen Basistransformationen zu tun, welche einem Übergang zu einem neuen Wert $\psi^\prime$ des
Spinorfeldes $\psi$ entsprechen, welches in ($\ref{fieldequation_Weyl}$) erscheint

\begin{equation}
\psi=\psi_\varphi \varphi+\psi_\chi \chi \rightarrow \psi^\prime=\psi_\varphi \varphi^\prime+\psi_\chi \chi^\prime.
\label{fundamentalfield_basisrepresentation}
\end{equation}
Dieser Wert ist äquivalent zu dem alten Wert in Bezug auf den nicht-trivialen Zusammenhang $A_\mu^{\alpha\beta}$.

Das Vierbein, welches oben definiert wurde ($\ref{tetrad}$), führt für alle Spinoren zu einer Metrik, welche proportional zur
Minkowski-Metrik ist. Um allgemeine Metriken zu erhalten, muss ein verallgemeinertes Vierbein wie folgt definiert werden

\begin{eqnarray}
e^m_\mu=\frac{1}{2}
\left(\begin{matrix}
\bar \varphi \sigma^m \varphi+\bar \chi \sigma^m \chi\\
\bar \varphi \sigma^m \chi+\bar \chi \sigma^m \varphi\\
i\bar \varphi \sigma^m \chi-i\bar \chi \sigma^m \varphi\\
\bar \varphi \sigma^m \varphi-\bar \chi \sigma^m \chi
\end{matrix}\right)+\frac{\bar \varphi \sigma^m \partial_\mu \chi-\bar \chi \sigma^m \partial_\mu
\varphi}{\epsilon_{\alpha\beta}\varphi^\alpha \chi^\beta},
\label{general_tetrad}
\end{eqnarray}
wobei ein zusätzlicher Term erscheint, welcher äquivalent zum Zusammenhang ausgedrückt mit einem Minkowski-Raum-Index anstatt
zweier Spinindizes ist. Für den Fall einer konstanten Basis von Spinorfeldern $\varphi$ und $\chi$, welche einem
verschwindenden Spinzusammenhang entsprechen, verschwindet der zweite Term und das Vierbein ($\ref{general_tetrad}$) reduziert
sich zu ($\ref{tetrad}$), was auf den Spezialfall einer flachen Minkowski-Metrik führt.

Entsprechend ($\ref{general_tetrad}$) und ($\ref{metric}$) kann aus der Basis der Spinoren ein Vierbeinfeld
$e_\mu^m(\varphi,\chi)$ konstruiert werden, beziehungsweise ein metrisches Feld $g_{\mu\nu}(\varphi,\chi)$, welches aufgrund der
vorausgesetzten symplektischen Struktur die Signatur $(+ ,- ,- ,- )$ trägt und somit wird man auf ein Gravitationsfeld
geführt. Da die Metrik mit Hilfe des Zusammenhangs beziehungsweise der Basis von Spinorfeldern konstruiert wurde und da diese in
der hier vorliegenden Theorie die metrische Struktur der Raum-Zeit konstituieren sollen, steht das allgemeine Kovarianzprinzip
der Allgemeinen Relativitätstheorie in Beziehung zu den beliebigen Koordinatentransformationen innerhalb des Spinorraumes.
In Übereinstimmung mit der gewöhnlichen Allgemeinen Relativitätstheorie ist die hier eingeführte kovariante Ableitung in einer
solchen Weise definiert, dass sie die beiden Spinorfelder ($\ref{spinor_basis}$) konstant hält, welche eine unabhängige Basis
von Spinoren an jedem Raum-Zeit-Punkt definieren. Da der Spinzusammenhang gemäß ($\ref{general_tetrad}$) zu einem Vierbein
führt und zu einer Metrik mit Minkowski-Signatur gemäß ($\ref{metric}$), könnte das Faktum, dass Materie aus Spinorfeldern
zusammengesetzt ist, als der Grund angesehen werden, warum die Raum-Zeit eine Lorentz-Struktur aufweist und somit die kovariante
Ableitung ($\ref{covariant_derivative}$) auch das Vierbein und die Metrik konstant hält. Das bedeutet, dass man es so
ausdrücken könnte, dass das Spinorfeld den Ursprung der metrischen Struktur darstellt.

\begin{figure}[ht]
\centering
\epsfig{figure=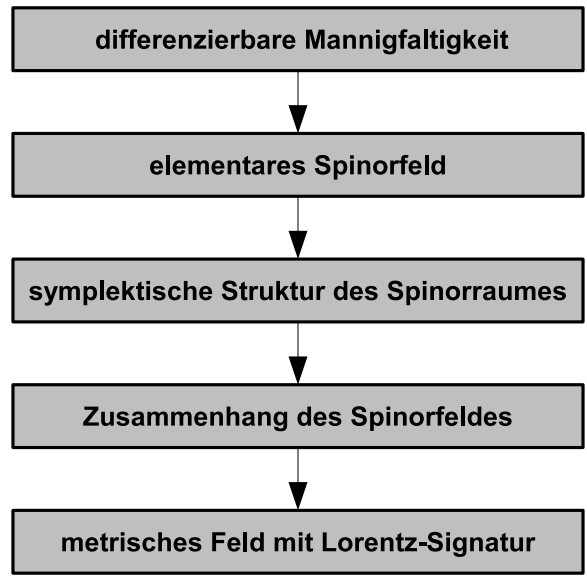,width=10cm}
\caption{\label{Raum-Zeit-Struktur} Hierarchie der Raum-Zeit-Struktur: Am Anfang wird nur die Struktur einer differenzierbaren
Mannigfaltigkeit vorausgesetzt. Auf dieser Mannigfaltigkeit, welche die Raum-Zeit darstellt, ist ein Spinorfeld definiert und
die metrischen Eigenschaften der Raum-Zeit sind  aus den Eigenschaften dieses abstrakten Raumes des Spinorfeldes hergeleitet.}
\end{figure}
Diese Hierarchie, welche den Ursprung der Raum-Zeit Struktur betrifft, in der das Vierbein und die Metrik aus einer
dahinterliegenden Spinstruktur eines fundamentalen Materiefeldes abgeleitete Größen darstellen, wird in Abbildung
($\ref{Raum-Zeit-Struktur}$) illustriert. Die Idee, dass ein Zusammenhang fundamentaler als das Vierbein beziehungsweise das
metrische Feld ist, wird auch bereits in Ashtekars neuer Fassung der Hamiltonschen Formulierung der Gravitation
\cite{Ashtekar:1986yd},\cite{Ashtekar:1987gu},\cite{Ashtekar:1989ju} und dem entsprechenden Ansatz zur Quantisierung der
Gravitation vertreten. Aber dort wird der Zusammenhang nicht mit einem fundamentalen Materiefeld assoziiert, dass durch
Spinoren beschrieben wird. 

\subsection{Wirkung des Materiefeldes und Beziehung zur eichtheoretischen Beschreibung der Gravitation}

In der gewöhnlichen Eichbeschreibung der Gravitation, wie sie in \textbf{[3.2]} thematisiert wurde,
beginnt man mit einer Lagrange-Dichte eines Materiefeldes auf der Minkowski-Raum-Zeit, welche invariant unter globalen
Transformationen der Poincar\'{e}-Gruppe ist und postuliert Invarianz unter den entsprechenden lokalen Transformationen. In
\cite{Dehnen:1986mx} wurde eine Spineichtheorie der Gravitation mit der Eichgruppe $SU(2) \otimes U(1)$ vorgeschlagen. Gemäß
diesem Ansatz beginnt man mit einem Spinorfeld und postuliert Invarianz unter lokalen Transformationen innerhalb des Raumes der
Weyl-Spinoren. Dies führt zu einem Spinzusammenhang und von diesem ausgehend kann man eine Feldstärke definieren und somit
erlangt man eine Theorie, welche äquivalent zu der linearen Näherung der Allgemeinen Relativitätstheorie ist. Aber diese
Theorie ist nicht hintergrundunabhängig. Sie nimmt noch nicht einmal an, dass Störungen der flachen Raum-Zeit-Metrik
existieren. Gemäß der Theorie in \cite{Dehnen:1986mx} ist die Gravitation ein Phänomen, welches völlig unabhängig von der
Raum-Zeit-Metrik ist und stattdessen auf einem a priori gegebenen Minkowski-Hintergrund formuliert ist. Außerdem setzt sie
keine einheitliche Theorie der Spinoren voraus und somit ist die Beschreibung von Wechselwirkungsfeldern, welche durch Spin
$1$ Teilchen vermittelt werden, nicht miteinbezogen. 

Im Gegensatz zu Heisenbergs Theorie \cite{Heisenberg:1957, Heisenberg:1967, Heisenberg:1974du} und der Spineichbeschreibung der
Gravitation \cite{Dehnen:1986mx} wird hier zu Beginn keine Minkowski-Struktur der Raum-Zeit und kein metrischen Feld
vorausgesetzt. Es wird lediglich von der Existenz eines Spinorfeldes $\psi(x^\mu)$ ausgegangen, dass auf einer vierdimensionalen
differenzierbaren Mannigfaltigkeit definiert ist, welche die Raum-Zeit beschreibt, und aus diesem Spinorfeld sind alle anderen
Felder zusammengesetzt. Dies führt zu einer hintergrundunabhängigen Theorie der Gravitation in einem strikten Sinne, weil das
Gravitationsfeld eine Konsequenz des Spinzusammenhangs des fundamentalen Materiefeldes darstellt. Da in dieser Theorie keine a
priori gegebene metrische Struktur vorausgesetzt wird, kann man nicht zuerst eine Wirkung definieren und dann das Eichprinzip
benutzen. Es muss zunächst die allgemeine metrische Struktur mit Hilfe der Eigenschaften des Spinorraumes definiert werden und
dann kann die Wirkung aufgestellt werden. Somit muss die fundamentale Wirkung, welche einer allgemein relativistischen Fassung
von ($\ref{fieldequation_Weyl}$) entspricht, direkt unter Einbeziehung der kovarianten Ableitung ($\ref{covariant_derivative}$)
formuliert werden, welche den sich auf das Spinorfeld beziehenden Zusammenhang enthält und auf einem Hintergrund
$e_m^{\mu}(\varphi,\chi)$ beziehungsweise $g_{\mu\nu}(\varphi,\chi)$, welcher aus diesem Zusammenhang hergeleitet ist,
beziehungsweise aus den Feldern $\varphi$ und $\chi$, welche direkt in Relation zum Zusammenhang ($\ref{covariant_derivative}$)
stehen. Somit wird von einer Wirkung für das Spinorfeld von der folgenden Gestalt ausgegangen

\begin{eqnarray}
S_{m}&=&\int d^4 x \sqrt{-g(\varphi,\chi)}\left(i\bar \psi \sigma^m e^\mu_m(\varphi,\chi) \nabla_\mu \psi
\pm \frac{l^2}{2} \bar \psi \sigma^\mu \psi \bar \psi \sigma_\mu \psi\right),
\label{matteraction}
\end{eqnarray}
wobei $g(\varphi,\chi)=\mbox{det}\left[g^{\mu\nu}(\varphi,\chi)\right]$ und somit
$\sqrt{-g(\varphi,\chi)}=\mbox{det}\left[e^\mu_m\right(\varphi,\chi)]$. Es sei bemerkt,
dass gemäß ($\ref{covariant_derivative_basis}$) der Zusammenhang $A^{\alpha}_{\mu\beta}(\varphi,\chi)$ wie
$e_m^{\mu}(\varphi,\chi)$ und $g_{\mu\nu}(\varphi,\chi)$ innerhalb der kovarianten Ableitung $\nabla_\mu$ auch
von $\varphi$ und $\chi$ abhängt. Diese Wirkung und die entsprechende Lagrange-Dichte sind invariant unter
$GL(2,\mathbb{C})$ Transformationen.

\subsection{Wirkung des Gravitationsfeldes}

Die Wirkung des Gravitationsfeldes muss aus dem Spinzusammenhang oder dem entsprechenden Paar von Spinoren gebildet werden.
In Analogie zur gewöhnlichen Fassung der Allgemeinen Relativitätstheorie kann ein Feldstärketensor
$F_{\mu\nu}^{\alpha\beta}(\varphi,\chi)$ (welcher dem Riemann-Tensor entspricht) als der Kommutator der kovarianten
Ableitungen definiert werden 

\begin{eqnarray}
F_{\mu\nu}^{\alpha\beta}(\varphi,\chi)&=&-i[\nabla_\mu,\nabla_\nu]
\nonumber\\
&=&\partial_\mu A_\nu^{\alpha\beta}(\varphi,\chi)-\partial_\nu A_\mu^{\alpha\beta}(\varphi,\chi)\nonumber\\
&&+iA_\mu^{\alpha\gamma}(\varphi,\chi)A_\nu^{\gamma\beta}(\varphi,\chi)-iA_\nu^{\alpha\gamma}(\varphi,\chi)A_\mu^{\gamma\beta}
(\varphi,\chi),\nonumber\\
\label{fieldstrength}
\end{eqnarray}
wobei $[A,B] \equiv AB-BA$. Es sei angemerkt, dass der Zusammenhang $A_\mu^{\alpha\beta}$, welcher im Feldstärketensor
erscheint, über  ($\ref{spinor_basis}$) und ($\ref{connection}$) definiert ist. Hiervon ausgehend kann eine Gravitationswirkung
definiert werden, welche der Wirkung entspricht, welche im Rahmen der Spineichtheorie der Gravitation \cite{Dehnen:1986mx}
formuliert wurde. Wie bereits erwähnt, ist diese Theorie nicht hintergrundunabhängig und sie enthält kein
selbstwechselwirkendes Spinorfeld, das als fundamental angenommen wird. Wie dem auch sei, in der hier vorgeschlagenen Theorie
wird die gleiche Dynamik für das Gravitationsfeld postuliert, welche zumindest in der linearen Näherung derjenigen der
gewöhnlichen Einsteinschen Fassung äquivalent ist, was in \cite{Dehnen:1986mx} gezeigt wurde. Dies führt auf die folgende
Gravitationswirkung

\begin{eqnarray}
S_{g}=\frac{1}{g}\int d^4 x \sqrt{-g(\varphi,\chi)}
g^{\mu\rho}(\varphi,\chi)g^{\nu\sigma}(\varphi,\chi)
F_{\mu\nu}^{\alpha\beta}(\varphi,\chi)F_{\rho\sigma\alpha\beta}(\varphi,\chi),
\label{gravityaction}
\end{eqnarray}
wobei $g$ eine fundamentale Konstante darstellt, welche die Stärke der Gravitation definiert und proportional zur gewöhnlichen
Gravitationskonstante ist, mit der sie über folgende Relation verknüpft ist 

\begin{equation}
g=32\pi G.
\label{relationGg}
\end{equation}
Wie in der Materiewirkung tauchen hier die Spinorbasisfelder $\varphi$ und $\chi$ aus dem Spinzusammenhang in der
Gravitationswirkung auf, welche eine Basis im Raum des fundamentalen Materiefeldes bilden sowie das Vierbeinfeld und damit das
metrische Feld definieren. Durch Verwendung von ($\ref{fieldstrength}$) in ($\ref{gravityaction}$) kann die Gravitationswirkung
ausführlicher geschrieben werden

\begin{eqnarray}
S_{g}&=&\frac{1}{g}\int d^4 x \sqrt{-g(\varphi,\chi)}g^{\mu\rho}(\varphi,\chi)g^{\nu\sigma}(\varphi,\chi)
\nonumber\\&&\quad\quad\quad
\left(2 \partial_{[\mu} A_{\nu]}^{\alpha\beta}(\varphi,\chi)\partial_{\rho}
A_{\sigma\alpha\beta}(\varphi,\chi)
+4i \partial_{[\mu} A_{\nu]}^{\alpha\beta}(\varphi,\chi)
A_{\rho\alpha\gamma}(\varphi,\chi)A_{\sigma\gamma\beta}(\varphi,\chi)
\right.\nonumber\\&&\quad\quad\quad\left.
-2A_{[\mu}^{\alpha\gamma}(\varphi,\chi)A_{\nu]}^{\gamma\beta}(\varphi,\chi)
A_{\rho\alpha\delta}(\varphi,\chi)A_{\sigma\delta\beta}(\varphi,\chi)\right).
\label{gravity_action_connection}
\end{eqnarray}
Die Klammern $[\mu\nu]=\mu\nu-\nu\mu$ bezeichnen eine Antisymmetrisierung in Bezug auf $\mu$ und $\nu$.
Das dynamische Verhalten des Vierbeinfeldes $e^\mu_m(\varphi,\chi)$ und des metrischen Feldes $g_{\mu\nu}(\varphi,\chi)$
ist also vollkommen durch ($\ref{gravityaction}$) determiniert, da diese durch $\varphi$ und $\chi$ definiert sind. Wenn die
Wechselwirkung des Gravitationsfeldes mit der Materie in Betracht gezogen werden soll, welches gemäß dem hier vorgestellten
Ansatz durch das fundamentale Spinorfeld beschrieben wird, so muss die entsprechende Lagrange-Dichte ($\ref{matteraction}$)
miteinbezogen werden. Somit stellt sich die gesamte Wirkung als die Summe der Materiewirkung des fundamentalen Spinorfeldes
($\ref{matteraction}$) und der Gravitationswirkung ($\ref{gravityaction}$) dar, welche auf den Spinzusammenhang beziehungsweise
die entsprechende Basis an Spinorfeldern bezogen ist. Mit dieser Annahme sieht die fundamentale Wirkung der Natur wie folgt aus

\begin{eqnarray}
S&=&\int d^4 x \sqrt{-g(\varphi,\chi)}
\left(\frac{1}{g}g^{\mu\rho}(\varphi,\chi)g^{\nu\sigma}(\varphi,\chi)
F_{\mu\nu}^{\alpha\beta}(\varphi,\chi)F_{\rho\sigma\alpha\beta}(\varphi,\chi) \right.\nonumber\\
&&\left.\quad+i\bar \psi \sigma^m e^\mu_m(\varphi,\chi) \nabla_\mu \psi \pm \frac{l^2}{2}
\bar \psi \sigma^\mu \psi \bar \psi \sigma_\mu \psi\right).
\label{action}
\end{eqnarray}
Es ist offensichtlich, dass außer den fundamentalen Konstanten der Speziellen Relativitätstheorie und der Quantentheorie,
namentlich der Lichtgeschwindigkeit $c$ und dem Planckschen Wirkungsquantum $h$, welche wie gewöhnlich gleich eins gesetzt
werden, die Konstante $l$ und die Konstante $g$, welche eine Hierarchie zwischen der Wirkung des fundamentalen Materiefeldes
$\psi$ und der Wirkung des Gravitationsfeldes definieren,
als die einzigen fundamentalen Konstanten in dieser Theorie erscheinen. Somit muss die Hierarchie zwischen der elektroschwachen
und der Planck-Skala als eine Konsequenz des Verhältnisses zwischen $l$ und $g$ betrachtet werden, wobei $l$ die Rolle einer
fundamentale Massenskala für Teilchen zu spielen scheint und $g$ zur Gravitationskonstante $G$ durch ($\ref{relationGg}$)
in Relation gesetzt ist. Aufgrund der Beziehung ($\ref{relationGg}$) steht die Planck-Länge, welche die untere Grenze für Längen
darstellt, in direkter Beziehung zu $g$. Somit kann $l$ gemäß der einheitlichen Spinorfeldtheorie Heisenbergs nicht als eine
kleinste Länge interpretiert werden. Da die metrische Struktur der Raum-Zeit eine Konsequenz der Eigenschaften des Spinorraumes
der Materie darstellt, spiegelt das Faktum, dass die Wirkung ($\ref{action}$) invariant unter beliebigen Transformationen der
$GL(2,\mathbb{C})$ ist, die allgemeine Kovarianz der Materie- und der Gravitationswirkung wider. Variation von ($\ref{action}$)
nach $\bar \psi$ führt auf die fundamentale Feldgleichung für die Materie

\begin{equation}
i\sigma^m e^\mu_m(\varphi,\chi) \nabla_\mu \psi \pm l^2 \sigma^\mu \psi \bar \psi \sigma_\mu \psi=0
\end{equation}
und Variation nach $\bar \varphi$ und $\bar \chi$ führt auf die fundamentalen Gleichungen für das Gravitationsfeld,
welche den Einsteinschen Gleichungen analog sind

\begin{eqnarray}
&&\left\{2 g^{\mu\rho}(\varphi,\chi)g^{\nu\sigma}(\varphi,\chi) \frac{\partial  F_{\mu\nu}^{\alpha\beta}(\varphi,\chi)}
{\partial \bar \varphi}F_{\rho\sigma\alpha\beta}(\varphi,\chi)
+2 \frac{\partial g^{\mu\rho}(\varphi,\chi)}{\partial \bar \varphi}g^{\nu\sigma}(\varphi,\chi)  
F_{\mu\nu}^{\alpha\beta}(\varphi,\chi)F_{\rho\sigma\alpha\beta}(\varphi,\chi)
\right. \nonumber\\&&\left.-e_{n \nu}(\varphi,\chi)\cdot
\frac{\partial e^{n \nu}(\varphi,\chi)}{\partial \bar \varphi}
g^{\mu\rho}(\varphi,\chi)g^{\nu\sigma}(\varphi,\chi)F_{\mu\nu}^{\alpha\beta}(\varphi,\chi)
F_{\rho\sigma\alpha\beta}(\varphi,\chi) \right\} \nonumber\\
&&=g\left\{-i\bar \psi \sigma^m \frac{\partial e^\mu_m(\varphi,\chi)}{\partial \bar \varphi} \nabla_\mu \psi
+e_{n \nu}(\varphi,\chi)\frac{\partial e^{n \nu}(\varphi,\chi)}{\partial \bar \varphi}
\cdot\left[i\bar \psi \sigma^m e^\mu_m(\varphi,\chi)\nabla_\mu \psi \pm l^2 \bar \psi \sigma^\mu \psi 
\bar \psi \sigma_\mu \psi\right]\right\},\nonumber\\
\nonumber\\
&&\left\{2 g^{\mu\rho}(\varphi,\chi)g^{\nu\sigma}(\varphi,\chi) \frac{\partial 
F_{\mu\nu}^{\alpha\beta}(\varphi,\chi)}{\partial \bar \chi}F_{\rho\sigma\alpha\beta}(\varphi,\chi)
+2 \frac{\partial g^{\mu\rho}(\varphi,\chi)}{\partial \bar \chi}g^{\nu\sigma}(\varphi,\chi)
F_{\mu\nu}^{\alpha\beta}(\varphi,\chi)F_{\rho\sigma\alpha\beta}(\varphi,\chi)
\right. \nonumber\\&&\left.-e_{n \nu}(\varphi,\chi)
\cdot \frac{\partial e^{n \nu}(\varphi,\chi)}{\partial \bar \chi}g^{\mu\rho}(\varphi,\chi)g^{\nu\sigma}(\varphi,\chi)
F_{\mu\nu}^{\alpha\beta}(\varphi,\chi)F_{\rho\sigma\alpha\beta}(\varphi,\chi)\right\} \nonumber\\
&&=g\left\{-i\bar \psi \sigma^m \frac{\partial e^\mu_m(\varphi,\chi)}{\partial \bar \chi} \nabla_\mu \psi
+e_{n \nu}(\varphi,\chi)\frac{\partial e^{n \nu}(\varphi,\chi)}{\partial \bar \chi}
\cdot\left[i\bar \psi \sigma^m e^\mu_m(\varphi,\chi) \nabla_\mu \psi \pm l^2 \bar \psi \sigma^\mu
\psi \bar \psi \sigma_\mu \psi\right]\right\}.\nonumber\\
\label{equations_gravity}
\end{eqnarray}
Hierbei wurde benutzt, dass $\delta \sqrt{-g}=\delta \mbox{det} \left[e^\mu_m \right]
=-\mbox{det} \left[e^\mu_m\right]e_{\nu n}\delta e^{\nu n}$ und

\begin{equation}
\delta_{\bar \varphi} e^\mu_m(\varphi,\chi)=\frac{\partial e^\mu_m(\varphi,\chi)}{\partial \bar \varphi}\delta \bar
\varphi,\quad \delta_{\bar \chi} e^\mu_m(\varphi,\chi)=\frac{\partial e^\mu_m(\varphi,\chi)}{\partial \bar \chi}\delta \bar\chi,
\end{equation}
wobei $\delta_{\bar \varphi}$ Variation nach $\bar \varphi$ und $\delta_{\bar \chi}$ Variation nach $\bar \chi$ bedeutet.
Die Terme, welche von der Materiewirkung herrühren, stehen auf den rechten Seiten der Gleichungen (\ref{equations_gravity})
geschrieben. Somit kann die Summe ihrer rechten Seiten, $\frac{\delta S_{matter}}{\delta \bar \varphi}
+\frac{\delta S_{matter}}{\delta \bar \chi}$ namentlich, als das Analogon zum Energie-Impuls-Tensor angesehen werden,
welcher in den Einsteinschen Gleichungen erscheint. Das dynamische Verhalten von $e^\mu_m(\varphi,\chi)$ und
$g^{\mu\nu}(\varphi,\chi)$ ist indirekt durch die oberen Gleichungen für $\varphi$ und $\chi$ definiert, was bedeutet

\begin{eqnarray}
\partial_\lambda e^\mu_m(\varphi,\chi)=\frac{\partial e^\mu_m(\varphi,\chi)}{\partial \varphi}\partial_\lambda {\varphi}
+\frac{\partial e^\mu_m(\varphi,\chi)}{\partial \chi}\partial_\lambda {\chi},\\ \partial_\lambda
g^{\mu\nu}(\varphi,\chi)=\frac{\partial g^{\mu\nu}(\varphi,\chi)}{\partial \varphi}\partial_\lambda {\varphi}
+\frac{\partial g^{\mu\nu}(\varphi,\chi)}{\partial \chi}\partial_\lambda {\chi}.
\end{eqnarray}

\subsection{Interpretation und begriffliche Grundlagen}

Aus konzeptioneller oder philosophischer Sicht stellt die Hintergrundunabhängigkeit, welche zur Diffeomorphismeninvarianz in
engem Zusammenhang steht, die entscheidende Eigenschaft der Allgemeinen Relativitätstheorie dar, was bereits in \textbf{[4.2]}
thematisiert wurde. Es ist eine der zentralen Aufgaben bezüglich der Suche nach einer Quantentheorie der Gravitation, eine
allgemein relativistische Fassung relativistischer Quantenfeldtheorien zu finden, welche diesem zentralen Prinzip Rechnung
trägt. Dieses Thema wurde in der Einleitung des Kapitels \textbf{[12.1]} angesprochen und wird ausführlich zum Beispiel in
\cite{Rovelli:1999hz},\cite{Rovelli:2006yt} diskutiert. Die Gravitationstheorie, welche hier vorgeschlagen wird, ist in einem
noch grundsätzlicheren Sinne hintergrundunabhängig. In der gewöhnlichen Fassung der Allgemeinen Relativitätstheorie stellt die
metrische Struktur der Raum-Zeit keine absolute Struktur mehr dar wie in der Speziellen Relativitätstheorie. Sie wird selbst zu
einer dynamischen Entität. Da alle Materiefelder auf der Raum-Zeit leben, sind sie alle in der gleichen Weise an die Gravitation
gekoppelt. Hierin liegt der Ursprung des Äquivalenzprinzips. Aber begrifflich sind sie dennoch vom Gravitationsfeld
unterschieden. Es sind im Prinzip beliebige Typen von Feldern denkbar, welche in beliebiger Weise miteinander in Wechselwirkung
stehen, die auf der Raum-Zeit definiert sein könnten, deren Struktur durch die Allgemeine Relativitätstheorie beschrieben wird.
Zusammenhänge auf der Raum-Zeit sind über das metrische Feld definiert, welches die Gravitation repräsentiert und seine
Wechselwirkung mit der Materie beschreibt, aber begrifflich dennoch von ihr völlig unabhängig ist. In der Theorie, wie sie hier
vorgeschlagen wird, steht am Anfang nur das fundamentale Spinorfeld. Der Zusammenhang dieses Materiefeldes beziehungsweise die
entsprechende Basis an Spinorfeldern ist der Grund der Existenz des Vierbeinfeldes und somit auch des metrischen Feldes und es
existiert kein metrisches Feld, welches a priori definiert wäre. Dies ist der Grund warum hier das Prinzip der
Hintergrundunabhängigkeit in einer noch grundsätzlicheren Weise in Erscheinung zu treten scheint. Die metrische Struktur der
Raum-Zeit spiegelt die Eigenschaften der elementaren Materie wider und steht nicht bloß in einer dynamischen Beziehung zu ihr
gemäß der Feldgleichung für das metrische Feld. Dies konstituiert zusätzlich zur durch die Diffeomorphismeninvarianz der
Allgemeinen Relativitätstheorie implizierte relationalistische Anschauung in Bezug auf die Natur der Raum-Zeit als
differenzierbarer Mannigfaltigkeit nun auch eine auf Materie im Sinne von Elementarteilchen bezogene Natur des
Gravitationsfeldes und damit der metrischen Struktur der Raum-Zeit. Die Gravitation kann gemäß der hier vorgestellten Theorie
als Eichtheorie in Bezug auf das fundamentale Spinorfeld angesehen werden, welche keine metrische Struktur der Raum-Zeit
voraussetzt, sondern den Ursprung einer solchen Struktur darstellt. Wie bereits erwähnt steht dies im Gegensatz zu den
gewöhnlichen eichtheoretischen Beschreibungen der Gravitation, zumindest in ihrer üblichen Interpretation.

Gemäß ($\ref{gravityaction}$) kann das Gravitationsfeld an sich in der hier vorgestellten Theorie auf einer fundamentalen Ebene
als eine Theorie der Spinoren beschrieben werden. Deshalb hat nicht nur die Materie sondern auch die Gravitation selbst ihren
Ursprung in einer Spinorformulierung. Dies führt auf eine Vereinheitlichung in einem sehr radikalen Sinne, welche der
ursprünglichen Heisenbergschen Theorie noch den in der ursprünglichen Fassung nicht berücksichtigten Aspekt einer aus der
hier formulierten Fassung der Theorie hervorgehenden hintergrundunabhängigen Gravitationsbeschreibung hinzufügt. Im Sinne
von ($\ref{general_tetrad}$) und ($\ref{metric}$) kann die Vierbeinbeschreibung des Gravitationsfeldes, obwohl sie selbst die
Konsequenz einer sogar noch fundamentaleren Spinorbeschreibung darstellt, trotzdem als fundamentaler angesehen werden als die
metrische Beschreibung.

Unabhängig von diesen Betrachtungen muss die Struktur der Raum-Zeit als einer vierdimensionale Mannigfaltigkeit im Rahmen des
in dieser Arbeit vorgeschlagenen Ansatzes als eine basale Annahme vorausgesetzt werden. Wie bereits erwähnt, ist in die Struktur
der Raum-Zeit als einer (3+1)-dimensionalen Mannigfaltigkeit in dem Penroseschen Twistoransatz selbst mit einer
darunterliegenden Spinstruktur verbunden. Aber hier wird die Frage nach einer einheitlichen Beschreibung der Materie übergangen.
In der von Weizsäckerschen Rekonstruktion der Physik \cite{Weizsaecker:1985}, die im dritten Teil dieser Dissertation behandelt
wurde, ist sogar die Existenz einer (3+1)-dimensionalen Raum-Zeit aus der Quantentheorie der Ur-Alternativen hergeleitet,
welche ebenfalls auf Spinoren führt.

\section{Programm für eine Quantisierung des Gravitationsfeldes}

\subsection{Quantisierung der fundamentalen Spinorfelder der Gravitation}

Im Rahmen der kanonischen Quantisierung der Gravitation, wie sie in \textbf{[11.3]} beschrieben wurde,
wird eine Zerlegung der Raum-Zeit durchgeführt $\Sigma \times \mathcal{R}$, indem eine raumartige Hyperfläche $\Sigma$ gewählt
und somit eine zeitartige Richtung $\mathcal{R}$ separiert wird. Dann wird eine induzierte Metrik $h^{ab}$ auf der
drei-dimensionalen Untermannigfaltigkeit $\Sigma$ eingeführt, wobei $a$ und $b$ räumliche Koordinaten beschreiben.
Die Definition der Zeitkoordinate erlaubt die Definition einer kanonisch konjugierten Variablen $\pi^{ab}$, welche $h^{ab}$
korrespondiert, indem sich bei dieser Definition auf die gewöhnliche Einstein-Hilbert-Wirkung bezogen wird, welche allerdings
durch die neuen Variablen ausgedrückt wird und somit wird man auf eine Hamilton-Funktion $\mathcal{H}$ für die Gravitation
geführt, welche mit Hilfe der Variablen $h^{ab}$ und $\pi^{ab}$ formuliert wird. Im Ansatz der Quantengeometrodynamik werden
auf diesen Größen basierend Vertauschungsrelationen zwischen $h^{ab}$ und $\pi^{ab}$ postuliert und somit werden $h^{ab}$ und
$\pi^{ab}$ zu Operatoren, welche auf Quantenzustände $\Psi\left[h^{ab}(x)\right]$ wirken, welche von $h^{ab}(x)$ abhängen.
Nach dieser Quantisierungsprozedur müssen Zwangsbedingungen für die Zustände $\Psi\left[h^{ab}(x)\right]$ eingeführt werden, um
den Hilbert-Raum der wirklichen physikalischen Zustände zu erhalten. Im dem Sinne, wie es Dirac vorgeschlagen hat, werden die
Zwangsbedingungen in der Weise auf die Zustände übertragen, dass die in ihnen enthaltenen Größen durch die entsprechenden
Operatoren ersetzt werden und der sich ergebende Ausdruck auf die Zustände $\Psi\left[h^{ab}(x)\right]$ angewandt wird, sodass
sich für diesen eine Gleichung ergibt. Im Falle der dynamischen Zwangsbedingung ergibt sich eine Gleichung, welche als eine
Schrödinger-Gleichung zu interpretieren ist. Da in dem hier vorgestellten Ansatz der Spinzusammenhang des fundamentalen
Materiefeldes als für die Gravitation fundamental angenommen wird und das Vierbein beziehungsweise die Metrik eine Konsequenz
der Kombination der Spinoren sind, welche eine Basis im Spinorraum repräsentieren, muss die Quantisierung des Vierbeinfeldes
und des metrischen Feldes eine Konsequenz der Quantisierung der fundamentaleren Spinzusammenhangsstruktur darstellen, welche
in direktem Zusammenhang mit den Spinorfeldern $\varphi$ und $\chi$ steht. Im Gegensatz zur Quantisierung gemäß der
Quantengeometrodynamik ist es nicht sinnvoll, eine induzierte Metrik zu betrachten, welche sich auf die Untermannigfaltigkeit
bezieht, welche den räumlichen Teil der Raum-Zeit nach der Zerlegung der Raum-Zeit beschreibt. Der Grund besteht darin, dass
die Felder $\varphi$ und $\chi$ die vollständige Metrik determinieren wobei die Komponente der Metrik mit dem positiven
Vorzeichen sich auf die gewählte Zeitrichtung bezieht. Deshalb kann eine Zeitrichtung $t$ ausgezeichnet werden, ohne dass eine
Aufspaltung der Metrik erfolgt. Somit werden die Quantisierungsbedingungen für $\varphi$ und $\chi$ und die ihnen
korrespondierenden kanonisch konjugierten Variablen quantentheoretische Eigenschaften implizieren, welche sich auf die
komplette Metrik beziehen. Die kanonisch konjugierten Variablen müssen gemäß ($\ref{gravityaction}$) in der gewöhnlichen Weise
mit Bezug auf die gewählte Zeitkoordinate definiert werden

\begin{equation}
\Pi_\varphi=\frac{\delta S_{g}}{\delta \partial_t \varphi} \quad,\quad \Pi_\chi=\frac{\delta S_{g}}{\delta \partial_t \chi}.
\label{canonical_momenta}
\end{equation}
Die Wahl der Zeitkoordinate kann gemäß der gewöhnlichen Formulierung der Hamiltonschen Fassung der Gravitation durchgeführt
werden, weil die Felder $\varphi$ und $\chi$, welche den Zusammenhang der Spinorfelder darstellen, eine metrische Struktur als
Konsequenz haben, in Bezug auf welche eine Zerlegung in eine raumartige Hyperfläche und eine Zeitrichtung durchgeführt werden
kann. Es sei angemerkt, dass der Zeitkoordinate in diesem Zusammenhang keine absolute Bedeutung zukommt. Sie entspricht dieser
Zerlegung der Raum-Zeit in Cauchy-Hyperflächen. Die Kovarianz der Allgemeinen Relativitätstheorie ist aufgrund der Tatsache
gewährleistet, dass allen möglichen Zerlegungen dieser Art der gleiche Status zukommt (siehe zum Beispiel \cite{Kiefer:2004}).

Es wird nun die Frage aufgeworfen, wie die Felder $\varphi$ und $\chi$ quantisiert werden müssen, mit Vertauschungs- oder
Antivertauschungsrelationen. Wenn man sich daran erinnert, dass $\varphi$ und $\chi$ eine Spinorbasis des Materiefeldes $\psi$
in ($\ref{fieldequation_Weyl}$) darstellen, was impliziert, dass $\psi$ gemäß ($\ref{fundamentalfield_basisrepresentation}$)
als eine Linearkombination von ihnen ausgedrückt werden kann, muss man Antivertauschungsrelationen postulieren, um die 
Gültigkeit der Antivertauschungsrelationen des Materiefeldes ($\ref{quantization_spinorfield}$) aufrechtzuerhalten. Somit
nehmen die Quantisierungsregeln folgende Gestalt an

\begin{eqnarray}
\left\{\varphi^\alpha(x,t),\varphi^\beta(x^\prime,t)\right\}
&=&\left\{\Pi_\varphi^\alpha(x,t),\Pi_\varphi^\beta(x^\prime,t)\right\}=0\quad,\quad
\left\{\varphi^\alpha(x,t),\Pi_\varphi^\beta(x^\prime,t)\right\}
=i\delta^{\alpha\beta}\delta^3(x-x^\prime),\nonumber\\
\left\{\chi^\alpha(x,t), \chi^\beta(x^\prime,t)\right\}
&=&\left\{\Pi_\chi^\alpha(x,t),\Pi_\chi^\beta(x^\prime,t) \right\}=0\quad,\quad
\left\{\chi^\alpha(x,t),\Pi_\chi^\beta(x^\prime,t)\right\}
=i\delta^{\alpha\beta}\delta^3(x-x^\prime),\nonumber\\
\left\{\varphi^\alpha(x,t),\chi^\beta(x^\prime,t)\right\}
&=&\left\{\varphi^\alpha(x,t),\Pi_\chi^\beta(x^\prime,t)\right\}=
\left\{\Pi_\varphi^\alpha(x,t),\chi^\beta(x^\prime,t)\right\}
=\left\{\Pi_\chi^\alpha(x,t),\Pi_\varphi^\beta(x^\prime,t) \right\}=0.\nonumber\\
\label{anticommutation_relation}
\end{eqnarray}
Dies impliziert Quantenzustände $\Psi\left[\varphi(x),\chi(x)\right]$, welche von $\varphi$ und $\chi$ abhängen, auf welche
$\varphi$ und $\chi$ genauso wie die kanonisch konjugierten Variablen $\Pi_\varphi$ und $\Pi_\chi$ als Operatoren wirken.
Die Hamiltonschen Zwangsbedingungen können spezifiziert werden, indem mit ($\ref{gravityaction}$) eine Hamilton-Dichte für
$\varphi$ und $\chi$ aufgestellt wird 

\begin{equation}
\mathcal{H}(\varphi,\chi)=\Pi_\varphi \partial_t \varphi+\Pi_\chi \partial_t \chi-\mathcal{L}(\varphi,\chi), 
\label{Hamiltonian}
\end{equation} 
wobei die Lagrange-Dichte durch ($\ref{gravityaction}$) gemäß $S_g=\int d^4 x \mathcal{L}$ definiert ist und die folgende 
Gestalt aufweist

\begin{eqnarray}
\mathcal{L}(\varphi,\chi)=
\frac{1}{g}\sqrt{-g(\varphi,\chi)} g^{\mu\rho}(\varphi,\chi)g^{\nu\sigma}(\varphi,\chi)
F_{\mu\nu}^{\alpha\beta}(\varphi,\chi)F_{\rho\sigma\alpha\beta}(\varphi,\chi).
\label{Lagrangian}
\end{eqnarray}
Mit Hilfe von ($\ref{Hamiltonian}$) kann man nun die dynamischen Zwangsbedingungen gemäß dem Heisenberg-Bild der Dynamik der
Quantentheorie formulieren, welche durch den Kommutator der Feldoperatoren mit dem Hamilton-Operator beschrieben wird 
(siehe zum Beispiel \cite{Weinberg:1995}) 

\begin{eqnarray}
\partial_t \varphi=i\left[\mathcal{H},\varphi \right] \quad, \quad \partial_t \chi=i\left[\mathcal{H},\chi \right].
\label{dynamics_Heisenberg}
\end{eqnarray}
Die Dynamik des Vierbeinfeldes und des metrischen Feldes, welche aus aus $\varphi$ und $\chi$ zusammengesetzt sind, ist auch
durch die Relationen ($\ref{dynamics_Heisenberg}$) determiniert und somit durch die Hamilton-Dichte ($\ref{Hamiltonian}$). Dies
führt auf die Gleichungen

\begin{eqnarray}
\partial_t e^\mu_m(\varphi,\chi)=i\left[\mathcal{H},e^\mu_m(\varphi,\chi)\right]\quad,\quad
\partial_t g^{\mu\nu}(\varphi,\chi)=i\left[\mathcal{H},g^{\mu\nu}(\varphi,\chi)\right].
\end{eqnarray}
Das quantentheoretische Verhalten der Operatoren, welche das Vierbeinfeld $e^\mu_m(\varphi,\chi)$ beziehungsweise das
metrische Feld $g_{\mu\nu}(\varphi,\chi)$ beschreiben müssen als aus den Quantisierungsregeln für $\varphi$ und $\chi$
abgeleitet betrachtet werden.

\subsection{Linearisierte Näherung}

Die komplette Gravitationswirkung ausgedrückt durch $\varphi$ und $\chi$ und die entsprechenden kanonisch konjugierten
Variablen haben eine sehr komplizierte mathematische Struktur. Deshalb wird der Fall einer linearen Näherung betrachtet 
werden, innerhalb derer die Felder $\varphi$ und $\chi$ als beinahe freie Felder ohne Selbstwechselwirkung angesehen werden.
In solch einer linearen Näherung der Gravitation kann man die fundamentalen Spinorfelder $\chi$ und $\varphi$ als beinahe
konstant ansehen. In geeigneten Koordinaten bedeutet dies

\begin{equation}
\varphi \approx \begin{pmatrix} 1\\0 \end{pmatrix}=\mbox{const} \quad,\quad \chi \approx \begin{pmatrix} 0\\1
\end{pmatrix}=\mbox{const}.
\end{equation}
Mit dieser Annahme kann man auch $g_{\mu\nu}(\varphi,\chi)=e_\mu^m(\varphi,\chi) e_{\nu m}(\varphi,\chi)$ als näherungsweise
konstant ansehen und somit kann man in ($\ref{gravityaction}$) $g_{\mu\nu}(\varphi,\chi) \approx \eta_{\mu\nu}$ setzen und die
Lagrange-Dichte ausgedrückt durch den Zusammenhang $A_{\mu}^{\alpha\beta}$, welche in ($\ref{gravity_action_connection}$)
erscheint und ($\ref{Lagrangian}$) korrespondiert, lautet in solch einer linearen Näherung

\begin{equation}
\mathcal{L}(\varphi,\chi)=\frac{2}{g} \partial_\mu A_\nu^{\alpha\beta}(\varphi,\chi)
\partial^{[\mu}A^{\nu ]}_{\alpha\beta}(\varphi,\chi).
\label{Lagrangian_gravity_linear}
\end{equation}
Um die kanonisch-konjugierten Variablen zu berechnen, muss man den expliziten Ausdruck für die Gravitationswirkung
ausgedrückt durch $\varphi$ und $\chi$ betrachten. Man erlangt diesen Ausdruck, indem man ($\ref{connection}$) in
($\ref{Lagrangian_gravity_linear}$) verwendet. Dies führt zu dem folgenden Ausdruck für die Lagrange-Dichte der Gravitation
ausgedrückt durch $\varphi$ und $\chi$

\begin{eqnarray}
&&\frac{1}{g}F_{\mu\nu}^{\alpha\beta}(\varphi,\chi)F^{\mu\nu}_{\alpha\beta}(\varphi,\chi)\nonumber\\
&=&\frac{2}{g}\left[\frac{\partial_\nu \chi^\alpha \partial^{[\nu} 
\chi_\alpha \partial_\mu \varphi^\beta \partial^{\mu]} \varphi_\beta
-\partial_\mu \chi^\beta \partial^{[\nu} \chi_\beta \partial_\nu \varphi^\alpha \partial^{\mu]}
\varphi_\alpha}{\varphi_\epsilon \chi^\epsilon}
-\frac{\partial_\mu (\varphi_\gamma \chi^\gamma)\partial^{[\mu} (\varphi_\delta \chi^\delta)\partial_\nu \chi^\alpha
\partial^{\nu]} \varphi_\alpha}{\left(\varphi_\epsilon \chi^\epsilon\right)^3}
\right.\nonumber\\&-&\left.
\frac{\partial^{[\mu}(\varphi_\gamma \chi^\gamma)\left(
\partial_\nu \chi^\alpha \partial_\mu \varphi^\beta \partial^{\nu]} \chi_\alpha \varphi_\beta 
+\partial_\nu \varphi^\alpha \partial_\mu \chi^\beta \partial^{\nu]} \varphi_\alpha \chi_\beta\right)}{(\varphi_\epsilon
\chi^\epsilon)^3}\right.
\nonumber\\
&+&\left.\frac{\partial^{[\mu}(\varphi_\gamma \chi^\gamma)\left(
\partial_\nu \chi^\alpha \partial_\mu \varphi^\beta \partial^{\nu]} \chi_\alpha \chi_\beta
+\partial_\nu \varphi^\alpha \partial_\mu \chi^\beta \partial^{\nu]} \varphi_\alpha \varphi_\beta\right)}{(\varphi_\epsilon
\chi^\epsilon)^3}\right].\nonumber\\
\label{gravity_action_spinor_linearized}
\end{eqnarray}
Hieraus kann man die kanonisch-konjugierten Variablen $\Pi_\varphi$ und $\Pi_\chi$ ($\ref{canonical_momenta}$) erhalten

\begin{eqnarray}
\Pi_\varphi^\alpha
&=&\frac{2}{g}\left[\frac{4 \theta^\alpha_\varphi(\varphi_\epsilon \chi^\epsilon)^2-2\omega_\varphi \chi^\alpha
-\partial_\mu (\varphi_\gamma \chi^\gamma)\partial^{[\mu} (\varphi_\delta \chi^\delta)\partial^{0]} \chi^\alpha}
{(\varphi_\epsilon \chi^\epsilon)^3}\right.\nonumber\\
&&\quad\quad\left.\frac{+\chi^\alpha(\theta_\varphi^\beta \chi_\beta+\theta_\chi^\beta \varphi_\beta-\theta_\varphi^\beta
\varphi_\beta-\theta_\chi^\beta \chi_\beta)}
{(\varphi_\epsilon \chi^\epsilon)^3}
\frac{+\lambda_\varphi(\chi^\alpha-\varphi^\alpha)+2\omega_\varphi^{\alpha\beta}(\varphi_\beta-\chi_\beta)}
{(\varphi_\epsilon \chi^\epsilon)^3}\right],\nonumber\\
\nonumber\\
\Pi_\chi^\alpha
&=&\frac{2}{g}\left[\frac{4 \theta^\alpha_\chi(\varphi_\epsilon \chi^\epsilon)^2-2\omega_\chi \varphi^\alpha
-\partial_\mu (\varphi_\gamma \chi^\gamma)\partial^{[\mu} (\varphi_\delta \chi^\delta)\partial^{0]} \varphi^\alpha}
{(\varphi_\epsilon \chi^\epsilon)^3}\right.\nonumber\\
&&\quad\quad\left.\frac{+\varphi^\alpha(\theta_\varphi^\beta \chi_\beta+\theta_\chi^\beta \varphi_\beta-\theta_\varphi^\beta
\varphi_\beta-\theta_\chi^\beta \chi_\beta)}{(\varphi_\epsilon \chi^\epsilon)^3}
\frac{+\lambda_\chi(\varphi^\alpha-\chi^\alpha)+2\omega_\chi^{\alpha\beta}(\chi_\beta-\varphi_\beta)}
{(\varphi_\epsilon \chi^\epsilon)^3}\right].\nonumber\\
\label{canonical_momenta_linearized}
\end{eqnarray}
Benutzung dieser kanonisch-konjugierten Variablen in ($\ref{anticommutation_relation}$) führt auf die Quantisierungsregeln der
linearen Näherung. Die entsprechende Hamilton-Dichte kann durch einsetzen von ($\ref{gravity_action_spinor_linearized}$) und
($\ref{canonical_momenta_linearized}$) in ($\ref{Hamiltonian}$) erhalten werden und sie lautet

\begin{eqnarray}
\mathcal{H}&=&\frac{2}{g}\left[\frac{4 \theta^\alpha_\varphi(\varphi_\epsilon \chi^\epsilon)^2-2\omega_\varphi \chi^\alpha
-\partial_\mu (\varphi_\gamma \chi^\gamma)\partial^{[\mu} (\varphi_\delta \chi^\delta)\partial^{0]} \chi^\alpha}
{(\varphi_\epsilon \chi^\epsilon)^3}\right.\nonumber\\
&&\quad\quad\left.\frac{+\chi^\alpha(\theta_\varphi^\beta \chi_\beta+\theta_\chi^\beta \varphi_\beta-\theta_\varphi^\beta
\varphi_\beta-\theta_\chi^\beta \chi_\beta)}{(\varphi_\epsilon \chi^\epsilon)^3}
\frac{+\lambda_\varphi(\chi^\alpha-\varphi^\alpha)+2\omega_\varphi^{\alpha\beta}(\varphi_\beta-\chi_\beta)}
{(\varphi_\epsilon \chi^\epsilon)^3}\right] \partial_0 \varphi_\alpha\nonumber\\
&+&\frac{2}{g}\left[\frac{4 \theta^\alpha_\chi(\varphi_\epsilon \chi^\epsilon)^2-2\omega_\chi \varphi^\alpha
-\partial_\mu (\varphi_\gamma \chi^\gamma)\partial^{[\mu} (\varphi_\delta \chi^\delta)\partial^{0]} \varphi^\alpha}
{(\varphi_\epsilon \chi^\epsilon)^3}\right.\nonumber\\
&&\quad\quad\left.\frac{+\varphi^\alpha(\theta_\varphi^\beta \chi_\beta+\theta_\chi^\beta \varphi_\beta-\theta_\varphi^\beta
\varphi_\beta-\theta_\chi^\beta \chi_\beta)}{(\varphi_\epsilon \chi^\epsilon)^3}
\frac{+\lambda_\chi(\varphi^\alpha-\chi^\alpha)+2\omega_\chi^{\alpha\beta}(\chi_\beta-\varphi_\beta)}
{(\varphi_\epsilon \chi^\epsilon)^3}\right] \partial_0 \chi_\alpha\nonumber\\&-&\frac{2}{g}\left[\frac{\partial_\nu \chi^\alpha
\partial^{[\nu} \chi_\alpha \partial_\mu \varphi^\beta \partial^{\mu]} \varphi_\beta-\partial_\mu \chi^\beta \partial^{[\nu}
\chi_\beta \partial_\nu \varphi^\alpha \partial^{\mu]}\varphi_\alpha}{\varphi_\epsilon \chi^\epsilon}\right. \nonumber\\
&-&\left.\frac{\partial_\mu (\varphi_\gamma \chi^\gamma)\partial^{[\mu} (\varphi_\delta \chi^\delta)\partial_\nu \chi^\alpha
\partial^{\nu]} \varphi_\alpha}{\left(\varphi_\epsilon\chi^\epsilon\right)^3}
-\frac{\partial^{[\mu}(\varphi_\gamma \chi^\gamma)\left(\partial_\nu \chi^\alpha \partial_\mu \varphi^\beta
\partial^{\nu]} \chi_\alpha \varphi_\beta +\partial_\nu \varphi^\alpha \partial_\mu \chi^\beta \partial^{\nu]} 
\varphi_\alpha \chi_\beta\right)}{(\varphi_\epsilon \chi^\epsilon)^3}\right.\nonumber\\
&+&\left.\frac{\partial^{[\mu}(\varphi_\gamma \chi^\gamma)\left(\partial_\nu \chi^\alpha \partial_\mu \varphi^\beta
\partial^{\nu]} \chi_\alpha \chi_\beta+\partial_\nu \varphi^\alpha \partial_\mu \chi^\beta \partial^{\nu]} \varphi_\alpha
\varphi_\beta\right)}{(\varphi_\epsilon \chi^\epsilon)^3}\right].\nonumber\\
\label{Hamiltonian_linearized}
\end{eqnarray}
In ($\ref{canonical_momenta_linearized}$) und ($\ref{Hamiltonian_linearized}$) wurden die folgenden Definitionen eingeführt

\begin{eqnarray}
\omega_\varphi^{\alpha\beta}&\equiv& \partial^{[\mu}(\varphi_\gamma \chi^\gamma)\partial_\mu
\chi^{\beta}\partial^{0]}\varphi^{\alpha}\quad,\quad
\omega_\chi^{\alpha\beta}\equiv \partial^{[\mu}(\varphi_\gamma \chi^\gamma)\partial_\mu
\varphi^{\beta}\partial^{0]}\chi^{\alpha},\nonumber\\
\omega_\varphi&\equiv& \partial^{[0}(\varphi_\gamma \chi^\gamma)\partial_\mu
\chi^{\alpha}\partial^{\mu]}\varphi_{\alpha}\quad,\quad
\omega_\chi\equiv \partial^{[0}(\varphi_\gamma \chi^\gamma)\partial_\mu
\varphi^{\alpha}\partial^{\mu]}\chi_{\alpha},\nonumber\\
\theta_\varphi^{\alpha}&\equiv& \partial_\mu \chi^\beta \partial^{[\mu} \chi_\beta \partial^{0]} \varphi^{\alpha}
\quad,\quad
\theta_\chi^{\alpha}\equiv \partial_\mu \varphi^\beta \partial^{[\mu} \varphi_\beta \partial^{0]} \chi^{\alpha},
\nonumber\\
\lambda_\varphi&\equiv& \partial^{[0}(\varphi_\alpha \chi^{\alpha})\partial_\mu \chi^\beta \partial^{\mu]}\chi_\beta
\quad,\quad
\lambda_\chi\equiv \partial^{[0}(\varphi_\alpha \chi^{\alpha})\partial_\mu \varphi^\beta
\partial^{\mu]}\varphi_\beta.
\end{eqnarray}
Diese Quantisierungsprozedur führt auch zu Antivertauschungsrelationen für das Vierbeinfeld
$\left[e^m_\mu(x,t),e^n_\nu(x^\prime,t)\right]\not= 0$ beziehungsweise das metrische Feld
$\left[g^{\mu\nu}(x,t)g^{\rho\sigma}(x^\prime,t)\right]\not= 0$ und somit zu einem Quantenzustand, welcher von der Metrik
abhängt: $\Psi \left[g^{\mu\nu}(x)\right]=\Psi \left[g^{\mu\nu}(\varphi(x),\chi(x))\right]$. Dies bedeutet, dass die
quantentheoretische Beschreibung des Gravitationsfeldes zur Quantisierung des fundamentalen Spinorfeldes in Beziehung steht.

\section{Abschließende Betrachtungen}

Es wurde vorgeschlagen, dass die Raum-Zeit-Struktur der Allgemeinen Relativitätstheorie eine Konsequenz des Zusammenhangs
eines fundamentalen selbstwechselwirkenden Spinorfeldes sein könnte, welches auf einer vierdimensionalen Mannigfaltigkeit
definiert ist, welche die Raum-Zeit vor der Einführung der Gravitation und einer entsprechenden metrischen Struktur
repräsentiert. Im Rahmen solch einer Beschreibung scheinen die Eigenschaft der Hintergrundunabhängigkeit und der
Diffeomorphismeninvarianz noch grundsätzlicheren Charakter zu erhalten, als dies bereits in der gewöhnlichen Fassung der
Allgemeinen Relativitätstheorie der Fall ist, denn das Gravitationsfeld, welches die metrische Struktur der Raum-Zeit
darstellt, steht direkt mit den Eigenschaften der Spinstruktur des Materiefeldes in Beziehung, welches als fundamental
angenommen wird. In diesem Sinne könnte man behaupten, dass das Gravitationsfeld und somit die metrische Struktur der Raum-Zeit
nicht als so fundamental anzusehen sind wie Materiefelder, sondern vielmehr eine Konsequenz der Existenz dieser Felder
darstellen.
Als eine Konsequenz des Faktums, dass die Dynamik des Zusammenhangs durch zwei Spinorfelder ausgedrückt wird, welche als
fundamental für die Gravitation angesehen werden, ist das dynamische Verhalten des metrischen Feldes aus einer fundamentaleren
Wirkung hergeleitet, welche sich auf diese Spinorfelder bezieht. Die Quantisierung des Gravitationsfeldes erscheint ebenfalls
in einer völlig neuen Weise, weil die quantentheoretische Beschreibung des Gravitationsfeldes aus fundamentaleren kanonischen
Quantisierungsregeln des Paares an Spinorfeldern hergeleitet ist, welches zu der Darstellung des fundamentalen Spinorfeldes in
Beziehung steht, das die Materie beschreibt. Somit ist die Quantisierung der Gravitation direkt mit der Quantisierung des
fundamentalen Spinorfeldes verbunden, welches somit gewissermaßen den Ursprung aller anderen Felder bildet. Insgesamt stellt
diese Theorie damit in Bezug auf das Verhältnis der Beschreibung der metrischen Struktur der Raum-Zeit einerseits und der
Elementarteilchen andererseits sowie der Sparsamkeit der Voraussetzung von a priori gegebener räumlicher Struktur eine wichtige
Errungenschaft dar.

\chapter{Programm für die Beschreibung der Allgemeinen Relativitätstheorie im Rahmen der Quantentheorie der Ur-Alternativen}

\section{Grundsätzliche Betrachtungen}

Im letzten Kapitel wurde eine Beschreibung der Gravitation geliefert, innerhalb derer diese aus einer
einheitlichen Quantenfeldtheorie eines fundamentalen Spinorfeldes hervorgeht, aus dessen Eigenschaften, wie der in dem
entsprechenden Spinorraum ausgezeichneten symplektischen Struktur, sich die Eigenschaften des metrischen Feldes konstituieren,
welches die Gravitation beschreibt. Diese Theorie ist als ein erster Schritt in Richtung einer Naturbeschreibung zu sehen, bei
der so wenig räumliche Struktur wie möglich vorausgesetzt wird, was in \textbf{[11.3]} in Anlehnung an die Betrachtungen in
\textbf{[4.2]} und \textbf{[7.4]} als wesentliches Ziel bei der Formulierung einer Quantentheorie der Gravitation herausgestellt
wurde. Im dritten Teil der Dissertation wurde die von Weizsäckersche Quantentheorie der Ur-Alternativen vorgestellt, deren
wesentlicher Begriff der quantentheoretisch beschriebenen Alternative ohne jeden Bezug auf konkrete physikalische Begriffe wie
den physikalischen Ortsraum auskommt. In \textbf{[9.2]},\textbf{[9.3]} und \textbf{[9.4]} wird dann geschildert, wie sich die Existenz
des physikalischen Raumes mit der Zeit zu einer Raum-Zeit verbunden aus der Quantentheorie der Ur-Alternativen jedoch begründen
lässt. Damit scheint es höchst sinnvoll und aussichtsreich, eine quantentheoretische Beschreibung der Gravitation im Rahmen der
Quantentheorie der Ur-Alternativen anzustreben. Denn auf diese Weise würde man genau zu einer Quantentheorie der Gravitation
gelangen, die in ihren Grundbegriffen überhaupt keine a priori gegebene räumliche Struktur mehr voraussetzt. Da aufgrund der
Diskretheit der Basiszustände im Tensorraum, wie bereits in \textbf{[8.3]} geäußert wurde, weiter die Aussicht besteht, die
bei der Formulierung einer Quantenfeldtheorie üblicherweise auftretenden Divergenzen in natürlicher Weise zu vermeiden, wird
hier die Vermutung aufgestellt, dass eine mit Hilfe der Ur-Alternativen begründete Quantentheorie der Gravitation zu einer
richtigen Beschreibung der Natur führen würde. Die Aufgabe besteht also nun darin, eine solche Theorie zu formulieren. Hierzu
können bisher nur interessante Grundgedanken geschildert werden. Daher soll das vorliegende Kapitel dazu dienen, ein wenig
Klarheit darüber zu gewinnen, welche Rolle der Gravitation in der Quantentheorie der Ur-Alternativen zukommen würde und in welcher
Weise man versuchen könnte, ein Gravitationsfeld zu konstruieren. Bezüglich dessen ist auch in $\cite{Lyre:1996ep}$ ein erster
Ansatz zu finden.

\section{Aufbau des freien Gravitationsfeldes aus Ur-Alternativen}

Zunächst ist natürlich wichtig, dass das Gravitationsfeld und daher auch die durch das Gravitationsfeld beschriebene
metrische Struktur des Raumes eine dynamische Entität darstellt. Damit weist es aber bestimmte Freiheitsgrade auf, die
Information enthalten. Da sich aber alle physikalische Information in der Quantentheorie der Ur-Alternativen in
Ur-Alternativen aufspalten lässt, muss das Gravitationsfeld natürlich auch durch Ur-Alternativen beschrieben werden.
Das bedeutet, dass es sich im Tensorraum der Ur-Alternativen darstellen lassen muss. Dies legt nun eine Behandlung in Analogie
zu den Betrachtungen in \textbf{[10.1]} nahe. Da das Gravitationsfeld im Rahmen einer quantenfeldtheoretischen 
Beschreibungsweise gewöhnlich als aus Spin $2$ Teilchen zusammengesetzt angesehen wird, könnte man versuchen, den Zustand
eines solchen Teilchens und aus diesem einen entsprechenden einem Quantenfeld entsprechenden Vielteilchenzustand in Anlehnung
an \textbf{[10.1]} zu konstruieren. Die Beschreibung des Spin $2$ entspricht gemäß der Reihe in
($\ref{Darstellungen_Lorentz-Gruppe}$) der folgenden Darstellung der Lorentz-Gruppe:
$\left[\left(\frac{1}{2},0\right) \otimes \left(0,\frac{1}{2}\right)\right]\otimes
\left[\left(\frac{1}{2},0\right) \otimes \left(0,\frac{1}{2}\right)\right]$. Damit kann man der Definition des Zustandes
eines Spin $1/2$ Teilchens in ($\ref{Zustand_Teilchen_Spin}$) folgen und einen Gravitonenzustand $|\Psi\rangle_G$ durch das
Tensorprodukt eines Zustandes im Tensorraum der symmetrischen Produktzustände von Ur-Alternativen mit vier weiteren einzelnen
Ur-Alternativen definieren, die mit $u_A, u_B, u_C, u_D$ bezeichnet seien, wobei $u_B$ und $u_D$ als Anti-Ur-Alternativen
formuliert werden

\begin{eqnarray}
|\Psi\rangle_G&=&\sum_{N_1}\sum_{N_2}\sum_{N_3}\sum_{N_4}c(N_1,N_2,N_3,N_4)|N_1,N_1,N_2,N_4\rangle\nonumber\\
&&\otimes \left(\begin{matrix} u_{A1}\\u_{A2}\end{matrix}\right)
\otimes \left(\begin{matrix} u_{B2}^{*}\\-u_{B1}^{*}\end{matrix}\right)
\otimes \left(\begin{matrix} u_{C1}\\u_{C2}\end{matrix}\right)
\otimes \left(\begin{matrix} u_{D2}^{*}\\-u_{D1}^{*}\end{matrix}\right).
\label{Zustand_Graviton}
\end{eqnarray}
Der Spinfreiheitsgrad, der durch die vier Ur-Alternativen $u_A, u_B, u_C, u_D$ ausgedrückt wird, kann in direkter Weise durch
eine Metrik $g_{\mu\nu}$ dargestellt werden. Wenn man zunächst die beiden Dirac-Spinoren $\psi$ und $\chi$ definiert

\begin{equation}
\psi=\left(\begin{matrix}u_{A1}\\u_{A2}\\u_{B2}^{*}\\-u_{B1}^{*}\end{matrix}\right)\quad,\quad
\chi=\left(\begin{matrix}u_{C1}\\u_{C2}\\u_{D2}^{*}\\-u_{D1}^{*}\end{matrix}\right),
\end{equation}
wobei deren rechts- und linkshändige Komponenten nun natürlich jeweils unabhängige Freiheitsgrade aufweisen, so können diese
anschließend in der folgenden Weise auf eine Metrik abgebildet werden 

\begin{equation}
g^{\mu\nu}=\bar \psi \gamma^\mu \psi \bar \chi \gamma^\nu \chi
+\bar \psi \gamma^\nu \psi \bar \chi \gamma^\mu \chi.
\end{equation}
Dadurch ergibt sich für den Zustand ($\ref{Zustand_Graviton}$)

\begin{equation}
|\Psi\rangle_G=\sum_{N_1}\sum_{N_2}\sum_{N_3}\sum_{N_4}c(N_1,N_2,N_3,N_4)|N_1,N_2,N_3,N_4
\rangle \otimes g^{\mu\nu}\left(u_A, u_B, u_C, u_D\right).
\label{Zustand_Graviton_2}
\end{equation}
Ein solcher Zustand gehorcht natürlich gemäß ($\ref{Klein-Gordon-Gleichung_Tensorraum_Ur-Alternativen}$) einer
Wellengleichung

\begin{equation}
P^\mu P_\mu |\Psi\rangle_G=0,
\end{equation}
wobei $P_\mu$ gemäß ($\ref{Impulsoperatoren}$) definiert ist. Diese Gleichung entspricht der linearisierten Einsteinschen
Gleichung ($\ref{linearisierte_Einsteinsche_Gleichung}$) für einen verschwindenden Energie-Impuls-Tensor $T_{\mu\nu}$. Da auf
dieser Ebene der Definition eines freien Gravitonenfeldes im Tensorraum der symmetrischen Produktzustände von Ur-Alternativen
noch keine Wechselwirkung erscheinen kann, sondern gemäß \textbf{[10.2.3]} durch Beziehungen zwischen Zuständen der
Parabose-Ordnung $p=1$ innerhalb eines Zustandes mit Parabose-Ordnung $p>1$ beschrieben wird, kann hier noch keine nichtlineare
Gleichung auftreten, wodurch nur zu erwarten ist, dass sich eine lineare Näherung der vollständigen das Gravitationsfeld
beschreibenden Gleichung ergibt.
Zu einer quantenfeldtheoretischen Beschreibungsweise gelangt man nun, indem man die Beschreibung von Quantenfeldern
in \textbf{[10.1.3]} nun auf ($\ref{Zustand_Graviton}$) beziehungsweise ($\ref{Zustand_Graviton_2}$) überträgt. Demgemäß kann
man also Erzeugungs- und Vernichtungsoperatoren für Gravitonen im Sinne von $(\ref{Operatoren_Ur-Alternativen_Teilchen}$)
einführen und ein Zustand eines Gravitonenfeldes $|\Phi\rangle_G$ lautet demgemäß

\begin{eqnarray}
&&|\Phi\rangle_G=\sum_{N_1}\sum_{N_2}\sum_{N_3}\sum_{N_4}\sum_{\bar N}c(\bar N)
\left[\prod_{i=1}^4\left(a_i^{\dagger}\right)^{N_i}|0\rangle
\otimes \sum_{j=1}^{4}u_A(1,2,3,4,\bar N)\bar a_j^{\dagger}|0\rangle\right.\nonumber\\&&\left.
\otimes \sum_{k=1}^{4}u_B(1,2,3,4,\bar N)\bar a_k^{\dagger}|0\rangle
\otimes \sum_{l=1}^{4}u_C(1,2,3,4,\bar N)\bar a_l^{\dagger}|0\rangle
\otimes \sum_{m=1}^{4}u_D(1,2,3,4,\bar N)\bar a_m^{\dagger}|0\rangle\right]^{\bar N},\nonumber\\
\label{Quantenfeld-Zustand_Graviton}
\end{eqnarray}
wobei gilt: $\bar N=\bar N(N_1,N_2,N_3,N_4)$, und die Operatoren $a_i,a_i^{\dagger}$ beziehungsweise
$\bar a_i,\bar a_i^{\dagger}$ die folgenden Vertauschungsrelationen beziehungsweise Antivertauschungsrelationen erfüllen

\begin{equation}
\left[a_i^{\dagger},a_j\right]=\delta_{ij}\quad,\quad\left\{\bar a_i^{\dagger},\bar a_j\right\}=\delta_{ij}.
\end{equation}
Natürlich wird im Rahmen der Beschreibung des Gravitationsfeldes durch
($\ref{Zustand_Graviton}$), ($\ref{Zustand_Graviton_2}$) und ($\ref{Quantenfeld-Zustand_Graviton}$) noch nicht klar, warum
gerade das Gravitationsfeld als physikalische Entität existiert. Das Bestreben muss letztlich darin bestehen, aus der
Quantentheorie der Ur-Alternativen zu erklären, warum sich gerade das Gravitationsfeld mit seinen spezifischen Eigenschaften
ergibt. Das gleiche gilt auch für seine Wechselwirkung mit anderen Feldern.

\section{Die gravitative Wechselwirkung und das Gravitationsfeld bei der Konstituierung der Struktur der Raum-Zeit}

Die Frage nach der Art und Weise der Wechselwirkung des Gravitationsfeldes mit den anderen Feldern, setzt eine weitergehende
konkrete Beschreibung der Wechselwirkung an sich voraus, wie sie in der Quantentheorie der Ur-Alternativen auftritt. An den
Betrachtungen in \textbf{[10.2.3]} kann man bereits ersehen, dass es sich hierbei um eine sehr schwierige Aufgabe handelt.
In jedem Falle wird sich die Behandlung der Beschreibung der Wechselwirkung des Gravitationsfeldes mit anderen Feldern
demgemäß nur unter Einbeziehung der Frage nach den anderen Wechselwirkungen in der Natur realisieren lassen. Es muss in diesem
Zusammenhang dann auf der Ebene der Ur-Alternativen auch der Grund dafür gesucht werden, dass die Gravitation im Gegensatz zu
den anderen Wechselwirkungen auf alle Objekte gleich wirkt, weshalb sie ja gerade als geometrische Eigenschaft der Raum-Zeit
beschrieben werden kann. Das Äquivalenzprinzip (siehe \textbf{[2.3.2]}) muss sich also aus der Quantentheorie der Ur-Alternativen
konstituieren. Interessant ist zudem die Tatsache, dass man um ein Elementarteilchen im Kosmos bis auf die Compton-Wellenlänge
zu lokalisieren, in etwa $10^{40}$ Ur-Alternativen benötigt. Hieraus kann unter anderem die der tatsächlich geschätzten Zahl
der Nukleonen im Kosmos bezüglich der Größenordnung entsprechende Zahl von $10^{80}$ Nukleonen näherungsweise hergeleitet
werden, was sehr erstaunlich ist. In diesem Zusammenhang ist aber nun weiter wichtig, dass um den schwachen Isospin eines
Teilchens zu determinieren, lediglich eine einzige Ur-Alternative nötig ist, da der Isospin nur zwei mögliche Einstellungen
hat. Das Verhältnis der Stärke der Gravitation als Eichtheorie der Raum-Zeit-Translationen zur elektroschwachen Wechselwirkung,
welche als Eichtheorie in Bezug auf den schwachen Isospin formuliert werden kann, entspricht nun ziemlich genau dem umgekehrten
Verhältnis, nämlich $1$ zu $10^{40}$. Diese Tatsache ist höchst bemerkenswert, wenn sie auch bisher nicht genauer verstanden
werden konnte.  

\begin{figure}[h]
\centering
\epsfig{figure=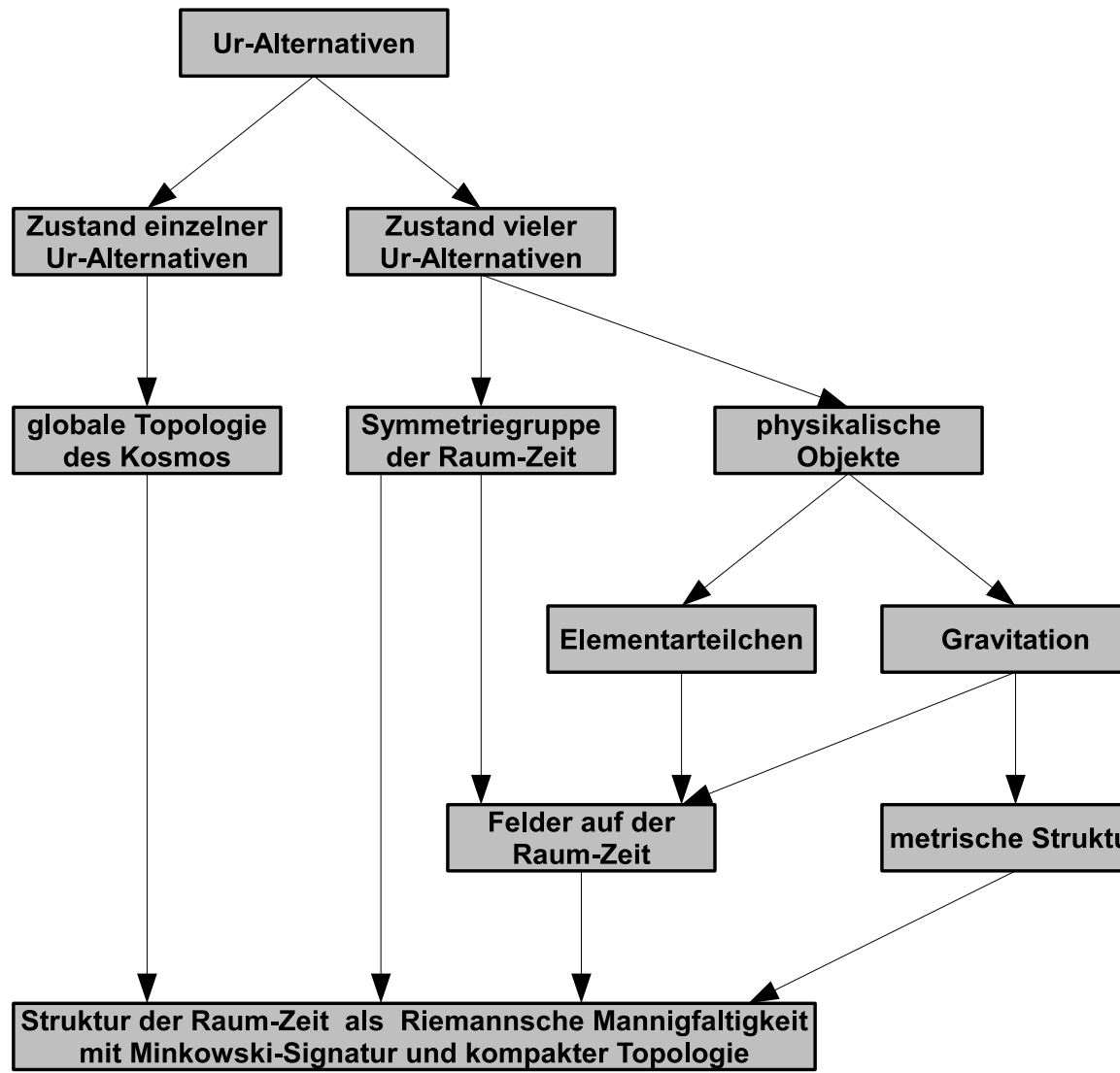,width=19cm}
\caption{\label{Ur-Alternativen_Raum-Zeit_Gravitation} Dieses Schaubild veranschaulicht die Art und Weise, wie sich die
Struktur der Raum-Zeit unter Einbeziehung der Gravitation in der Quantentheorie der Ur-Alternativen konstituiert. Aus dem Begriff
der Ur-Alternative gehen die lokale und die globale Struktur der Raum-Zeit hervor, sowie die Beschreibung physikalischer Objekte
und ihrer Wechselwirkungen, unter anderem der Gravitation, welche als allgemeinste Wechselwirkung die metrische Struktur der
Raum-Zeit konstituiert. Hieraus ergibt sich schließlich die Raum-Zeit als Riemannsche Mannigfaltigkeit mit Minkowski-Signatur
und einer kompakten Topologie. Die in diesem Schaubild nicht explizit miteinbezogene spezifische Rolle der Zeit unterscheidet
sich natürlich nicht gegenüber derjenigen in ($\ref{Ur-Alternativen_Raum-Zeit}$).}
\end{figure}

Es ist nun weiter wichtig zu erwähnen, dass der Gravitation im Rahmen der Quantentheorie der Ur-Alternativen keine konstitutive
Bedeutung bei der Beschreibung der Struktur des Kosmos als Ganzem zukommt. Denn gemäß \textbf{[9.3]} ergibt sich die topologische
Struktur des Kosmos bereits aus den mathematischen Eigenschaften des Zustandsraumes einer einzelnen Ur-Alternative, ist damit
also bereits vor dem Begriff des Gravitationsfeldes definiert. Das Gravitationsfeld bestimmt also lediglich lokal die metrische
Struktur der Raum-Zeit, nicht aber deren globale Topologie.
Aus dem Begriff der Ur-Alternative geht sowohl die globale Struktur der Raum-Zeit im Sinne eines topologisch kompakten Kosmos
hervor, als Konsequenz des Zustandsraumes einzelner Ur-Alternativen, als auch die Symmetriegruppe der Raum-Zeit, welche mit
der lokalen Struktur als Minkowski-Raum-Zeit zusammenhängt, als Konsequenz des Tensorraumes der Zustände vieler
Ur-Alternativen. Aus den Ur-Alternativen sollen durch Kombination physikalische Objekte dargestellt werden, die schließlich auf
Elementarteilchen und deren Wechselwirkungen, insbesondere die Gravitation führen, wobei letztere
ihrerseits die metrische Struktur der Raum-Zeit konstituiert. Die Elementarteilchen und ihre Wechselwirkungen einschließlich 
der Gravitation können aufgrund der sich aus den Zustandsräumen der Ur-Alternativen ergebenden Struktur der Raum-Zeit
schließlich als Felder in jener Raum-Zeit dargestellt werden. Schließlich ergibt sich hieraus die Beschreibung der Physik durch
Objekte auf einer Raum-Zeit mit der mathematische Struktur einer Riemannschen Mannigfaltigkeit mit Minkowski-Signatur, deren
globale Topologie kompakt ist. Es erscheint daher sinnvoll, ein gegenüber dem in \textbf{[9.4]} gegebenen Schaubild
($\ref{Ur-Alternativen_Raum-Zeit}$) erweitertes und das Gravitationsfeld miteinbeziehendes Schaubild bezüglich der
Konstituierung der Struktur der Raum-Zeit aufzuführen (siehe Abbildung ($\ref{Ur-Alternativen_Raum-Zeit_Gravitation}$)).
\newpage
\pagestyle{empty}

\chapter*{Zusammenfassung}
\addcontentsline{toc}{chapter}{Zusammenfassung}
\pagestyle{plain}

In der vorliegenden Dissertation wird die Frage der Vereinheitlichung der Quantentheorie mit der Allgemeinen
Relativitätstheorie behandelt, wobei entsprechend dem Titel der Arbeit der Beziehung der Grundbegriffe der beiden Theorien
die entscheidende Bedeutung zukommt. Da das Nachdenken über Grundbegriffe in der Physik sehr eng mit philosophischen
Fragen verbunden ist, werden zur Behandlung dieser Thematik zunächst in einem Kapitel, das die vier jeweils drei Kapitel
umfassenden Teile vorbereitet, die Entwicklung der Theoretischen Physik betreffende wissenschaftstheoretische
Betrachtungen sowie einige wesentliche Gedanken aus der Klassischen Philosophie vorgestellt, welche für die weitere
Argumentation wichtig sind. Bei letzteren geht es neben einer kurzen Schilderung der Platonischen Ideenlehre in Bezug auf
ihre Relevanz für die Physik insbesondere um die Kantische Auffassung von Raum und Zeit als a priori gegebenen
Grundformen der Anschauung, deren Bezug zur Evolutionären Erkenntnistheorie ebenfalls thematisiert wird.
In den beiden ersten Teilen werden die wesentlichen Inhalte der Allgemeinen Relativitätstheorie und der Quantentheorie
vorgestellt, wobei der Deutung der beiden Theorien jeweils ein Kapitel gewidmet wird. In Bezug auf die Allgemeine
Relativitätstheorie wird diesbezüglich die Bedeutung der Diffeomorphismeninvarianz herausgestellt und in Bezug auf die
Quantentheorie wird zunächst die Grundposition der Kopenhagener Deutung verdeutlicht, die im Mindesten als eine notwendige
Bedingung zum Verständnis der Quantentheorie angesehen wird, um anschließend eine Analyse und Interpretation des
Messproblems und vor allem entscheidende Argumente für die grundlegende Nichtlokalität der Quantentheorie zu geben.
Im dritten Teil der Arbeit wird die seitens Carl Friedrich von Weizsäcker in der zweiten Hälfte des letzten Jahrhunderts
entwickelte Quantentheorie der Ur-Alternativen beschrieben, in welcher die universelle Gültigkeit der allgemeinen Quantentheorie
begründet und aus ihr die Existenz der in der Natur vorkommenden Entitäten hergeleitet werden soll, auf deren
Beschreibung die konkrete Theoretische Physik basiert. Es werden sehr starke Argumente dafür geliefert, dass diese Theorie von
den bislang entwickelten Ansätzen zu einer einheitlichen Theorie der Natur, welche die heute bekannte Physik in sich enthält,
die vielleicht aussichtsreichste Theorie darstellt und damit die Aussicht bietet, auch für das Problem der Suche nach einer
Quantentheorie der Gravitation den richtigen begrifflichen Rahmen zu bilden. Ihre große Glaubwürdigkeit erhält sie durch eine
die Klassische Philosophie miteinbeziehende philosophische Analyse der Quantentheorie. Dieses Urteil behält seine
Gültigkeit auch dann, wenn die Quantentheorie der Ur-Alternativen aufgrund der ungeheuren Abstraktheit der Begriffsbildung
innerhalb der Theorie und der sich hieraus ergebenden mathematischen Schwierigkeiten bisher noch nicht zu einer vollen
physikalischen Theorie entwickelt werden konnte.
Die alles entscheidende Kernaussage dieser Dissertation besteht darin, dass aus einer begrifflichen Analyse der Quantentheorie
und der Allgemeinen Relativitätstheorie mit nahezu zwingender Notwendigkeit zu folgen scheint,
dass die physikalische Realität auf fundamentaler Ebene nicht-räumlich ist. Dies bedeutet, dass die These vertreten
wird, dass es sich bei dem physikalische Raum, wie er gewöhnlich schlicht vorausgesetzt wird, wenn auch in unterschiedlicher
Struktur, in Wahrheit nur um eine Darstellung dahinterstehender dynamischer Verhältnisse nicht-räumlicher Objekte handelt.
Diese These stützt sich auf die Diffeomorphismeninvarianz in der Allgemeinen Relativitätstheorie und in noch höherem Maße auf
die Nichtlokalität in der Quantentheorie, welche sich wiederum nicht nur in konkreten für die Quantentheorie konstitutiven
Phänomenen, sondern dazu parallel ebenso im mathematischen Formalismus der Quantentheorie manifestiert. In Kombination mit der
Kantischen Behandlung von Raum und Zeit ergibt sich damit ein kohärentes Bild in Bezug auf die eigentliche Natur des Raumes.
Die Quantentheorie der Ur-Alternativen ist diesbezüglich als einzige derzeit existierende Theorie konsequent, indem sie auf der
basalen Ebene den Raumbegriff nicht voraussetzt und rein quantentheoretische Objekte als fundamental annimmt, aus deren
Zustandsräumen sie die Struktur der Raum-Zeit allerdings zu begründen in der Lage ist. Damit befinden sich diese fundamentalen
durch Ur-Alternativen beschriebenen Objekte nicht in einem vorgegebenen Raum, sondern sie konstituieren umgekehrt den Raum.
Dies ist eine Tatsache von sehr großer Bedeutung.
Im vierten Teil wird schließlich die vorläufige Konsequenz aus diesen Einsichten gezogen. Nach einer kurzen Behandlung der
wichtigsten bisherigen Ansätze zu einer quantentheoretischen Beschreibung der Gravitation, wird die Bedeutung der Tatsache,
dass die Allgemeine Relativitätstheorie und die Quantentheorie eine relationalistische Raumanschauung nahelegen, nun konkret
in Bezug auf die Frage der Vereinheitlichung der beiden Theorien betrachtet.
Das bedeutet, dass das Ziel also letztlich darin besteht, einen Ansatz zu einer quantentheoretischen Beschreibung der Gravitation
zu finden, bei der so wenig räumliche Struktur wie möglich vorausgesetzt wird. In Kapitel 12 wird diesbezüglich ein von mir
entwickelter Ansatz vorgestellt, um zumindest eine Theorie zu formulieren, bei der die metrische Struktur der Raum-Zeit nicht
vorausgesetzt sondern in Anlehnung an die Eigenschaften eines fundamentalen Spinorfeldes konstruiert wird, das im Sinne
der Heisenbergschen einheitlichen Quantenfeldtheorie die Elementarteilchen einheitlich beschreiben soll.
Dieser Ansatz geht bezüglich der Sparsamkeit der Verwendung von a priori vorhandener räumlicher Struktur über die bisherigen
Ansätze zu einer Quantentheorie der Gravitation hinaus. Er ist aber dennoch nur als ein erster Schritt zu verstehen.
Die konsequente Weiterführung dieses Ansatzes würde in dem Versuch bestehen, eine Verbindung zur von Weizsäckerschen Quantentheorie
der Ur-Alternativen herzustellen, die überhaupt keine räumliche Struktur mehr voraussetzt. Hierzu konnten bisher nur
aussichtsreiche Grundgedanken formuliert werden.
Es wird allerdings basierend auf den in dieser Dissertation dargelegten Argumentationen die Vermutung aufgestellt, dass es im
Rahmen der von Weizsäckerschen Quantentheorie der Ur-Alternativen möglich ist, eine konsistente quantentheoretische Beschreibung der
Gravitation aufzustellen. In jedem Falle scheint die Quantentheorie der Ur-Alternativen die einzige Theorie zu sein, die aufgrund
ihrer rein quantentheoretischen Natur in ihrer Begriffsbildung grundsätzlich genug ist, um eine Aussicht zu bieten,
diejenige Realitätsebene zu erfassen, in welcher die Dualität zwischen der Quantentheorie und der Allgemeinen Relativitätstheorie
zu einer Einheit gelangt.

\bibliographystyle{plain}

\chapter*{Danksagung}
\addcontentsline{toc}{chapter}{Danksagung}

Zunächst danke ich der Adolf Messer Stiftung für die Gewährung eines großzügigen Promotionsstipendiums für zwei Jahre. Ich
danke weiter meinem Betreuer Marcus Bleicher sowie Horst Stöcker für ihre Unterstützung bezüglich meiner Bewerbung bei der Adolf
Messer Stiftung und dafür, dass sie es mir ermöglicht haben, während meiner Promotion meine eigenen Forschungsideen
durchzuführen. Weiter danke ich Benjamin Koch für wichtige Anmerkungen in Bezug auf die in Kapitel 12 vorgestellte Theorie
und Piero Nicolini für eine sehr gute wissenschaftliche Zusammenarbeit während meiner Promotionszeit, deren Ergebnisse
allerdings nicht Gegenstand dieser Dissertation sind. Am meisten danke ich jedoch meinen Eltern für ihre immerzu währende
vielseitige moralische und praktische Unterstützung.

\end{document}